\documentclass[structabstract]{aa}  
\usepackage{graphicx}
\usepackage{txfonts}
\usepackage{natbib}
\bibpunct{(}{)}{;}{a}{}{,}
\usepackage{amssymb}
\usepackage{xcolor}
\usepackage{lipsum}
\usepackage{textcomp}
\usepackage{multirow}
\usepackage{enumerate}
\usepackage{url}
\usepackage{lscape}

\newcommand{\msun}{M$_{\sun}$}

\newcommand{\xmm}{XMM-{\it Newton}}
\newcommand{\chandra}{{\it Chandra}}

\newcommand{\suzaku}{{\it Suzaku}}

\newcommand{\ROSAT}{ROSAT}
\newcommand{\ASCA}{\textit{ASCA}}
\newcommand{\einstein}{{\it Einstein}}

\newcommand{\siiha}{[\ion{S}{ii}]/H$\alpha$}
\newcommand{\Halpha}{\ensuremath{{\rm H}\alpha}\xspace}
\newcommand{\Hone}{\ion{H}{i}\xspace}

\newcommand{\eg}{\mbox{e.\,g.}\xspace}

\newcommand{\ie}{\mbox{i.\,e.}\xspace}

\newcommand{\src}{\texttt{SRC}\xspace}
\newcommand{\bg}{\texttt{BG}\xspace}

\newcommand{\ccIa}{$N_{\mathrm{CC}}/N_{\mathrm{Ia}}$}

\newcommand{\NOB}{\ensuremath{N_{{\rm OB}}}\xspace}

\newcommand{\RD}{\multicolumn{1}{c}{RD92}}
\newcommand{\abund}[1]{\makebox[2em]{#1:}}

\newcommand{\revision}[1]{\normalfont{#1}}

\defcitealias{2011ApJ...733L..35P}{P11}
\defcitealias{2012ApJ...752..103D}{D12}
\defcitealias{2013ApJ...764...11H}{H12}
\defcitealias{1999A&AS..139..277H}{HP99}
\defcitealias{1966MNRAS.131..371W}{WM66}
\defcitealias{1973ApJ...180..725M}{MC73}
\defcitealias{1981ApJ...248..925L}{LHG81}
\defcitealias{1983ApJS...51..345M}{MFD83}
\defcitealias{1984ApJS...55..189M}{MFD84}
\defcitealias{1984PASAu...5..537T}{TM84}
\defcitealias{1985ApJS...58..197M}{MFT85}
\defcitealias{1993ApJ...414..213C}{CMG83}
\defcitealias{1994AJ....108.1266S}{SCM94}
\defcitealias{1995AJ....109.1729C}{CDS95}
\defcitealias{1997PASP..109..554C}{CKS97}
\defcitealias{2004AJ....127..125L}{LCG04}
\defcitealias{2006ApJS..165..480B}{BGS06}
\defcitealias{2007MNRAS.378.1237B}{BFP07}
\defcitealias{2008MNRAS.383.1175P}{PFW08}
\defcitealias{2010MNRAS.407.1301B}{BMD10}
\defcitealias{2010ApJ...725.2281K}{KPS10}
\defcitealias{2012A&A...539A..15G}{GSH12}
\defcitealias{2012MNRAS.420.2588B}{BFC12a}
\defcitealias{2012RMxAA..48...41B}{BFC12b}
\defcitealias{2012A&A...540A..25D}{DFB12}
\defcitealias{2012A&A...546A.109M}{MHB12}
\defcitealias{2013A&A...549A..99K}{KSP13}
\defcitealias{2013MNRAS.432.2177B}{BFC13}
\defcitealias{2014A&A...561A..76M}{MHK14}
\defcitealias{2014MNRAS.439.1110B}{BKM14}
\defcitealias{2014xru..confE.333W}{WKS14}
\defcitealias{2015A&A...573A..73K}{KSB15a}

\defcitealias{2012A&A...548L...3M}{M12}

\begin{document}

   \title{The population of X-ray supernova remnants\\ in the Large 
Magellanic Cloud\,\thanks{Based on observations
obtained with \xmm, an ESA science mission with instruments and contributions
directly funded by ESA Member States and NASA.}}

   \subtitle{}

	\titlerunning{The population of X-ray supernova remnants in the 
Large Magellanic Cloud}

	\author{P. Maggi    \inst{1,2}
	\and F.    Haberl   \inst{1}
	\and P. J. Kavanagh \inst{3}
	\and M.    Sasaki \inst{3}
	\and L. M. Bozzetto \inst{4}
	\and M. D. Filipovi\'c \inst{4}
	\and G.    Vasilopoulos \inst{1}
	\and\\ W.  Pietsch \inst{1}
	\and S. D. Points \inst{5}
	\and Y.-H. Chu \inst{6}
	\and J.    Dickel \inst{7}
	\and M.    Ehle \inst{8}
	\and R.    Williams \inst{9}
	\and J.    Greiner \inst{1}
	}

   \institute{Max-Planck-Institut f\"ur extraterrestrische Physik, Postfach
	1312, Giessenbachstr., D-85741 Garching, Germany\\ 
\email{pierre.maggi@cea.fr}
	\and
	Laboratoire AIM, CEA-IRFU/CNRS/Universit\'e Paris Diderot, Service 
	d'Astrophysique, CEA Saclay, F-91191 Gif sur Yvette, France
	\and
	Institut f\"ur Astronomie und Astrophysik T\"ubingen, Universit\"at
	T\"ubingen, Sand 1, D-72076 T\"ubingen, Germany
	\and
	University of Western Sydney, Locked Bag 1797, Penrith South DC, NSW 
	1797, Australia
	\and
	Cerro Tololo Inter-American Observatory, National Optical
	Astronomy Observatory, Cassilla 603 La Serena, Chile 
	\and
	Institute of Astronomy and Astrophysics, Academia Sinica, No.1, Sec. 4, 
	Roosevelt Rd, Taipei 10617, Taiwan, Republic of China
	\and
	Physics and Astronomy Department, University of New Mexico, MSC 07-4220,
	Albuquerque, NM 87131, USA
	\and
	\xmm\ Science Operations Centre, ESA/ESAC, PO Box 78, 28691 Villanueva 
	de la Ca\~nada, Madrid, Spain
	\and
	Columbus State University Coca-Cola Space Science Center, 701 Front 
	Avenue, Colombus, GA 31901, USA
    }

   \date{Received 10 July 2015\,/\,Accepted 28 September 2015}

  \abstract
{}
{We present a comprehensive X-ray study of the population of supernova remnants 
(SNRs) in the Large Magellanic Cloud (LMC). Using primarily \xmm\ observations, 
we conduct a systematic spectral analysis of LMC SNRs to gain new insight into
their evolution and the interplay with their host galaxy.}
  {We combined all the archival \xmm\ observations of the LMC with those of our 
Very Large Programme LMC survey. We produced X-ray images and spectra of 51 
SNRs, out of a list of 59 objects compiled from the literature and augmented 
with newly found objects.
  Using a careful modelling of the background, we consistently 
analysed all the X-ray spectra and measure temperatures, luminosities, and 
chemical compositions.
  The locations of SNRs are compared to the distributions of stars, cold gas, 
and warm gas in the LMC, and we investigated the connection between the SNRs 
and their local environment, characterised by various star formation histories. 
We tentatively typed all LMC SNRs, in order to constrain the ratio of 
core-collapse to type Ia SN rates in the LMC.
  We also compared the column densities derived from X-ray spectra to 
\ion{H}{I} maps, thus probing the three-dimensional structure of the LMC.
}
{This work provides the first homogeneous catalogue of the X-ray spectral 
properties of SNRs in the LMC. It offers a complete census of LMC remnants 
whose X-ray emission exhibits Fe~K lines ($13\%$ of the sample), or 
reveals the contribution from hot supernova ejecta ($39\%$), which both 
give clues to the progenitor types.
  The abundances of O, Ne, Mg, Si, and Fe in the hot phase of the LMC 
interstellar medium are found to be between 0.2 and 0.5 times the solar values 
with a lower abundance ratio {[}$\alpha$/Fe{]} than in the Milky Way.
  The current ratio of core-collapse to type Ia SN rates in the LMC is 
constrained to \ccIa~$=1.35(_{-0.24}^{+0.11})$, which is lower than in local SN 
surveys and galaxy clusters.
  Our comparison of the X-ray luminosity functions of SNRs in Local Group 
galaxies (LMC, SMC, M31, and M33) reveals an intriguing excess of bright 
objects in the LMC. 
  Finally, we confirm that 30 Doradus and the LMC Bar are offset from the main 
disc of the LMC to the far and near sides, respectively.
  }    
{}

   \maketitle

\section{Introduction}
\label{introduction}

Supernova remnants (SNRs) are the imprints of stars that died in supernova (SN) 
explosions on the interstellar medium (ISM). SNRs return nucleosynthesis 
products to the ISM, enriching and mixing it with freshly produced heavy 
elements. A core-collapse (CC) SN is the explosion of a massive star, and it
produces large quantities of $\alpha$-group elements (e.g. O, Ne, Mg, Si, S). 
Thermonuclear (or type Ia) SNe mark the disruption of a carbon-oxygen white 
dwarf (WD) that reached the Chandrasekhar limit, \revision{although recent 
models suggest that sub-Chandrasekhar WDs may also explode as type Ia SNe 
\citep{2010ApJ...714L..52S,2010ApJ...722L.157V,2011ApJ...734...38W}}. The 
thermonuclear burning front in a type Ia SN incinerates most of the progenitor 
to Fe-group elements. Despite the essential role of type Ia SNe in cosmology as 
standard candles, leading to the discovery that the expansion of the Universe 
is accelerating \citep{1998AJ....116.1009R,1999ApJ...517..565P}, the exact 
nature of the progenitor system as either a white dwarf accreting from a 
companion or a merger of two white dwarves is still hotly debated 
\citep[see][for a review]{2012PASA...29..447M}.

SNe of either type instantaneously release a tremendous amount of 
\revision{kinetic} energy ($\sim 10^{51}$~erg) in the ISM and consequently have 
a profound and long-lasting impact on their surrounding environment. SN ejecta 
are launched to velocities in excess of $10^4$~km~s$^{-1}$, producing shock 
waves that heat the ISM and ejecta up to X-ray emitting temperatures ($> 10^6$ 
K). SNe are the main source of energy for the ISM, in the form of kinetic energy 
\revision{and turbulence \citep[\eg][and references 
therein]{2004RvMP...76..125M}} or in the form of cosmic rays that are 
accelerated at SNR shock fronts.

X-ray observations are a powerful tool for studying SNRs \citep[see \eg the 
review of][]{2012A&ARv..20...49V}. While some SNRs exhibit non-thermal 
\revision{X-ray} emission, originating in synchrotron-emitting electrons 
accelerated up to 100~TeV \citep[see][and references 
therein]{1995Natur.378..255K,2002ApJ...581.1116R,2005ApJ...621..793B}, 
\revision{most X-ray emitting SNRs have thermal spectra} dominated by highly 
ionised species of C, N, O, Ne, Mg, Si, S, and Fe. At the typical electron 
temperatures of SNR shocks ($kT \sim$~0.2~--~5~keV), all these astrophysically 
abundant elements have emission lines in the range accessible to X-ray space 
observatories. Thus, the thermal X-ray spectrum of an SNR encrypts precious 
information about the temperature, ionisation state, and chemical composition of 
the hot plasma \citep{2014IAUS..296..226S}. This, in turn, provides clues to the 
evolutionary state of the remnant, ambient density (of the inter- or 
circum-stellar medium), age, explosion energy, and the type of supernova 
progenitor. The distribution of these parameters, the impact of the environment 
on them, and their interrelations (\eg temperature vs. size/age, luminosity vs. 
ambient density) are valuable information to understand the evolution of SNRs 
and their role in the hydrodynamical and chemical evolution of galaxies.

Furthermore, SNRs are \revision{observable} for a few tens of thousands of 
years. Thus, even though SNe are rare events in a galaxy (typically one per 
century or less), there will be tens or hundreds of SNRs for us to access. In 
our own Galaxy, the Milky Way (MW), 294 SNRs are known 
\citep{2014BASI...42...47G}. However, studies of Galactic SNRs are plagued by 
the large distance uncertainties towards sources in the Galactic plane. In 
addition, many important X-ray lines of O, Ne, Mg, and Fe are emitted at 
energies $kT <$ 2 keV and are readily absorbed by the high column densities in 
front of Galactic sources.

On the other hand, the Large Magellanic Cloud (LMC), our closest neighbour 
galaxy, offers an ideal laboratory for such (X-ray) studies: First, the distance 
towards the LMC is relatively small \citep[50~kpc,][]{2013Natur.495...76P} and 
very well studied \citep{2014AJ....147..122D}. Second, the moderate inclination 
angle \citep[between 25\textdegree\ and 40\textdegree, 
\eg][]{2014ApJ...781..121V} and the small line-of-sight depth of the LMC 
\citep[between 0.3~kpc and 1.5~kpc,][]{2002AJ....124.2639V} mean that we can 
assume all LMC sources to be at a very similar distance. Third, the interstellar 
absorption by gas in the foreground is much smaller towards the LMC ($N_H < 
10^{21}$ cm$^{-2}$) than towards the Galactic plane ($N_H > 10^{22}$ cm$^{-2}$), 
allowing detection of photons even in the soft X-ray regime, below 1~keV. 
Finally, a wealth of data is available for the LMC, allowing for easier 
detection and multi-wavelength analysis of SNRs. For all these reasons, we aim 
to discover and study the \emph{complete} sample of SNRs in the LMC.

\revision{While several studies exist analysing the sample of LMC remnants as a 
whole (see references in Sect.\,\ref{compiling_literature})}, they focus either 
on surveys with a particular instrument (at a particular wavelength, \eg 
infrared or ultraviolet), or on some specific aspects (\eg the size distribution 
of SNRs). In X-rays, \citet{1981ApJ...248..925L} used the \einstein\ survey of 
the LMC to detect 26 SNRs. Later, \citet{1999ApJS..123..467W} compiled a list of 
37 SNRs, amongst which they studied the X-ray morphology of 31 objects with 
ROSAT. Since 2000, more than twenty new remnants were discovered or confirmed, 
primarily through \xmm\ observations. However, the X-ray spectral analyses of 
LMC SNRs were presented in a wide collection of individual papers with little 
consistency in the instruments, spectral models, and analysis methods used. 
Furthermore, several known SNRs were observed for the first time with modern 
X-ray instrumentation during our \xmm\ LMC survey (see 
Sect.\,\ref{observations_observations}) and their spectral properties are as yet 
unpublished \revision{(see Sect.\,\ref{results_spectra_general} and 
Appendix~\ref{appendix_spectra}}). Because of these limitations, it is not 
feasible to study the spectral properties of the whole population of LMC 
remnants with a mere survey of the available literature.

The main ambition of this work is to alleviate these limitations and provide for 
the first time an up-to-date study of the X-ray emission of LMC SNRs, using 
primarily \xmm\ observations. To that end, we performed a systematic and 
homogeneous X-ray spectral analysis of \emph{all} LMC SNRs for which \xmm\ data 
are available. This allows meaningful comparisons of remnants at various 
evolutionary stages, and provides a complete census of various spectral 
features, such as Fe~K or SN ejecta emission. In turn, SNRs are used as probes 
of their surroundings, thanks to which one can derive the chemical abundances in 
the hot phase of the LMC ISM, and compare those to abundances measured in older 
populations (globular clusters and red giant stars).

In addition, we take advantage of the availability of star formation history 
(SFH) maps of the LMC, based on spatially resolved stellar photometry, to 
investigate the connection between LMC SNRs and their local environment, 
characterised by different SFHs. Doing so, we devise a method to tentatively 
type all LMC SNRs, which can then be used to retrieve the ratio of core-collapse 
to type Ia SN rates in the LMC. Then, via their X-ray luminosity function,  we 
compare SNR populations in galaxies of the Local Group (M31, M33, LMC, SMC), 
which have different metallicities and SFHs. Finally, we study the spatial 
distribution of SNRs in the LMC with respect to cool gas, star-forming regions, 
and stars.

This work is organised as follows: We start in Sect.\,\ref{observations} by 
presenting the X-ray observations and their processing, along with supplementary 
data. In Sect.\,\ref{compiling}, we compile a complete, clean sample of LMC SNRs 
which is used throughout the rest of the paper. The details of our data analysis 
methods are given in Sect.\,\ref{data}. The following Sections present the 
results of the systematic spectral analysis of LMC SNRs 
(Sect.\,\ref{results_spectra}), the SNR typing and measurement of the ratio of 
core-collapse to type Ia SN rates (Sect.\,\ref{results_sfh}), the comparative 
study of the X-ray luminosity functions of Local Group SNRs 
(Sect.\,\ref{results_XLF}), and the spatial distribution of SNRs in the LMC 
(Sect.\,\ref{results_distribution}). Finally, we summarise our findings and 
offer our conclusions in Sect.\,\ref{summary}.

\section{Observations and data reduction}
\label{observations}

\begin{figure}[t]
    \centering
\includegraphics
[width=0.999\hsize]{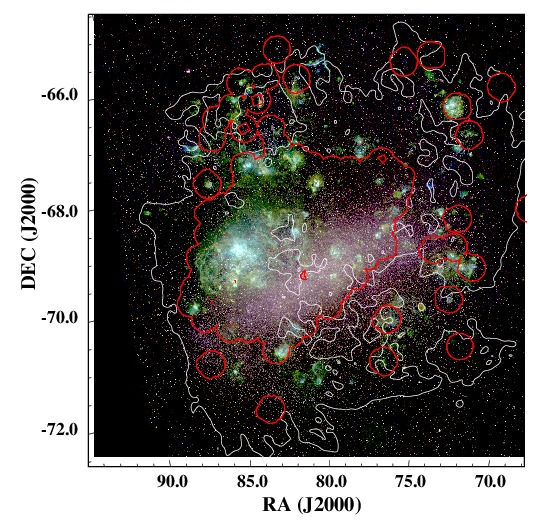}
\caption{\revision{The LMC in the light of [\ion{S}{ii}] (red), H$\alpha$ 
(green), and [\ion{O}{iii}] (blue), all data from MCELS (see 
Sect.\,\ref{observations_supplementary}). The red contours delineates the 
coverage of the LMC with \xmm, combining archival data and observations of our 
large survey (see Sect.\,\ref{observations_observations}). The white contours 
outline a LMC \ion{H}{I} column density of $1 \times 10^{21}$~cm$^{-2}$
\citep[data from][]{2003ApJS..148..473K}.}
}
\label{fig_observations_survey}
\end{figure}

\subsection{XMM-Newton observations of the LMC}
\label{observations_observations}

The \xmm\ space observatory \citep{2001A&A...365L...1J,2012OptEn..51a1009L} was 
placed in a 48~hours highly eccentric orbit by an Ariane-V on 1999 December 10. 
It carries three identical X-ray telescopes, each consisting of 58 gold-coated 
nested Wolter-I mirrors with a focal length of 7.5~m. Three CCD imaging cameras 
are placed at the focal points of each telescope. Two of them have Metal Oxide 
Semi-conductor (MOS) CCD arrays \citep{2001A&A...365L..27T} and the third uses 
pn-CCDs \citep{2001A&A...365L..18S}. Together, they form the European Photon 
Imaging Camera (EPIC). Other instruments are the two Reflection Grating 
Spectrometers \citep[RGS,][]{2001A&A...365L...7D} for high-resolution 
spectroscopy of bright on-axis point sources, and the optical monitor 
\citep[OM,][]{2001A&A...365L..36M}, a 30~cm Ritchey-Chr\'etien telescope 
observing simultaneously the central field of view in optical and ultraviolet 
light. However, data from RGS and OM were not used in this work.

About 200 \xmm\ observations of the LMC were performed since the ``first light'' 
image of the observatory, of the 30 Doradus region \citep{2001A&A...365L.202D}. 
In most cases, one specific object is placed at the focus of the telescopes 
(regular or ``target of opportunity'' observations). Some fields were observed 
several times, yielding very deep exposures. For instance, the SNR N132D is used 
as a calibration source and regularly observed; also SN~1987A is frequently 
monitored \citep{2008ApJ...676..361H,2012A&A...548L...3M}. In these two regions, 
the combined exposure reaches $10^6$~s. We have also carried out dedicated \xmm\ 
observations of SNR candidates found in ROSAT, radio, and optical data 
\citep[][]{2012A&A...539A..15G,2014MNRAS.439.1110B,2015A&A...579A..63K}.

Other programmes are raster surveys: The most ambitious such project was the 
survey of the LMC, proposed as a Very Large Programme (VLP) for \xmm\ (PI: Frank 
Haberl). The survey comprises 70 pointings chosen to fill the gaps between all 
existing observations. This provides a contiguous field in the central region of 
the LMC, a strategy similar to the \xmm\ survey of the SMC 
\citep{2012A&A...545A.128H,2012PhDT......ppppS,2013A&A...558A...3S}. 
\revision{The LMC coverage with \xmm, including both the 70 observations of that 
survey \emph{and} the archival data, is shown in 
Fig.\,\ref{fig_observations_survey} on an optical image of the galaxy}. Because 
the LMC is closer and has a larger extent on the sky than the SMC, all the 
observations combined still cover less than half of the total extent of the 
galaxy. 

\subsection{Data reduction}
    \label{observations_reduction}
The processing of all available \xmm\ data in the LMC region, and those of the
VLP survey in particular, was done with the data reduction pipeline developed in
our research group over several years. This pipeline was already used for the
surveys of M31 \citep{2005A&A...434..483P,2011A&A...534A..55S} and M33
\citep{2004A&A...426...11P,2006A&A...448.1247M}. It was then enhanced for the
analysis of the SMC survey by Richard Sturm (\citeyear{2012PhDT......ppppS}).
The data reduction pipeline is similar in essence to that used for the \xmm\
Serendipitous Source Catalogue \citep{2009A&A...493..339W}, with the advantage
of a better spatial accuracy (thanks to astrometric boresight corrections), and
dedicated source screenings and cross-identifications. It is a collection of 
tasks from the \xmm\ Science Analysis Software\,\footnote{SAS, 
\url{http://xmm.esac.esa.int/sas/}}, organised in \texttt{bash} scripts, 
together with other tools, in particular the FITS-file manipulation tasks of the 
\texttt{FTOOLS} package\,\footnote{\url{http://heasarc.gsfc.nasa.gov/ftools/}} 
(Blackburn \citeyear{1995ASPC...77..367B}). We summarise below the important 
steps of the pipeline.
\paragraph{Preparing the data:} To point to the Current Calibration Files (CCFs)
corresponding to each observation, a CCF index file (CIF) is created with the
SAS task \texttt{cifbuild}. Then, using the task \texttt{odfingest}, the ODF
summary file is extended with data extracted from the instrument housekeeping
datasets. The instrument mode is also determined based on the CIF.

\paragraph{Creating event lists:} The meta-tasks \texttt{epchain} and
\texttt{emchain} produce EPIC-pn and MOS event lists, respectively. Raw events 
are first extracted from each exposure and CCD chip. Bad pixels are flagged. In 
the case of EPIC-pn, the task \texttt{epreject} corrects shifts in the energy 
scale of some pixels induced by high-energy particles hitting the detector while 
the offset map is calculated. Raw events are then assigned pattern and detector 
position information. EPIC-pn events are corrected for gain variation and charge 
transfer inefficiency (CTI). The calibrated events are (tangentially) projected 
on the sky using the task \texttt{attcalc} and an attitude history file (AHF), 
which records the attitude of the spacecraft during the observation. The AHF is 
created by the task \texttt{atthkgen} that is automatically ran before the main 
chain (unless the AHF already exists). EPIC-pn event times are randomised within 
their read-out frame. Finally, event lists from all CCDs are merged in the final 
list by \texttt{evlistcomb}.

\paragraph{Time filtering:} Times that are useful for analysis are known as good 
time intervals (GTIs). In particular, periods of high background must be 
filtered out. The pipeline identifies the background-GTIs as times when the 
count rate in the (7~--15)~keV band is below a threshold of $8 \times 
10^{-3}$~cts\,s$^{-1}$\,arcmin$^{-2}$ and $2.5 \times 
10^{-3}$~cts\,s$^{-1}$\,arcmin$^{-2}$ for EPIC-pn and EPIC-MOS, respectively. 
Soft proton flares affect all detectors, so only the GTIs \emph{common} to pn 
and MOS are used. When one instrument starts earlier or observes longer, this 
interval is added to the GTIs. For instance, EPIC-pn calculates an offset map 
before an exposure. Thus, pn exposures usually start later than those of MOS, 
but these times should not be vetoed, unless the background in MOS is above the 
threshold.

\paragraph{Images creation:} The pipeline then produces images from the
calibrated, cleaned, and background-filtered event lists. The image pixels have
a size of 2\arcsec$\times$~2\arcsec.  All single to quadruple-pixel
(\texttt{PATTERN} = 0 to 12) events with \texttt{FLAG = 0} from the MOS
detectors are used. From the pn detector single and double-pixel events
(\texttt{PATTERN} = 0 to 4) with \texttt{(FLAG \&\& 0xf0000) = 0} (including
events next to bad pixels or bad columns) are used. Below 500~eV, only
single-pixel events are selected to avoid the higher detector noise contribution
from the double-pixel events. Exposure maps taking into account the telescope
vignetting (which is energy-dependent) are created with the task
\texttt{eexpmap}. Images and exposure maps are extracted in various energy 
bands for all three cameras. Out-of-time (OoT) images are created from the 
EPIC-pn OoT event lists, scaled by the corresponding OoT fraction 
$f_{\mathrm{OoT}}$\,\footnote{Values taken from the \xmm\ Users Handbook.}, and
subtracted from the source+background images. MOS and pn images are then merged, 
smoothed with a 10\arcsec\ full width at half maximum (FWHM) Gaussian kernel, 
and finally divided by the vignetted exposure maps.

Detector-background images are also created, by using \xmm\ filter wheel closed
(hereafter FWC) data, obtained with the detectors shielded from astrophysical
and soft-proton backgrounds by a 1.05~mm-thick aluminium filter. FWC data are
collected several times per year, and the merged event lists of these
observations are made available by the \xmm\ Science Operations 
Centre\,\footnote{
\url{http://xmm2.esac.esa.int/external/xmm_sw_cal/background/filter_closed/}}. 
The detector corners are always shielded from the X-ray telescopes, and the 
count rate in the corners is used to estimate the contribution of the 
instrumental background $f_{\mathrm{FWC}}$ to the science image. The FWC image 
is scaled by $f_{\mathrm{FWC}}$ and removed from the science image to create 
the background-subtracted image.

\paragraph{Source detection:} X-ray source detection is performed simultaneously
using source+background images in all available energy bands of all three 
instruments with the SAS meta-task \texttt{edetectchain}. Although this work is 
concerned with SNRs, \ie extended sources, detecting point sources is highly 
desirable: it allows us to excise unrelated point sources from spectral 
extraction regions and to look for central compact objects or pulsar wind 
nebulae inside SNRs.

\paragraph{Fine-tuning for SNRs:} Several scripts for the analysis of the LMC 
SNRs were produced. For imaging purposes, all observations of an SNR are 
combined to produce an image centred on the source. The smoothing of the images 
(using the SAS task \texttt{asmooth}) is performed both in constant and adaptive 
mode. In the latter, the task calculates a library of Gaussian kernels such that 
the resulting images reached a minimum (Poissonian) signal-to-noise ratio of 5 
everywhere. Regions of good statistics (\eg bright sources) will be smoothed 
with a 10\arcsec\ FWHM kernel (the chosen minimum value), whereas fainter 
regions (diffuse emission, rims of the field of view) will be smoothed with 
wider kernels. The (minimum) kernel size for (adaptive) smoothing was chosen 
depending on the available data and brightness of the SNR under investigation. 
Moderately bright and faint SNRs (\ie most of the sample) have smoothing kernel 
sizes of $\gtrsim 10$\arcsec\ or $\gtrsim 20$\arcsec. The bright objects and 
SNRs in very deep fields (\eg the field around SNR~1987A) only need shallow 
smoothing (kernels $\gtrsim 3$\arcsec\ or $\gtrsim 6$\arcsec).

Images were produced in a set of energy bands tailored to the thermal spectrum 
of SNRs: A soft band from 0.3~keV to 0.7~keV includes strong lines from oxygen; 
a medium band from 0.7~keV to 1.1~keV comprises Fe L-shell lines as well as 
Ly$\alpha$ lines from \ion{Ne}{IX} and \ion{Ne}{X}; and a hard band 
(1.1~--~4.2~keV) which includes lines from Mg, Si, S, Ca, Ar, and possibly 
non-thermal continuum. Thus, the composite images of SNRs provide a visual 
evaluation of their temperature: evolved objects with a relatively cool plasma 
(0.2~keV~$\lesssim kT\lesssim$~0.4~keV) are most prominent in the soft band, 
those with higher temperatures (0.4~keV~$\lesssim kT\lesssim$~1~keV) in the 
medium band. Only (young) SNRs with a much hotter component or a non-thermal 
continuum will have significant emission in the hard band as well.

\subsection{Supplementary data}
\label{observations_supplementary}

Various non-X-ray data were used to supplement the \xmm\ observations. They 
allow us \eg to assess the relation between the population of SNRs and large 
scale structure of the LMC (Sect.\,\ref{results_distribution}), or to evaluate 
doubtful candidates in the sample compilation (Sect.\,\ref{compiling}). Here, 
we present those data briefly.

\paragraph{Optical data:} The Magellanic Clouds Emission Line Survey 
\citep[MCELS, \eg][]{2000ASPC..221...83S} was carried out at the Cerro Tololo 
Inter-American Observatory (CTIO). It is a spatially complete, flux-limited 
survey with the 0.6/0.9~m Curtis Schmidt telescope of the University of 
Michigan. A 8\degr~$\times$~8\degr\ region centred on the LMC was imaged with
three narrow-band filters [\ion{S}{ii}]$\lambda\lambda$6716,\,6731 \AA,
\Halpha\,\footnote{the \Halpha filter included the 
[\ion{N}{ii}]$\lambda\lambda$6548,\,6584 \AA\ doublet in its bandpass.}, and 
[\ion{O}{iii}]$\lambda$5007 \AA. Observations with green and red broad-band 
filters centred at 5130~\AA\ and 6850~\AA\ were obtained to subtract stellar 
continua. The pixel size of the mosaicked data is 2\arcsec~$\times$~2\arcsec. 

For optical photometry, we used results of the Magellanic Clouds Photometric 
Survey \citep[MCPS,][]{2004AJ....128.1606Z}, a $UBVI$ survey of 24 million stars 
in the central $\sim 64$~deg$^2$ of the LMC down to $V\sim 20 -21$~mag 
(depending on crowding). Additionally, we used optical images (red continuum and 
\Halpha) from the Southern H-Alpha Sky Survey Atlas 
\citep[SHASSA][]{2001PASP..113.1326G}.

\paragraph{Radio\,:}
The neutral hydrogen (\Hone) content and structure of the LMC has been studied
(at 21~cm) by \citet{2003MNRAS.339...87S} and \citet{2003ApJS..148..473K}. The
former used data from the 64-m single-dish Parkes radio-telescope, sensitive
to large-scale structures (200~pc to 10~kpc). They show the distribution of
\Hone in a well-defined disc and three ``arms'' interpreted as tidal features.
Several \Hone holes (the largest ones) are associated to supergiant shells
(SGS).
In \citet{2003ApJS..148..473K}, the Parkes data are merged with data from the 
Australia Telescope Compact Array (ATCA) interferometer, which provides a view 
of the smaller structures (15 pc to 500 pc). The resulting map (which we used in 
this work) reveal the clumpiness of the \Hone distribution, or in their words, 
``the filamentary, bubbly, and flocculent structures of the ISM in the LMC''. 
Finally, the molecular content of the LMC is assessed by the $\sim 30$~deg$^2$ 
survey with the NANTEN telescope in the \element[][12][]{CO}~$(J = 1 - 0)$ line 
\citep{2008ApJS..178...56F}, from which we borrowed the velocity-integrated CO 
map.

\paragraph{Star formation history map of the LMC\,:}
\label{observations_supplementary_SFH}
The first studies of the LMC's stellar content in the 1960s suggested a 
different SFH than for the Milky Way 
\citep{1960ApJ...131..351H,1961ApJ...133..413H}. Most of the early studies used
age-dating of LMC clusters. The most striking feature they revealed was the
``Age Gap'', \ie the lack of clusters between ages of $\sim 5$~Gyr and $\sim
12$~Gyr \citep[\eg][]{1991IAUS..148..183D}. Studies of \emph{field star}
populations \citep[\eg with 
\textit{HST},][]{1999AJ....118.2262H,2002ApJ...566..239S} reveal essentially the
same results, \ie a dearth of star formation between an initial burst ($\gtrsim
12$~Gyr) and a second episode 4--5~Gyr ago.

The first truly global analysis of the LMC's SFH was conducted by
\citet{2009AJ....138.1243H}. They used the results from the MCPS to perform
colour-magnitude diagram fitting. They obtained a reconstruction of the star
formation rate (SFR, in \msun\, yr$^{-1}$) in 13 time bins and four metallicity
bins, for 1380 cells, most of them having a size of 12\arcmin $\times$
12\arcmin. Although poorly sensitive to old ages because the survey does not
reach the main-sequence turn-off (MSTO) in the crowded fields\,\footnote{in the
Bar the old ($\gtrsim 4$~Gyr) SFH is constrained to match that obtained with
\textit{HST}.}, the SFH obtained is extremely useful to study the recent and
intermediate-age star formation episodes, and to compare the integrated SFH of
small- and medium-scale regions. We used the SFH map to compare the 
local stellar populations around LMC SNRs in Sect.\,\ref{results_sfh}.

\section{Compiling a complete sample of LMC SNRs}
\label{compiling}
Obtaining a complete and clean census of LMC remnants is a complex task, for
several reasons\,:\\
\indent \textbullet\ \emph{Classification\,:} different authors may use
different criteria to classify an object as a definite SNR.\\
\indent \textbullet\ \emph{Literature size\,:} with the exception of the early
studies, the discovery of most new objects was reported in separate papers,
building up a vast literature. \\
\indent \textbullet\ \emph{Nomenclature\,:} an additional problem related to the
previous point is the inconsistencies in the naming convention for LMC SNRs. The
common names of many remnants used in the literature, especially those
discovered first, are an unruly collection of various surveys and catalogues in
specific wavelengths. Some are referred to after the \ion{H}{ii} complex within
which they are located (\eg ``SNR in N44''), or worse, a nearby \ion{H}{ii} 
region (\eg DEM~L109, though it is most likely unrelated to the remnant). Other 
names use B1950 coordinates, with little to no consistency in the coordinates 
convention. Consequently, some objects were mistakenly listed twice in SNR 
compilations (Sect.\,\ref{compiling_cleaning}).

To bypass these shortcomings, we performed a complete literature survey to build
a list of LMC SNRs, combining all papers that either \emph{i)} report the
discovery or classification of one or more SNRs, \emph{ii)} give a list of LMC
SNRs, or \emph{iii)} present new candidates
(Sect.\,\ref{compiling_literature}). The list is then cleaned from the wrongly 
identified or misclassified objects (Sect.\,\ref{compiling_cleaning}). 
Unconfirmed candidates, particularly in light of new X-ray observations, are 
also removed. For the naming of all SNRs in the Magellanic Clouds, we made use 
of the acronym ``MCSNR'', which was pre-registered to the International 
Astronomical Union by R. Williams et al., who maintain the Magellanic Cloud 
Supernova Remnants online database\,\footnote{MCSNR, 
\url{http://www.mcsnr.org/Default.aspx}}. This ensures a consistent and general 
naming system. Therefore, all SNRs are assigned the identifier ``MCSNR 
JHHMM+DDMM'', although we also retained the old ``common names'' from the 
literature for easy cross-identifications.

    \subsection{Literature survey}
        \label{compiling_literature}

The first extragalactic supernova remnants were found in the LMC in the 1960s. 
Combining Parkes observations with \Halpha\ photographs, 
\citet*{1963Natur.199..681M} first identified N49 as an SNR, to which 
\citet{1966MNRAS.131..371W} soon added N63A and N132D. Less than ten years 
later, \citet{1973ApJ...180..725M}, using the same method, had already 
discovered 12 new SNRs\,\footnote{Counting the two distinct shells they 
identified in N135 (the remnants to be known as DEM~L316A and DEM~L316B) and 
including the two objects in the 30 Doradus region that they identified as 
candidates.}. The survey with \einstein\ allowed \citet{1981ApJ...248..925L} to 
list 26 SNRs detected in X-rays, confirming many previously suggested candidates 
(based on optical or radio data). \citet{1983ApJS...51..345M} provided a 
catalogue of 25 SNRs with radio, optical, and X-ray results. With more 
observations, \citet{1984AuJPh..37..321M} and 
\citet{1984ApJS...55..189M,1985ApJS...58..197M} increased the size of the sample 
to 32.

In the 1990s, several new SNRs were discovered with ROSAT pointed observations
\citep{1993ApJ...414..213C,2000AJ....119.2242C,1994AJ....108.1266S}, sometimes
aided by optical spectroscopy \citep{1995AJ....109.1729C,1997PASP..109..554C}.
Since then, about twenty new remnants were discovered or confirmed in a
collection of papers. Some discoveries stemmed from new radio observations
\citep[\eg][]{2012MNRAS.420.2588B,2012RMxAA..48...41B,2012A&A...540A..25D}. The
majority, though, used \xmm\ observations, either optically selected candidates
\citep{2010ApJ...725.2281K}, ROSAT-selected candidates
\citep[][Kavanagh et al., in prep.]{2012A&A...539A..15G,2014MNRAS.439.1110B}, or
serendipitously observed during the LMC VLP survey
\citep{2012A&A...546A.109M,2014A&A...561A..76M} or other programmes 
\citep{2014A&A...567A.136W}.

Several groups compiled lists of SNRs in the (Large) Magellanic Cloud(s), the 
purpose being to analyse some of their global properties. 
\citet{1998A&AS..130..421F} used Parkes surveys to study the radio spectral 
index and luminosityy distribution of 34 confirmed and 24 probable LMC SNRs. 
\citet{1999ApJS..123..467W} were the first to study the X-ray morphology of all 
known LMC SNRs at that time. They showed ROSAT images for 31 out of their list 
of 37 SNRs. \citet[][hereafter 
\citetalias{2006ApJS..165..480B}]{2006ApJS..165..480B} compiled a sample of 39 
SNRs in the LMC which was observed with the \textit{Far Ultraviolet 
Spectroscopic Explorer (FUSE)} satellite. The goal was to study UV emission from 
SNRs, in particular in the light of highly ionised oxygen (\ion{O}{vi}~$\lambda 
1032$). A sample of 52 confirmed and 20 candidate radio-selected SNRs was 
observed spectroscopically in \citet{2008MNRAS.383.1175P}, but the exact list 
was not given. Instead, they reported the results for the 25 objects which were 
detected. \citet{2010AJ....140..584D} studied the triggering of star formation 
by SNRs. To that end, they examined the young stellar objects and molecular 
clouds associated to LMC SNRs. Their census resulted in a list of 45 objects. A 
total of 54 SNRs was used by \citet[][hereafter 
\citetalias{2010MNRAS.407.1301B}]{2010MNRAS.407.1301B} to study their size 
distribution. The difference in numbers stems from their including objects from 
unpublished sources (\ie online catalogues). 
\citet{2008PASJ...60S.453S,2013ApJ...779..134S} combined \textit{AKARI} and 
\textit{Spitzer} observatories to survey the infrared emission of LMC SNRs. They 
presented a list of 47 SNRs, warning that some sources in 
\citetalias{2010MNRAS.407.1301B} still needed confirmation.

    \subsection{Cleaning the sample: Objects not included}
        \label{compiling_cleaning}

To build the final list of LMC SNRs, we combined objects from the older 
catalogues \citep{1973ApJ...180..725M,1981ApJ...248..925L,1983ApJS...51..345M, 
1984ApJS...55..189M,1985ApJS...58..197M} with those reported in individual 
studies since then. We also included all sources present in the various 
compilations described in the previous Section. After removing all multiple 
occurrences of the same object, we ``cleaned'' the sample, searching for\,:\\
\indent \textbullet\ \textit{Misclassification:} the object is something else 
than an SNR, \eg a superbubble. \revision{The X-ray properties of non-SNR 
extended sources that can be found in the field of the LMC were described in 
\citet{2014A&A...561A..76M}}.\\
\indent \textbullet\ \emph{Unconfirmed candidates:} new data obtained since 
the classification as an SNR/candidate argue against this interpretation. 
\revision{This includes mainly candidates observed with \xmm\ for the first 
time in our VLP survey. The absence of coincident X-ray emission strongly 
disfavours an SNR nature, unless radio and optical emission typical of SNRs is 
found.}\\
\indent \textbullet\ \emph{Misidentification:} spurious source due to confusion 
(of the coordinates or nomenclature) in the literature. 

Below, we describe the objects erroneously classified as SNRs or candidates and 
the evidence motivating the decision. These objects are listed in 
Table~\ref{table_compiling_rejected} and were not included in our final sample.

\paragraph{[BGS2006b] J0449$-$693:} This object was observed in the UV by
\citet{2006ApJS..165..480B} and in optical by \citet{2008MNRAS.383.1175P},
although the latter used a different location, further to the south-east than
the former. None of these studies gave conclusive evidence of an SNR nature
(no UV lines detected, moderate \siiha\ ratio). \citet{2010ApJ...725.2281K}
used MCELS and \xmm\ to identify the true SNR in that region, that they named
SNR0449$-$6921, now registered as [BMD2010]~SNR~J0449.3$-$6920 in Simbad. The
X-ray emission originates from an optical shell clearly distinct from the
position given for [BGS2006b] J0449$-$693. In \citetalias{2010MNRAS.407.1301B}, 
both sources are listed, although only [BMD2010]~SNR~J0449.3$-$6920 
(SNR0449$-$6921) is the true source. This is an example of a misidentification 
due to coordinate confusion.

\paragraph{LHA 120$-$N~185:} \citet{2006ApJS..165..480B} could not detect UV 
emission from this source (that they incorrectly listed as SNR~0453-672). It was 
not included in the compilations from \citet{2010AJ....140..584D} and 
\citet{2013ApJ...779..134S}. Only \citetalias{2010MNRAS.407.1301B} classified 
the source as an SNR. X-ray emission is detected, surrounded by the large, 
bright optical shell N~185. However, the nature of the source remains uncertain. 
Most likely, N~185 is actually a superbubble, and not the remnant of a single 
supernova \citep{2014ApJ...792...58Z,2014AJ....148..102R}.

\paragraph{SNR J051327$-$691119:} This source is located north-westwards of SNR~
B0513$-$692 (which has the name MCSNR J0513$-$6912 in our list).
\citet{2007MNRAS.378.1237B} present the optical and radio observations of this
region, identifying the large (4.1\arcmin~$\times$~3.3\arcmin) shell of MCSNR
J0513$-$6912. They detected a strong unresolved radio source at its
north-western edge, that they classified as an unrelated \ion{H}{ii} region or
background galaxy \citep[GH 6$-$2, see references in][]{2007MNRAS.378.1237B}.

In addition, they observed a faint optical shell seen in both MCELS
[\ion{S}{ii}] and AAO/UKST deep H$\alpha$ images. Follow-up optical spectroscopy
revealed distinct, higher \siiha\ ratios from this faint shell, prompting
\citep{2007MNRAS.378.1237B} to classify this shell as a new candidate SNR,
J051327$-$691119. This region was covered by the \xmm\ survey, revealing in
details the X-ray emission of MCSNR J0513$-$6912 
(Sect.\,\ref{results_spectra}, Appendix~\ref{appendix_spectra} \& 
\ref{appendix_images}). On the other hand, the candidate J051327$-$691119 lacks 
any X-ray feature. The small extent of the source (40\arcsec\ diameter in 
H$\alpha$) would suggest a young, X-ray bright SNR, easily detectable in 
observations of the \xmm\ survey. With only weak optical evidence, a confused 
field in the radio, and a stringent non-detection in X-rays, one is forced to 
conclude that J051327$-$691119 is \emph{not} an SNR.

\begin{figure}[t]
    \centering
\includegraphics[angle=0,width=0.87\hsize]
{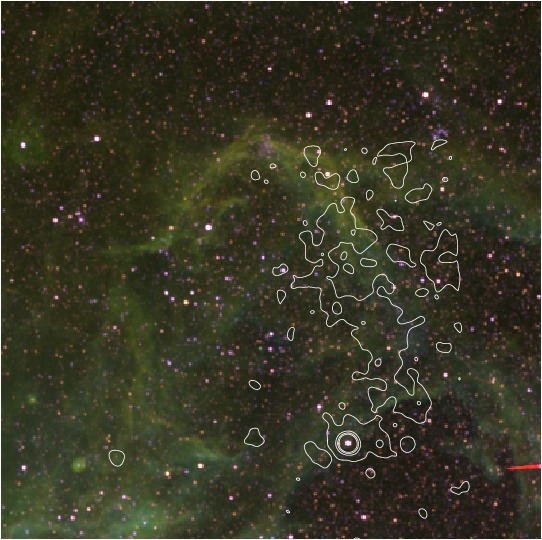}
\caption[\ The rejected SNR candidate in DEM L203 in optical lines with soft
X-ray contours]{The rejected SNR candidate in DEM L203 in optical lines
([\ion{S}{ii}] (red), H$\alpha$ (green), and [\ion{O}{iii}] (blue), data from
MCELS), with soft X-ray contours (from \xmm) overlaid in white. The image spans
20\arcmin\ across. The bright star seen in X-rays (lower right corner) is the
Galactic star HD~269602.
}
    \label{fig_compiling_DEML203}
\end{figure}

\paragraph{LHA 120$-$N 204:} It is only listed as an SNR in the compilation of
\citetalias{2010MNRAS.407.1301B}. It was selected from the radio observations of
\citet{2008MNRAS.383.1175P} where it appeared for the first time in the
literature. Therefore, it was selected from radio catalogues. The ``SNR'' lies
within the large (diameter of 14\arcmin) optical shell N~204, although a size of
1\arcmin\ was given in \citet{2008MNRAS.383.1175P}. The field 61 of the \xmm\
survey covered this region, detecting no extended X-ray emission. With the small
size of this source, bright emission is expected. Instead, an X-ray point
source is detected in projection in N~204, which correlates with a mid-IR
selected AGN \citep[MQS J052749.08$-$703641.7,][]{2012ApJ...746...27K}. The
background AGN is most likely the origin of the radio emission which led to the
misclassification of the target as an SNR candidate.

\begin{table*}[t]
\caption{LMC objects erroneously classified as SNRs or candidates, not included
in the final sample.}
\begin{center}
\label{table_compiling_rejected}
\begin{tabular}{@{\hspace{1em}}l @{\hspace{1em}} c @{\hspace{1em}} c
@{\hspace{1em}} c @{\hspace{1em}}}
\hline\hline
\noalign{\smallskip}
  \multicolumn{1}{c}{Name} &
  \multicolumn{1}{c}{Alternative name} &
  \multicolumn{1}{c}{Category} &
  \multicolumn{1}{c}{Ref. code} \\
\noalign{\smallskip}
\hline
\noalign{\smallskip}
{[}BGS2006b{]} J0449$-$693 & B0450$-$6927 & Wrong identification & BGS06\\

LHA 120$-$N 185 & N185 & Wrong classification (superbubble) & PWF08\\

SNR J051327$-$691119 & DEM L109 & Unconfirmed candidate & BFP07\\

LHA 120$-$N 204 & B0528$-$7038 & Wrong identification & PWF08\\

{[}BMD2010{]} SNR J0529.1$-$6833 & DEM L203 & Unconfirmed candidate & BMD10\\
\noalign{\smallskip}
\multirow{2}{*}{RX J0533.5$-$6855} & X-ray arc around  &
\multirow{2}{*}{Unconfirmed candidate} & \multirow{2}{*}{LCG04}\\
   &  RX J053335$-$6854.9 & \\
\noalign{\smallskip}

30 DOR C & {[}BGS2006b{]} J0536$-$692 & Wrong classification (superbubble) &
MFT85\\

SNR B0538$-$69.3 & {[}BGS2006b{]} J0538$-$693 & Unconfirmed candidate & MFD84\\
%
%
%
\noalign{\smallskip}
\hline
\end{tabular}
\end{center}
\tablefoot{See text in Sect.\,\ref{compiling_cleaning} for a description of 
each object. Reference codes\,:
\citetalias{1984ApJS...55..189M}: \citet{1984ApJS...55..189M};
\citetalias{1985ApJS...58..197M}: \citet{1985ApJS...58..197M};
\citetalias{2004AJ....127..125L}: \citet{2004AJ....127..125L};
\citetalias{2006ApJS..165..480B}: \citet{2006ApJS..165..480B};
\citetalias{2007MNRAS.378.1237B}: \citet{2007MNRAS.378.1237B};
\citetalias{2008MNRAS.383.1175P}: \citet{2008MNRAS.383.1175P};
\citetalias{2010MNRAS.407.1301B}: \citet{2010MNRAS.407.1301B}.
}
\end{table*}

\paragraph{[BMD2010] SNR J0529.1$-$6833:} The classification as an SNR candidate
(in the MCSNR online database) stems from the detection of radio emission
correlating with the large optical shell DEM L203. This object is however in the
compilation of ``confirmed'' SNRs of \citetalias{2010MNRAS.407.1301B}. Again, 
X-ray observations can shed light on the nature of the source. DEM L203 has no 
X-ray counterpart in the ROSAT catalogue. More importantly, \xmm\ covered the 
object on three occasions during the LMC survey. Combining $\sim 35$~ks of EPIC 
data, only unrelated large-scale diffuse emission is detected, without any
correlation with the optical shell, as shown in 
Fig.\,\ref{fig_compiling_DEML203}. A very old age, as indicated by the large 
extent, might explain the lack of X-ray emission, although \xmm\ can and did 
detect the largest SNRs, such as MCSNR J0450$-$7050 \citep[5.7\arcmin\ 
diameter,][]{2009SerAJ.179...55C} or J0506$-$6541 
\citep[6.8\arcmin][]{2010ApJ...725.2281K}. Furthermore, the MCELS image reveals
no clear enhanced [\ion{S}{ii}] emission, and the source was not 
spectroscopically observed by \citet{2008MNRAS.383.1175P}. In light of this and
the absence of X-ray emission, we do not confirm the classification of this
object as an SNR and did not include it in the final sample.

\paragraph{RX J0533.5$-$6855:} \citet{2004AJ....127..125L} used ROSAT to study
the X-ray diffuse emission around the point source RX J053335$-$6854.9
(referenced as RX J0533.5$-$6855 in Simbad) and concluded that the X-ray arc
seen was a large SNR candidate; they classified the X-ray point source as a
dwarf M2-M3 star in the Solar neighbourhood. This region was covered in the
\xmm\ survey. The diffuse emission detected with ROSAT is found to be part of
larger scale structures from the hot phase of the LMC ISM. There is \emph{no}
large SNR around RX J0533.5$-$6855.

\paragraph{30 DOR C:} This is a large shell seen in X-rays with a non-thermal 
spectrum \citep{2004ApJ...602..257B,2015A&A...573A..73K}. Its nature as a 
superbubble rather than a standard SNR was already recognised by 
\citet{1985ApJS...58..197M}. It was however listed as an SNR in \citet[][with 
the identifier {[}BGS2006b{]} J0536$-$692]{2006ApJS..165..480B} and 
\citetalias[][as {[}BMD2010{]} SNR J0536.2$-$6912]{2010MNRAS.407.1301B}. 
Interestingly, there \emph{is} an SNR (in projection) in 30 DOR C \citep[MCSNR 
J0536$-$6913,][]{2015A&A...573A..73K}, but it was revealed only later and is 
most likely distinct from the non-thermal shell.

\paragraph{SNR B0538$-$69.3:} The first classification as an SNR dates back to
\citet{1984ApJS...55..189M}, based on radio and weak optical evidence.
\citetalias{2010MNRAS.407.1301B} included that source with the wrong J2000
coordinates. \citet{2006ApJS..165..480B} used the correct position but did not 
detect UV emission from the object. B0538$-$69.3 is unusually bright in radio 
(Miroslav Filipovi\'c 2014, personal communication) considering the general 
lack of X-ray and optical emission. \citet{1984ApJS...55..189M} noted that the 
absence of X-ray emission might be due to the high $N_H$ towards this region of 
the LMC. However, other SNRs are found in that region (\eg MCSNR J0536$-$6913, 
DEM L299, the Honeycomb nebula), so a negative result with \xmm\ is puzzling. 
This objects remains at best an SNR \emph{candidate}.

    \subsection{The final sample}
        \label{compiling_final}
Our compilation results in a list of 59 definite SNRs. In 
Table~\ref{appendix_table_snrs_sample} we list the final sample of LMC SNRs 
used in this work. Basic information is given for each object: MCSNR 
identifier and old name, position, X-ray data available, and reference. In 
addition, we added columns with X-ray results: X-ray luminosity 
(Sect.\,\ref{results_spectra} and \ref{results_XLF}), X-ray size 
(Sect.\,\ref{data_imaging}), and $N_H$ fraction 
(Sect.\,\ref{results_distribution}). Finally, we give for each SNR the values of 
the two metrics used to assess the local stellar environment described in 
Sect.\,\ref{results_sfh}. See text in Appendix~\ref{appendix_sample} for 
detailed description of each column.

This work focuses on the X-ray emission of LMC SNRs. Therefore, there are only 
confirmed SNRs in the final sample (no candidate). The resulting list provides 
the most complete sample of SNRs in the LMC, \emph{as far as X-rays are 
concerned}: \xmm\ observations exist for 51 SNRs out of the list of 59 SNRs 
defined here. Out of the eight objects without \xmm\ data available, three were 
covered with \chandra, and two only by ROSAT. Only three objects have not any 
X-ray information available (yet), though their radio and optical properties 
warrant their classifications as SNR. In Sect.\,\ref{results_XLF} and 
Sect.\,\ref{summary}, we discuss the total number of LMC SNRs and the overall 
completeness of the sample.

\section{Data analysis}
\label{data}

\subsection{X-ray imaging}
\label{data_imaging}

For each SNR, we combined the smoothed images in the soft, medium, and hard 
bands (obtained as described in Sect.\,\ref{observations_reduction}) into X-ray 
composite images. These are shown in Appendix~\ref{appendix_images}. The same 
images are used to obtain X-ray contours to help in defining regions for 
spectral extraction (Sect.\,\ref{data_spectra_extraction}).

The study of the size distribution of SNRs provides clues to small-scale 
structures in galaxies and the energy and matter cycles in the ISM. The sample 
of LMC SNRs (at various levels of completeness) has been already used for such 
studies \citep[\eg][\citetalias{2010MNRAS.407.1301B}]{1983ApJS...51..345M}. An 
SNR can appear to have different sizes depending on the wavelength (\eg a larger 
radius in radio than in X-rays), or can have an asymmetric morphology that 
complicates the definition of its ``size''. To help future studies of the size 
distribution, we provide in this work the \emph{maximal} extent of each SNR in 
X-rays, which we measured from the X-ray images and contours. The values are 
listed in Table~\ref{appendix_table_snrs_sample}. The size distribution of LMC 
remnants, combining measurements at various wavelengths, is presented and 
discussed in Bozzetto et al. (in prep.).

\subsection{X-ray spectra}
\label{data_spectra}

\subsubsection{Analysis method}
\label{data_spectra_method}
SNRs are \emph{extended} X-ray sources, and many of those in our sample have a 
low surface-brightness. Consequently, the analysis of their spectra is 
challenging. A careful treatment of the background, both instrumental and 
astrophysical, is utterly important in order to obtain meaningful fits and 
extract the purest possible information from the source. It is not desirable to 
simply subtract a background spectrum extracted from a nearby region, because of 
the different responses and background contributions associated to different 
regions, and because of the resulting loss in the statistical quality of the 
source spectrum. An alternative method, which we used in this work, is to 
extract a nearby background spectrum, define a (physically motivated) model for 
the background and simultaneously fit the source and background spectra. Below, 
we explain the method in detail. Our account of the background is detailed in 
Appendix~\ref{appendix_background}. The only source for which a different method 
was used is SNR~1987A, as described in Sect.\,\ref{results_spectra_1987A}.

The spectral-fitting package XSPEC \citep{1996ASPC..101...17A} version 12.8.0m 
was used to perform the spectral analysis. Unless otherwise stated, spectra were 
rebinned with a minimum of 25 counts to allow the use of the $\chi 
^2$-statistic. Interstellar absorption was reproduced by two photoelectric 
absorption components (\texttt{phabs} and \texttt{vphabs} in XSPEC, where the 
previx ``v'' indicates that abundances can vary), one with a column density 
$N_{H\mathrm{\ Gal}}$ and solar abundances for the foreground Galactic 
absorption, and another one with $N_{H\mathrm{\ LMC}}$ and LMC elemental 
abundances \citep{1992ApJ...384..508R} for absorption within the LMC. 
Cross-sections for photoelectric absorption were set to those of 
\citet{1992ApJ...400..699B}. The foreground column density $N_{H\mathrm{\ Gal}}$ 
at the location of each analysed source is taken (and fixed) from the \ion{H}{I} 
maps of \citet[][available online on the HEASARC 
pages\footnote{\url{http://heasarc.gsfc.nasa.gov/cgi-bin/Tools/w3nh/w3nh.pl}}]{ 
1990ARA&A..28..215D}.

For the analysis of one extended source, two different regions are defined: 
\emph{i)} a source spectrum extraction region (hereafter \src region), and 
\emph{ii)} a background spectrum extraction region (hereafter \bg region).
Two spectra are extracted per instrument (pn, MOS1, and MOS2) from each region,
one from the event list of the science observation, the second from the FWC
data. The FWC spectra must be extracted at the same \emph{detector} position as
in the science observation, because of the strong position-dependency of the
instrumental background for both pn and MOS. The \src and \bg regions are best 
defined in World Coordinates System (WCS, \ie sky position\footnote{This is 
more practical, in particular when several observations of a source with 
different pointings exist.}). Therefore, we first project the FWC data at the 
same sky position as the science observation, using its attitude history file 
and the SAS task \texttt{attcalc}. One can then use the same extraction regions 
to select FWC spectra.

The four spectra are fitted simultaneously \revision{over the 0.3~keV~--~12~keV 
range}. The instrumental background model is constrained by the FWC data, and 
included (with tied parameters) in the spectra from the science observation. The 
science spectrum in the \bg region therefore allows the parameters of the 
astrophysical X-ray background (AXB) to be determined. It is assumed that the 
temperature of the thermal components and the surface brightness of the thermal 
and non-thermal components do not vary significantly between the \src and \bg 
regions. Thus, the appropriate temperature and normalisation parameters are 
linked. All background components are then accounted for, and one can explore 
the intrinsic emission from the source using several emission models 
(Sect.~\ref{data_spectra_models}).

Regarding which instruments are used, several configurations are possible, 
depending on the data present. Ideally, one would use all EPIC instruments 
(pn+MOS1+MOS2) together. However, our analysis method requires FWC data. Those 
are available for all read-out modes of pn, but only for the full-frame mode of 
MOS, limiting the use of MOS data in some cases (\eg MOS in Small Window mode). 
It also happens that the SNR is outside the MOS field of view, if it is too far 
off-axis or on one of the damaged chips of MOS1 \citep{2006ESASP.604..943A}. In 
these cases only the pn spectrum is used for analysis. The contrary (only MOS 
spectra available) occurs in rare cases.

About 80\,\% of the SNRs in the sample were observed only once. A few were
observed twice in overlapping survey observations; the deep field centred on
SNR~1987A contains four SNRs in total, and a plethora of \xmm\ data are at hand
for those. To keep the analysis the same for most sources, we restricted the 
number of observations analysed simultaneously to two for the latter cases. If 
more than two observations are available, we selected the two deepest datasets 
(\ie longest flare-filtered exposure times) for analysis. Finally, N132D is a
calibration target and frequently observed. It is however too bright for the
full-frame mode; only Small Window and Large Window modes have been used and
thus we only used the deepest pn dataset.

It was found more efficient to pre-fit the instrumental and astrophysical 
background of each SNR. That is, we first fitted the (FWC + AXB) EPIC-pn 
spectra alone and FWC MOS spectra alone. If the pre-fitting of the background 
components was satisfactory, their best-fit parameters were used as starting 
points in the final fit, which includes the SNR emission model. Doing so speeds 
up the process of analysing the SNR spectrum alone. It also helps, by visual 
examination of the background fits, to identify problematic cases, as 
described in Appendix~\ref{appendix_background}.

\begin{figure}[t]
    \centering
\includegraphics
[width=0.99\hsize]{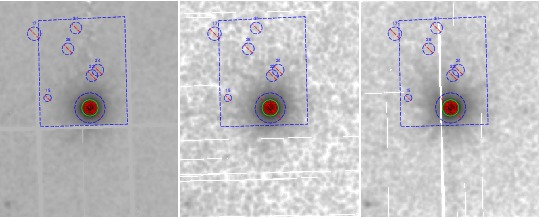}

\includegraphics
[width=0.99\hsize]{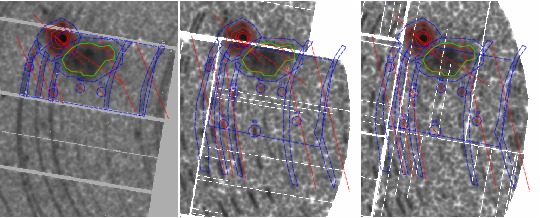}
\caption[\ Extraction regions used to extract spectra of MCSNR
J0519$-$6902 and J0547$-$6943]{\emph{Top:} Extraction regions used to extract
spectra of MCSNR J0519$-$6902 from EPIC pn, MOS1, and MOS2 detectors (left to
right). For the scale, we remind the reader that pn chips are 4.4\arcmin-wide. 
The X-ray contours (in red) are used to outline the boundary of the remnant 
emission and set the radius of the circular \src region (in green). The \bg 
regions are shown by the blue dashed rectangle. The barred blue circles show
detected point sources and a ``buffer'' area around the \src region. Those are
excluded from the \bg region.\\
\emph{Bottom:} Same for MCSNR J0547$-$6943, outlined by the green polygonal
region. The barred blue arcs are excluded to avoid single-reflections from LMC
X-1.
}
\label{fig_data_spectra_extraction}
\end{figure}

\subsubsection{Extraction of the spectra}
\label{data_spectra_extraction}

The first step of the analysis is to extract spectra for each SNR of the
sample, as well as corresponding background spectra from nearby regions (using
the same observation). Due to the spread in morphology and size of the SNRs, 
unequal properties of their background (diffuse emission and point source 
crowding), and their varying location on the EPIC detectors, \src and \bg 
regions cannot be created automatically. Therefore, extraction regions were 
manually defined for each SNR. For the \src region, the constraint was simply 
to include all the remnant's emission and exclude unrelated point sources that 
might be located in projection. We used the contours taken from the X-ray 
image (see Sect.\,\ref{data_imaging}), combining all observations of each 
remnant, to identify the boundaries of the SNR emission. If the morphology
of the object requires it, an arbitrary shape (polygonal region) is used
instead of a circle or ellipse.

The \bg regions are chosen from different locations on the pn and MOS detectors 
if needed, in order to be on the same CCD chip as (most of) the SNR emission. In
most cases where the remnant was the target of the observation (\ie was observed
on-axis), the same \bg region defined for pn can also be used for MOS data,
because of the chip configuration of the latter with one central chip and six
peripheral chips. Detected point sources are also excluded from the \bg
regions.

Two examples are shown in Fig.\,\ref{fig_data_spectra_extraction}. In the simple 
case (that of MCSNR J0519$-$6902), we used a circular \src region and the same 
\bg regions for all EPIC detectors. In the more complex case of MCSNR 
J0547$-$6943 (or DEM L316B), we used a polygonal \src region; the \bg region is 
narrower for pn than for MOS to fit on a single CCD chip. In addition to point 
sources, we excluded arc-shaped regions which are affected by instrumental 
artefacts (single-reflections from LMC X-1). Extraction regions for all LMC 
SNRs analysed in this work are shown in Appendix~\ref{appendix_images}. 

Because of the telescope vignetting, the effective area is not constant across
the extent of SNRs. To take this into account, all spectra (source and
background) are extracted from \emph{vignetting-weighted} event lists. These are
created with the SAS task \texttt{evigweight}, applied on the calibrated event 
lists produced by our data reduction pipeline (as described in 
Sect.\,\ref{observations_reduction}). It assigns a weight to each event of 
energy $E_j$ at detector coordinates $(detx_j,dety_j)$, which is the inverse of 
the ratio of the effective area at that position to the central effective area 
(at the same energy):
\begin{equation}
w_j \; = \; \frac{A_{0,0}(E_j)}{A_{detx_j,dety_j}(E_j)} 
\end{equation}
The corrected event lists are equivalent to that obtained with a flat 
instrument. For spectral analysis, a flat response file with the 
on-axis effective area must then be used\,\footnote{The extraction radii for 
the three smallest SNRs (except SNR~1987A which is handled as a point source) 
result in PSF losses less than 15\,\%.}. Instrumental background spectra were 
extracted from FWC data at the same detector position as the \src and \bg 
regions following the method described above.

\subsubsection{Spectral models}
\label{data_spectra_models}
The model of the SNR emission is built iteratively, in increasing order of 
complexity. First, a one-component collisional ionisation equilibrium (CIE) 
model (\emph{vapec} in XSPEC) is tried. The elemental abundances are initially 
set to the values measured by \citet{1992ApJ...384..508R}. Their abundance of 
silicon is highly uncertain, however, and therefore we use an initial value of 
half solar for Si. Analysis of the residuals and goodness-of-fit reveals if and 
how the model can be improved, by either thawing some elemental abundances, or 
switching to a non-equilibrium ionisation (NEI) model (\emph{vpshock} in XSPEC). 
Abundance effects are manifest when all lines from one element are over- or 
underestimated compared to the continuum or other elements, while the signatures 
of NEI are in the residuals from different ions of the same element (\eg 
relative strengths of \ion{O}{VII}/\ion{O}{VIII} lines). We evaluate the 
significance of the fit improvements (if any) with F-tests. We refer to these 
one-component models as ``1T'' (one temperature).

A second component is added if needed; those are the ``2T'' SNRs. Again we start 
with CIE, then assess whether there is need for NEI or free abundances in the 
second component. For several SNRs, the analysis of X-ray colour images already 
hints at the presence of two components, with \eg a different temperature, 
$N_H$, or abundance pattern. This iterative process is done until a satisfactory 
fit is achieved, at which point 90\,\%~C.\.L. errors are computed for all free 
parameters. More complicated models may be applied/needed for a handful of SNRs, 
particularly amongst the brightest ones. These cases are presented in 
Sects.\,\ref{results_spectra_brightest} and \ref{results_spectra_1987A}.


\section{The X-ray spectral properties of LMC SNRs}
    \label{results_spectra}

\subsection{General properties}
    \label{results_spectra_general}

Out of the sample of 59 SNRs and 51 with \xmm\ data available, 45 are fitted
with 1T or 2T models, while 6, amongst the brightest, are fitted with more 
complex models (see Sects.\,\ref{results_spectra_brightest} and 
\ref{results_spectra_1987A}). \revision{Six previously known SNRs have been 
covered by our VLP survey, giving the opportunity to study their spectra with 
\xmm\ for the first time. Among these, MCSNR J0534$-$6955 and J0547$-$7025 were 
observed with \chandra\ \citep{2003ApJ...593..370H}. \citet{2008AJ....136.2011R} 
analysed only MOS data of MCSNR J0518$-$6939 from an archival observation, while 
we now have EPIC-pn data. We show the \xmm\ spectra from these SNRs in 
Appendix~\ref{appendix_spectra}.}

The results of the spectral analysis for the 1T/2T 
sample are given in the Appendix~\ref{appendix_tables}  
(Table~\ref{appendix_table_spectra_all}). All relevant parameters are listed 
with their 90\,\%~C.\.L. uncertainties: The fitted LMC absorption column 
density (column 2), plasma temperature $kT$ (3), ionisation age $\tau$ (4), 
emission measure EM (5), and abundances (6). When a second component is used, 
its parameters are given in columns (7)~--~(11). The first component is the one 
with higher EM. The quality of the fits are evaluated by the $\chi^2 / \nu$ of 
column (12), where $\nu$ is the number of degrees of freedom. The reduced  
$\chi^2$ values ($\chi ^2 _{\mathrm{red}}$) are also listed in column (12). 
The median $\chi ^2 _{\mathrm{red}}$ is 1.16. 90\,\% of the fitted objects have 
a reduced $\chi^2$ less than 1.4.

In 32 cases, the SNR is fitted with, or the available data only require, a one 
component model. Amongst these, 9 do not show significant NEI effects and are 
fitted with a CIE model. Using a NEI model for these did not result in a 
statistically significant improvement, neither in the goodness-of-fit sense, nor 
in terms of residuals. Moreover, the ionisation ages $\tau$ in these cases are 
high and poorly constrained. Therefore we list the CEI model parameters. In the 
23 remaining ``1T'' objects, better fits are obtained with an NEI model. The 
plasma temperature for the ``1T SNRs'' clusters in the 0.25~keV~--~0.45~keV 
range. The highest values of temperature (above 1~keV) are associated with the 
smallest ionisation ages. In at least some cases, this could be an artefact of 
the analysis due to insufficient data. The ionisation age $\tau$ of this sample 
is broadly distributed around a median value of $1.7 \times 
10^{11}$~s~cm$^{-3}$.

There are 13 SNRs in the 2T sample. Two objects are fitted with two CIE 
component models (MCSNR J0530$-$7008 and J0517$-$6759). The rest was fitted with 
two NEI components, although for three SNRs the ionisation age of one of the 
components was unconstrained and on the high end ($\tau \gtrsim 
10^{13}$~s~cm$^{-3}$), indicating a plasma close to or in CIE. The median $\tau$ 
of the main component (\ie that with higher emission measure) for the 2T sample 
is slightly higher (5~--~7~$\times 10^{11}$~s\,cm$^{-3}$) than that of the 1T 
sample, but low number statistics preclude a direct comparison. The temperature 
distribution is bimodal: one component has a median temperature of $kT = 
0.31$~keV, the second a higher median of 0.8~keV. In several cases the high-$kT$ 
component also requires a different abundance pattern, revealing SN ejecta 
(Sect.\,\ref{results_spectra_ejecta}).

For nine SNRs, the data did not require or allow to fit elemental abundances. 
For a few cases, this happens because the spectrum is contaminated by a bright 
pulsar (N157B and MCSNR J0540$-$6920), or by LMC X-1 (MCSNR J0540$-$6944), and 
the thermal emission is not well separated by \xmm. The other SNRs fitted with 
abundances from \citet[][RD92 in 
Table~\ref{appendix_table_spectra_all}]{1992ApJ...384..508R} are relatively 
faint. The limited available data therefore prevent the use of free abundances 
in the fits.

Oxygen and iron are the main contributors to the 0.5~keV~--~2~keV X-ray
emission for the relevant plasma temperatures. Consequently, they are the first
elements for which abundances can be fitted. Out of the 45 1T/2T SNRs, 35 have
at least free O and Fe abundances. Neon and magnesium also have prominent lines
routinely detected below 2~keV, and their abundances were fitted in 33 and 30
SNRs, respectively. Silicon is detected and its abundance fitted, in
23 SNRs. This subset has a higher median temperature ($kT \sim 0.6$~keV) than
the whole sample, as expected. Indeed, Si emission becomes prominent for higher
temperatures than, say, O, Ne, or Fe. While obvious and fitted in all the
brightest SNRs, which are younger/hotter, lines of sulphur are not detected in
most 1T/2T SNRs. Only a handful (MCSNR J0534$-$6955, J0547$-$7025, N63A) allow 
to fit the S abundances. All have plasma temperatures in excess of 0.8~keV. The 
fitted abundance patterns can be used to type the supernova progenitor, if 
ejecta are detected (Sect.\,\ref{results_spectra_ejecta}), or to measure 
metallicity of the LMC ISM (Sect.\,\ref{results_spectra_abundance}).

\subsection{The analysis of the brightest SNRs}
    \label{results_spectra_brightest}
For six of the brightest SNRs, the simple 1T/2T models approach was clearly
insufficient to satisfactorily model the spectra. This is expected, because on
the one hand, the exquisite statistical quality of these spectra imply that
even a two-component model is not adequate to reproduce the complex multi-phase
structure in these objects. On the other hand, the very young SNRs, in addition
to a small ambient medium contribution, are dominated by ejecta. Because of
stratification of the ejecta heated by the reverse shock, elements synthesised
at different radii in the SN explosion can have distinct spectral properties.

All the ``bright SNR'' sample was observed in individual \xmm\ and \chandra\
pointings. Detailed results are published in several papers (references are
given below), with which our results were never at odds. Here, we used 
multi-temperature empirical models to reproduce the spatially integrated
spectra. This allows \emph{i)} to derive accurate X-ray fluxes, so that the
luminosity function (Sect.\,\ref{results_XLF}) is complete at the bright end, 
\emph{ii)} to measure the properties of the Fe~K emission, if present (see
Sect.\,\ref{results_spectra_FeK}), and \emph{iii)} to obtain spectral 
properties (\eg $N_{H}$, $kT$, $\tau$) for statistical studies and comparison of 
their distributions for various sub-samples (Sect.\,\ref{results_XLF}). The 
adopted models are described below. The spectral parameters are given in
Table~\ref{appendix_table_spectra_brightest} and 
Table~\ref{appendix_table_spectra_1987A}.


\paragraph{DEM~L71 (MCSNR J0505$-$6753):}
DEM~L71 is a notorious type Ia SNR, owing to the detection of iron-rich
ejecta \citep[\eg][]{1995ApJ...444L..81H}. \citet{2003A&A...406..141V}
presented the \xmm\ EPIC and RGS results for this remnant, and
\citet{2003ApJ...582L..95H} those obtained with \chandra\ observations.
Different conditions are measured in the shell and central regions. It is then
unsurprising that a 2T model as used for other SNRs did not produce acceptable
fits. Instead, we obtained satisfactory results with three components: Two
components (``Fe-low $kT$'' and ``Fe-high $kT$'') had Si, S, and Fe (the main 
nucleosynthesis products of Ia SNe) freed and common to the two components,
while other metals were set to zero. These components account for the
ejecta-rich emission, as well as the Si, S, and Fe contribution of the ISM. A
third component, with O, Ne, and Mg abundances free and Si, S, Fe set to zero,
accounts for the bulk of ISM emission.

In addition, Fe~K emission is clearly detected, pointing to the presence of
very hot ejecta ($kT > 2$~keV). The statistical weight of this feature remains
small. Therefore, instead of adding another thermal component, we modelled the
line with a Gaussian. The parameters of the Fe~K line are used in comparison
with other SNRs in Sect.\,\ref{results_spectra_FeK}. The ejecta components have 
best-fit temperatures of $\sim 0.4$~keV and $\sim0.9$~keV 
(Table~\ref{appendix_table_spectra_brightest}). The ionisation age of the cooler 
component is twice that of the hotter one. The ISM component has a temperature 
of $kT = 0.46$~keV, the same as measured with \chandra\ 
\citep{2003ApJ...582L..95H}, and in between the two temperatures used for the 
shell emission by \citet{2003A&A...406..141V}.

\paragraph{N103B (MCSNR J0509$-$6844):}
The spectrum of N103B is remarkable because of the numerous lines from highly
ionised metals: \ion{Si}{XII} and \ion{Si}{XIV}, \ion{S}{XV} and (marginally)
\ion{S}{XVI}, \ion{Ar}{XVII}, and \ion{Ca}{XIX}. A strong Fe~K blend is also
detected. We fit the spectrum with the same three-temperature model as for
DEM~L71. One component had abundances fixed to RD92, accounting for the ISM
emission. Two components with different $kT$ and $\tau$ were used to reproduce
the (dominating) ejecta emission. All relevant elements (O, Ne, Mg, Si, S, Ar,
Ca, and Fe) were freed, but common to both components. A Gaussian was also
included to fit the Fe~K~feature.

With this model, the spectrum of N103B is well reproduced across the whole
0.3~keV~--~8~keV band. The results are comparable to those of
\citet[][focusing on \xmm\ data]{2002A&A...392..955V} and
\citet[][with \chandra]{2003ApJ...582..770L}, especially regarding: \emph{i)}~
the column density $N_H \sim 3 \times 10^{21}$~cm$^{-2}$; \emph{ii)}~the
presence of one high ionisation age component (at $kT \sim 0.7$~keV) and a
hotter (1.6~keV) underionised component. Because the Fe~K blend is modelled
separately with a Gaussian, the fitted temperature of the hottest component is
lower than in the previous references; \emph{iii)}~high abundances of S, Ar, and
Ca.

\paragraph{N132D (MCSNR J0525$-$6938):}
\citet{2001A&A...365L.242B} presented the \xmm\ observations of N132D from the
Performance Verification programme. Results of the \chandra\ ACIS-S
observations can be found in \citet{2007ApJ...671L..45B}. Both instruments
spatially resolve the SNR into regions with different spectral properties.
Therefore, though a three-temperature model can reproduce the main features of
the spectrum (thus allowing to measure accurately the integrated flux of
the remnant), strong residual structures are seen between 0.5~keV and 1~keV,
where the strongest variations are observed (lines of O, Ne, Fe).

The best fit is obtained with a cool ($\sim 0.5$~keV) component with abundances
close to the normal LMC values (\ie it represents a blast wave component) that
dominates the soft emission (below 1.5~keV). A second component with $kT \sim
1$~keV is characterised by enriched levels of O, Ne, and Mg, as well as a
higher column density ($\sim 10^{22}$~cm$^{-2})$. This component thus describes
the bulk of the ejecta emission, and accounts for most of the Si and S emission.
Finally, the presence of highly ionised iron is evident from the 
\revision{$6.69$~keV line (see 
Table~\ref{table_results_spectra_FeK}), corresponding to the K$\alpha$ energy 
of \ion{Fe}{XXV}}. This indicates a third,
very hot component ($\sim 5$~keV). In this component only Fe, Ar, and Ca are
included. The two latter elements improve the residuals around 3.1~keV
(\ion{Ar}{XVII}), and 3.9/4.1~keV (\ion{Ca}{XIX} and \ion{Ca}{XX}). These K
lines were already mentioned in the early \xmm\ results
\citep{2001A&A...365L.242B}.

\paragraph{0519$-$69.0 (MCSNR J0519$-$6902):}
The SNR was observed early in the \chandra\ and \xmm\ missions. In addition, the
LMC survey covered the source, at an off-axis angle of $\sim 9$\arcmin, adding
23~ks and 27~ks to the existing 8~ks and 46~ks of full-frame pn and MOS data,
respectively. Spectra from the two observations were fitted simultaneously.
0519$-$69.0 exhibits strong lines of Si, S, Ar, and Ca, as well as prominent
Fe~L and K blends. To reproduce the spectra we used the multi-component approach
of \citet{2010A&A...519A..11K}, who extensively studied the \xmm\ and \chandra\
data.

First, one NEI component with LMC abundances accounts for circumstellar medium 
(CSM) or ISM emission. Then, one NEI component for each (group of) element(s) 
having detected lines: oxygen, silicon and sulphur, argon and calcium, and iron. 
In the latter case two NEI components with distinct parameters are used, as the 
spectrum evidently includes both medium temperature and very hot iron. Due to 
the low count rate, and therefore statistical weight, of the Fe~K blend, the hot 
iron component was driven to fit lower energy lines instead. To alleviate this 
issue we fitted the high-energy part of the spectrum separately with this 
component, then froze the best-fitting parameters in the global fits. Residuals 
around 0.72~keV (lines of \ion{Fe}{XVII}) were fitted with an additional 
Gaussian line.

\paragraph{0509$-$67.5 (MCSNR J0509$-$6731):}
\xmm\ observed the SNR for $\approx 40$~ks in 2000, with pn operated in Large
Window mode. This dataset is presented in \citet{2008A&A...490..223K}, while
\citet{2004ApJ...608..261W} reported the spectral and imaging analysis of a
\chandra\ observation. Finally, \citet{2008ApJ...680.1149B} attempted to
reproduce spectra from both instruments using a grid of hydrodynamical models
and an X-ray emission code. Inconsistencies between pn and MOS spectra were
found, with lines in the pn spectrum (red-)shifted relative to those in MOS
spectra by about 1~\%. This is likely a gain issue of the pn instrument. We
discarded spectra from the MOS instruments, as they were operated in Small
Window mode, for which no FWC data are available. To get the spectral model to
match the observed energies of atomic lines, we freed the ``redshift'' parameter 
available in XSPEC models, which allows an \emph{ad hoc} change of the
energy scale. Satisfying results were obtained for a shift of $\approx 1$~\%,
which is the measured pn/MOS discrepancy \citet{2008A&A...490..223K}.

As for J0519$-$6902, lines from heavy elements are prominent, and we used a
multi-component model. Si, S, and Ar were grouped in a NEI component, and shared 
the same temperature and ionisation age. Another NEI component modelled the
continuum+lines emission from the CSM/ISM. No Si, S, Ar, or Ca were included in 
this component. Iron was included in two NEI components, one with a medium 
temperature ($\sim~1.4$~keV) and a high-$kT$ one ($\sim 11$~keV) that
reproduces the strong Fe~K line. The latter component also includes calcium.
Even with this model, residuals remained around Fe lines (0.72~keV and 
1.22~keV), which we fitted with two Gaussian lines.
\revision{The very high temperature of the second Fe component is atypical for 
SNRs, but was also suggested in \chandra\ data by \citet{2004ApJ...608..261W}.  
The SNR exhibits a high-energy continuum tail, which previous studies tried 
to reproduce with non-thermal models. This tail can also be reproduced with the 
Bremsstrahlung continuum of a high-$kT$ thermal model, driving our fit to 
temperatures above 10~keV. Furthermore, we already noted the energy shift issue 
of the pn spectra, that results in a small centroid energy of the Fe K line for 
a given fitted $kT$. Given this caveats, it remains unclear whether the 11~keV 
plasma is physical.
}

\begin{table*}[ht]
\caption{Fe~K line properties of LMC SNRs}
\label{table_results_spectra_FeK}
\centering
\begin{tabular}{l l c c c c c}
\hline\hline
\noalign{\smallskip}
MCSNR & Alt. name & type & \multicolumn{2}{c}{Energy centroid (eV)} &
\multicolumn{2}{c}{Line luminosity ($10^{42}$~ph s$^{-1}$)} \\
 & & & \xmm & \suzaku & \xmm & \suzaku \\
\noalign{\smallskip}
\hline
\noalign{\smallskip}
J0509$-$6731 & B0509$-$675 & Ia & 6432$_{-27}^{+29}$ & 6425$_{-15}^{+14}$ &
0.87$\pm0.21$ & 0.96$\pm0.12$ \\
\noalign{\smallskip}
J0505$-$6753 & DEM L71 & Ia & 6494$\pm58$ & --- & 0.26$_{-0.09}^{+0.08}$ & ---
\\
\noalign{\smallskip}
J0509$-$6844 & N103B & Ia & 6514$_{-32}^{+31}$ & 6545$\pm6$ & 5.10$\pm0.87$ &
6.43$\pm0.30$ \\
\noalign{\smallskip}
J0519$-$6902 & B0519$-$690 & Ia & 6543$_{-31}^{+28}$ & 6498$_{-8}^{+6}$ &
1.71$\pm0.45$ & 2.78$\pm0.15$ \\
\noalign{\smallskip}
J0526$-$6605 & N49 & CC & --- & 6628$_{-26}^{+29}$ & $<$ 4.75\tablefootmark{a} &
0.54$\pm0.12$ \\
\noalign{\smallskip}
J0535$-$6916 & SNR~1987A\tablefootmark{b} & CC & 6635$\pm70$ &
6646$_{-54}^{+55}$ & 0.64$\pm0.18$
& 0.57$\pm0.24$ \\
\noalign{\smallskip}
J0535$-$6602 & N63A & CC & 6683$_{-99}^{+88}$ & 6647$_{-17}^{+16}$ &
2.36$_{-1.08}^{+1.03}$ & 2.57$\pm0.36$ \\
\noalign{\smallskip}
J0525$-$6938 & N132D & CC & 6685$_{-14}^{+15}$ & 6656$\pm9$ & 4.58$\pm0.58$ &
5.47$\pm0.51$ \\
\noalign{\smallskip}
\hline
\end{tabular}
\tablefoot{\suzaku\ results are from  \citet{2014ApJ...785L..27Y}.
\tablefoottext{a}{$3\sigma$ upper limit.}
\tablefoottext{b}{The quoted numbers are average values over the last six
epochs, and the uncertainties are the RMS scatter. Note that we found the 
energy centroid to evolve rapidly at recent epochs (see 
Sect.\,\ref{results_spectra_1987A}).}
}
\end{table*}

\subsection{Update on the monitoring of SNR~1987A}
\label{results_spectra_1987A} SN~1987A, the nearest supernova in almost 
400~years, was discovered in the LMC on 23 February 1987. It is exceptional in 
many ways and has been extensively studied ever since. We have the unique 
opportunity to follow the early evolution of a supernova \emph{remnant} (hence 
the use of the identifier ``SNR~1987A''\,\footnote{In our nomenclature SNR~1987A 
is also given the name MCSNR J0535$-$6916.}). Results from the many existing 
\xmm\ observations of SNR~1987A are presented in \citet{2006A&A...460..811H}, 
\citet{2008ApJ...676..361H}, and \citet{2010A&A...515A...5S}. \citet[hereafter 
\citetalias{2012A&A...548L...3M}]{2012A&A...548L...3M} analysed data from the 
2007--2011 monitoring, focusing on the rapid evolution of the X-ray light curve 
and the properties and evolution of the Fe~K lines, which were detected 
unambiguously for the first time.  However, the spectral parameters (except 
fluxes and Fe~K line properties) were not given in 
\citetalias{2012A&A...548L...3M}. We take advantage of this work to give these 
detailed results, and include an unpublished observations (ObsID 0690510101) 
performed on December 2012, after \citetalias{2012A&A...548L...3M} was released.

All spectra from SNR~1987A were extracted from a circular region centred on the 
source, with a radius of 25\arcsec. The use of spatially integrated spectra
is dictated by the small radius of the source \citep[still less than
1\arcsec,][]{2013ApJ...764...11H}, which is completely unresolved by \xmm. The 
background spectra were extracted from a nearby point-source-free region common 
to all observations. Only single-pixel events (\texttt{PATTERN} = 0) from the 
pn 
detector were selected. Contrary to all other SNRs in this work, the background 
spectra were not modelled but \emph{subtracted} from the source spectra. We 
used 
the same three-component plane-parallel shock model as in 
\citetalias{2012A&A...548L...3M}, with one fixed-temperature component ($kT = 
1.15$~keV) and free abundances of N, O, Ne, Mg, Si, S, and Fe. EPIC-pn spectra 
from all seven epochs of the monitoring are fitted simultaneously between 
0.2~keV and 10~keV, with common abundances and N$_{H\mathrm{\ LMC}}$. To 
characterise the Fe~K line, we performed separate fits in the range (5--8)~keV 
on the \emph{non-rebinned} spectra using the C-statistic 
\citep{1979ApJ...228..939C}. We used a Bremsstrahlung model for the continuum 
and a Gaussian for the Fe~K line complex.

The simultaneous (broad-band) fit was satisfactory, with $\chi ^2 = 5114.2$ for 
4109 degrees of freedom (reduced $\chi ^2 _r~=~1.24$). Spectral results are the 
same as in \citetalias{2012A&A...548L...3M}. We give the best-fit parameters for 
all seven epochs in Table\,\ref{appendix_table_spectra_1987A}. 
We list soft (0.5~keV~$-$~2~keV) and hard (3~keV~$-$~10~keV) X-ray fluxes at all 
epochs, with $3\sigma$ uncertainties (99.73\,\% confidence level, C.\,L.) in 
Table~\ref{appendix_table_spectra_1987A}. Echoing the findings of 
\citetalias{2012A&A...548L...3M}, we see that the soft X-ray flux keeps 
increasing after the 25$^{{\rm th}}$ anniversary of SNR~1987A. Since 2011, 
however, the rate of increase has dropped below 10\,\% per year, showing that 
subsequent observations of SNR~1987A with \xmm\ are highly desirable to follow 
the evolution of the X-ray flux and to identify the turn-over point.

The central energy, $\sigma$-width, total photon flux and equivalent width (EW) 
of the Fe~K feature are also listed in 
Table~\ref{appendix_table_spectra_1987A}. Up to December 2011 the results are 
the same as in \citetalias{2012A&A...548L...3M}. The new data point (December 
2012) reveals a line with roughly the same flux but a significantly higher 
central energy ($6.78_{-0.05}^{+0.06}$~keV) than previously ($6.60\pm0.01$~keV, 
averaging the earlier measurements). This hardening likely indicates an 
increased contribution from highly ionised iron (\ion{Fe}{xxvi}) that prior to 
2012 was either absent (as iron was in lower ionisation stages) or too weak to 
be detected. With the resolution of pn and the statistics in our hand, it is not 
possible to resolve the K-shell lines from various Fe ions, which will become 
possible with next-generation X-ray calorimeters onboard Astro-H 
\citep{2012SPIE.8443E..1ZT} or Athena \citep{2013arXiv1308.6784B}.

\subsection{Fe~K emission from LMC SNRs}
    \label{results_spectra_FeK}

\citet{2014ApJ...785L..27Y} used \suzaku\ to systematically search for Fe~K
emission from Galactic and LMC SNRs. Fe~K$\alpha$ emission was detected in 23
SNRs, including seven remnants in the LMC. Their essential finding is that the
centroid energy of the Fe~K emission, determined by the ionisation state of
iron, is a powerful tool for distinguishing progenitor types. Indeed, the Fe~K 
emission of type Ia remnants is significantly less ionised than in CC-SNRs. 
Furthermore, there is a positive correlation between the Fe~K$\alpha$ line 
luminosity and centroid energy \emph{within each progenitor group}.

Because the Fe~K blend is a promising typing tool, we extended the search for
Fe~K emission of \citet{2014ApJ...785L..27Y} to all LMC SNRs observed with
\xmm. Compared to the \suzaku\ sample, the coverage is more complete (\ie more
SNRs observed) and more sensitive (the EPIC-pn effective area is slightly
higher than that of \suzaku's XIS, even combining all detectors), and Fe~K can 
potentially be detected from more SNRs.

In Table~\ref{table_results_spectra_FeK}, we give the results for all LMC SNRs 
with detected Fe~K emission, ranked by increasing centroid energy. The \xmm\ and 
\suzaku\ results are consistent within the uncertainties. Strikingly, we found 
Fe~K emission undetected with \suzaku\ for only one source, DEM~L71. Its line 
luminosity is smaller than from any other LMC remnant. Likely, this fact and the 
smaller effective area of XIS explain why it was undetected in the 100~ks-long 
\suzaku\ observation of the remnant (Hiroya Yamaguchi 2014, personal 
communication). Furthermore, the second faintest Fe~K line from LMC SNRs is 
found in N49. With \xmm\ one does not formally detect the line. Including a 
Gaussian at the energy measured with \suzaku, the \xmm\ spectrum allows a line 
flux an order of magnitude above that actually detected. This is only a 
statistical issue. Indeed, there are less than 10~ks of EPIC-pn data available, 
which is no match to the 158~ks spent by \suzaku\ on N49 when detecting the Fe~K 
line.

The properties of the Fe~K emission from DEM~L71 fit well with its type Ia
nature. Furthermore, \citet[][their Figure~1, right]{2014ApJ...785L..27Y} used
simple (one-dimensional) theoretical models of type Ia SNe exploding in uniform
ambient media of various densities to predict the luminosity and energy of the
line. Even with this simplistic approach, they are able to reproduce all the
parameter space spanned by type Ia SNRs. In this context, the location of
DEM~L71 in the Fe~K luminosity~--~energy diagram is well reproduced by a
delayed-detonation model with a rather high explosion energy \citep[$1.4 \times
10^{51}$~erg, DDTa in][]{2003ApJ...593..358B,2005ApJ...624..198B}, in an ambient
medium of density $\rho = 2 \times 10^{-24}$~g~cm$^{-3}$, at age between 2000~yr
and 5000~yr. This is in line with the measured density and age of DEM~L71
\citep{2003A&A...406..141V,2003ApJ...590..833G}. Furthermore, the DDTa model
predicts a silicon-to-iron mass ratio of 0.08, close to that measured in X-rays
\citep[$\sim 0.15$,][]{2003ApJ...582L..95H,2003A&A...406..141V}. Since the hot,
K$\alpha$-emitting iron was previously overlooked, the $M_{{\rm Si}}/M_{{\rm
Fe}}$ ratio should be even lower, closer to the prediction of the DDTa model.

The dearth of Fe~K-emitting remnants, aside from the combined \xmm/\suzaku\
sample (eight objects), is somehow expected. Indeed, most of the SNRs have
plasma temperatures less than 1~keV (Sect.\,\ref{results_spectra_general}), 
which is too low to excite iron K-shell electrons, so that no emission is 
expected. \revision{MCSNR J0547$-$6973 is one case where a hotter spectral 
component ($kT \sim 2.2$~keV) is present but no Fe~K emission is detected, 
likely because the emission measure of that component is too low, so that it 
is not detectable with current instruments.}
Even if a spectrally unresolved hot iron component exists in more LMC 
remnants, a further issue is again detectability. The LMC SNRs of 
\citet{2014ApJ...785L..27Y} have hard X-ray (2~keV~--~8~keV) luminosities above 
$10^{35}$~erg~s$^{-1}$. There are only two other SNRs in the LMC above this 
level, MCSNR J0540$-$6920 and N157B, which are powered by a bright pulsar and 
pulsar wind nebula, respectively.

Despite these observational difficulties, it is very likely that the sample of
LMC Fe~K-emitting remnants \citep[of][plus DEM~L71]{2014ApJ...785L..27Y} is
complete, because all young SNRs ($\lesssim 5000$~yr old) are now known and
observed in X-rays. Translating the fraction of remnants with Fe~K emission in
the LMC (\mbox{$\approx 13$~\%}) to the Galactic population \citep[294
objects,][]{2014BASI...42...47G}, we expect more than 40 such sources in the
Milky Way. This number is a lower limit, since fainter line fluxes than in the
LMC can be reached. \citet{2014ApJ...785L..27Y} list 16 Galactic SNRs detected,
out of 56 objects observed with \suzaku\ \citep[][online
database\footnote{\url{http://www.physics.umanitoba.ca/snr/SNRcat/}}]{
2012AdSpR..49.1313F}. About 80 more SNRs were observed and detected with
\chandra\ or \xmm, and 150 have not been covered in X-rays. A systematic
analysis of all X-ray-detected SNRs and new/deeper observations of promising
candidates with more sensitive instruments (\eg \xmm\ vs. \chandra, future
missions such as Athena) will provide a better census of Fe~K lines in SNRs.
This will allow to type more remnants and to study the pre-SN evolution of their
progenitors.




%
%
%
%
%
%
%
%
%

\subsection{Detection of SN ejecta}
    \label{results_spectra_ejecta}
When SN ejecta give an observable contribution to the X-ray emission of an SNR,
the fitted abundances, or rather the fitted \emph{abundance ratios}, will
reflect the nucleosynthesis yields of either thermonuclear or CC SNe. To
identify SNRs with detected ejecta and the origin thereof, we computed abundance
ratios X/Fe, where X is O, Ne, Mg, or Si. The ratios are normalised with
respect to (X/Fe)$_{\mathrm{LMC}}$, the corresponding ratios with the LMC
abundances \citep[from][]{1992ApJ...384..508R}. As CC-SNRs produce large 
amounts of light-Z elements and little
iron, high (X/Fe)/(X/Fe)$_{\mathrm{LMC}}$ ratios (in excess of one) indicate a
massive star progenitor. On the contrary, the main product of thermonuclear SNe
is iron, and ejecta in type Ia SNRs (if detected), are expected to have
(X/Fe)/(X/Fe)$_{\mathrm{LMC}}$~$\ll 1$.

\begin{figure}[t]
    \begin{center}
\includegraphics[width=0.530\hsize]
{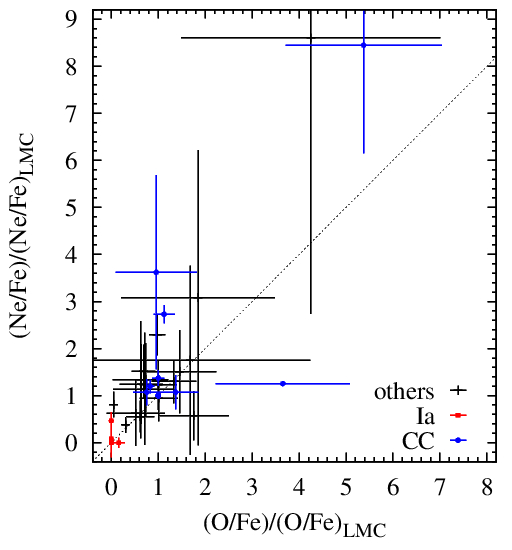}
\includegraphics[width=0.455\hsize]
{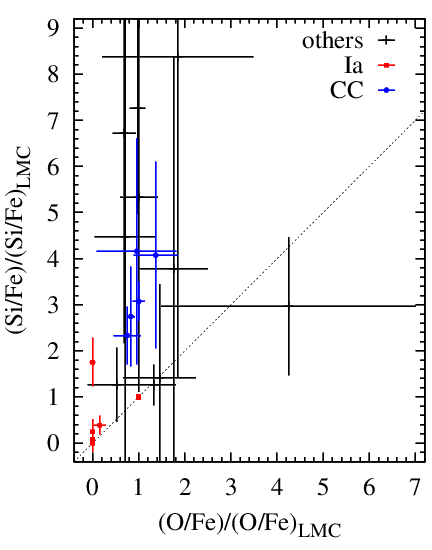}

\includegraphics[width=0.635\hsize]
{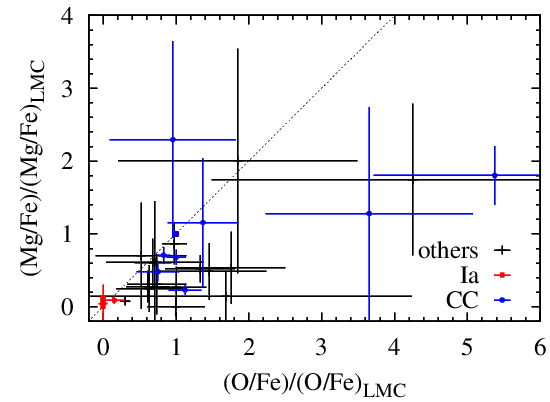}
    \end{center}
\caption[\ Abundance ratio diagrams of LMC SNRs with fitted
abundances]{Abundance ratio diagrams of LMC SNRs with fitted abundances. Sources
firmly classified as type Ia or CC-SNRs are plotted in red and blue,
respectively, \revision{and the rest of the sample in black}. See text 
(Sect.\,\ref{results_spectra_ejecta}) for details.
}
    \label{fig_spectra_scatter}
\end{figure}

In Fig.\,\ref{fig_spectra_scatter}, the abundance ratio diagrams of all SNRs
with corresponding fitted abundances are shown. The samples of SNRs with a 
secured CC or type Ia classification (as described in 
Appendix~\ref{appendix_secured}) are marked. Evidently, many of the known CC 
SNRs are located in regions of super-LMC X/Fe. The known type Ia SNRs are 
unsurprisingly in the (X/Fe)/(X/Fe)$_{\mathrm{LMC}}$~$\ll 1$ regions of the 
diagrams, because in most cases it is this very iron-enhancement that was used 
to classify them.

\begin{table}[t]
\caption{Constraints used for the identification of ejecta in SNR spectra.}
\begin{center}
\label{table_results_spectra_flags}
\vspace{-0.2cm}
\begin{tabular}{@{}c c c c c@{}}
\hline\hline
\noalign{\smallskip}
  X &
  \multicolumn{2}{c}{``high X/Fe'' flag} &
  \multicolumn{2}{c}{``low X/Fe'' flag} \\
 & (1) & (2) & (1) & (3) \\
\noalign{\smallskip}
\hline
\noalign{\smallskip}
O  & $> 1.0 $ & $> 0.83$ & $< 0.60$ & $< 0.83$ \\
Ne & $> 1.43$ & $> 1.30$ & $< 0.55$ & $< 1.30$ \\
Mg & $> 0.62$ & $> 0.48$ & $< 0.22$ & $< 0.48$ \\
Si & $> 2.70$ & $> 1.60$ & $< 1.30$ & $< 1.60$ \\
\noalign{\smallskip}
\hline
\end{tabular}
\end{center}
\vspace{-0.45cm}
\tablefoot{(1) Constraints on the ratio (X/Fe)/(X/Fe)$_{\mathrm{LMC}}$.
(2) Constraints on the lower limit (X/Fe)/(X/Fe)$_{\mathrm{LMC}} -
\Delta$(X/Fe), \revision{where $\Delta$(X/Fe) is the uncertainty of the 
abundance ratio}.
(3) Constraints on the upper limit (X/Fe)/(X/Fe)$_{\mathrm{LMC}} +
\Delta$(X/Fe).
}
\end{table}

\begin{table*}[t]
\caption[\ SNRs with detected ejecta]{1T/2T SNRs with detected ejecta (top
part), and used for measurements of ISM composition (bottom part)}
\vspace{-0.45cm}
\begin{center}
\label{table_results_spectra_flagged}
\begin{tabular}{l c c c c c c | c c c c}
\hline\hline
\noalign{\smallskip}
  \multicolumn{1}{c}{\multirow{2}{*}{MCSNR}} &
  \multicolumn{1}{c}{\multirow{2}{*}{Old name}} &
  \multicolumn{1}{c}{\multirow{2}{*}{SN type}} &
\multicolumn{4}{c}{High X/Fe flags} &
\multicolumn{4}{c}{Low X/Fe flags}  \\
& & &
  \multicolumn{1}{c}{O} &
  \multicolumn{1}{c}{Ne} &
  \multicolumn{1}{c}{Mg} &
  \multicolumn{1}{c}{Si} &
  \multicolumn{1}{c}{O} &
  \multicolumn{1}{c}{Ne} &
  \multicolumn{1}{c}{Mg} &
  \multicolumn{1}{c}{Si} \\
\noalign{\smallskip}
\hline
\noalign{\smallskip}
J0453-6829 & B0453-685 & CC & --- & --- & Y & Y & --- & --- & --- & ---\\
J0506-6541 & --- &   & --- & Y & --- & --- & --- & --- & --- & ---\\
J0506-7026 & [HP99] 1139 &   & --- & --- & --- & --- & Y & Y & Y & ---\\
J0508-6830 & --- & Ia & --- & --- & --- & --- & Y & --- & --- &---\\
J0508-6902 & [HP99] 791 & Ia & --- & --- & --- & --- & Y & --- & --- &---\\
J0511-6759 & --- & Ia & --- & --- & --- & --- & Y & --- & --- & ---\\
J0519-6926 & B0520-694 &   & --- & Y & Y & Y & --- & --- & --- & ---\\
J0523-6753 & N44 &   & Y & Y & Y & --- & --- & --- & --- & ---\\
J0525-6559 & N49B & CC & --- & Y & Y & Y & --- & --- & --- & ---\\
J0526-6605 & N49 & CC & Y & --- & Y & Y & --- & --- & --- & ---\\
J0529-6653 & DEM L214 &   & Y & --- & --- & --- & --- & --- & --- & ---\\
J0531-7100 & N206 &   & Y & --- & --- & Y & --- & --- & --- & ---\\
J0533-7202 & 1RXSJ053353.6-7204 & & --- & --- & --- & --- & --- & --- & Y &---\\
J0534$-$6955 & B0534$-$699 & Ia & --- & --- & --- & --- & Y & Y & Y & ---\\
J0534-7033 & DEM L238 & Ia & --- & --- & --- & --- & Y & Y & Y & Y\\
J0535-6602 & N63A & CC & Y & Y & --- & --- & --- & --- & --- & --- \\
J0535-6918 & Honeycomb &   & --- & --- & --- & Y & --- & --- & --- & ---\\
J0536-6735 & DEM L241 & CC & Y & Y & Y & --- & --- & --- & --- & ---\\
J0536-6913 & B0536-6914 & CC & Y & --- & --- & --- & --- & --- & --- & ---\\
J0536-7039 & DEM L249 & Ia & --- & --- & --- & --- & Y & Y & Y & Y\\
J0537-6628 & DEM L256 &   & --- & --- & --- & --- & --- & --- & Y & ---\\
J0547-6941 & DEM L316A & Ia & --- & --- & --- & --- & Y & Y & Y & Y\\
J0547-7025 & B0548-704 & Ia & --- & --- & --- & --- & Y & Y & Y & ---\\
\hline\hline
\noalign{\smallskip}
& & & \multicolumn{8}{c}{ISM abundance} \\
& & & \multicolumn{2}{c}{O \& Fe} &
  \multicolumn{2}{c}{Ne} &
  \multicolumn{2}{c}{Mg} &
  \multicolumn{2}{c}{Si} \\
\noalign{\smallskip}
\hline
\noalign{\smallskip}
  J0450-7050 & B0450-709 &   & \multicolumn{2}{c}{Y} & \multicolumn{2}{c}{Y} &
\multicolumn{2}{c}{Y} & \multicolumn{2}{c}{---}\\
  J0453-6655 & N4 &   & \multicolumn{2}{c}{Y} & \multicolumn{2}{c}{Y} &
\multicolumn{2}{c}{Y} & \multicolumn{2}{c}{---}\\
  J0453-6829 & B0453-685 & CC & \multicolumn{2}{c}{Y} & \multicolumn{2}{c}{Y} &
\multicolumn{2}{c}{---} & \multicolumn{2}{c}{---}\\
  J0454-6626 & N11L &   & \multicolumn{2}{c}{Y} & \multicolumn{2}{c}{Y} &
\multicolumn{2}{c}{Y} & \multicolumn{2}{c}{---}\\
  J0505-6802 & N23 & CC & \multicolumn{2}{c}{Y} & \multicolumn{2}{c}{Y} &
\multicolumn{2}{c}{Y} & \multicolumn{2}{c}{Y}\\
  J0514-6840 & --- &   & \multicolumn{2}{c}{Y} & \multicolumn{2}{c}{---} &
\multicolumn{2}{c}{---} & \multicolumn{2}{c}{---}\\
  J0518-6939 & N120 &   & \multicolumn{2}{c}{Y} & \multicolumn{2}{c}{Y} &
\multicolumn{2}{c}{Y} & \multicolumn{2}{c}{Y}\\
  J0519-6926 & B0520-694 &   & \multicolumn{2}{c}{Y} & \multicolumn{2}{c}{---} &
\multicolumn{2}{c}{---} & \multicolumn{2}{c}{---}\\
  J0527-6912 & B0528-692 &   & \multicolumn{2}{c}{Y} & \multicolumn{2}{c}{Y} &
\multicolumn{2}{c}{Y} & \multicolumn{2}{c}{Y}\\
  J0528-6727 & DEM L205 &   & \multicolumn{2}{c}{Y} & \multicolumn{2}{c}{---} &
\multicolumn{2}{c}{---} & \multicolumn{2}{c}{---}\\
  J0531-7100 & N206 & CC & \multicolumn{2}{c}{---} & \multicolumn{2}{c}{Y} &
\multicolumn{2}{c}{Y} & \multicolumn{2}{c}{---}\\
  J0532-6732 & B0532-675 &   & \multicolumn{2}{c}{Y} & \multicolumn{2}{c}{Y} &
\multicolumn{2}{c}{Y} & \multicolumn{2}{c}{Y}\\
  J0533-7202 & 1RXSJ053353.6-7204 &   & \multicolumn{2}{c}{Y} &
\multicolumn{2}{c}{Y} & \multicolumn{2}{c}{---} & \multicolumn{2}{c}{---}\\
  J0535-6918 & Honeycomb &   & \multicolumn{2}{c}{Y} & \multicolumn{2}{c}{Y} &
\multicolumn{2}{c}{Y} & \multicolumn{2}{c}{---}\\
  J0543-6858 & DEM L299 &   & \multicolumn{2}{c}{Y} & \multicolumn{2}{c}{Y} &
\multicolumn{2}{c}{Y} & \multicolumn{2}{c}{Y}\\
  J0547-6943 & DEM L316B & & \multicolumn{2}{c}{Y} & \multicolumn{2}{c}{Y} &
\multicolumn{2}{c}{Y} & \multicolumn{2}{c}{Y}\\
\noalign{\smallskip}
\hline
\end{tabular}
\end{center}
\vspace{-0.5cm}
\tablefoot{The classification given (type Ia or core-collapse) is described in
Appendix~\ref{appendix_secured}.
}
\end{table*}

Several sources without previous classification are located in the high- and 
low-ratio regions of the diagrams. For typing purpose, we assign ``high X/Fe'' 
and ``low X/Fe'' flags to these objects, using the following scheme: For each 
element X, we plot the cumulative distribution of the ratio 
(X/Fe)/(X/Fe)$_{\mathrm{LMC}}$. We then assign a ``high X/Fe'' flag to an object 
if its ratio is above the 68$^{\mathrm{th}}$ percentile ($\sim 1\sigma$) of the 
cumulative distribution. Symmetrically, a ``low X/Fe'' flag is given if the 
ratio is below the 32$^{\mathrm{th}}$ percentile. Since the uncertainties in the 
fitted abundances can be large, it is necessary to put a second constraint using 
the uncertainty of the ratio, \revision{noted $\Delta$(X/Fe)}: A ``high X/Fe'' 
flag is only given if the lower limit (\ie ratio minus the uncertainty) is above 
the median of the cumulative distribution. For a ``low X/Fe'' flag the upper 
limit must be below the median. This excludes all cases where the ratios are 
elevated (or much smaller than one) but highly uncertain. Though some of the 
criteria for ``low X/Fe'' flags may seem high, the selected SNRs have actual 
ratios well below half the LMC average (well below 0.2 times the average for 
Mg). The criteria for ``high X/Fe'' and ``low X/Fe'' flag are given in 
Table~\ref{table_results_spectra_flags}. There are 23 SNRs in the 1T/2T sample 
with high or low abundance ratio flags, as listed in 
Table~\ref{table_results_spectra_flagged}. These flags are used in 
Sect.\,\ref{results_sfh} to help the typing of all LMC SNRs.

\subsection{Metal abundances of the LMC ISM}
    \label{results_spectra_abundance}
When no SN ejecta is detected, the X-ray emission is dominated by the ISM
swept-up by the SN blast wave. Therefore, the fitted abundances in these cases
provide us with measurements of the chemical composition of the gas phase of
the LMC ISM. \citet{1992ApJ...384..508R} and \citet{1998ApJ...505..732H} have
used samples of SNRs to obtain the abundance of some elements (using optical
and X-ray observations, respectively), but the smaller sample of known SNRs and
sensitivity of the X-ray instrument used (\ASCA) at the time limited the number
of SNRs eligible to measure LMC abundances.

\begin{figure*}[t]
    \begin{center}
    \includegraphics[width=0.995\hsize]
    {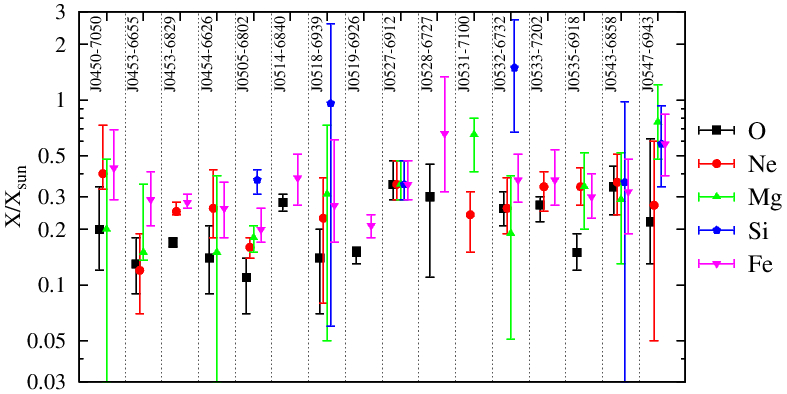}
    \end{center}
\caption[\ LMC ISM abundances (relative to solar values) measured in a sample of
16 X-ray SNRs]{LMC ISM abundances, relative to the solar values of
\citet{2000ApJ...542..914W}, measured in a sample of 16 X-ray SNRs. The
selection of the sample and the measurements of abundance are described in
Sect.\,\ref{results_spectra_abundance}.}
    \label{fig_spectra_ISMabund}
\end{figure*}

We first selected all 1T/2T SNRs with fitted abundances but no high or low
abundance ratios. To increase that sample, we included SNRs where some 
abundances are enhanced but others can still be used. For example in MCSNR 
J0453$-$6829, the spectrum is enhanced in Mg and Si, but the fitted values for 
O, Ne, and Fe, are still (assumed to be) reflecting the LMC ISM abundance. 
Furthermore, if the abundance of a given element is too uncertain, then the 
SNR is not used to measure the average abundance of that element. This limits 
in particular the size of the SNR sample allowing the abundance of silicon to 
be measured.

In Table~\ref{table_results_spectra_flagged} we give the list of SNRs used to 
measure the abundance of O, Ne, Mg, Si, and Fe, or a subset of these elements. 
The measured abundances for this sample are plotted relative to solar values in 
Fig.\,\ref{fig_spectra_ISMabund}. The final LMC abundances are obtained by 
taking the average values from all SNRs where an element is used; the errors 
given are the RMS scatter amongst the SNRs used. This method is similar to that 
of \citet{1998ApJ...505..732H}. Resulting abundances range from $\sim 0.2$ solar 
for oxygen to $\sim 0.7$ solar for silicon. The results are listed in 
Table~\ref{table_results_spectra_abundance}. The absolute abundances, in the 
form 12 + log(X/H) (by number), are given, in comparison with results from 
\citet{1992ApJ...384..508R} and \citet{1998ApJ...505..732H}. Abundances of Fe 
and Si measured with \xmm\ are in good agreement with the results measured for a 
different sample of SNRs by 
\citet{1998ApJ...505..732H}\,\footnote{\revision{They used six CC SNRs 
(J0453-6829, N23, N49, N49B, N63A, and N132D) and one type Ia SNR (DEM~L71).}}. 
More recent studies of abundances in the LMC, using large samples of field 
stars \citep{2005AJ....129.1465C,2008A&A...480..379P,2012ApJ...761...33L, 
2013A&A...560A..44V}, can be used to evaluate our results. The metallicity 
distributions {[Fe/H]}\footnote{using the conventional notation: {[X/Y]}$ = 
\log\left(\mathrm{X/Y}\right) - \log \left(\mathrm{X/Y}\right)_{\sun}$.} peak at 
about $-$0.5~dex for most field star samples \citep{2012ApJ...761...33L}, 
\revision{corresponding to 12 + log(Fe/H) $= 7$. This matches very well our 
value based on \xmm\ SNRs (6.97$_{-0.18} ^{+0.13}$)}, indicating no 
metallicity difference between field stars and gas-phase ISM.

\begin{table}[t]
\caption{LMC abundances}
\begin{center}
\label{table_results_spectra_abundance}
\small
\begin{tabular}{@{\hspace{0.05cm}} c @{\hspace{0.1cm}} @{\hspace{0.15cm}} c 
@{\hspace{0.15cm}} @{\hspace{0.15cm}} c @{\hspace{0.15cm}} @{\hspace{0.15cm}} c 
@{\hspace{0.15cm}} @{\hspace{0.10cm}} c @{\hspace{0.10cm}} @{\hspace{0.15cm}} c 
@{\hspace{0.15cm}} @{\hspace{0.15cm}} c @{\hspace{0.05cm}}}
\hline\hline
\noalign{\smallskip}
  Element & X/X$_{\sun}$ & N & RMS & 12 + log(X/H) & Hughes et &  RD92 \\
  & (1) & (2) & (3) & & al. (1998) & \\
\noalign{\smallskip}
\hline
\noalign{\smallskip}
O  & 0.21 & 15 & 0.08 & 8.01$_{-0.21} ^{+0.14}$ & 8.21$\pm 0.07$ & 8.35$\pm0.06$
\\ \noalign{\smallskip}
Ne & 0.28 & 13 & 0.08 & 7.39$_{-0.15} ^{+0.11}$ & 7.55$\pm 0.08$ & 7.61$\pm0.05$
\\ \noalign{\smallskip}
Mg & 0.33 & 11 & 0.19 & 6.92$_{-0.37} ^{+0.20}$ & 7.08$\pm 0.07$ & 7.47$\pm0.13$
\\ \noalign{\smallskip}
Si & 0.69 & 6  & 0.42 & 7.11$_{-0.41} ^{+0.20}$ & 7.04$\pm 0.08$ &
7.81\tablefootmark{a}\\ \noalign{\smallskip}
Fe & 0.35 & 15 & 0.12 & 6.97$_{-0.18} ^{+0.13}$ & 7.01$\pm 0.11$ & 7.23$\pm0.14$
\\ \noalign{\smallskip}
\hline
\end{tabular}
\end{center}
\tablefoot{
(1) Abundance relative to the solar value of \citet{2000ApJ...542..914W}. (2)
Number of SNRs used to measure X/X$_{\sun}$. (3) RMS scatter amongst the N SNRs.
\tablefoottext{a}{Silicon abundance was quoted as highly uncertain in
\citet{1992ApJ...384..508R}}
}
\end{table}

However, the abundances of light $\alpha$-elements tend to be lower (by
$\sim$0.15~dex~--~0.2~dex) compared to \citet{1998ApJ...505..732H}, although the
results for Mg and Ne might still be reconciled given the larger uncertainties. 
Still, we measured a ratio {[}O/Fe{]} of $-$0.21 while \ASCA\ SNRs gave $-$0.06. 
The likely explanation is two-fold. First, the $\alpha$-elements abundance has 
an intrinsic scatter \citep[about 0.05~dex~--~0.08~dex at the relevant 
metallicity,][]{2013A&A...560A..44V} that can partly explain the discrepancy. 
The second reason is the sample used by \citet{1998ApJ...505..732H}: 
\revision{N132D, N49, and N49B contain regions that have been (since then) found 
to be enhanced in low-Z elements (\eg this work, and references in 
Appendix~\ref{appendix_secured}), while the bright central regions of the type 
Ia SNR DEM L71 are enriched in iron. These contributions from ejecta to the 
integrated spectra measured with \ASCA\ affect the measured LMC abundances. Six 
out of the seven SNRs used by \citet{1998ApJ...505..732H} are well-established 
CC-SNRs, and this bias is likely to explain their higher {[}O/Fe{]} (or more 
generally {[}$\alpha$/Fe{]}).} On the contrary, the \xmm\ sample used here is 
explicitly cleaned of SNRs with abnormal abundance patterns (\ie those with 
ejecta detected), resulting in a purer sample better suited to the measurement 
of the ISM composition. However, this sample comprises SNRs fainter than used in 
previous studies, and the abundances thus obtained are consequently
relatively uncertain.

\begin{figure}[t]
    \begin{center}
    \includegraphics[width=0.95\hsize]
    {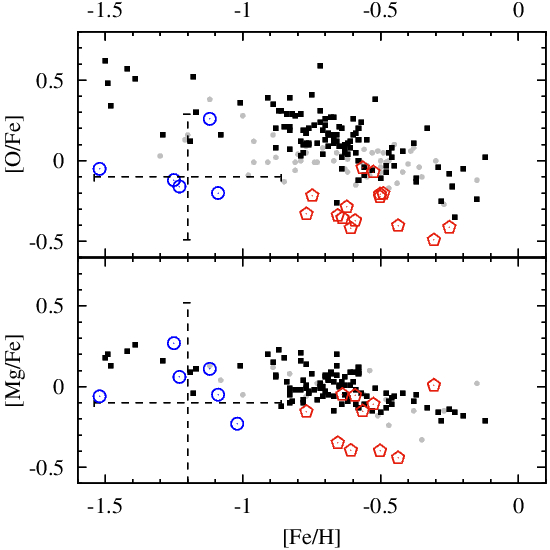}
    \end{center}
\vspace{-0.3cm}
\caption[\ {[}O/Fe{]} and {[}Mg/Fe{]} vs. {[}Fe/H{]} diagrams for various LMC
populations]{[O/Fe] and [Mg/Fe] vs. [Fe/H] diagrams for various LMC
populations: Abundances measured in SNRs (ISM gas phase, this work) are shown
with red pentagons. The crosses indicate median error bars. Blue open circles
are the old globular clusters from \citet[][ages~$\sim
10$~Gyr]{2006ApJ...640..801J}. Chemical abundances of Bar and disc stars are
marked by black squares and grey dots, respectively \citep[from][ages~$\gtrsim
1$~Gyr]{2013A&A...560A..44V}.
}
    \label{fig_spectra_ISMabund_vsFe}
\end{figure}

The abundance pattern of metals should reflect the past history of chemical 
enrichment, and in particular the relative number of CC and Ia SNRs (hereafter 
\ccIa), because their metal yields are markedly different. In 
Fig.\,\ref{fig_spectra_ISMabund_vsFe} we show the [O/Fe] and [Mg/Fe] vs. [Fe/H] 
diagrams. Abundances measured with SNRs (\ie that of the ISM gas phase) are 
compared with that measured in older populations: old globular clusters from 
\citet[][ages~$\sim 10$~Gyr]{2006ApJ...640..801J} and Bar and disc field red 
giant stars from \citet[][ages $\gtrsim 1$~Gyr]{2013A&A...560A..44V}. 
\revision{SNRs are only found in the higher metallicity ([Fe/H]) range. There is 
also a clear trend for SNRs to be at lower {[}$\alpha$/Fe{]} ratios, although 
uncertainties from X-ray spectral fitting are large (particularly for [Mg/Fe]). 
A larger sample and more data would be desirable to demonstrate this result 
definitely. Nevertheless, this trend should} reflect the continued enrichment 
by type Ia SNe in the last $\sim 1$~Gyr, which inject large amounts of Fe back 
in the ISM and drive younger populations towards the bottom right corner of the 
[$\alpha$/Fe] -- [Fe/H] diagrams.

There are SNR-to-SNR variations in the abundances, but the metallicity scatter 
in the ISM gas phase is less than for field stars. In particular there is no 
metal-poor population \revision{([Fe/H]~$\lesssim -\, 0.9$)}. We also checked 
that there is no clear correlation between the location of an SNR in the 
[$\alpha$/Fe] -- [Fe/H] diagrams and the SFH around the SNR. For instance, SNRs 
with relatively high [$\alpha$/Fe] are not necessarily in regions with increased 
recent SF, which would produce massive stars that release low-Z elements. 
Despite the uncertainties and the limited size of the sample, this lack of 
correlation likely indicates that SNe-produced elements are well mixed in the 
ISM. In other words, the ISM is quickly homogenised, at least at the spatial 
scales over which SFH is measured ($\sim 200$~pc).

After LMC abundances were measured \citep[\eg][]{1992ApJ...384..508R},
\citet{1995MNRAS.277..945T} found with chemical evolution models that the
deficit of light $\alpha$-elements of the MCs (\ie lower {[$\alpha$/Fe]} for a
given {[}Fe/H{]}) compared to the Galaxy must be explained by a \emph{smaller}
\ccIa\ (more type Ia SNe). They measured a Galactic ratio of 6.7, but
\ccIa$\sim4-5$ and $\sim 3.3$ for the LMC and SMC, respectively. Our results for
the LMC ISM abundance suggest an even lower ratio \ccIa, because the deficit of
light $\alpha$-elements is wider than previously assumed by
\citet{1995MNRAS.277..945T}. By tentatively typing all LMC remnants, we show in
Sect.\,\ref{results_sfh} that indeed \ccIa\ is particularly low, compared to 
previous measurements of the ratio in the LMC or inferred from galaxy cluster 
X-ray observations, and discuss likely explanations.

\section{Measuring the ratio of CC to type Ia SNe in the LMC using 
``SFH-typing''}
\label{results_sfh}

\subsection{The typing of SNRs in general}
\label{results_sfh_general}
Because the two flavours of SNe deposit a similar amount of energy in the ISM, 
they produce remnants which become increasingly hard to type as they age. The 
most secured typing methods are the study of SN \emph{optical} light echoes
(\citealt{2005Natur.438.1132R,2008ApJ...680.1137R}; \emph{infrared} light
echoes can be used to probe the ISM dust, see e.g. 
\citealt{2012ApJ...750..155V}), the measurement of the nucleosynthesis products 
in the ejecta \citep[e.g. ][]{1995ApJ...444L..81H}, or the association with a 
neutron star/pulsar wind nebula. Optical spectroscopy can also be used: In some 
cases the fast-moving ejecta are detected in optical lines with highly elevated
abundances of oxygen. Those so-called \emph{oxygen-rich SNRs} have massive star 
progenitors, see for instance \citet{1978ApJ...223..109L}, 
\citet{1979ApJ...233..154C}, \citet{1995AJ....109.2104M}, and references
therein. On the contrary, some SNRs have prominent Balmer lines of hydrogen,
but absent or weak [\ion{S}{ii}] and [\ion{O}{iii}] lines. These 
Balmer-dominated optical spectra are interpreted as non-radiative shocks
overtaking (partially) neutral gas 
\citep{1978ApJ...225L..27C,1980ApJ...235..186C}. A type Ia SN progenitor is
consistent with the presence of neutral gas, as massive stars would ionise their 
surrounding. A sample of optically bright Balmer-dominated SNRs was detected in 
the LMC by \citet{1982ApJ...261..473T}.

These methods work best for relatively young remnants ($\lesssim 5000$ yr), 
leaving a significant fraction of the SNR population untyped. However, several 
\emph{evolved} SNRs have been discovered (in X-rays) in the Magellanic Clouds 
with an iron-rich, centrally bright emission 
\citep{2001PASJ...53...99N,2003ApJ...593..370H,2004A&A...421.1031V,
2006ApJ...640..327S,2006ApJ...652.1259B,2014MNRAS.439.1110B, 
2014A&A...561A..76M}, naturally leading to their classification as type Ia 
remnants.

In addition, studies of the X-ray and infrared morphologies of SNRs 
\citep{2009ApJ...706L.106L,2013ApJ...771L..38P} suggest that, as a class, type 
Ia and CC SNRs have distinct symmetries: type Ia remnants are statistically 
more spherical and mirror symmetric than the CC SNRs. However, this method 
cannot give definite results for \emph{individual} objects: Prominent 
counterexamples include SNR 1E 0102.2$-$7219, a textbook CC SNR which is highly 
symmetric \citep{2004ApJ...605..230F}, and MCSNR J0547$-$7025, which is a type 
Ia SNR \citep[based on its X-ray spectrum][this work]{2003ApJ...593..370H} with 
``anomalous'' ejecta distribution \citep{2009ApJ...706L.106L}.

For a decent fraction of the LMC remnants, these various methods give a secured 
classification. This ``secured-type'' sample is presented in 
Appendix~\ref{appendix_secured} and listed in 
Table~\ref{appendix_table_securedSNRs}. To tentatively type the rest of the 
sample, we devise \revision{a new way to quantify the local stellar environment 
of LMC, as described in Sect.\,\ref{results_sfh_environment}}. 
This method is calibrated with the ``secured-type'' sample and applied in
Sect.\,\ref{results_sfh_typing}. We can then discuss the measured ratio
of CC to type Ia SNRs in the LMC and its implications in 
Sect.\,\ref{results_sfh_ratio}.

\subsection{Evaluating the local stellar environment}
    \label{results_sfh_environment}
We devised two metrics to assess the local stellar environment of LMC SNRs. Both
ultimately stem from the same set of data \citep[the MCPS catalogue
of][]{2004AJ....128.1606Z}. Although connected, they still measure two distinct
properties and are therefore complementary, as we discuss below. The two 
metrics are given for each SNR in Table~\ref{appendix_table_snrs_sample}.

\begin{figure}[t]
    \includegraphics[width=0.495\hsize]
{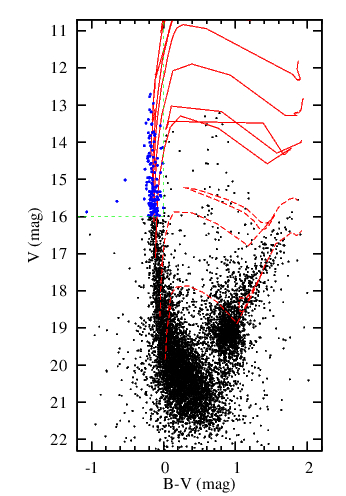}
    \includegraphics[width=0.495\hsize]
{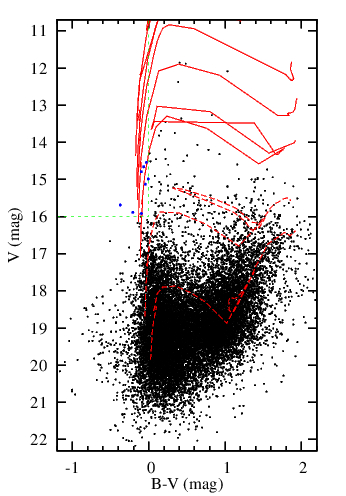}
\caption{Colour-magnitude diagram (CMD) of the MCPS stars
\citep{2004AJ....128.1606Z} within 100~pc ($\sim$6.9\arcmin) of the central
position of two remnants, MCSNR J0528$-$6727 (left) and MCSNR J0534$-$6955 
(right). Geneva stellar evolution tracks \citep{2001A&A...366..538L} are shown 
as red lines, for metallicity of 0.4~Z$_{\sun}$ and initial masses of 3, 
5~M$_{\sun}$ (dashed lines) and 10, 15, 20, 25, and 40~M$_{\sun}$ (solid lines), 
from bottom to top. The green dashed line shows the criteria used to identify 
the OB stars ($V < 16$ and $B-V< 0$). Stars satisfying these criteria are shown 
as blue dots.}
    \label{fig_Nob_cmd}
\end{figure}

\begin{figure}[t]
    \begin{center}
    \includegraphics[width=0.995\hsize]
    {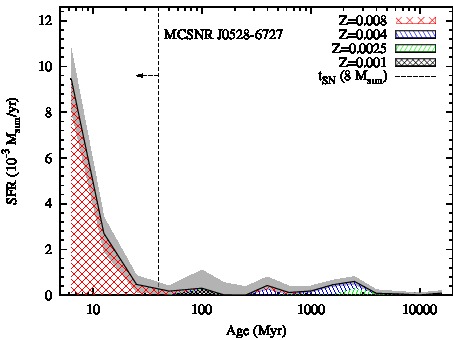}

    \includegraphics[width=0.995\hsize]
    {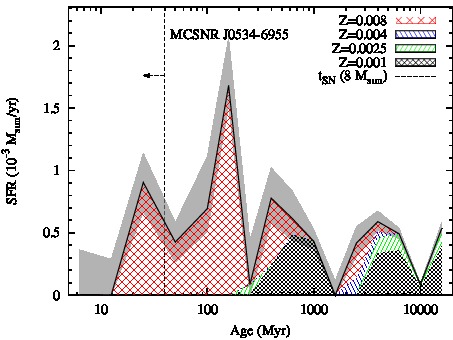}
    \end{center}
\caption[\ Star formation history around MCSNR J0528$-$6727 and
J0534$-$6955]{Star formation history around MCSNR J0528$-$6727 (top) and
J0534$-$6955 (bottom). Data are taken from \citet{2009AJ....138.1243H}. The star
formation rate in four metallicity bins are plotted against lookback time. The
errors (combining all metallicities) are shown by the grey shading. The vertical
dashed line at 40 Myr indicates the maximal lifetime of a CC SN progenitor. Note
the changing vertical scale.}
    \label{fig_sfh_examples}
\end{figure}

\textbullet\ \textbf{\boldmath$N_{{\rm OB}}$, the number of blue early-type 
stars in the immediate vicinity of the remnant:}\\
To obtain this number, we constructed a $V$ vs. ($B-V$) colour-magnitude diagram 
(CMD) of all stars whose projected position lies within 100~pc 
($\sim$6.9\arcmin) of each SNR. This value corresponds to the drift distance for 
a star of age $10^7$~yr at a velocity of 10~km~s$^{-1}$ and was used by 
\citet{1988AJ.....96.1874C}. The upper main-sequence of stars in the LMC was 
identified by adding the stellar evolutionary tracks of 
\citet{2001A&A...366..538L}, for Z $=$ 0.4~Z$_{\sun}$ and initial masses from 
3~M$_{\sun}$ to 40~M$_{\sun}$. We assumed a distance modulus of 18.49 and an 
extinction $A_V = 0.5$ \citep[the average extinction for ``hot'' 
stars,][]{2004AJ....128.1606Z}. From there, we used the criteria of $V < 16$ and 
$B - V < 0$ to identify OB stars. In Fig.\,\ref{fig_Nob_cmd} we show two example 
CMDs of the regions around MCSNR J0528$-$6727 (left) and MCSNR J0534$-$6955 
(right). In the former case, a prominent upper-main sequence is obvious and the 
number of OB stars \NOB $= 142$. By contrast, the region around 
MCSNR~J0534$-$6955 is devoid of young massive stars. For this remnant, \NOB is 
only 8. \revision{Note that \NOB is likely a lower limit on the actual number 
of massive stars due to stellar crowding (the typical seeing of the MCPS is 
1.5\arcsec). This issue affects chiefly regions with higher \NOB.}

\medskip

\begin{figure*}[t]
    \begin{center}
    \includegraphics[width=0.33\hsize]
    {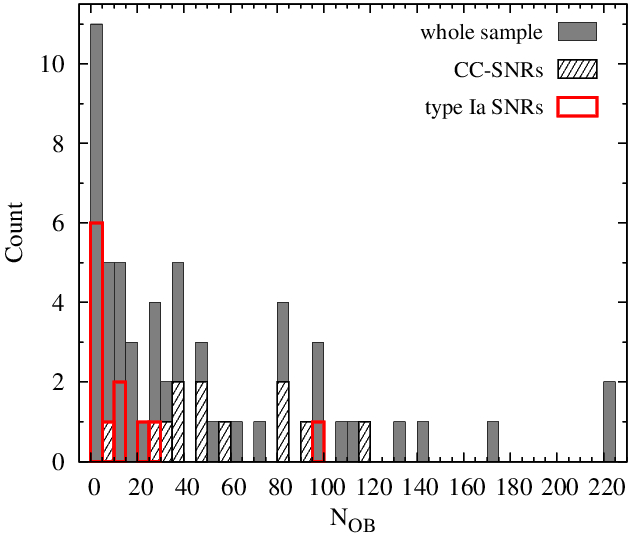}
    \includegraphics[width=0.33\hsize]
    {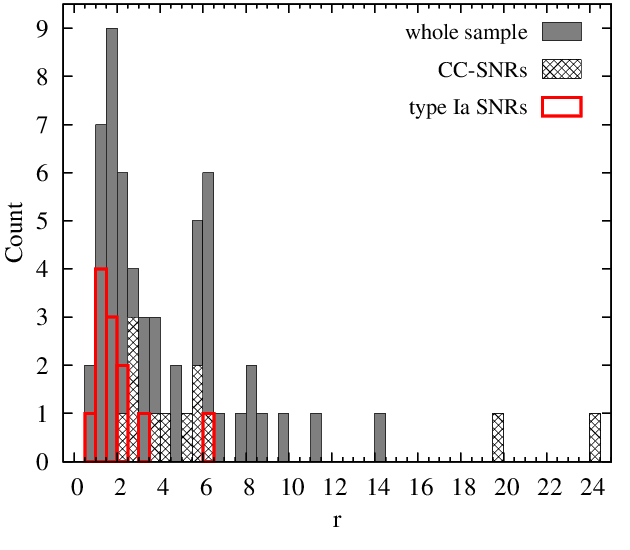}
    \includegraphics[bb=0 0 455 400, clip,width=0.330\hsize]
    {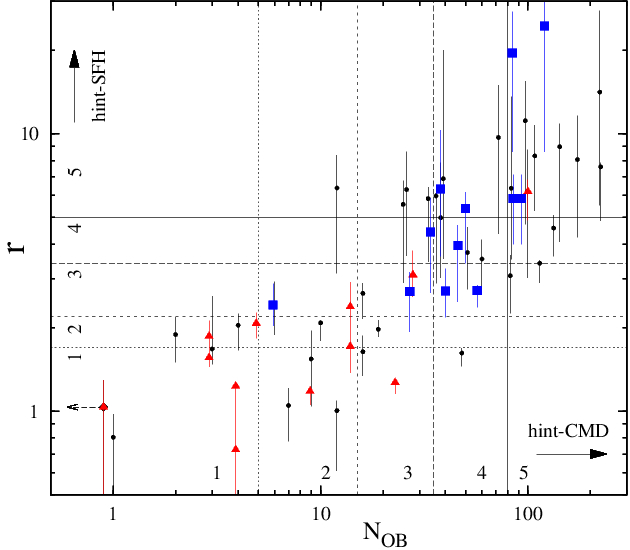}
    \end{center}
\caption{\emph{Left and middle panel:} Count distribution of LMC SNRs as 
function of \NOB and $r$. The distribution for the SNRs with a secured CC 
classification is shown with the hatched boxes; that for type Ia SNRs is 
outlined in red. The whole sample is shown by the solid grey histograms.
\emph{Right panel:} $r$--\NOB diagram of LMC SNRs. Secured Ia and CC SNRs are 
marked by red triangles and blue squares, respectively; the rest of the sample 
is shown with black dots. The arrow in the lower left corner indicates an SNR 
with \NOB $=$ 0. The regions corresponding to different ``Hint-SFH'' and 
``Hint-CMD'' are marked by the gridding.}
    \label{fig_typing_calibration}
\end{figure*}

\textbullet\ \textbf{\boldmath$r=N_{\rm CC}/N_{\rm Ia}$,
the ratio of CC SNe to thermonuclear SNe expected from the observed distribution
of stellar ages in the neighbourhood of the remnants:}\\
This number is obtained via the spatially resolved SFH map of \citet[][see
Sect.\,\ref{observations_supplementary_SFH}]{2009AJ....138.1243H}. For each SNR 
we plot the SFR of the cell including the remnant as a function of lookback 
time and metallicity. Two example SFHs are shown in Fig.\,\ref{fig_sfh_examples} 
for the same SNRs of Fig.\,\ref{fig_Nob_cmd}. They are strikingly different: The 
SFR around J0528$-$6727 soared in the last 20~Myr, when the numerous early-type 
stars in the vicinity of the remnant were formed, while the star formation 
around J0534$-$6955 peaked (at a lower absolute rate) about 125~Myr ago and was 
shut down in the most recent 20~Myr.

Because stars might drift away from their birth place, one potentially important
caveat is that the SFH of a cell hosting an SNR may be derived from stars having
no physical connection with the SNR progenitor. For a detailed discussion on the
relevance of local stellar populations to the study of progenitors, we point to
\citet{2009ApJ...700..727B}. However, we stress that most of the information
that can be gained from the study of the local SFHs, in the context of typing
remnants, is contained in the most recent time bins. Namely, the presence of
recent star formation episode is a strong necessary (but not sufficient)
condition to tentatively type a remnant as having a CC origin. Conversely, the
lack of recent star-forming activity favours a thermonuclear origin.

To approach this question in a quantitative way, we did the following: We used
the delay time distribution (DTD) $\Psi _{i} (\tau)$, the SN rate at time $\tau$
following a star formation event, measured by \citet{2010MNRAS.407.1314M} in the
Magellanic Clouds, with $i=1$, 2, and 3 designating the time intervals they used
($t <$ 35~Myr, 35~Myr $< t <$ 330~Myr, and 330~Myr $< t < 14$~Gyr,
respectively). From timescale arguments it is reasonably assumed that $\Psi
_1$ will correspond to the CC-SN rate, whilst $\Psi _2$ and $\Psi _3$ will be
that of SNe Ia (regardless of their ``prompt'' or ``delayed'' nature). The SFR
is integrated to obtain $M_{i}$, the stellar mass formed in each time interval.
The SFH of \citet{2009AJ....138.1243H} is only given at $t = 25$~Myr and $t =
50$~Myr. To obtain $M_1$, the mass formed at $t < 35$~Myr, we approximate $M (25
< t < 35)$ as half that formed between 25~Myr and 50~Myr (the second half is
included in $M_2$). Likewise, we split the mass formed between $t = 250$~Myr and
$t = 400$~Myr in two and include a half in both $M_2$ and $M_3$.

Then, we compute $r=N_{\rm CC}/N_{\rm Ia}$ as the ratio of the \emph{rates} of 
CC and Ia SNe, since the visibility times are the same for both types, \ie:
\begin{equation}
\label{eq_r}
    r = \frac{\Psi_1 M_1}{\Psi_2 M_2 + \Psi_3 M_3}
\end{equation}

Over the visibility time of a remnant --- taking 100 kyr as a very conservative
limit --- the stars in the SFH cell including the remnant will not drift away.
In other words, the distribution of stellar ages observed \emph{now} is the same
as that when the SN exploded. $r$ is therefore a measure of the relative size
of the pool of possible progenitors of both types. Using the same example SNRs
as in Fig.\,\ref{fig_sfh_examples}, a value of $r = 9.0_{-4.9}^{+1.9}$ is 
obtained for J0528$-$6727\,\footnote{
The uncertainty given for $r$ solely includes that of the mass formed $M_i$,
which is computed from uncertainties of the SFR given in
\citet{2009AJ....138.1243H}. The uncertainties on $\Psi_2$ and $\Psi_3$ are
larger, but are the same for all SNRs in the sample, allowing to use $r$ in a
comparative fashion. we adopted $\Psi_2 = 0.26$ SNe yr$^{-1}$ 
($10^{10}$\msun)$^{-1}$ and $\Psi_3 < 0.0014$ SNe  yr$^{-1}$ 
($10^{10}$\msun)$^{-1}$. Note that because $\Psi_3$ is an upper limit, $r$ is
formally a lower limit.
}
while for J0534$-$6955 it is only $r = 1.2\pm0.1$.

\subsection{``SFH-typing'' all LMC SNRs}
\label{results_sfh_typing}
We now proceed to give a tentative type to the whole sample of SNRs in the
LMC, using \NOB and $r$. We assign two numbers called ``Hint--CMD'' and 
``Hint--SFH'', depending on the \NOB and $r$-value obtained for each SNR,
respectively. The numbers range from 1 meaning ``strongly favours a type Ia SN
origin'', to 5 meaning ``strongly favours a CC-SN origin''. We used the 
distribution of \NOB and $r$ for the sample of ``secured-type'' SNR to establish
the correspondence between their values and the hints.

\begin{table}[t]
\begin{center}
\caption{\normalsize{Criteria and ``Hints'' attributed to SNRs as function of 
\NOB and $r$.}}
\label{table_results_sfh_hints}
\small
\begin{tabular}{@{}c @{\hspace{0.0cm}} c @{\hspace{1em}} c 
@{\hspace{1em}} c @{}}
\hline\hline
\noalign{\smallskip}
  \multicolumn{1}{c}{Value} &
  \multicolumn{1}{c}{Hint-CMD} &
  \multicolumn{1}{c}{Hint-SFH} &
  \multicolumn{1}{c}{Meaning} \\
\noalign{\smallskip}
\hline
\noalign{\smallskip}
1 & \NOB $< 5$           & $r < 1.7$       & Strongly favours a type Ia\\
2 & $5 \leq$ \NOB $< 15$ & $1.7 < r < 2.2$ & Moderately favours a type Ia\\
3 & $15 \leq$ \NOB $< 35$& $2.2 < r < 3.4$ & Undecided \\
4 & $35 \leq$ \NOB $< 80$& $3.4 < r < 5$   & Moderately favours a CC\\
5 & $80 \leq$ \NOB       & $5 < r$         & Strongly favours a CC\\
\noalign{\smallskip}
\hline
\end{tabular}
\normalsize
\end{center}
\end{table}

This method is conceptually similar to that used by \citet{1988AJ.....96.1874C},
albeit with several improvements: Firstly, the sample in this work is twice the
size of that available to \citeauthor{1988AJ.....96.1874C}. Secondly, many
($\sim 25$) SNRs have now a secured type (Appendix~\ref{appendix_secured}) and 
can be used to calibrate the method and evaluate the rate of erroneous 
classification. Then, the completeness of the census of early-type stars in the 
vicinity of the remnants is higher, owing to the use of the MCPS catalogue. 
Finally, the spatially resolved SFH reconstruction was simply unavailable before
\citet{2009AJ....138.1243H}.

\begin{figure*}[ht]
    \includegraphics[width=0.495\hsize]{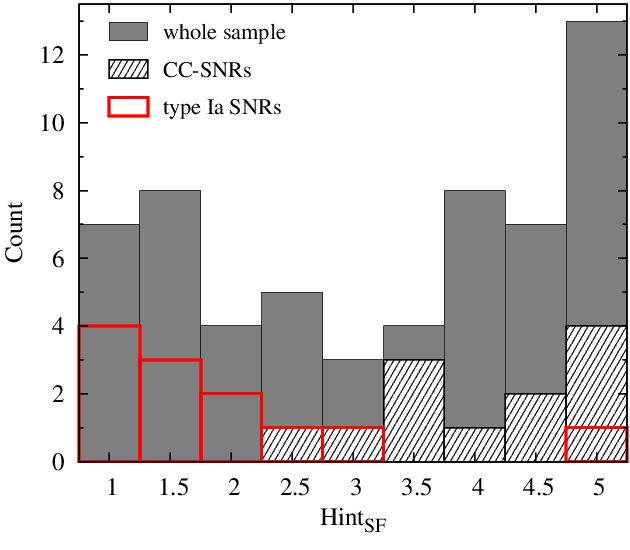}
    \includegraphics[width=0.495\hsize]{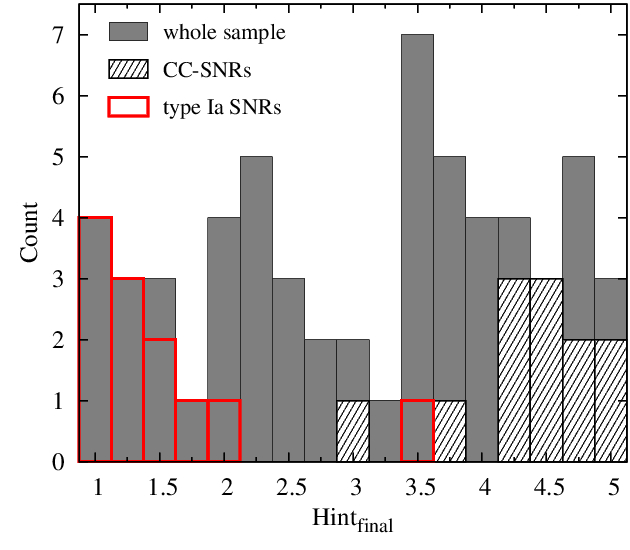}
\caption{\emph{Left:} Count distribution of the LMC SNRs as function of 
``Hint-SF'', combining \NOB and $r$.
\emph{Right:} Count distribution of the LMC SNRs as function of 
``Hint--final'', 
combining spectral \emph{and} SFH information (see text for details). Hatching 
and colours as in Fig.\,\ref{fig_typing_calibration}.}
    \label{fig_hint_sf}
\end{figure*}

\paragraph{Calibration of the ``SFH-typing'':}
The number of OB stars in the vicinity of the secured type Ia and CC SNRs is
shown in Fig.\,\ref{fig_typing_calibration} (left panel). The two samples are 
rather well separated: The majority of type Ia SNRs have less than 20 
early-type stars in their neighbourhood, while most of the CC-SNRs have \NOB~$> 
30$. The single major type Ia outlier is N103B (\NOB~$= 99$), which is known to 
be in a region with a vigorous recent star formation activity
\citep[\eg][]{2009ApJ...700..727B}. MCSNR J0453$-$6829 is the only CC-SNRs to
have a moderate \NOB ($< 25$). The choice of ``Hint-CMD'' is given in 
Table~\ref{table_results_sfh_hints} to reflect this distribution: \NOB less than 
5 (less than 15) strongly (moderately) favours a type Ia classification, while 
\NOB in excess of 80 (35) strongly (moderately) favours the CC-SN case.

Intuitively, any value $r > 1$ should favour a CC SN origin (conversely for a
thermonuclear origin). However, an important caveat in interpreting $r$ is that 
the rates of \citet{2010MNRAS.407.1314M}, especially $\Psi_2$ and $\Psi_3$, are
quite uncertain, due to the still limited sample of SNRs. Specifically, $\Psi_2$
has a value that changes by a factor of four depending on the tracer used to
constrain the SNR visibility time. To provide a better feeling on what $r$-value
to expect in either case (and to decide where is the separation), we show the
count distribution of secured type Ia and CC SNRs in the $r$-domain in
Fig.\,\ref{fig_typing_calibration} (middle panel). There is a stronger overlap 
of both types in the intermediate range ($2.2 \lesssim r \lesssim 3.5$) than 
with \NOB. However, the lower end ($r < 2.2$) still includes most of the type Ia 
SNRs, without contamination by the other type. N103B is again the only outlier 
at $r = 6.2$; at $r > 3.4$ only CC-SNRs are found. In view of this observed 
distribution, the ratio $r=N_{\rm CC}/N_{\rm Ia}$ is still a useful tool to 
assign a type to SNRs using the observed local SFH, and should be valid in a 
comparative and statistical sense. The ``Hints-SFH'' attributed to the sample 
based on $r$ are listed in Table~\ref{table_results_sfh_hints}. $r$ and \NOB are
also displayed as a scatter plot for secured Ia and CC SNRs (right panel of 
Fig.\,\ref{fig_typing_calibration}). There, the regions corresponding 
to different ``Hints'' are marked.

\paragraph{Caveat on the complementarity of \NOB and $r$\,:}
It is clear that the two metrics are connected. Both are based on the MCPS
catalogue; the early-type stars detected in a cell drive the fitting of the
most recent time bins in the SFH reconstruction of \citet{2009AJ....138.1243H}.
However, the $r$-value of a cell can be moderate even though \NOB is high, as
evident from the scatter along the horizontal axis in 
Fig.\,\ref{fig_typing_calibration} (right panel). That is because $r$ is a 
\emph{relative} measure of the recent SFR compared to that at earlier epochs, 
while \NOB gives a measure of the \emph{absolute} strength of the recent star 
formation. In the (high \NOB -- moderate $r$) case, there are many available 
progenitors of both CC- and type Ia SN; these are typically cases where the 
classification is inconclusive.

\begin{table*}[t]
\caption{``Hint-spec'' attributed to SNRs as function of spectral results.}
\begin{center}
\label{table_results_sfh_hints_spec}
\begin{tabular}{@{}c c @{}}
\hline\hline
\noalign{\smallskip}
\noalign{\smallskip}
  \multicolumn{1}{c}{Hint-spec} &
  \multicolumn{1}{c}{Criteria}
  \\
\noalign{\smallskip}
\noalign{\smallskip}
\hline
\noalign{\smallskip}
1   & at least three ``low X/Fe'' flags, no ``high X/Fe'' flag \\
1.5 & two ``low X/Fe'' flags or low O/Fe, no ``high X/Fe'' flag \\
2   & one ``low X/Fe'' flag (except O/Fe), no ``high X/Fe'' flag \\
2.5 & low Si/Fe, no ``high X/Fe'' flag \\
3   & ISM abundances, unfitted abundances, or no \xmm\ data \\
3.5 & high Si/Fe, no ``low X/Fe'' flag \\
4   & one `high X/Fe'' flag (except O/Fe), no ``low X/Fe'' flag \\
4.5 & two ``high X/Fe'' flags or high O/Fe, no ``low X/Fe'' flag \\
5   & at least three ``high X/Fe'' flags, no ``low X/Fe'' flag; pulsar/PWN
detected \\ 
\noalign{\smallskip}
\hline
\end{tabular}
\end{center}
\end{table*}

\paragraph{Results for the whole sample:} The count distributions for all LMC
SNRs in the \NOB and $r$ spaces are shown in Fig.\,\ref{fig_typing_calibration} 
(grey histograms, left and middle panels), and as a scatter plot in the right 
panel. They are similar, with larger numbers, to the distributions of the 
secured-type SNRs. About twenty remnants are in regions with a low number of 
early-type stars (\NOB $< 15$) and not dominated by recent SF ($r \lesssim 2$). 
There is a peak at $r \sim 6$ with a dozen remnants. Those are SNRs in 
star-forming regions which are widely spread across the LMC. They are often 
associated with giant \ion{H}{II} complexes (\eg LHA-120 N4, N11, N44). The 
objects with extreme values for $r$ ($\gtrsim 8$) also have the largest \NOB. 
Those are located in the two most intensively star-forming regions of the LMC: 
30 Doradus, and the rim of the supergiant shell LMC 4 \citep[which embeds the 
``Constellation III'' region,][]{2008PASA...25..116H,2009AJ....138.1243H}.

To combine the two ``Hints'' into one, we computed the arithmetic mean of
Hint-CMD and Hint-SFH. The resulting ``star-formation Hint'' (Hint-SF) again
ranges from 1 to 5. Its distribution for the whole sample and the secured-type
SNRs is shown in Fig.\,\ref{fig_hint_sf}. There are 19 remnants with Hint-SF 
$\leq 2$; they most likely all result from a type Ia SN. We call this sample 
``likely-Ia''. Likewise, the 28 objects above Hint-SF $\geq 4$ are probably 
most of the CC-SNR population. They form the ``likely-CC'' sample.

The single type Ia SNR contaminating the sample (N103B) allows to estimate a
false-positive rate of 5~\%~--~10~\%. The false-positive rate of the
``likely-Ia'' sample is probably lower: The massive stars formed at (roughly)
the same time as the progenitor of a CC-SN can hardly be missed by photometric
survey, because they would form the bright end of the population.

There are 12 SNRs in between $2.5 \leq \mathrm{Hint_{SF}} \leq 3.5$, for which
the local stellar environment cannot be used to decisively type the origin;
they form the ``SFH-untyped'' sample. Interestingly, two and five of these
remnants can be classified from other indicators as type Ia (the iron-rich MCSNR
J0508$-$6830 and DEM L71) or CC-SNRs (\eg the oxygen-rich N132D or MCSNR
J0453$-$6829, which hosts a PWN), respectively.

\paragraph{Including the spectral results for typing purpose}
The spectral analysis of Sect.\,\ref{results_spectra} revealed the presence of 
ejecta-enhanced plasma in almost half of the sample 
(Tables~\ref{table_results_spectra_flags} and 
\ref{appendix_table_spectra_brightest}). One should take advantage of this for 
the typing of the remnant, in combination with the SFH-based method we just 
presented. We assign another number, ``Hint--spec'', which depends on the high- 
or low-abundance flags of each SNR (Sect.\,\ref{results_spectra_ejecta}). The 
numbers range from 1 (strongly favouring a type Ia origin) if ``low X/Fe'' flags 
are raised (\ie the SNR is iron-rich), to 5 (strongly favouring a CC origin) 
when ``high X/Fe'' flags are raised (\ie CC nucleosynthesis pattern is 
detected). 

The choice of ``Hint-spec'' is given in 
Table~\ref{table_results_sfh_hints_spec}. Note that a bigger impact is given to 
the low or high O/Fe ratio, as these elements are the most abundant. Therefore, 
this ratio is easier to fit, more reliable, and sometimes the only one 
available. A value of 5 is also attributed to remnants where a pulsar/PWN is 
detected in the remnant. \revision{The values of ``Hint--spec'' for each SNR are
given in Table~\ref{appendix_table_snrs_sample}.} We combined ``Hint--SF'' and 
``Hint--spec'' by taking their arithmetic mean. The distribution of the 
resulting ``Hint--final'' is shown in Fig.\,\ref{fig_hint_sf} (right panel). The 
contamination (\ie misclassification of N103B) is slightly alleviated, while a 
better separation of the ``secured-type'' SNRs is evident. There are 23 SNRs 
with ``Hint--final'' $\leq 2.5$ which are likely of type Ia, and 31 SNRs where 
``Hint--final'' $\geq 3.5$ which are likely attributed to CC, although N103B 
(Hint--final$=$3.5) is still contaminating the sample. There are five sources 
with inconclusive ``Hint--final'', including one secured-CC (N23).

\subsection{Ratio of CC to type Ia SNe and implications}
\label{results_sfh_ratio}

The observed number of SNRs of both types provides a measurement of the ratio
of CC to type Ia SNe that exploded in the LMC over the last few $10^4$~yr, \ie
very close to the current ratio of CC/Ia SN \emph{rates}. Based on the ``star
formation Hint'', the numbers of SNRs in the ``likely Ia'' and ``likely CC''
samples translate in \ccIa~$=$~1.47 (28/19). Assuming all ``SFH-untyped'' SNRs
which do not have a secured type are of type Ia then the ratio \ccIa is 1.27
(33/26). Conversely, if the `SFH-untyped'' are all CC, the ratio is 1.81
(38/21). Even correcting for N103B, \ccIa is conservatively in the range 1.2 to
1.8.

Including the spectral results (detection of ejecta in X-rays), we have a ratio
\ccIa~$= 1.35$ (31/23). Correcting for N103B and N23 (with wrong and uncertain
classifications), the ratio CC:Ia based on SFH \emph{and} X-ray spectroscopy is 
between 1.11 (31/28) and 1.46 (35/24), depending on the type assigned to the 
remaining four objects. This range is compatible with that derived from the 
``SFH-typing'' alone, albeit narrower because of the greater amount of 
information included in the calculation.

This ratio can be compared to two kinds of measurements: First, to the observed
ratio of current rates, obtained from SNe search. For instance,
\citet{2011MNRAS.412.1441L} measured a ratio of about 3:1 in a volume-limited
sample. Second, to the ratio \ccIa\ derived from intracluster medium (ICM) 
abundances. Galaxy clusters retain all the metals produced by SNe. The X-ray 
spectrum of the ICM reveals the elemental abundances, which are used to 
constrain the \emph{time-integrated} numbers of CC and Ia SNe. From \suzaku\ 
observations of a small sample of clusters and groups of galaxies, 
\citet{2007ApJ...667L..41S} estimated \ccIa~$\sim 3.5$ (ranging between 2 and 4, 
depending on the type Ia SN model assumed). With \xmm\ and a larger cluster 
sample, \citet{2007A&A...465..345D} measured a \ccIa\ between 1.7 and 3.5, again 
depending on the adopted SN Ia models. However, none of the explored SN models 
could reproduce the Ar/Ca ratio. \citet{2011A&A...528A..60L} derived 
\ccIa~$\sim1.5-3$. Therefore, \textbf{the current ratio of CC/Ia SNe in the LMC 
is significantly lower than that measured in local SN surveys and in galaxy 
clusters.}

One possible caveat could be that we are missing CC-SNRs. For instance, SNe
exploding in superbubbles \citep[SBs, see \eg][for a 
review]{2008IAUS..250..341C} will not be directly recognised as SNRs. 
\citet{1991ApJ...373..497W} and \citet{2001ApJS..136..119D} found a dozen LMC
SBs with an X-ray luminosity, measured with \einstein\ and ROSAT, brighter than
theoretically expected for a wind-blown bubble, and possibly energised by
interior SNRs. The limited spatial resolution of the instruments used may result
in \emph{distinct} SNRs to have been overlooked and the X-ray emission of the SB
overestimated (\eg MCSNR J0523$-$6753, near the \ion{H}{II} region/SB N44 in
\citealt{1991ApJ...373..497W}, see also \citealt{2011ApJ...729...28J}). With
a dozen extra CC SNRs, the ratio \ccIa\ is pushed to $\sim 1.5 - 2$. However,
the number of type Ia SNRs currently known in the LMC is also expected to be
below the actual number (see Sect.\,\ref{results_XLF} for a discussion on sample 
completeness). Therefore, it is unlikely that the ratio \ccIa\ is significantly 
underestimated. Furthermore, the abundance pattern of the LMC, with its low 
{[}$\alpha$/Fe{]} (Sect.\,\ref{results_spectra_abundance}), lends support to
such a low \ccIa. This should be lower than the value of \ccIa$\sim4-5$
estimated by \citet{1995MNRAS.277..945T} using metallicity alone.

The low \ccIa\ ratio measured in the LMC therefore has to be a consequence of
the different SFH of the Cloud compared to that in other nearby galaxies or
galaxy clusters. The local SN rates depend on the summed SFH of galaxies
included in the SN surveys. The higher ratio measured by \eg 
\citet{2011MNRAS.412.1441L} simply indicates that many star-forming galaxies are
included in the local volume explored. The SFHs of galaxy clusters are
relatively simple, with short episodes of star-formation at a redshift of $z
\sim 3$ \citep{2008ApJ...684..905E}, so that the integrated numbers of type Ia
and CC SNe inferred from X-ray observations correspond to the fractions of stars
formed that end their lives as SN of either type.

In the LMC, star formation occurred during several episodes. In addition to many
regions with recent or ongoing star formation where, unsurprisingly, the CC-SNRs
are found (see Sect.\,\ref{results_distribution}), the LMC had enhanced
star formation episodes 100~Myr, 500~Myr, and 2~Gyr ago as well 
\citep{2009AJ....138.1243H}. The SN Ia DTD follows fairly well a $t^{-1}$ power 
law for delays $t > 1$~Gyr, and appear to keep increasing below 1~Gyr \citep[for 
a review of SN Ia DTD measurements, see][]{2012PASA...29..447M}. The majority 
of type Ia SNe explode within 2~Gyr after star-forming episodes.  We are 
therefore coincidentally observing the LMC at a time when the rate of type Ia SN 
from the stellar populations formed 500~Myr to 2~Gyr ago is high. Integrated 
over an SNR lifetime (a few $10^4$~yr), it results in the relatively large 
number of type Ia SNRs. It is not possible to use \ccIa\ to estimate $\eta$, the 
fraction of stars that eventually explode as Ia SNe \citep{2008MNRAS.384..267M}, 
because of the complex SFH of the LMC: stars exploding now (as either SN types) 
were created at different epochs. Furthermore, $\eta$ is also dependent on the 
initial mass function (IMF), over which one has little freedom, since the 
SFH-reconstruction already assumes a particular form (the Salpeter IMF).

\revision{Currently, the only other galaxy with which to compare the \ccIa\ of 
the LMC is the SMC}. In our own Milky Way, there remain too many untyped SNRs. 
More problematic are the distance uncertainties and line-of-sight confusion that 
prevent associating remnants to regions of star formation (\eg spiral arms). In 
the Local Group (M31, M33) and beyond \citep[\eg M83,][]{2010ApJ...710..964D}, 
the problem is again the lack of secured typing methods, and generally the 
absence of spatially resolved SFH. The situation is likely to improve in the 
near future with more sensitive X-ray observatories (\eg \emph{Athena}), and 
large observing programmes of M31 and M33 with \emph{Hubble} which allow to 
build SFH maps \citep[so far, this was done in the few archival field 
available,][]{2012ApJ...761...26J,2014ApJ...795..170J}. The SMC is the only 
obvious target remaining where a similar study can be currently performed, 
although the smaller sample of SNRs and inclination of the galaxy (and 
corresponding line-of-sight confusion) might complicate direct comparisons to 
the LMC.

\section{X-ray luminosity function of SNRs in the Local Group}
\label{results_XLF}

X-ray luminosity functions (XLFs) are valuable tools for the study of X-ray 
sources and comparisons between populations. The \xmm\ dataset is ideally 
suited to derive the XLF of LMC SNRs. Out of the 59 objects in the sample, 
\xmm\ covered 51 of them, to which we fitted spectral models 
(Sect.\,\ref{results_spectra}). For all these, the X-ray fluxes in various 
bands are obtained from the best-fit models 
(Tables~\ref{appendix_table_spectra_all} and 
\ref{appendix_table_spectra_brightest}) with the XSPEC command
\texttt{flux}. The final results are presented in the ``broad'' band, from
0.3~keV to 8~keV (the effect of including the high-energy part is very minor
and discussed below).

Three SNRs have been covered with \chandra\ but not \xmm: For MCSNR J0454$-$6713 
(N9), we used the spectral results of \citet{2006ApJ...640..327S} to measure the 
flux. MCSNR J0459$-$7008 (N186D) was covered in the \chandra\ observation of the 
SB DEM~L50. \citet{2011ApJ...729...28J} published the results from these data. 
We used their best-fit NEI model for the SNR emission, which is spatially 
resolved from the SB, to derive the X-ray flux. Finally for MCSNR J0550$-$6823, 
we used the spectral parameters given in the entry of the \chandra\ SNR
catalogue\,\footnote{Maintained by Fred Seward\,:
\url{http://hea-www.cfa.harvard.edu/ChandraSNR/index.html}}.

Two SNRs have neither \xmm\ nor \chandra\ observations available, but were
covered with ROSAT. \citet{1999ApJ...514..798W} present a spectral analysis of
MCSNR J0455$-$6839 (in N86). We obtained the X-ray flux of the SNR using their
best-fit model. MCSNR J0448$-$6700 corresponds to the ROSAT PSPC source
{[HP99]}~460, with a count rate of $1.41 \times 10^{-2}$~cts~s$^{-1}$
\citep{1999A&AS..139..277H}. With the multi-mission count rate simulator 
WebPIMMS\,\footnote{
\url{http://heasarc.gsfc.nasa.gov/cgi-bin/Tools/w3pimms/w3pimms.pl}}, we 
calculated the flux in several energy bands for various temperatures of an APEC 
model, assuming a total absorbing column $N_H  = 7 \times 10^{20}$~cm$^{-2}$ 
towards the source and a mean abundance of 0.4~solar. The observed hardness 
ratios could be reproduced for $kT = 0.97$~keV. These spectral parameters and
normalisation can be converted into fluxes in the same bands as used for the
rest of the sample.

\begin{figure}[t]
    \centering
\includegraphics
[width=0.999\hsize]{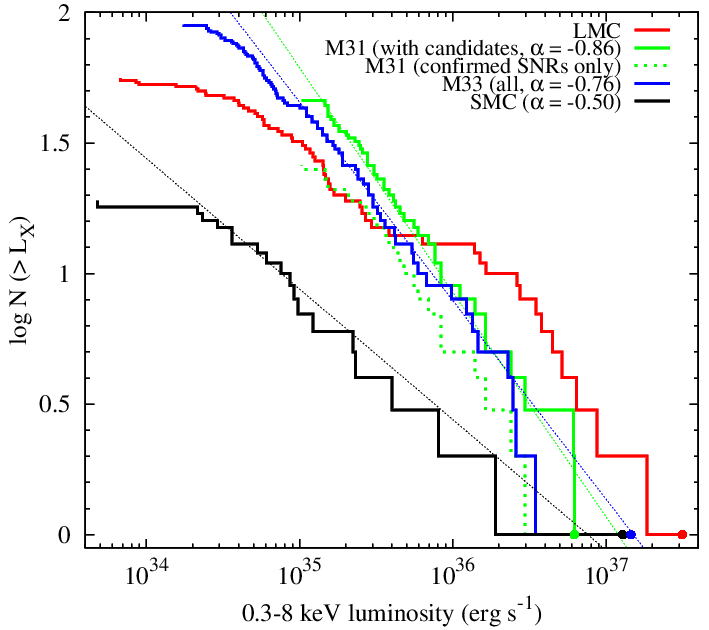}
\caption[\ Cumulative X-ray luminosity function of SNRs in Local Group
galaxies]{Cumulative X-ray luminosity function of SNRs in Local Group galaxies.
See text for details and references on how $L_X$ was measured for each sample.
The brightest SNR in each galaxy is marked by a dot. The thin dotted lines are
nonlinear least-square fits of a power law ($N (> L_X) \propto L_X\ ^{\alpha}$).
Slopes $\alpha$ are given in the legend. These fits are only used to 
characterise the slopes and illustrate the differences between galaxies; they 
do 
not represent a physical fit of the population.
}
\label{fig_results_XLF}
\end{figure}

In total, 56 objects have well-defined X-ray fluxes and make it into the XLF. 
The adopted values are listed in Table~\ref{appendix_table_snrs_sample}. Only 
three SNRs have no X-ray data available. The cumulative XLF is shown in 
Fig.\,\ref{fig_results_XLF}. The sample spans almost four orders of magnitude in 
luminosity, from the brightest (N132D) at $L_X = 3.15 \times 
10^{37}$~erg~s$^{-1}$ down to $\sim 7\times 10^{33}$~erg~s$^{-1}$. The $L_X$ 
used is the observed value, uncorrected for (LMC) absorption, because the fitted 
column densities can be quite uncertain, in particular in the faint end. 
\revision{We checked the $N_{H\mathrm{\ LMC}}$-corrected XLF and found no 
significant variations in its shape compared to that of 
Fig.\,\ref{fig_results_XLF}, except the shift to higher $L_X$.}

\paragraph{SNR XLF from other Local Group galaxies:}
The LMC XLF can be best compared to other Local Group galaxies such as M31,
M33, and the SMC. \citet{2012A&A...544A.144S} studied M31 SNRs and candidates
identified in the \xmm\ Large Programme survey of the Andromeda galaxy
\citep{2011A&A...534A..55S}. They converted EPIC count rates into
0.35~keV~--~2~keV luminosities assuming a thermal (APEC) spectrum with $kT =
0.2$~keV, $N_{H\ \mathrm{M31}} = 10^{21}$~cm$^{-2}$, and $N_{H\ \mathrm{MW}} =
0.7 \times 10^{21}$~cm$^{-2}$. The quoted values, however, are corrected for
$N_{H\ \mathrm{MW}}$, while for the LMC we give the observed luminosities. For 
consistency with the LMC XLF, we re-included $N_{H\ \mathrm{MW}} = 0.7 \times
10^{21}$~cm$^{-2}$ in the results of \citet{2012A&A...544A.144S} and converted
the luminosity in the 0.3~keV~--~8~keV by scaling their $L_X$ by 0.577 (a factor
derived from simulating their assumed spectrum with and without $N_{H\
\mathrm{MW}}$). Note that the effect of the foreground absorption should be
very minor, since $N_{H\ \mathrm{MW}}$ values are very similar in the directions 
of M31, M33, and the LMC ($5 - 7 \times 10^{21}$~cm$^{-2}$). 26 objects were 
confirmed SNRs in \citet{2012A&A...544A.144S}, and another 20 were candidate 
SNRs.

\citet{2010ApJS..187..495L} present a large catalogue of 82 confirmed SNRs in 
M33, based on the \chandra\ ACIS survey of M33 
\citep[ChASeM33,][]{2011ApJS..193...31T}. They give $L_X$ in the 
0.35~keV~--~2~keV band, converted from ACIS count rates, assuming a thermal 
plasma at 0.5 solar, $kT = 0.6$~keV, and total $N_{H} = 10^{21}$~cm$^{-2}$ (\ie 
the spectrum found for the brightest source). We obtained the corresponding 
0.3~keV~--~8~keV luminosity by scaling up the values of 
\citet{2010ApJS..187..495L} by 4~\%.

Recently, \citet{2015ApJS..218....9W} published the results of a deep \xmm\ 
survey of M33 with a larger coverage than ChASeM33, up to the $D_{25}$ isophote 
of the galaxy. They recovered most of the SNRs of \citet{2010ApJS..187..495L}, 
except in the central region where source confusion is an issue for \xmm, and 
detected or confirmed eight new SNRs, including three sources far in the 
outskirts of M33. We converted the unabsorbed 0.2--4.5~keV EPIC flux of the new 
inclusions of \citet{2015ApJS..218....9W}, obtained assuming a power-law 
spectrum, to the 0.3--8~keV luminosity for the same spectrum as the SNRs of 
\citet{2010ApJS..187..495L}. The final, concatenated list of M33 SNRs thus 
comprises 90 objects.

Converting count rate to luminosity in different energy bands assuming a single
temperature might affect the slope of the XLF. For instance, from a count rate
in the 0.35~keV~--~2~keV band, the luminosity in the broad band is 25~\% higher
with $kT = 0.6$~keV than if it is 0.2~keV. The two studies have limited
knowledge of the actual spectrum of each remnant, because the larger distances
prohibit more complex spectral fits, and they have to assume a particular
spectrum, regardless of luminosity. This is not the case in the LMC. We indeed 
found a trend for brighter remnants to have higher plasma temperatures
(Sect.\,\ref{results_spectra_general}). Quantitatively, the median temperatures 
are $kT = 0.31$~keV for luminosities less than $10^{35}$~erg~s$^{-1}$, 0.55~keV 
between $10^{35}$~erg~s$^{-1}$ and $10^{36}$~erg~s$^{-1}$, and 0.8~keV above 
$10^{36}$~erg~s$^{-1}$. The luminosities of M31 SNRs were given assuming $kT = 
0.2$~keV; we scaled the 0.3~keV~--~8~keV luminosity up by 1.05, 1.20, and 1.35 
for sources with $L_X < 10^{35}$, $10^{35} < L_X < 10^{36}$, and $L_X > 
10^{36}$~erg~s$^{-1}$, respectively. M33 SNRs were assumed to have a higher 
temperature (0.6~keV), which means that the luminosity of objects below $\sim 
10^{35}$~erg~s$^{-1}$ was overestimated by about 15~\%, while for those above 
$10^{36}$~erg~s$^{-1}$ it was underestimated by $\sim 8$~\%. Correcting for 
this effect ensures a meaningful comparison between M31, M33, and the LMC.

The SMC SNR population is comparatively smaller. \citet{2004A&A...421.1031V}
presented an X-ray spectral analysis of all SNRs in the SMC known at that time.
We used their best-fit models to measure the observed (\ie absorbed) X-ray 
luminosity in the same 0.3~keV~--~8~keV band\,\footnote{The luminosity given in
\citet{2004A&A...421.1031V}, Table~3, for IKT~22 (1E0102$-$7219, the brightest
SMC SNR) was mistyped. Instead of the $150\times10^{27}$~W, it should read
$1500\times10^{27}$~W ($1.5 \times 10^{37}$~erg~s$^{-1}$).}, except for IKT~16. 
For this SNR we used results from \citet{2011A&A...530A.132O}, which included 
more data from subsequent \xmm\ observations. Three additional SNRs were 
covered with \xmm; the results were published in \citet{2008A&A...485...63F}, 
from which we borrowed the best-fit spectral models. The latter study also 
reported a new SNR, HFPK~334. For this one, we used the best-fit model from 
\citet{2014AJ....148...99C}, which combined \xmm\ and \chandra\ observations. 
Also included is the SNR XMMU~J0056.5$-$7208 identified during the SMC survey
\citep{2012A&A...545A.128H,2012PhDT......ppppS}. Finally, the Be/X-ray binary
pulsar SXP~1062 was found to be associated to an SNR, of which it is most
likely the progenitor \citep{2012MNRAS.420L..13H}. The thermal emission from
the SNR was studied by \citet{2012A&A...537L...1H}. This sample of 19 SMC SNRs
is the most up to date.

\paragraph{Comparative study of SNR XLFs:}
The cumulative XLFs of M31 and M33 in the 0.3~keV~--~8~keV band, corrected for
the $kT$~--~$L_X$ trend, are shown along that of the SMC and LMC in 
Fig.\,\ref{fig_results_XLF}. \textbf{In terms of depth}, the LMC XLF dominates. 
There is a single SNR at $L_X < 2\times 10^{34}$~erg~s$^{-1}$ in M33 and in the 
SMC, but the bright interior pulsar in the SMC case (SXP~1062) makes the 
measurement of the thermal emission luminosity uncertain. In contrast, there are 
eight SNRs with $L_X \lesssim 2\times 10^{34}$~erg~s$^{-1}$ in the LMC, of which 
seven were discovered or confirmed thanks to \xmm\ observations.

\textbf{In terms of number}, the largest population so far is found in M33 (90
SNRs in X-rays), probably owing to the depth of the \chandra\ survey
(using 100~ks pointings) in the central 15\arcmin, the overlap with a deep 
\xmm\ survey up to the $D_{25}$ isophote, and the favourable (face-on) 
orientation of M33. However, the population of M31 SNRs is larger than any other 
at $L_X \lesssim 5\times 10^{35}$~erg~s$^{-1}$ and is only limited by the depth 
of the survey ($\sim 10^{35}$~erg~s$^{-1}$). The ratio of M31-to-M33 SNRs in the
$10^{35}$~--~$10^{36}$~erg~s$^{-1}$ range is at most 1.5, \ie substantially
smaller than the mass ratio of the galaxies 
\citep[10--20,][]{2003MNRAS.342..199C,2014MNRAS.443.2204P}. This shows the
effect of the higher (recent) SFR in M33 compared to M31 
\citep[0.45~\msun~yr$^{-1}$ vs. 0.27~\msun~yr$^{-1}$,][]{2009A&A...493..453V,
2010A&A...517A..77T} leading to a larger production of CC SNRs in M33. In the 
same luminosity range, the number of LMC SNRs is comparable to that in M33. This 
is expected because the LMC is only slightly less massive than M33. Furthermore, 
the recent SFR of the LMC is high, 0.3--0.4~\msun~yr$^{-1}$ in the last 40 Myr 
\citep{2009AJ....138.1243H}. This conspires with the high current type Ia SN 
rate (Sect.\,\ref{results_sfh_ratio}) to build up the large population of
SNRs in the LMC. Finally, the ``feather-weight'' SMC (about ten times less
massive than the LMC, \citep{2004ApJ...604..176S,2006AJ....131.2514H} has a
smaller, yet decent population of remnants, likely owing to its recent star
formation activity \citep[0.08--0.3~\msun~yr$^{-1}$,][]{2004AJ....127.1531H}.

\textbf{In terms of shape}, the XLF of M31 SNRs is the most uniform, following a 
power law ($N (> L_X) \propto L_X\ ^{\alpha}$) with $\alpha = -0.86\pm0.04$ down 
to $\sim 2\times 10^{35}$~erg~s$^{-1}$. This holds with or without including the 
candidates, which means that most are indeed bona-fide SNRs. The M33 remnants 
follow mostly the same distribution, with $\alpha = -0.76\pm0.05$. Towards the 
faint end, the M33 XLF flattens and diverges from the power law below 
$10^{35}$~erg~s$^{-1}$, indicating incompleteness. \citet{2010ApJS..187..495L} 
concluded that no SNR brighter than $4\times 10^{35}$~erg~s$^{-1}$ was missed 
across the surveyed field. It is likely that they were over-conservative and 
that missing SNRs are only those which have luminosity below 
$10^{35}$~erg~s$^{-1}$. The combined ChASeM33 and \xmm\ surveys cover the total 
extent of the galaxy \citep{2008ApJS..174..366P,2015ApJS..218....9W}, so the 
missing SNRs are either too X-ray-faint (below the surveys' detection limits), 
or absent/undetected at radio and optical wavelengths, precluding their 
identifications as SNRs.

We performed Kolmogorov-Smirnov (KS) tests to compare the different populations. 
Using a bootstrapping method, we produced 1000 luminosity functions from the 
original data. We checked that similar results were obtained when increasing 
that number to $10^6$. Restricting the analysis to SNRs brighter than $3\times 
10^{35}$~erg~s$^{-1}$ to ensure completeness of the samples, we found that the 
XLFs of M31 and M33 SNRs follow the same distribution at the $3\sigma$ 
confidence level. There was a marginal indication that the M33 distribution was 
steeper than that of M31 \citep{2012A&A...544A.144S}, but this difference 
essentially disappears once the $kT$--$L_X$ trend is taken into account.

In the  SMC, although the population is limited to about 20 objects, the 
distribution is relatively uniform. The XLF is however flatter ($\alpha = 
-0.5\pm0.05$), and KS tests confirm that the SMC population is different from 
those of M31 and M33. This might indicate that SMC remnants evolve faster (and 
fade earlier) than in M31 and M33, due to a lower ISM density. The lower 
metallicity in the SMC \citep[about 0.2 solar,][]{1992ApJ...384..508R} may also 
participates in the lower luminosities of the SMC SNR, \revision{since the 
emissivity of hot plasmas is smaller for lower metallicities}.

In contrast to the other galaxies, the luminosity function of SNRs in the LMC 
exhibits a complex behaviour and does not obey a smooth power-law distribution 
over most of the dynamical range. Regardless of the lower luminosity cut used, 
the KS tests reject the null hypothesis that LMC SNRs have the same XLF than 
those in M31, M33, or the SMC.

The most striking and robust result is the very prominent bright end of the LMC 
XLF. There are 13 SNRs with $L_X > 10^{36}$~erg~s$^{-1}$, more than in M31 and 
M33. Amongst these, there are two SNRs hosting bright pulsars/PWNe and a harder 
non-thermal spectrum. Even restricting the XLF to the soft band or excluding 
these two objects, the population of bright LMC SNRs is still above the other 
ones. This bright population is not a clearly distinct group. In particular, it 
is not made up of remnants from only one SN type. There are four type Ia SNRs 
and nine CC-SNRs, so the \ccIa\ ratio is higher than overall 
(Sect.\,\ref{results_sfh_ratio}), but not exceedingly so. Higher luminosities 
are expected from SNRs interacting with denser ISM. We compared the average LMC 
\ion{H}{I} column density \citep[from the map of][]{2003ApJS..148..473K} around 
the position of remnants in various luminosity bins, but no trend could be 
found. However, the line-of-sight integrated column density might not be a good 
indicator of the ISM density \emph{local} to the remnant, considering that the 
SNR could be in front of or behind the regions where most of the neutral 
hydrogen is (see Sect.\,\ref{results_distribution}).

A possible explanation for the population of bright SNRs in the LMC stems from
its lower metallicity. Massive stars lose a considerable amount of mass in the
form of winds \citep[\eg][]{2000ARA&A..38..613K}. The stellar winds blow
low-density cavities, bordered by dense shells, around the stars that eventually 
explode as (core-collapse) SNe. The interaction of the SN shocks with the 
modified CSM results in a different evolution compared to that in a
constant-density ISM. \citet[][and references
therein]{2005ApJ...630..892D,2007ApJ...667..226D} explored the evolution of
remnants in wind-blown cavities. It was shown that it critically depends on one
parameter (coined $\Lambda$), the ratio of the mass of the dense shell to that
of the ejected material. For low values ($\Lambda < 1$) the X-ray luminosity
increases sharply when the shock reaches the dense shell early on ($t < 
10^3$~yr). If instead the shell is more massive compared to the ejecta material,
the shock propagates in the very low density of the (much larger) bubble,
producing less X-ray emission. The increase of X-ray luminosity upon impact
(after a few thousand years) is also smaller than in the low-$\Lambda$ case
\citep[][his Figs. 7 and 12]{2005ApJ...630..892D}. The properties of the
cavities around massive stars are determined by the mass loss rate $\dot{M}$
during their various evolutionary stages. This in turn is affected by the
elemental abundance (\ie metallicity), because the main driving mechanism of
stellar winds is the transfer of momentum from photons to the star atmospheric
gas by line interactions\,\footnote{The product abundance $\times$ ionisation
fraction $\times$ number of available lines for metals is comparable to that of
hydrogen and helium.} \citep{2000ARA&A..38..613K,2001A&A...369..574V}. By
measuring mass-loss rates of early-type stars in the Galaxy, LMC, and SMC,
\citet{2007A&A...473..603M} could quantify the dependence of $\dot{M}$ on the
metallicity as $\dot{M} \propto Z^{0.83}$. It is therefore expected that in
lower metallicity environment (\eg LMC) massive stars explode in wind-blown
cavities with lower $\Lambda$, and are more likely to produce 
\revision{\emph{young}} remnants that are brighter in X-rays.
\revision{Since the SMC has the lowest metallicity of our sample of galaxies 
(0.2 solar), one would also expect an excess of bright sources. However, the 
SMC does not host SNRs as young (less than a few thousand years) as the LMC 
\citep{2004A&A...421.1031V}, and it is likely that the smaller wind-blown 
bubbles do not affect the X-ray luminosities of more evolved remnants. 
Furthermore, the small number of SNRss in the SMC hinders conclusions regarding 
a possible excess of bright sources.}

Finally, there are also four type Ia SNRs amongst the bright end of the
population, to which the explanation discussed above does not apply. If we
exclude, these however, there is still an excess. Because they are prominently
young (three are less than a thousand years old), it appears that the high
current type Ia SN rate in the LMC (Sect.\,\ref{results_sfh_ratio}) will also 
contribute to a larger population of bright remnants.

Between $\sim 1\times 10^{35}$~erg~s$^{-1}$ and $5\times 10^{35}$~erg~s$^{-1}$, 
where many SNRs reside (a third of the sample), the LMC XLF is comparable in 
shape to the M31 and M33 XLFs, with a power-law distribution (consistent with 
$\alpha$ between $-1$ and $-0.8$), and in number to M33 (M31 begins to have 
more sources below $\sim 8\times 10^{35}$~erg~s$^{-1}$).
Towards the fainter end, the LMC XLF is again remarkable via its significant
flattening. It is unlikely that this represents an overall flatter distribution 
(at least not as strongly as in the SMC), because it would imply that a lot of
SNRs with $L_X \sim$(5--8)$\times 10^{35}$~erg~s$^{-1}$ (thus easy to identify) 
have been missed. It is more plausible that the flattening of the XLF is almost 
exclusively due to incompleteness. The majority of the remnants at $L_X < 
8\times 10^{34}$~erg~s$^{-1}$ (15/22) were identified/confirmed thanks to 
(pointed or serendipitous) \xmm\ observations. Though many were already 
detected with ROSAT, the combination of the large effective area and resolution 
of \xmm\ is usually needed to confirm the extent and thermal emission of 
candidates. Even with the VLP survey, the area of the LMC covered by \xmm\ is 
less than 20 square degrees, \ie about a third of the whole galaxy. 
\revision{In particular, X-ray coverage to the south-west of the LMC Bar is 
sparse (see Fig.\,\ref{fig_observations_survey})}. 
Extending the covered fraction warrants to find the missing remnants. The M31 
survey with \xmm\ exemplifies how a full coverage results in a high 
completeness: the M31 SNR XLF is uniform down to the sensitivity limit of
the survey, which fully covers the $D_{25}$ ellipse of M31 
\citep{2012A&A...544A.144S}.
In the LMC, the situation could easily be improved with more X-ray
observations. We briefly discuss possible strategies in Sect.\,\ref{summary}.

\section{3D spatial distribution}
\label{results_distribution}
\subsection{Comparison with other wavelengths}
\label{results_distribution_comparison}

\begin{figure}[t]
    \centering
\includegraphics
[width=0.999\hsize]{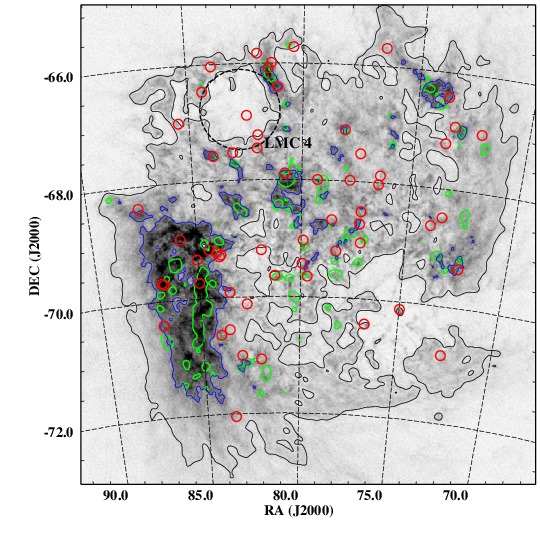}
\caption[\ Positions of LMC SNRs on an \ion{H}{I} column density
map]{Positions of LMC SNRs (red circles) on the \ion{H}{I} column
density map of \citet{2003ApJS..148..473K}, displayed on a linear scale ranging
from 0 to $6 \times 10^{21}$~cm$^{-2}$. Black and blue contours indicate levels
of 1 and $3 \times 10^{21}$~cm$^{-2}$, respectively. The green contours are the
3$\sigma$ level (1.2~K~km~s$^{-1}$) of the velocity-integrated map of $^{12}$CO
 $(J = 1 - 0)$ from the NANTEN survey \citep{2008ApJS..178...56F}. The position
of the SGS LMC~4 is marked with a dashed black circle.
}
\label{fig_results_distribution_HI}
\end{figure}

The positions of SNRs in the LMC are plotted on the \ion{H}{I} column density 
map of \citet{2003ApJS..148..473K}, showing the LMC gas disc 
(Fig.\,\ref{fig_results_distribution_HI}). The population exhibits correlations 
with neutral hydrogen structures. The most striking example is the many SNRs (a 
dozen) around the supergiant shell (SGS) LMC~4 \citep[][SGS~11 in the notation 
of \citealt{1999AJ....118.2797K}]{1980MNRAS.192..365M}. SGSs are formed by the 
combined action of multiple generations of massive star formation. Their 
expansions shock and sweep up the ISM, which can trigger further star formation 
along the SGS rims \citep[][and references therein]{1998ASPC..148..150E}. The 
impact of SGSs on star formation, particularly in the LMC, was demonstrated by 
\citet{2001ApJ...553L.185Y,2001PASJ...53..959Y}. They found that the 
concentration of molecular clouds and young star clusters is enhanced by a 
factor of 1.5--2 near the SGS rims, and most of these clusters are on the side 
of the molecular clouds facing the interior of the SGSs. 
\citet{2009AJ....137.3599B} added massive young stellar objects and \ion{H}{II} 
regions/OB associations to the list of tracers of recent star formation that are 
well correlated with the shell peripheries.

\begin{figure*}[ht]
    \centering
\includegraphics
[width=0.495\hsize]{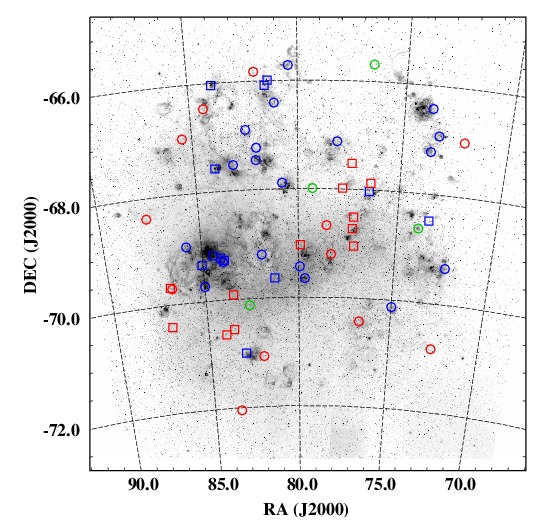}
\includegraphics
[width=0.495\hsize]{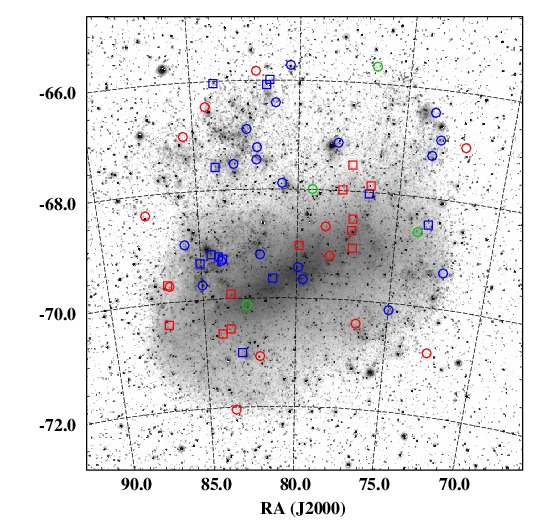}
\caption[\ Location of LMC SNRs relative to H$\alpha$ and red continuum
emission.]{
\emph{Left:} Location of LMC SNRs on the MCELS H$\alpha$ mosaic, displayed 
logarithmically in greyscale. ``Likely-Ia'' and ``secured-Ia'' SNRs are marked 
by red circles and squares, respectively, while ``likely-CC'' and 
``secured-CC'' SNRs are shown in blue. Green circles are SNRs with undecided 
type. \emph{Right:} Same as left on a red continuum image from the SHASSA 
survey.
}
\label{fig_results_distribution_optical}
\end{figure*}

Because (core-collapse) SNRs are themselves very good indicators of recent star
formation, the distribution of many SNRs around the edge of LMC~4 is a further
sign of the important role played by SGSs in triggering star formation. In
turn, this could be used to look for \emph{new} SNRs: The high number of
remnants around LMC~4 is explained in part by the large size of the SGS
($\sim 1.2$~kpc), but also by the good X-ray coverage (only two out of twelve 
SNRs around LMC~4 were not observed with \xmm). Exploring SGSs less well 
studied, \eg in the west and south-west regions of the LMC, is promising, as we 
discuss in Sect.\,\ref{summary}.

Another prominent \ion{H}{I} feature is the density enhancement in the east
that extends southwards into ``arms B and E'' \citep[see Fig.~1 of 
][]{2003MNRAS.339...87S}, which are interpreted as tidal features. Most of the 
SNRs in the south-east of the LMC are associated to the 30~Doradus complex 
(which itself might be a manifestation of tidal shear). Only a handful of 
sources are known in the regions of the B and E arms \citep[and a single SNR is 
confirmed south of a declination of $-71$\textdegree,][]{2013MNRAS.432.2177B}. 
The southern region of the LMC is poorly studied in X-rays, preventing 
conclusions regarding the dearth of SNRs observed there. However, it could be an 
interesting target for future studies (Sect.\,\ref{summary}).

In Fig.\,\ref{fig_results_distribution_optical}, we show the position of SNRs 
relative to H$\alpha$ (left, MCELS data), and to a red continuum image from the 
SHASSA survey \citep{2001PASP..113.1326G}. \revision{The association of many 
CC-SNRs with large \ion{H}{II} regions, which trace regions of active star 
formation, is evident: Out of 31 secured or ``likely'' CC SNRs, 25 correlate 
with strong H$\alpha$ emission, three with moderate H$\alpha$ emission, and only 
three are not associated to large optical nebulosities.} In contrast, many SNRs 
are not associated to H$\alpha$ emission, \eg in the Bar, or south-east and 
north-west of it. These are the type Ia SNRs\,: \revision{22 out of 24 secured 
or ``likely'' Ia SNRs have no coincident H$\alpha$ emission (except the 
remnant's emission itself). Only N103B is spatially associated to strong 
H$\alpha$ emission, although it is plausibly a projection effect, with the SNR 
on the far side of the LMC (as suggested by its large ``$N_H$ fraction'', see 
Sect.\,\ref{results_distribution_adding}). This would resolve the long-standing 
issue of the association of a type Ia SNR with a region of intense star 
formation \citep{1988AJ.....96.1874C,2009ApJ...700..727B}. The type Ia SNRs are} 
in regions of relatively high stellar density (\eg the Bar, as traced in the 
red continuum image) but are also present in more isolated, less active 
regions, where intermediate- and old-age stellar populations dominate.

\subsection{Adding the third dimension} 
\label{results_distribution_adding}

So far, we discussed the 2-D distribution of SNRs, projected on the sky. It is 
possible to gain a rudimentary sense of depth, by comparing the absorbing 
column density derived from X-ray observations (hereafter $N_{H} ^{\ X}$), to 
the line-of-sight integrated \ion{H}{I} column density, derived from 21~cm 
observations (hereafter $N_{H} ^{\mathrm{\,21\,cm}}$). We recall that $N_{H} 
^{\ X}$ is an \emph{equivalent} neutral hydrogen column density assuming a given 
chemical composition\,\footnote{X-rays are absorbed not only by \ion{H}{I}, but 
also by molecular hydrogen, helium, and metals \citep{2000ApJ...542..914W}.}. 
The ratio $N_{H} ^{\ X} / N_{H} ^{\mathrm{\,21\,cm}}$ (hereafter ``$N_H$ 
fraction'') is a measurement of how deep an SNR is with respect to the 
\ion{H}{I} structure. Interpreting the $N_H$ fraction is made easier by the 
favourable orientation of the LMC. Neutral hydrogen is mainly distributed in a 
nearly circular disc at a moderate inclination angle\,\footnote{
Measurements of the orientation of the disc, \ie inclination $i$ (with 
0~\textdegree\ defined as face-on) and position angle of line of nodes $\Theta$ 
(the intersection of the disc and sky planes, measured eastwards of north), are 
widely scattered but are in the range 25~\textdegree$< i <$ 40~\textdegree\ and 
120~\textdegree$< \Theta <$ 155~\textdegree\ 
\citep[][]{1997macl.book.....W,2013A&A...552A.144S,2014ApJ...781..121V}.
}
, with a thickness of $\sim 360$~pc \citep{1999AJ....118.2797K}. Small $N_H$ 
fractions ($\lesssim 0.3$, \eg when $N_{H} ^{\ X}$ is consistent with zero) 
indicate that the SNR is well in front of the disc; intermediate values (0.3 to 
0.8) are expected from sources within the disc; high fractions (0.8--1.2; a 
value of 1.23 is expected when including contributions of neutral and 
singly ionised helium, \citealt{1999ApJ...510..806A}) are associated to 
remnants 
on the far side, or behind, the disc. Values significantly above 1.2 are 
discussed below.

\begin{figure}[t]
\includegraphics
[width=0.95\hsize]{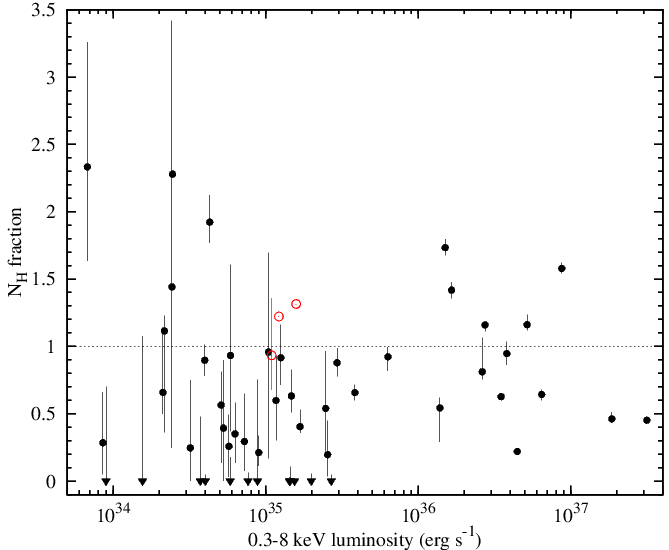}
\caption[\ $N_H$ fraction $= N_{H} ^{\ X} / N_{H} ^{\mathrm{\,21\,cm}}$ as
function of broad-band X-ray luminosity]{$N_H$ fraction $= N_{H} ^{\ X} / N_{H}
^{\mathrm{\,21\,cm}}$ as function of broad-band X-ray luminosity (see text for
details). Downward pointing arrows indicate upper limits, for objects with
$N_{H} ^{\ X}$ consistent with 0. SNRs covered with \chandra\ are shown in red.
}
    \label{fig_results_distribution_nH_Lx}
\end{figure}

\begin{figure*}[ht]
    \centering
\includegraphics
[width=0.85\hsize]{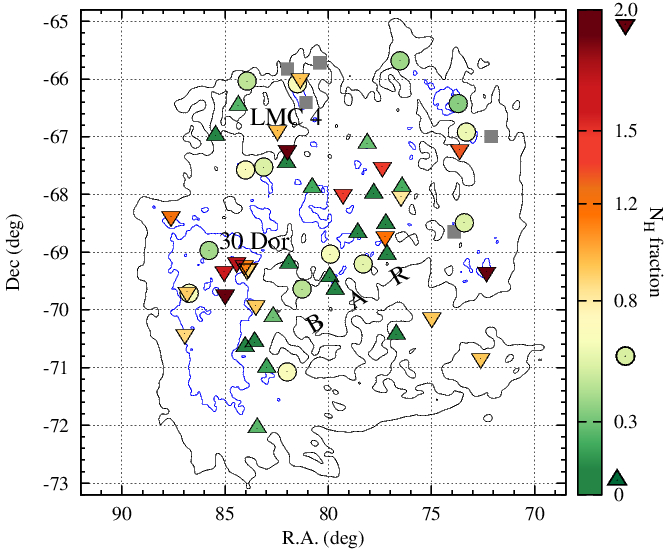}
\vspace{-0.2cm}
\caption[\ ``Pseudo-3D'' distribution of LMC SNRs, using $N_H$ fractions as
indicators of location along the line of sight]{``Pseudo-3D'' distribution of
LMC SNRs, using $N_H$ fractions (quantified by the colour bar) as indicators of
location along the line of sight. Objects ``in front of the disc'' ($N_H$
fraction $< 0.3$) are marked by upward pointing triangles; downward pointing
triangles are used for those ``behind the disc ($N_H$ fraction $> 0.8$). Objects
within the disc (0.3 to 0.8) are marked by dots. The black and blue contours
delineate \ion{H}{I} column densities of 1 and $3 \times 10^{21}$~cm$^{-2}$,
respectively (same as in Fig.\,\ref{fig_results_distribution_HI}). Prominent 
LMC structures are labelled.
}
\label{fig_results_distribution_nH}
\end{figure*}

$N_{H} ^{\ X}$ is taken from the spectral results of 
Sect.\,\ref{results_spectra}. For the 1T/2T sample, the adopted value is simply 
that in Table~\ref{appendix_table_spectra_all}. Only two 2T remnants have two 
different absorption components: For MCSNR~J0517$-$6759 we used the higher 
value. For MCSNR~J0535$-$6602 (N63A), the highly absorbed component is 
ejecta-rich and has a lower EM; we therefore adopted the (lower) $N_{H}$ of the 
ISM component, which is more representative. For the brightest SNRs, we adopted 
the best-fit values given in Table~\ref{appendix_table_spectra_brightest} and in 
Sect.\,\ref{results_spectra_1987A} (for SNR~1987A). For three SNRs with 
\chandra\ data only, we obtained $N_{H} ^{\ X}$ from the same references as in 
Sect.\,\ref{results_XLF}. Five remaining SNRs have either no or only ROSAT data 
available, and are not used in this analysis.

$N_{H} ^{\mathrm{\,21\,cm}}$ is measured from the map of 
\citet{2003ApJS..148..473K}, by averaging the column density around each SNR
over a 5\arcmin\ radius (the resolution of the map is about 1\arcmin, \ie 
$\sim 15$~pc). We checked
that using a smaller averaging radius, closer to the typical SNRs size, gave
essentially the same results. We then computed the ratio, propagating only the
uncertainties on $N_{H} ^{\ X}$ since they should dominate the error budget in
most cases. $N_H$ fractions are plotted against $L_X$ in 
Fig.\,\ref{fig_results_distribution_nH_Lx}. No correlation is evident, as 
expected: $L_X$ depends mostly on the evolutionary state of the remnant, while 
the depth within the LMC does not. At lower luminosities, however, there are 
more remnants with only upper limits on $N_{H} ^{\ X}$ (and thus on the $N_H$ 
fraction). This likely stems from the difficulty of deriving $N_{H} ^{\ X}$ from 
limited X-ray statistics. For the same reason, the error bars are larger in the 
handful of cases below a few $10^{34}$~erg~s$^{-1}$, and the sense of depth 
provided by the $N_H$ fraction becomes blurry.

In Fig.\,\ref{fig_results_distribution_nH}, the $N_H$ fraction is projected on 
the sky, on the same field of view as showed in 
Figs.\,\ref{fig_results_distribution_HI} and 
\ref{fig_results_distribution_optical}. Prominent LMC structures are labelled, 
including the LMC Bar: The bar is traced by the stars, both in young and 
intermediate-age populations 
\citep[][respectively]{1972VA.....14..163D,2001AJ....122.1827V}. It was found to 
be on the near side of the LMC, ``floating'' $\sim 0.5$~kpc to 5~kpc above the 
plane of the disc, as evidenced from near-infrared star count maps and distances 
to Cepheids, red clump, and RR Lyrae stars 
\citep{2000ApJ...545L..35Z,2004ApJ...601..260N,
2009AJ....138....1K,2012AJ....144..106H}. This interpretation is challenged by
red clump stars distance measurements of \citet{2009ApJ...703L..37S} and 
\citet{2013A&A...552A.144S}. \citet{2004ApJ...614L..37Z} proposed an alternative 
model, where the Bar is a stellar bulge (with a $z$ scale height of 2.5--3~kpc) 
whose south-eastern part is obscured by the gas disc. Consequently, the 
photometric centre is offset (in the plane of the sky), and distance 
measurements are biased to stars in the near side of the bulge. As can be seen 
in Fig.\,\ref{fig_results_distribution_nH}, SNRs in the Bar region are primarily 
on the near side (low $N_H$ fraction). Since some of these remnants must 
originate from the stellar population of the Bar, this lends support to previous 
findings that the Bar is indeed ``floating'' in front of the disc. One advantage 
of our method is that it does not need distance measurements of both disc and 
Bar objects; it directly gives locations \emph{relative} to the disc. In the 
bulge model of \citet{2004ApJ...614L..37Z}, SNRs in the Bar, but behind the 
disc\footnote{The obscuring effect by the disc on X-rays is moderate, not 
sufficient to mask SNRs as it does on stars in optical surveys.}, should have 
large $N_H$ fractions, while some scatter should be found along the line of 
nodes, where the disc and bulge intersect. Unfortunately, there are too few SNRs 
known in the Bar region to adequately test this alternative model.

The remnants in the 30 Doradus region and directly south of it 
(MCSNR~J0540$-$6920 and J0540$-$6944) are the most absorbed, both in absolute 
and relative terms (largest $N_{H} ^{\ X}$ and largest $N_H$ fractions). From 
distance measurements with red clump stars, \citet{2009AJ....138....1K} found 
that 30 Dor was further away, although it was noted that this could be an effect 
of 30 Dor being next to the Bar floating in front of the disc. With our analysis 
it is confirmed that not only does 30 Dor lie at a larger distance compared to 
neighbouring features, but is indeed \emph{behind} the plane of the gas disc.

Finally, it is striking from Figs.~\ref{fig_results_distribution_nH_Lx} and
\ref{fig_results_distribution_nH} that a few SNRs have an $N_H$ fraction in 
excess of 1.2, and up to 2.3. The extra absorption is likely to come from 
molecular hydrogen in front of the object \citep{1999ApJ...510..806A}. We show 
in Fig.\,\ref{fig_results_distribution_HI} CO contours from the NANTEN
survey \citep{2008ApJS..178...56F}. CO is used as a tracer of molecular
hydrogen. In the east of the LMC there are large regions of molecular gas,
following the peak density in \ion{H}{I}. In most cases with large $N_H$
fractions, we find nearby (less than a few arcmin away in projection) CO clouds, 
using either the NANTEN catalogue or the higher resolution MAGMA survey 
\citep[][respectively]{2008ApJS..178...56F,2011ApJS..197...16W}. We stress that
this does not imply that the remnants and the molecular clouds are physically
connected, but can be merely a projection effect, with the remnant behind, and 
not interacting with, the molecular cloud. However, physical interactions can 
happen, as exemplified by the case of MCSNR J0517$-$6759, where secondary 
evidence hints at a physical connection \citep{2014A&A...561A..76M}.

\section{Summary and outlooks}
\label{summary}
We have studied the X-ray emission of the rich population of SNRs in the LMC, 
using data from the \xmm\ observatory. We compiled a sample of 59 definite SNRs, 
cleaned of misclassified objects and doubtful candidates. \xmm\ data are 
available for the vast majority (51 SNRs) of the sample, which called for a 
homogeneous re-analysis of the X-ray spectra of the entire population. This 
alleviates the inconsistencies in spectral models and analysis methods used, and 
allows meaningful comparisons of, \eg, temperature, chemical composition, and 
luminosity of SNRs. The main outcomes of this systematic spectral analysis are 
the following:
\begin{itemize}
\item First, it provides the best census of LMC remnants with an Fe~K
line ($\approx 13$~\% of the sample), which is a powerful tool to retrieve the
type of SN progenitor.
\item Second, it reveals the contribution to the X-ray emission by hot SN
ejecta for 23 SNRs ($\approx 39$~\% of the sample). Since the abundance ratios
measured in the ejecta components reflect the nucleosynthesis yields of either
type Ia and CC SNe, this is of great help for the typing of a substantial
fraction of the sample.
\item And third, it allows us to select 16 SNRs ($\approx 27$~\% of the sample)
where the X-ray emission is dominated by swept-up ISM. In these objects, the
fitted abundances provide a measurement of chemical abundances in the gas phase
of the LMC ISM. A metallicity of {[Fe/H]} $= -0.46(_{-0.18}^{+0.13})$~dex is
found based on \xmm\ SNRs. Light $\alpha$-elements (O, Ne, Mg) have lower
abundance ratios {[}$\alpha$/Fe{]} than in the Milky Way. Although this general
result was previously known, one can now study abundance ratios within the LMC
as function of age. In comparison to old clusters ($\sim 10$~Gyr) and red giant
stars (1~Gyr and older), the relatively young gas phase ISM ($\lesssim 100$~Myr)
has a higher metallicity [Fe/H] and lower {[}$\alpha$/Fe{]} (in particular
[O/Fe]). This reflects the continued enrichment by type Ia SNe in the last $\sim
1$~Gyr, which injected large amounts of Fe back in the ISM.
\end{itemize}

We devised a \revision{quantitative way} to tentatively type all LMC SNRs, based 
on their local SFHs and stellar environments, combined with spectral information 
(\ie detection of SN ejecta, when present). We calibrated this method with SNRs 
having a well-established type based on robust indicators. The resulting ratio 
of CC to type Ia SNe that exploded in the LMC over the last few $10^4$~yr (\ie 
very close to the current ratio of CC/Ia \emph{rates}) is \ccIa~$= 
1.35(_{-0.24}^{+0.11})$. This is lower than the ratio typically measured in 
local SNe surveys and in galaxy clusters. After arguing that SNRs of both types 
might be absent from the sample (\ie the current sample is not biased towards 
one type only), we concluded that the low \ccIa\ ratio is a consequence of the 
specific SFH of the LMC, and particularly the enhanced star formation episodes 
that occurred 500~Myr and 2~Gyr ago. Because the majority of type Ia SNe explode 
within 2~Gyr after star-forming episodes, we are coincidentally observing the 
LMC at a time when the type Ia SN rate is high. Integrated over an SNR lifetime, 
this results in the relatively low \ccIa\ observed.

We also assessed the spatial distribution of SNRs with respect to cool gas 
(traced by \ion{H}{I} and molecular emission), star-forming regions (H$\alpha$), 
and stars (red continuum). A concentration of SNRs around the edge of the SGS 
LMC~4 exemplifies the role of SGSs in triggering star formation. The column 
density $N_{H} ^{\ X}$ obtained during the X-ray spectral analysis of the whole 
sample, when compared to the \ion{H}{I} column density, provides a measurement 
of the position of each SNR relative to the \ion{H}{I} structure. Since most of 
the neutral gas lies in a well-defined thin disc seen at a moderate inclination 
angle, the fraction $N_{H} ^{\ X} / N_{H} ^{\mathrm{\,21\,cm}}$ is a good 
indicator of the depth along the line-of-sight, revealing the ``pseudo-3D'' 
distribution of SNRs in the LMC. Previous studies found that the Bar is 
``floating'' in front of the disc, but this statement was challenged by some 
authors. Our analysis shows that SNRs in the Bar regions are primarily on the 
near side (low $N_H$ fraction), lending support to the foreground location of 
the Bar.

Finally, we compared the populations of SNRs in Local Group galaxies via their
X-ray luminosity function. The XLF of SNRs in the SMC, M31, and M33 are
relatively homogeneous over all the observed luminosity range, although that of
the SMC is flatter. The LMC XLF is remarkable by its prominent bright end. The
largest population of SNRs brighter than $L_X > 10^{36}$~erg~s$^{-1}$ is found
in the LMC (13 SNRs vs. 8 and 7 in M31 and M33, respectively). This is possibly
an effect of the lower metallicity in the LMC: Massive stars have smaller mass
loss rates (less heavy elements to drive stellar winds) and the interaction of
SN ejecta with less massive CSM shells produce brighter remnants. 
The number of SNRs brighter than $10^{35}$~erg~s$^{-1}$ in the LMC is
comparable to that in M31 and M33, likely owing to its high recent SFR and high
current type Ia SN rate. The LMC XLF flattens significantly because of
incompleteness: Many X-ray-faint SNRs have been missed so far, due to the
incomplete coverage of the LMC with sensitive X-ray instruments (\ie\ \chandra\
or \xmm).

This work presents the state of the art on X-ray emission of SNRs in the LMC.
However, it is clear that the \emph{current} sample is incomplete, as evidenced 
by the flattening of the X-ray luminosity function of LMC SNRs 
(Sect.\,\ref{results_XLF}). In the last 15 years, new SNRs were confirmed or 
discovered in the LMC at an almost constant rate (one or few per year), 
principally using X-ray observations. There is no indication that this trend 
will stop in the near future, so that more observations of the LMC will 
increase the sample of SNRs.

Nevertheless, the observing time of major observatories is limited and
expensive. We conclude this work by offering several strategies to maximise the 
chance of finding ``missing'' SNRs:
\begin{itemize}
    \item As shown in Sect.\,\ref{results_distribution}, star formation is 
intense around the SGS LMC~4, and the edges of the shell abound in SNRs. Many 
LMC SGS have not been (fully) surveyed by \xmm, for instance \citep[in the 
notation of][]{1999AJ....118.2797K} SGS~3 and 6 in the north, SGS~2 and 5 in 
the west, and SGS~4 in the \revision{south}. Targeting in particular SGSs 
associated to star formation (\eg with \ion{H}{II} region along the rims) 
warrants successful SNR searches.
    \item The follow-up observations of X-ray-selected candidates (usually 
ROSAT sources) with \xmm\ have been extremely successful. Such programmes should 
be continued until completion of the list of candidates.
    \item Even the ROSAT (targeted) survey of the LMC was not covering the LMC
up to its outskirts. To find SNRs in these regions, the future \emph{eROSITA}
survey \citep{2012arXiv1209.3114M} will be most useful, covering the full sky in
the 0.5~keV~--~8~keV band. The LMC is located close to the South Ecliptic Pole
and will be observed with a deeper exposure than the rest of the sky. Looking
for new SNR candidates, especially evolved X-ray-only SNRs, will be of special
interest.
\end{itemize}

Even in \emph{existing} data, some SNRs might be as yet unrecognised. There is
significant diffuse emission from large-scale structures of the hot ISM in the
LMC \citep{2002A&A...392..103S}, which is seen in greater spatial and spectral
detail by \xmm. By looking for ejecta-enhancement, it might be possible to 
distinguish old SNRs with low surface brightness hiding in the diffuse emission.

Finding new SNRs is desirable. Individual objects of special interest are often 
found serendipitously, without prior knowledge of their exciting nature. The 
evolved type Ia SNRs presented in \citet{2014A&A...561A..76M} and 
\citet{2014MNRAS.439.1110B}, are good examples; the discovery of the SNR around 
the Be/X-ray binary SXP~1062 is another one 
\citep{2012MNRAS.420L..13H,2012A&A...537L...1H}. Furthermore, as demonstrated in 
this work, SNRs are powerful probes of the ISM of their host galaxies. With more 
SNRs where metallicity can be measured, we will obtain a more accurate knowledge 
of the chemical composition of the hot ISM or better assess its homogeneity.

\begin{acknowledgements}
\revision{We thank the anonymous referee for carefully reading this rather long 
manuscript and providing us with comments and suggestions to improve it.}
The \xmm\ project is supported by the Bundesministerium f\"ur Wirtschaft und 
Technologie\,/\,Deutsches Zentrum f\"ur Luft- und Raumfahrt (BMWi/DLR, FKZ 50 OX 
0001) and the Max-Planck Society.
P.\,M. acknowledges support by the Centre National d’\'Etudes Spatiales (CNES) 
and from the BMWi/DLR grant FKZ 50 OR 1201. M.\,S. acknowledges support by the 
Deutsche Forschungsgemeinschaft through the Emmy Noether Research Grant SA 
2131/1-1. P.\,K. acknowledges support from the BMWi/DLR grant FKZ 50 OR 1309.
This research has made use of Aladin, SIMBAD and VizieR, operated at the CDS,
Strasbourg, France.
\end{acknowledgements}


\appendix

\section{The EPIC background}
\label{appendix_background}
The signal recorded in an \xmm\ observation comprises many components, which 
can be separated into three main groups\,: the X-ray emission of the target of 
the observation, an astrophysical X-ray background (hereafter AXB, \ie X-rays 
from any source located \emph{in projection} near the target), and an 
instrumental background. In this Appendix we describe the last two components.

\paragraph{The instrumental background of EPIC:} It consists of three 
components. The first is an \textit{electronic noise}, in the form of hot 
pixels/columns or read-out noise. In the case of EPIC-pn, the read-out noise 
increases dramatically below energies of $\lesssim 300$ eV, especially if 
double-pixel events are used.

The second component is the \textit{particle-induced background}, the spectrum
of which includes both a continuum and many lines. The continuum part is due to 
the quiescent particle background (QPB). Cosmic rays deposit a large amount of 
energy ($\gg$ 10 keV) in many adjacent pixels and are easy to distinguish from 
valid X-ray events. However, the unrejected fraction of direct and 
Compton-scattered cosmic rays produces a remaining continuum with a rather
flat spectrum and a rate of $0.021 \pm 0.0022$ events~cm$^{-2}$\,s$^{-1}$ for
the MOS cameras, and about twice as much for pn \citep{2002A&A...389...93L}. The
continuum is both \mbox{chip-,} position-, and time-dependent \citep[at least 
for MOS where it has been extensively studied, see][]{2008A&A...478..575K}. 
These variations have to be taken into account in the spectral analysis.
The line part of the particle background is composed of many X-ray fluorescence 
lines produced by the interaction of high-energy particles with the material 
surrounding the detectors (Al, Ni, Cu, etc.). Due to this origin the 
fluorescence line component varies with time. This component is highly 
position-dependent, mirroring the distribution of the camera material around the
detectors \citep{2002A&A...389...93L,2004SPIE.5165..112F,2008A&A...478..575K}.

The third component is the so-called \textit{soft proton contamination} (SPC).
Low-energy protons, accelerated in the Earth magnetosphere and focused by the
X-ray telescopes onto the detectors, produce events that cannot be
distinguished from genuine X-ray events. The soft proton flux has a highly
time-variable, ``flaring'' nature. At times of the strongest flares, most of
the data are unusable anyway (except in the case of a very bright target). But
soft proton flares can occur on longer time scales, at lower amplitudes. These
time intervals are typically used for science, though they include a small but
potentially important contamination by soft protons (hence the term ``SPC'').
The flaring spectrum was found by \citet{2008A&A...478..575K} to be rather flat,
with a small roll-off at high energy. The same authors showed that the stronger
the flare, the flatter (\ie harder) the SPC spectrum.

The contribution of the instrumental background will be relatively higher in 
spectra of sources with low surface brightness and must be taken into account, 
i.e. modelled. To do so, we use the FWC data 
(Sect.\,\ref{observations_reduction}).
Several hundred kiloseconds worth of data are now available, providing a good 
knowledge of the spectrum of the instrumental background. As part of his PhD 
thesis, Richard Sturm (\citeyear{2012PhDT......ppppS}) developed an empirical 
model of the EPIC-pn FWC data. This includes an exponential decay (modified by a 
spline function), a power law, and a combination of Gaussian lines to account 
for the electronic noise, QPB, and instrumental lines, respectively. In 
addition, two smeared absorption edges to the continuum are included.

We extended his work and developed a similar model for the EPIC-MOS FWC 
spectra. This allows to analyse jointly the pn and MOS spectra of LMC SNRs and 
take advantage of the better spectral resolution of MOS. There is no low-energy 
noise as for EPIC-pn, so no exponential decay function is needed. Satisfactory 
results were obtained with a broken power law for the continuum, leaving the 
slope of the two segments as well as the energy of the break free. A smeared 
absorption edge around $E\approx 0.53$~keV (K edge of oxygen) improved the fit 
and was included. A set of Gaussian are used to model the fluorescence lines. 
The materials of the MOS and pn cameras are different, and so is the observed 
fluorescence line pattern. Both have a strong Al K line at 1.49~keV. MOS 
background also features a strong Si K line at 1.74~keV, as opposed to pn, 
where the strongest line is Cu K at 8.05~keV. Other lines from Au, Cr, Mn, Zn, 
Cu, Fe, and Ni are detected \citep{2002A&A...389...93L,2008A&A...478..575K} and 
included in the MOS detector background model.

Note that the same screening and filtering criteria used for the science data 
are applied to the FWC data, including the vignetting correction with 
\texttt{evigweight} (Sect.\,\ref{data_spectra_extraction}). Formally, the 
instrumental background is \emph{not} subjected to vignetting, which is an 
effect of the telescope on \emph{photons}. However, by applying a vignetting 
correction to the science data, one assigns weights to genuine X-ray events as 
well as to particle background events, since one cannot \emph{a priori} 
distinguish the two type of events. Therefore, one needs to vignetting-correct 
the FWC data to make sure that the FWC spectra, used for the modelling of the 
instrumental background contribution to the science data, have been processed 
in the same way as the latter. At a given position on the detector, 
\texttt{evigweight} will assign heavier weights to photons with higher 
energies, an effect that can be easily accounted for in the background model. 
To do so, we added a spline function to the pn and MOS instrumental background
models. This reproduces the effect of the vignetting correction, which
``overweights'' events above 5 keV, if they have been recorded at significant
off-axis angles.

\paragraph{The astrophysical X-ray background:} The AXB can usually be modelled 
with four or less components \citep{2008A&A...478..615S,2010ApJS..188...46K}. 
The soft part of the AXB ($E \lesssim 2$~keV) has mostly a thermal emission 
spectrum and originates from various regions/hot plasmas. The Local Hot Bubble 
(LHB) is a region in the solar neighbourhood filled with million-degree plasma 
\citep[$kT \approx 85$~eV,][]{2008ApJ...676..335H}. This component was modelled 
with an unabsorbed APEC model. Emission from the Galactic halo comprises a cool 
($kT \sim 0.1$~keV) and warm ($kT \sim 0.25$~keV) thermal component. Since the 
cool component is mostly absorbed by the foreground Galactic absorbing column, 
we did not include it in the AXB model to keep it as simple as possible. For 
the warm component we used an absorbed APEC model.

Above 2~keV, the background is mostly from the cosmic X-ray background (CXB), a
superposition of unresolved distant objects, in other words AGN. This component
has an absorbed power-law spectrum, with a photon index fixed at $\Gamma = 1.46$
\citep{1997MNRAS.285..449C}. To account for the non-uniformity of this
component, the normalisation is a free parameter in our background model.

\begin{figure}[t]
    \centering
\includegraphics
[angle=-90,width=0.995\hsize]{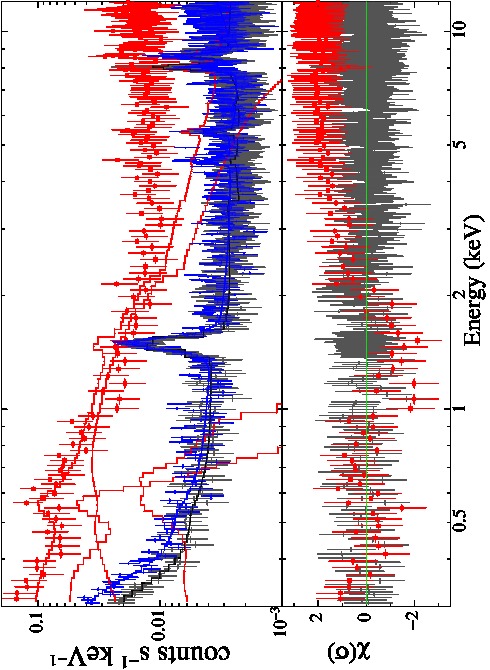}
\caption[\ Example of a strong SPC affecting the spectral analysis of MCSNR
J0529$-$6653]{Example of a strong soft proton contamination (SPC) affecting the
spectral analysis of MCSNR J0529$-$6653 (ObsID 0700381101). The instrumental
background, extracted from FWC data, is shown in the top panel in grey and blue.
The spectrum extracted from the \bg region is shown in red. The strong flat tail
above 2~keV is the SPC, which the instrumental + AXB model cannot account for.
}
\label{fig_appendix_background_SPC}
\end{figure}

\paragraph{Note: Cases with a problematic background} Several situations can
occur where the instrumental and X-ray backgrounds cannot be properly accounted
for at first, hindering the analysis of the SNR emission.\\
    \textbullet\ Bad extraction region: If the SNR is very bright, a background
spectrum extracted too close to the SNR will include contamination from the
telescope PSF wings. Alternatively, a nearby bright X-ray source can produce
artefacts such as singly reflected photons up to large angular distances
($\sim$1~\textdegree). When this happens, we selected the background in another
region better suited.\\
    \textbullet\ Anomalous MOS states: The model for the instrumental 
background of the MOS detectors was developed for data obtained in the normal 
state; \citet{2008A&A...478..575K} identified periods of ``anomalous'' 
background of EPIC-MOS. The instrumental background spectrum of this anomalous 
state is markedly stronger below 1~keV, which complicates the analysis of 
observations obtained during these epochs. If the SNR affected was faint, 
including the MOS data usually does not add much information, and only the pn 
spectrum was used. Conversely, if the SNR is bright, the error induced by 
fitting the anomalous spectra with the standard model is unimportant, because
the source count rate is much higher than the instrumental background, and this
issue can be discarded.\\ 
    \textbullet\ Soft proton contamination: Observations affected by a strong 
SPC are easily identified by fitting the FWC and astrophysical backgrounds 
spectra together. Indeed, the background above 2~keV is almost purely 
instrumental. Therefore, a residual component at high energy in the background 
spectrum (extracted from science observation) that is not present in the FWC 
spectrum betrays the SPC. An extreme example is shown in
Fig.\,\ref{fig_appendix_background_SPC}. An extra component needs 
to be added to the X-ray background and SNR models to account for the SPC.

\begin{table*}[ht]
\caption{LMC remnants with a secured SN classification}
\label{appendix_table_securedSNRs}
\centering
\begin{tabular}{@{\hspace{0.25cm}}l @{\hspace{0.25cm}} @{\hspace{0.25cm}} l 
@{\hspace{0.25cm}}
@{\hspace{0.25cm}} c @{\hspace{0.25cm}} @{\hspace{0.25cm}} c @{\hspace{0.25cm}}
@{\hspace{0.25cm}} c @{\hspace{0.25cm}}}
\hline\hline
\noalign{\smallskip}
MCSNR & Other name & Age & Evidence\tablefootmark{a} &
References\tablefootmark{b}\\
      &           & (yr)&          &     \\
\noalign{\smallskip}
\hline
\noalign{\smallskip}
\multicolumn{5}{c}{Core-collapse SNRs}\\
\noalign{\smallskip}
\hline
\noalign{\smallskip}
J0536$-$6916 & SNR~1987A & 28 & Historical & {\citet{2012A&A...548L...3M}}  \\
J0540$-$6920 & B0540$-$693 & $\sim$1600 & Pulsar & \citet{2001ApJ...546.1159K}\\
J0525$-$6938 & N132D & $\sim$3150 & Ejecta, morphology &
    \citet{2007ApJ...671L..45B} \\
J0535$-$6602 & N63A & 2000$-$5000 & Ejecta, Fe~K & \citet{2003ApJ...583..260W}\\
J0536$-$6913 & --- & 2200$-$4900 & Ejecta &  {\citet{2015A&A...573A..73K}} \\
J0505$-$6802 & N23  & $\sim$4600 & Ejecta, CCO, morphology &
    \citet{2006ApJ...645L.117H} \\
J0526$-$6605 & N49 & $\sim$4800 & Ejecta, SGR, Fe~K &
    \citet{2012ApJ...748..117P} \\
J0537$-$6910 & N157B & $\sim$5000\,\tablefootmark{c} & PWN  &
    \citet{2006ApJ...651..237C}  \\
J0525$-$6559 & N49B & $\sim$10000 & Ejecta, morphology &
    \citet{2003ApJ...592L..41P} \\
J0536$-$6735 & DEM L241 & $> 10^4$\,\tablefootmark{c} & HMXB &
    \citet{2012ApJ...759..123S} \\
J0453$-$6829 & B0453$-$685 & 12000$-$15000 & PWN, morphology &
    {\citet{2012A&A...543A.154H}} \\
J0531$-$7100 & N206 & $\sim$25000 & PWN candidate, morphology &
    \citet{2005ApJ...628..704W} \\
\noalign{\smallskip}
\hline
\noalign{\smallskip}
\multicolumn{5}{c}{Type Ia SNRs}\\
\noalign{\smallskip}
\hline
\noalign{\smallskip}
\multirow{2}*{J0509$-$6731} & \multirow{2}*{B0509$-$675} &
\multirow{2}*{400$\pm$120} & Light echo, ejecta &   
\multirow{2}*{\citet{2008ApJ...680.1137R}} \\
 & & & Fe~K, morphology & \\
\cline{4-4}\noalign{\smallskip}
J0509$-$6844 & N103B & 860 & Ejecta, Fe~K & \citet{1995ApJ...444L..81H} \\
J0519$-$6902 & B0519$-$690 & 600$\pm$200 & Ejecta, Fe~K, morphology &
    \citet{1995ApJ...444L..81H}  \\
J0505$-$6753 & DEM L71 & $\sim$4700 & Ejecta. Fe~K, morphology &
    \citet{1998ApJ...505..732H,2003ApJ...582L..95H} \\
J0547$-$7025 & B0548$-$704 & $\sim$7100 & Ejecta & \citet{2003ApJ...593..370H},
    this work \\
J0534$-$6955 & B0534$-$699 & $\sim$10000 & Ejecta, morphology &
    \citet{2003ApJ...593..370H}\\
J0534$-$7033 & DEM L238 & $\sim$13500 & Ejecta & \citet{2006ApJ...652.1259B} \\
J0536$-$7039 & DEM L249 & $\sim$15000\,\tablefootmark{c} & Ejecta &
    \citet{2006ApJ...652.1259B} \\
J0506$-$7026 & {[HP99] 1139} & 17000$-$21000 & Ejecta &
    Kavanagh et al. (in prep.), this work \\
J0508$-$6902 & {[HP99] 791} & 20000$-$25000 & Ejecta &
    \citet{2014MNRAS.439.1110B} \\
J0527$-$7104 & {[HP99] 1234} & $\sim$ 25000 & Ejecta &
   {\citet{2013A&A...549A..99K}}, this work \\
J0547$-$6941 & DEM L316A & $\sim$27000\,\tablefootmark{c} & Ejecta &
    \citet{2005ApJ...635.1077W} \\
J0511$-$6759  &  \multicolumn{1}{c}{---} & $\gtrsim 20000$\,\tablefootmark{c}  &
    Ejecta & {\citet{2014A&A...561A..76M}} \\
J0508$-$6830 & \multicolumn{1}{c}{---} & $\gtrsim 20000$\,\tablefootmark{c} &
    Ejecta & {\citet{2014A&A...561A..76M}} \\
\noalign{\smallskip}
\hline
\end{tabular}
\tablefoot{Ages for the first three type Ia SNRs are from light echo
measurements \citep{2005Natur.438.1132R}.\\
\tablefoottext{a}{Morphology: Typed from X-ray morphology by
\citet{2009ApJ...706L.106L,2011ApJ...732..114L}. Fe~K: Typed from the
properties of the Fe~K emission by \citet[][see also
Sect.\,\ref{results_spectra_FeK}]{2014ApJ...785L..27Y}.
CCO: Central compact object. PWN: Pulsar wind nebula. SGR: Soft gamma-ray
repeater.}\\
\tablefoottext{b}{Because of the multiple studies on most remnants, the
given references are ``see {[}...{]} and references therein''.}\\
\tablefoottext{c}{Uncertain age.}
}
\end{table*}

\section{Selection of SNRs with secured classifications}
    \label{appendix_secured}
The list of LMC remants with secured CC and type Ia classifications
is given in Table~\ref{appendix_table_securedSNRs}.

\subsection{Type Ia SNRs}
    \label{appendix_secured_typeIa}

\paragraph{The spectacular case of MCSNR J0509$-$6731:} One of the few SNRs
less than a thousand years old, this object was first typed as a type Ia remnant
by \citet{1982ApJ...261..473T} based on the Balmer-dominated optical spectrum.
This classification was confirmed by the analysis of the \ASCA\ spectrum,
revealing ejecta emission rich in nucleosynthesis products of thermonuclear SNe
\citep{1995ApJ...444L..81H}. Finally, light echoes from the SN, scattered off
interstellar dust, were detected around four LMC SNRs
\citep{2005Natur.438.1132R}. Optical spectroscopy of the light echoes of MCSNR
J0509$-$6731 allowed \citet{2008ApJ...680.1137R} to determine the SN spectral
type as an overluminous 1991T-like SN Ia.

\paragraph{Balmer-dominated SNRs with X-ray-detected ejecta:}
\citet{1982ApJ...261..473T} also detected Balmer-dominated emission from MCSNR
J0519$-$6902, J0505$-$6753 (DEM L71), and J0547$-$7025, concluding that they
were produced by type Ia events. In the two former cases, the X-ray spectra
clearly showed emission from the ejecta of thermonuclear SNe
\citep{1995ApJ...444L..81H,2003ApJ...582L..95H,2003A&A...406..141V}. For
J0547$-$7025, the \chandra\ spectra revealed ejecta but the O/Fe ratio was not
as decisive \citep{2003ApJ...593..370H}. Furthermore, this remnant was an
outlier in \citet{2009ApJ...706L.106L,2011ApJ...732..114L}, with a morphology
more consistent with the sample of CC-SNRs. The observations of the \xmm\ survey
confirm the iron-rich nature of J0547$-$7025
(Table~\ref{appendix_table_spectra_all}), and therefore secure a
type Ia classification consistent with the optical data.

\paragraph{Middle-aged to evolved iron-rich SNRs:} Several remnants with ages
exceeding $10^4$~yr revealed iron-rich X-ray spectra (observed with \xmm\ and
\chandra) that betrayed their type Ia nature. MCSNR J0534$-$6955 was first
identified as such with \chandra\ \citep{2003ApJ...593..370H} and \xmm\
observations give similar results
(Table~\ref{appendix_table_spectra_all}). Slightly more evolved,
MCSNR J0534$-$7033 and J0536$-$7039 (DEM L238 and L249, respectively) have a
more pronounced separation of the shell and central iron-rich plasma
\citep{2006ApJ...652.1259B}. The shell A of DEM L316 (MCSNR J0547$-$6941) has
striking spectral differences to the very close neighbour MCSNR J0547$-$6943
(DEM L316B): the former is also mostly exhibiting Fe L-shell emission, which
leads to the interpretation that it is another type Ia remnant
\citep{2001PASJ...53...99N,2005ApJ...635.1077W}.

Then come three more evolved (age $\gtrsim 15$~kyr) iron-rich SNRs that were 
presented in \citet{2014A&A...561A..76M} and \citet{2014MNRAS.439.1110B}. 
Finally, since these publications, we obtained \xmm\ follow-up observations of 
two remnants, which we classified as type Ia: \emph{i)} MCSNR J0506$-$7026 was a 
\ROSAT-selected candidate ({[}HP99{]}~1139), which revealed a remnant similar to 
DEM L238 and L249, about 17-21~kyr old and holding about 0.9~\msun~-~1~\msun\ in 
the central region (Kavanagh et al., in prep., 
Table~\ref{appendix_table_spectra_all}); \emph{ii)} MCSNR J0527$-$7104 was 
confirmed by our group in a multi-wavelength study \citep{2013A&A...549A..99K}. 
A subsequent observation (performed 2014 May 31) revealed yet another iron-rich 
core (with an unusual morphology, see Kavanagh et al., in prep.), so that this 
source completes the (currently known) sample of LMC SNRs with a secured type Ia 
origin.

\subsection{Core-collapse SNRs}
    \label{appendix_secured_CC}

\paragraph{Remnants hosting a compact object:} 
Several neutron stars have been detected inside LMC SNRs, mostly powering a
pulsar wind nebula (PWN). MCSNR J0540$-$6920 is the prototypical example: It
hosts the pulsar PSR B0540$-$69 and is known as a twin of the Crab nebula
\citep{2001ApJ...546.1159K}. MCSNR J0537$-$6910 (N157B) is also dominated by a
PWN around PSR J0537$-$6910 \citep{2006ApJ...651..237C}. Fainter, less obvious
PWNe have been found in MCSNR J0453$-$6829
\citep{2003ApJ...594L.111G,2012A&A...543A.154H} and in MCSNR J0535$-$6602
\citep{2005ApJ...628..704W}. In the latter, the case for a PWN is not as strong;
however, analysis of the X-ray morphology using a power-ratio method
\citep{2009ApJ...706L.106L} confirm the classification as CC-SNR.

\chandra\ observations \citep{2006ApJ...645L.117H} of MCSNR J0505$-$6802 (N23)
revealed, in addition to regions with enhanced O-group elements, a point source
in the centre of the remnant that shows properties similar to compact central
objects (CCOs) seen in other CC-SNRs, such as Cas A. Finally, a point source was
detected in the ``Head'' of MCSNR J0536$-$6735 \citep[DEM L241, 
see][]{2006A&A...450..585B} using \xmm\ observations and first classified as 
candidate PWN. With \chandra\ observations, \citet{2012ApJ...759..123S} could 
show that the source was not extended and identified the optical counterpart as 
an O5III(f) star. Based on this and the X-ray variability and spectrum, they 
concluded that the SNR was hosting an HMXB, akin to SXP1062 in the SMC 
\citep{2012MNRAS.420L..13H}.

\paragraph{Detection of the remains of massive star nucleosynthesis:}
MCSNR J0525$-$6938 (N132D), the brightest SNR in the LMC, belongs to the class
of oxygen-rich remnants. Many clumps of X-ray emitting O-ejecta are detected in
X-ray observations \citep{2007ApJ...671L..45B} and match the optical ejecta
morphology seen by \textit{Hubble}. \citet{2003ApJ...592L..41P} have revealed a
highly enhanced Mg abundance and derived a high mass of Mg ejecta from MCSNR
J0525$-$6559 (N49B), strongly suggesting a massive stellar progenitor. This
classification is supported by the X-ray morphology of the remnant
\citep{2009ApJ...706L.106L}.

The nearby MCSNR J0526$-$6605 (N49) is a more puzzling case. No compelling
evidence for overabundant O or Fe is found, but Si- and S-rich ejecta features
are detected by \chandra\ \citep{2003ApJ...586..210P,2012ApJ...748..117P}. These
can be interpreted as explosive O-burning or incomplete Si-burning deep inside a
CC SN explosion; however, the Si/S ejecta mass ratio favour a type Ia origin
\citep{2003ApJ...586..210P,2012ApJ...748..117P}. The soft gamma-ray repeater SGR
0526-66 lies in projection in the remnant, favouring the CC-SNR scenario,
although the physical association between the remnant and the SGR is uncertain
\citep{2001ApJ...559..963G,2001ApJ...556..399K}. The best evidence to terminate
the debate over the nature of N49 comes from the properties of its Fe~K
emission, which is clearly in the region occupied by CC-SNRs \citep[][
Sect.\,\ref{results_spectra_FeK}]{2014ApJ...785L..27Y}. MCSNR J0535$-$6916 
(N63A) is another case where ejecta features are detected but cannot yield a 
definite classification \citep{2003ApJ...583..260W}. As for N49, however, the 
Fe~K emission allows to include N63A in the sample of secured CC-SNRs.

\paragraph{SNR~1987A:} Last but not least comes the remnant with the most
secured core-collapse classification of all, as the SN itself was observed in 
detail and its progenitor identified in pre-explosion images.


\onecolumn

\begin{landscape}

\section{Sample of SNRs in the LMC}
\label{appendix_sample}

\begin{longtable}[c]{l c c l c c c c r r c c}
\caption{SNRs in the LMC}
\label{appendix_table_snrs_sample}\\
\hline\hline
\noalign{\medskip}
 \multicolumn{1}{c}{MCSNR} & RA & DEC & \multicolumn{1}{c}{Other name} & X-ray
& $L_X$ & X-ray size & $N_H$ fraction & \multicolumn{1}{c}{$N_{\mathrm{OB}}$} &
\multicolumn{1}{c}{$r$} & \multicolumn{1}{c}{Hint-spec} & ref.\\
  & \multicolumn{2}{c}{(J2000)} & & data & &  & & & & & \\
 \multicolumn{1}{c}{(1)} & \multicolumn{1}{c}{(2)} & \multicolumn{1}{c}{(3)} &
\multicolumn{1}{c}{(4)} & \multicolumn{1}{c}{(5)} & \multicolumn{1}{c}{(6)} &
\multicolumn{1}{c}{(7)} & \multicolumn{1}{c}{(8)} & \multicolumn{1}{c}{(9)} &
\multicolumn{1}{c}{(10)} & \multicolumn{1}{c}{(11)} & \multicolumn{1}{c}{(12)} 
\\
\noalign{\medskip}
\hline
\endfirsthead
%
%
\caption[]{(continued)}\\
\hline
\hline
\noalign{\medskip}
 \multicolumn{1}{c}{MCSNR} & RA & DEC & \multicolumn{1}{c}{Other name} & X-ray
& $L_X$ & X-ray size & $N_H$ fraction & \multicolumn{1}{c}{$N_{\mathrm{OB}}$} &
\multicolumn{1}{c}{$r$} & \multicolumn{1}{c}{Hint-spec} & ref.\\
  & \multicolumn{2}{c}{(J2000)} & & data & &  & & & & & \\
 \multicolumn{1}{c}{(1)} & \multicolumn{1}{c}{(2)} & \multicolumn{1}{c}{(3)} &
\multicolumn{1}{c}{(4)} & \multicolumn{1}{c}{(5)} & \multicolumn{1}{c}{(6)} &
\multicolumn{1}{c}{(7)} & \multicolumn{1}{c}{(8)} & \multicolumn{1}{c}{(9)} &
\multicolumn{1}{c}{(10)} & \multicolumn{1}{c}{(11)} & \multicolumn{1}{c}{(12)} 
\\\noalign{\medskip}
\hline
\endhead
%
\hline
\endfoot
%
\hline
\endlastfoot
\noalign{\smallskip} %
  J0448$-$6700 & 04:48:22 & -66:59:52 & [HP99] 460 & R & 0.46 & 220 & --- & 3 &
1.67$_{-0.20}^{+0.92}$ & 3 & BGS06\\
\noalign{\smallskip} %
  J0449$-$6920 & 04:49:20 & -69:20:20 &   & X & 0.07 & 162 (85) &
2.33$_{-0.70}^{+0.93}$ & 26 & 6.29$_{-2.66}^{+2.33}$ & 3 & KPS10\\
\noalign{\smallskip} %
  J0450$-$7050 & 04:50:27 & -70:50:15 & B0450-709 & X & 0.59 & 340 (85) &
0.93$_{-0.56}^{+0.68}$ & 2 & 1.89$_{-0.39}^{+0.29}$ & 3 & MFT85\\
\noalign{\smallskip} %
  J0453$-$6655 & 04:53:14 & -66:55:13 & N4 & X & 1.17 & 256 (122) &
0.60$_{-0.29}^{+0.33}$ & 83 & 6.36$_{-2.85}^{+7.24}$ & 3 & SCM94\\
\noalign{\smallskip} %
  J0453$-$6829 & 04:53:38 & -68:29:27 & B0453-685 & X & 13.85 & 120 &
0.54$_{-0.25}^{+0.08}$ & 5 & 2.42$_{-0.39}^{+0.46}$ & 5 & LHG81\\
\noalign{\smallskip} %
  J0454$-$6713 & 04:54:33 & -67:13:13 & N9 & C & 1.58 & 216 (5) &
1.32 & 72 & 9.69$_{-5.34}^{+5.26}$ & 3 & SCM94\\
\noalign{\smallskip} %
  J0454$-$6626 & 04:54:49 & -66:25:32 & N11L & X & 0.63 & 106 (50) &
0.35$_{-0.22}^{+0.23}$ & 33 & 5.84$_{-2.39}^{+0.58}$ & 3 & MC73\\
\noalign{\smallskip} %
  J0455$-$6839 & 04:55:37 & -68:38:47 & N86 & R & 1.42 & 366 & --- & 16 &
2.66$_{-0.51}^{+0.23}$ & 3 & MC73\\
\noalign{\smallskip} %
  J0459$-$7008 & 04:59:55 & -70:07:52 & N186D & C & 1.09 & 114 &
0.93$_{-0.25}^{+0.42}$ & 51 & 3.73$_{-0.98}^{+0.87}$ & 3 & MC73\\
\noalign{\smallskip} %
  J0505$-$6753 & 05:05:42 & -67:52:39 & DEM L71 & X & 44.59 & 76 (5) &
0.22$\pm 0.01$ & 13 & 2.38$_{-0.52}^{+0.53}$ & 1 & LHG81\\
\noalign{\smallskip} %
  J0505$-$6802 & 05:05:55 & -68:01:47 & N23 & MX & 26.25 & 96 &
0.81$_{-0.06}^{+0.25}$ & 26 & 2.7$_{-0.76}^{+0.47}$ & 3 & MC73\\
\noalign{\smallskip} %
  J0506$-$6541 & 05:06:05 & -65:41:08 & DEM L72 & X & 0.53 & 410 (170) &
0.39$_{-0.39}^{+0.50}$ & 12 & 1.00$_{-0.40}^{+0.08}$ & 4 & KPS10\\
\noalign{\smallskip} %
  J0506$-$7026 & 05:06:50 & -70:25:53 & [HP99] 1139 & X & 1.44 & 262 (10) &
0$(<0.11)$ & 82 & 3.08$_{-0.83}^{+0.57}$ & 1 & in prep.\\
\noalign{\smallskip} %
  J0508$-$6902 & 05:08:37 & -69:02:54 & [HP99] 791 & X & 0.37 & 304 (33) &
0$(<0.48)$ & 22 & 1.27$_{-0.11}^{+0.04}$ & 1 & BKM14\\
\noalign{\smallskip} %
  J0508$-$6830 & 05:08:50 & -68:30:50 & J0508-6830 & X & 0.09 & 138 (160) &
0$(<0.7)$ & 27 & 3.09$_{-0.51}^{+0.69}$ & 1 & MHK14\\
\noalign{\smallskip} %
  J0509$-$6844 & 05:08:59 & -68:43:35 & N103B & X & 51.7 & 30 &
1.16$_{-0.04}^{+0.08}$ & 99 & 6.18$_{-1.31}^{+0.63}$ & 2 & MC73\\
\noalign{\smallskip} %
  J0509$-$6731 & 05:09:31 & -67:31:17 & B0509-67.5 & X & 16.51 & 31.8 &
1.42$_{-0.06}^{+0.06}$ & 2 & 1.86$_{-0.41}^{+0.26}$ & 1 & LHG81\\
\noalign{\smallskip} %
  J0511$-$6759 & 05:11:11 & -67:59:08 &   & MX & 0.16 & 112 & 0$(<1.08)$ &
3 & 1.23$_{-0.43}^{+0.04}$ & 1 & MHK14\\
\noalign{\smallskip} %
  J0512$-$6707 & 05:12:27 & -67:07:18 & [HP99] 483 & X & 0.09 & 120 (45) &
0.28$_{-0.23}^{+0.38}$ & 25 & 5.56$_{-2.66}^{+1.2}$ & 3 & KSB15b\\
\noalign{\smallskip} %
  J0513$-$6912 & 05:13:14 & -69:12:20 & DEM L109 & X & 0.51 & 240 (155) &
0.56$_{-0.43}^{+0.25}$ & 16 & 1.64$_{-0.30}^{+0.23}$ & 3 & MFT85\\
\noalign{\smallskip} %
  J0514$-$6840 & 05:14:16 & -68:40:22 &   & MX & 0.4 & 220 & 0$(<0.05)$ & 7
& 1.05$_{-0.27}^{+0.16}$ & 3 & MHK14\\
\noalign{\smallskip} %
  J0517$-$6759 & 05:17:08 & -67:59:29 &   & X & 0.24 & 324 (30) &
1.44$_{-1.19}^{+1.98}$ & 12 & 6.37$_{-3.23}^{+2.01}$ & 3 & MHK14\\
\noalign{\smallskip} %
  J0518$-$6939 & 05:18:41 & -69:39:12 & N120 & MX & 0.88 & 148 (110) &
0$(<0.75)$ & 133 & 4.57$_{-0.96}^{+0.52}$ & 3 & MC73\\
\noalign{\smallskip} %
  J0519$-$6902 & 05:19:35 & -69:02:09 & B0519-690 & MX & 34.94 & 33.6 &
0.63$_{-0.03}^{+0.03}$ & 13 & 1.71$_{-0.34}^{+0.26}$ & 1 & LHG81\\
\noalign{\smallskip} %
  J0519$-$6926 & 05:19:44 & -69:26:08 & B0520-694 & X & 2.69 & 190 (140) &
0$(<0.05)$ & 114 & 3.41$_{-0.51}^{+0.09}$ & 5 & MFD83\\
\noalign{\smallskip} %
\pagebreak
\noalign{\smallskip} %
  J0521$-$6543 & 05:21:39 & -65:43:07 & DEM L142 & N & --- & 168 & --- & 36 &
5.97$_{-3.07}^{+0.30}$ & 3 & BGS06\\
\noalign{\smallskip} %
  J0523$-$6753 & 05:23:07 & -67:53:12 & N44 & X & 0.9 & 255 (90) &
0.21$_{-0.10}^{+0.13}$ & 99 & 6.17$_{-3.15}^{+2.61}$ & 5 & CMG93\\
\noalign{\smallskip} %
  J0524$-$6624 & 05:24:20 & -66:24:23 & DEM L175a & N & --- & 240 & --- & 39 &
6.88$_{-3.34}^{+13.1}$ & 3 & MFT85\\
\noalign{\smallskip} %
  J0525$-$6938 & 05:25:04 & -69:38:24 & N132D & X & 315.04 & 126 (60) &
0.45$_{-0.02}^{+0.02}$ & 56 & 2.72$_{-0.35}^{+0.13}$ & 5 & WM66\\
\noalign{\smallskip} %
  J0525$-$6559 & 05:25:25 & -65:59:19 & N49B & X & 38.03 & 170 &
0.95$_{-0.08}^{+0.09}$ & 33 & 4.42$_{-1.76}^{+1.43}$ & 5 & MC73\\
\noalign{\smallskip} %
  J0526$-$6605 & 05:26:00 & -66:04:57 & N49 & X & 64.37 & 84 &
0.64$_{-0.04}^{+0.04}$ & 37 & 6.33$_{-2.72}^{+3.99}$ & 5 & WM66\\
\noalign{\smallskip} %
  J0527$-$6912 & 05:27:39 & -69:12:04 & B0528-692 & MX & 1.99 & 198 (142) &
0$(<0.05)$ & 224 & 7.59$_{-2.74}^{+0.29}$ & 4.5 & MFD84\\
\noalign{\smallskip} %
  J0527$-$6550 & 05:27:54 & -65:49:38 & DEM L204 & N & --- & 282 & --- & 9 &
1.54$_{-0.50}^{+0.40}$ & 3 & LHG81\\
\noalign{\smallskip} %
  J0527$-$6714 & 05:27:56 & -67:13:40 & B0528-6716 & X & 0.25 & 270 (40) &
2.28 & 97 & 11.12$_{-6.42}^{+4.37}$ & 3 & TM84\\
\noalign{\smallskip} %
  J0527$-$7104 & 05:27:57 & -71:04:30 & [HP99] 1234 & X & 0.21 & 369 (155) &
0.66$_{-0.16}^{+0.48}$ & 6 & 2.4$_{-0.52}^{+0.52}$ & 1.5 & KSP13\\
\noalign{\smallskip} %
  J0528$-$6727 & 05:28:05 & -67:27:20 & DEM L205 & X & 0.58 & 324 (40) &
0$(<0.18)$ & 142 & 8.97$_{-4.9}^{+1.91}$ & 3 & MHB12\\
\noalign{\smallskip} %
  J0529$-$6653 & 05:29:51 & -66:53:28 & DEM L214 & X & 1.04 & 145 (140) &
0.96$_{-0.79}^{+0.74}$ & 222 & 14.12$_{-8.61}^{+13.7}$ & 4.5 & BFC12a\\
\noalign{\smallskip} %
  J0530$-$7008 & 05:30:40 & -70:07:30 & DEM L218 & X & 0.72 & 325 (50) &
0.29$_{-0.22}^{+0.36}$ & 19 & 1.97$_{-0.12}^{+0.16}$ & 3 & DFB12\\
\noalign{\smallskip} %
  J0531$-$7100 & 05:31:56 & -71:00:19 & N206 & X & 2.55 & 180 (90) &
0.20$_{-0.16}^{+0.25}$ & 49 & 5.36$_{-1.95}^{+0.79}$ & 4.5 & MC73\\
\noalign{\smallskip} %
  J0532$-$6732 & 05:32:30 & -67:31:33 & B0532-675 & X & 2.48 & 285 (145) &
0.54$_{-0.32}^{+0.43}$ & 173 & 8.07$_{-3.83}^{+3.56}$ & 3 & MFT85\\
\noalign{\smallskip} %
  J0533$-$7202 & 05:33:46 & -72:02:59 & 1RXSJ053353.6-7204 & X & 0.57 & 205 (85)
& 0.26$_{-0.23}^{+0.24}$ & 1 & 0.8$_{-0.37}^{+0.17}$ & 2 & BFC13\\
\noalign{\smallskip} %
  J0534$-$6955 & 05:34:02 & -69:55:03 & B0534-699 & X & 6.33 & 135 (35) &
0.92$_{-0.11}^{+0.07}$ & 8 & 1.18$_{-0.13}^{+0.05}$ & 1 & LHG81\\
\noalign{\smallskip} %
  J0534$-$7033 & 05:34:18 & -70:33:26 & DEM L238 & X & 1.55 & 186 (110) &
0$(<0.02)$ & 4 & 2.07$_{-0.24}^{+0.2}$ & 1 & LHG81\\
\noalign{\smallskip} %
  J0535$-$6916 & 05:35:28 & -69:16:11 & SNR1987A & MX & 27.39 & 1.62 &
1.16$_{-0.05}^{+0.03}$ & 84 & 5.85$_{-1.87}^{+1.28}$ & 4 & historical\\
\noalign{\smallskip} %
  J0535$-$6602 & 05:35:44 & -66:02:14 & N63A & X & 185.68 & 84 &
0.46$_{-0.03}^{+0.05}$ & 45 & 3.95$_{-1.47}^{+0.73}$ & 5 & WM66\\
\noalign{\smallskip} %
  J0535$-$6918 & 05:35:46 & -69:18:02 & Honeycomb & MX & 0.4 & 105 (155) &
0.90$\pm0.12$ & 108 & 8.32$_{-3.05}^{+2.40}$ & 3.5 & CDS95\\
\noalign{\smallskip} %
  J0536$-$6735 & 05:36:03 & -67:34:36 & DEM L241 & X & 3.84 & 310 (155) &
0.66$_{-0.06}^{+0.06}$ & 39 & 2.71$_{-0.53}^{+0.55}$ & 5 & MFT85\\
\noalign{\smallskip} %
  J0536$-$7039 & 05:36:07 & -70:38:37 & DEM L249 & MX & 1.43 & 200 (35) &
0$(<0.03)$ & 2 & 1.56$_{-0.10}^{+0.10}$ & 1 & LHG81\\
\noalign{\smallskip} %
  J0536$-$6913 & 05:36:17 & -69:13:28 & B0536-6914 & MX & 0.22 & 66 (90) &
1.11$_{-0.75}^{+0.11}$ & 92 & 5.85$_{-1.87}^{+1.28}$ & 4.5 & KSB15a\\
\noalign{\smallskip} %
  J0537$-$6628 & 05:37:27 & -66:27:50 & DEM L256 & X & 0.32 & 227 (42) &
0.25$_{-0.25}^{+0.50}$ & 48 & 1.62$_{-0.17}^{+0.07}$ & 2 & KPS10\\
\noalign{\smallskip} %
\pagebreak
\noalign{\smallskip} %
  J0537$-$6910 & 05:37:46 & -69:10:28 & N157B & MX & 15.0 & 120 &
1.73$_{-0.06}^{+0.06}$ & 83 & 19.56$_{-11.0}^{+8.02}$ & 5 & MC73\\
\noalign{\smallskip} %
  J0540$-$6944 & 05:39:59 & -69:44:02 & N159 & X & 0.43 & 92 (110) &
1.92$_{-0.15}^{+0.20}$ & 60 & 3.53$_{-0.85}^{+0.62}$ & 3 & CKS97\\
\noalign{\smallskip} %
  J0540$-$6920 & 05:40:11 & -69:19:55 & B0540-693 & X & 87.35 & 72 &
1.58$_{-0.04}^{+0.04}$ & 119 & 24.41$_{-15.8}^{+19.1}$ & 5 & MC73\\
\noalign{\smallskip} %
  J0541$-$6659 & 05:41:51 & -66:59:04 & [HP99] 456 & X & 0.77 & 300 & 0$(<0.06)$
& 10 & 2.08$_{-0.29}^{+0.04}$ & 3 & GSH12\\
\noalign{\smallskip} %
  J0543$-$6858 & 05:43:08 & -68:58:18 & DEM L299 & X & 1.68 & 330 (55) &
0.4$_{-0.05}^{+0.13}$ & 38 & 4.98$_{-1.95}^{+2.88}$ & 3 & LHG81\\
\noalign{\smallskip} %
  J0547$-$6943 & 05:46:59 & -69:42:50 & DEM L316B & X & 1.47 & 190 (95) &
0.63$_{-0.12}^{+0.19}$ & 0 & 1.03$_{-0.56}^{+0.26}$ & 3 & MC73\\
\noalign{\smallskip} %
  J0547$-$6941 & 05:47:22 & -69:41:26 & DEM L316A & X & 1.26 & 190 (170) &
0.92$_{-0.20}^{+0.24}$ & 0 & 1.03$_{-0.56}^{+0.26}$ & 1 & MC73\\
\noalign{\smallskip} %
  J0547$-$7025 & 05:47:49 & -70:24:54 & B0548-704 & X & 2.94 & 118 (75) &
0.88$_{-0.10}^{+0.11}$ & 3 & 0.72$_{-0.26}^{+0.20}$ & 1 & MFD83\\
\noalign{\smallskip} %
  J0550$-$6823 & 05:50:30 & -68:22:40 &   & C & 1.22 & 312 (90) &
1.22 & 4 & 2.04$_{-0.39}^{+0.20}$ & 3 & BFC12b\\
\noalign{\smallskip} %
\hline
\end{longtable}
\tablefoot{
Sample of SNRs in the LMC. The columns are the following:
\begin{itemize}
    \item[(1)] MCSNR identifier, in the form ``JHHMM$-$DDMM''.
    \item[(2)] Right ascension of the remnant, in J2000 equinox.
    \item[(3)] Declination in J2000 equinox.
    \item[(4)] Old ``common'' name used in the literature.
    \item[(5)] Flag coding the type of X-ray data available and used in this
work. ``X'' indicates that \xmm\ data are present, and ``MX'' that multiple
\xmm\ observation of the remnant exist. ``C'' or ``R'' are used when no \xmm\
observations are available but \chandra\ or ROSAT observations were used,
respectively. ``N'' means that no X-ray information was found.
    \item[(6)] $L_X$, the X-ray luminosity in the 0.3~keV~--~8~keV band, in
units of $10^{35}$~erg~s$^{-1}$, obtained as described in 
Sect.\,\ref{results_XLF}.
    \item[(7)] X-ray size in arcsec. Only the \emph{maximal} extent is given
(corresponding to the diametre in a circularly symmetric case). The number
in brackets gives the position angle (PA) of the maximal extent in the
non-symmetric case. The PA is measured in degrees, eastwards of north. Size was
measured from \xmm\ images whenever applicable (``X'' and ``MX'' flags). For
``C'' SNRs, the quoted value is taken from the entry in the \chandra\ SNR
catalogue. For the ROSAT-only SNRs (J0448$-$6700 and J0455$-$6839), we used the
value quoted in \citet{2010MNRAS.407.1301B} and \citet{1999ApJS..123..467W},
respectively.
    \item[(8)] $N_H$ fraction, as defined in
Sect.\,\ref{results_distribution}. Uncertainties are given at the 90\,\% C.L.
    \item[(9)] \NOB , the number of blue early-type stars within 100~pc of the
remnant (see Sect.\,\ref{results_sfh_environment}).
    \item[(10)] $r$, the ratio of CC SNe to thermonuclear SNe expected from the
observed distribution of stellar ages in the neighbourhood of the remnant, as
obtained by Eq.~\ref{eq_r} (see Sect.\,\ref{results_sfh} for details).
    \revision{\item[(11)] ``Hint-spec'', the number attributed to SNRs as 
function of spectral results, as described in Sect.\,\ref{results_sfh_typing} 
and Table~\ref{table_results_sfh_hints_spec}.}
    \item[(12)] Reference in which the SNR was first published\,: 
\citetalias{1966MNRAS.131..371W}: \citet{1966MNRAS.131..371W};
\citetalias{1973ApJ...180..725M}: \citet{1973ApJ...180..725M};
\citetalias{1981ApJ...248..925L}: \citet{1981ApJ...248..925L};
\citetalias{1983ApJS...51..345M}: \citet{1983ApJS...51..345M};
\citetalias{1984PASAu...5..537T}: \citet{1984PASAu...5..537T};
\citetalias{1984ApJS...55..189M}: \citet{1984ApJS...55..189M};
\citetalias{1985ApJS...58..197M}: \citet{1985ApJS...58..197M};
\citetalias{1993ApJ...414..213C}: \citet{1993ApJ...414..213C};
\citetalias{1994AJ....108.1266S}: \citet{1994AJ....108.1266S};
\citetalias{1995AJ....109.1729C}: \citet{1995AJ....109.1729C};
\citetalias{1997PASP..109..554C}: \citet{1997PASP..109..554C};
\citetalias{2006ApJS..165..480B}: \citet{2006ApJS..165..480B};
\citetalias{2010ApJ...725.2281K}: \citet{2010ApJ...725.2281K};
\citetalias{2012A&A...539A..15G}: \citet{2012A&A...539A..15G};
\citetalias{2012MNRAS.420.2588B}: \citet{2012MNRAS.420.2588B};
\citetalias{2012RMxAA..48...41B}: \citet{2012RMxAA..48...41B};
\citetalias{2012A&A...540A..25D}: \citet{2012A&A...540A..25D};
\citetalias{2012A&A...546A.109M}: \citet{2012A&A...546A.109M};
\citetalias{2013A&A...549A..99K}: \citet{2013A&A...549A..99K};
\citetalias{2013MNRAS.432.2177B}: \citet{2013MNRAS.432.2177B};
\citetalias{2014A&A...561A..76M}: \citet{2014A&A...561A..76M};
\citetalias{2014MNRAS.439.1110B}: \citet{2014MNRAS.439.1110B};
\citetalias{2015A&A...573A..73K}: \citet{2015A&A...573A..73K};
KSB15b: \citet{2015arXiv150906475K}.
\end{itemize}
}
\end{landscape}

\section{Spectra of SNRs observed with \xmm\ for the first time}
\label{appendix_spectra}

\begin{figure}[h]
  \centering
  \includegraphics[angle=-90,width=0.496\hsize]
  {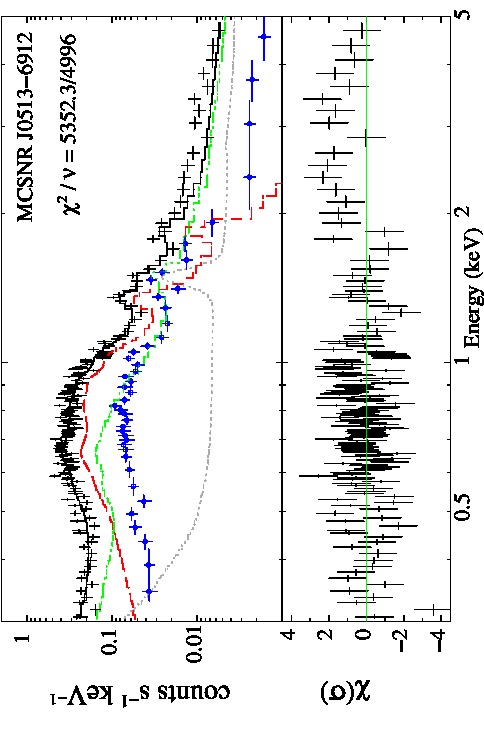}
  \includegraphics[angle=-90,width=0.496\hsize]
  {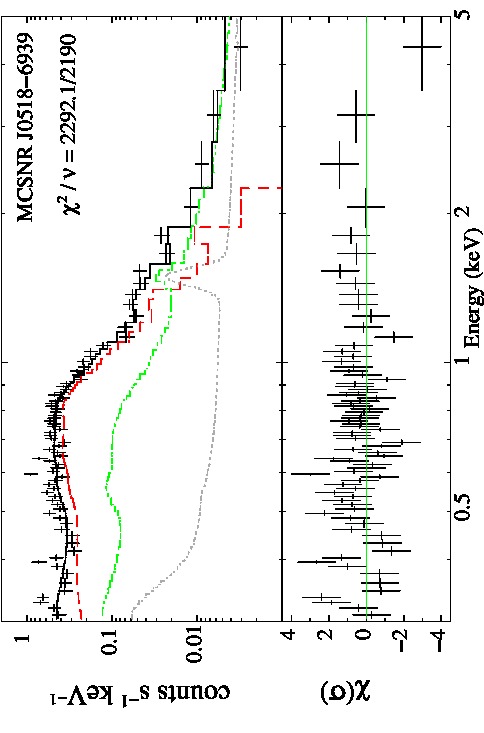}

  \vspace{0.45cm}
  \includegraphics[angle=-90,width=0.496\hsize]
  {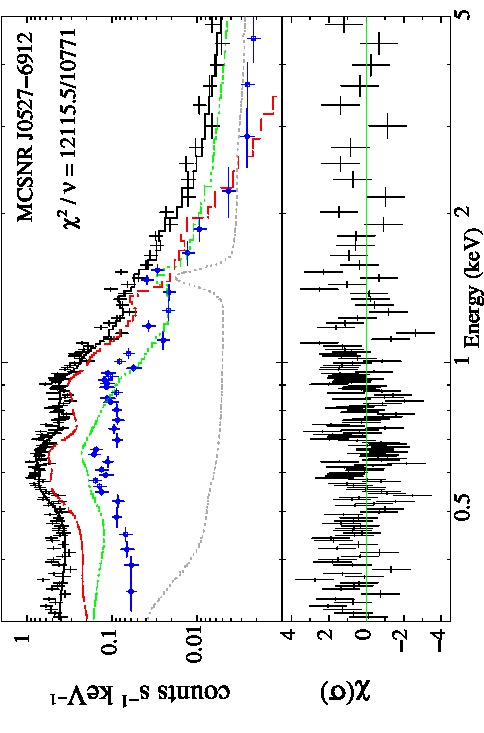}
  \includegraphics[angle=-90,width=0.496\hsize]
  {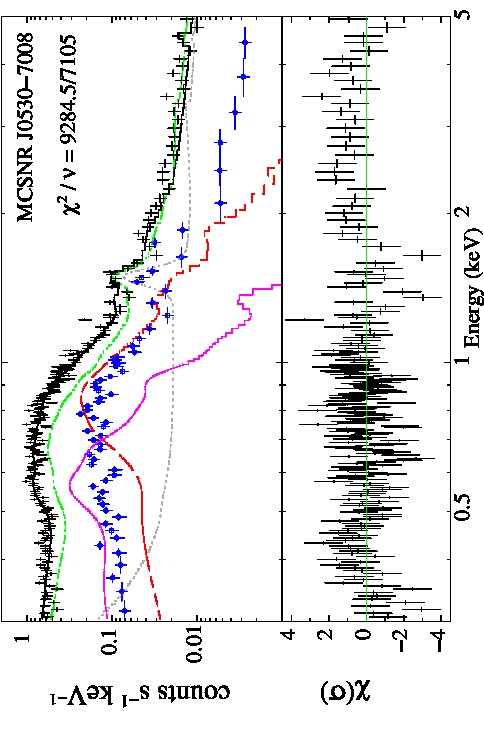}

  \vspace{0.45cm}
  \includegraphics[angle=-90,width=0.496\hsize]
  {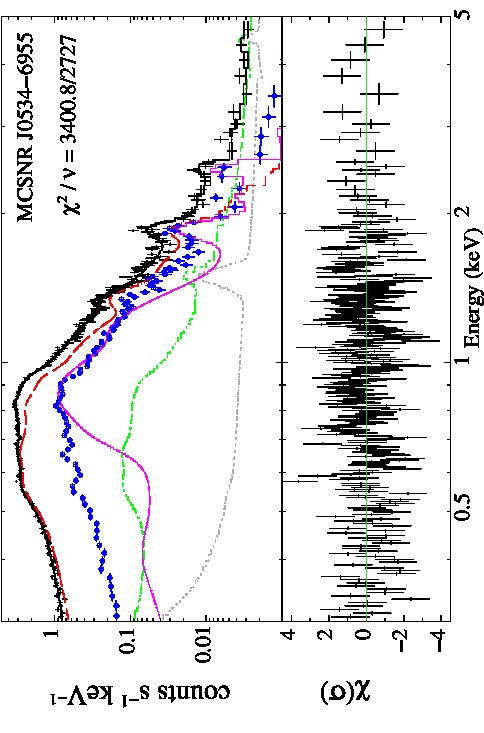}
  \includegraphics[angle=-90,width=0.496\hsize]
  {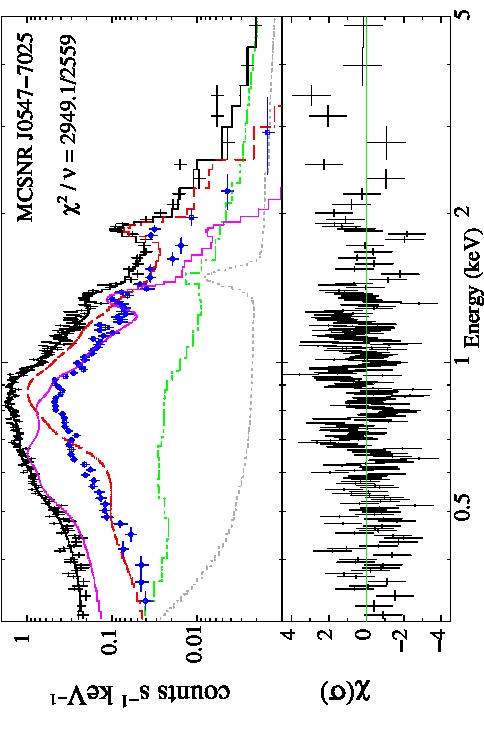}
  
  \vspace{0.3cm}
\caption{\revision{
\xmm\ EPIC spectrum of SNRs for which none have been published yet (see 
Sect.\,\ref{results_spectra_general}). The SNR names are labelled on each 
plot. The pn and MOS1 data are shown in black and blue points, respectively. 
MOS2 data are not shown for clarity. The total best-fit model (see parameters 
in Table~\ref{appendix_table_spectra_all}), convolved with the pn response, is 
shown as the solid black line. It includes:
\emph{i)} the instrumental background component with a dotted grey line (only 
pn shown);
\emph{ii)} the total AXB model (see Appendix~\ref{appendix_background}) with a
dot-dashed green line; and 
\emph{iii)} the SNR emission with a dashed red line. When a second component is 
used (Table~\ref{appendix_table_spectra_all}), it is shown with a solid magenta 
line.
The lower panels show the pn residuals (in terms of $\sigma$) for the total 
best-fit model. For MCSNR J0527$-$6712 data from two observations were used, 
though only one spectra is shown for clarity.
}
}
\label{fig_data_spectra_newSNRs}
\end{figure}

\begin{landscape}

\onltab{

\section{SNR X-ray spectral results}
\label{appendix_tables}


\begin{longtable}[c]{@{}l @{\hspace{0.0cm}} @{\hspace{0.0cm}} c
@{\hspace{0.0cm}} @{\hspace{0.0cm}} c @{\hspace{0.0cm}} @{\hspace{0.0cm}} c
@{\hspace{0.0cm}} @{\hspace{0.0cm}} c @{\hspace{0.0cm}} @{\hspace{0.0cm}} l
@{\hspace{-0.20cm}} @{\hspace{-0.20cm}} c @{\hspace{0.0cm}} @{\hspace{0.0cm}} c
@{\hspace{0.0cm}} @{\hspace{0.0cm}} c @{\hspace{0.05cm}} @{\hspace{0.05cm}} c
@{\hspace{-0.00cm}} @{\hspace{-0.0cm}} l @{\hspace{0.05cm}} @{\hspace{0.1cm}} r
@{\hspace{0.1cm}}}
\caption{X-ray spectral results of LMC SNRs.}
\label{appendix_table_spectra_all}\\
\hline\hline
\noalign{\medskip}
  & \multicolumn{5}{c}{Component 1:} &
\multicolumn{5}{c}{Component 2:} & \multirow{2}{*}{$\chi ^2 / \nu$ \ 
$\left(\chi ^2 _{\mathrm{red}}\right)$} \\
\noalign{\medskip}
\cline{2-6} \cline{7-11}
\noalign{\medskip}
 MCSNR & $N_{H\mathrm{\ LMC}}$ & $kT$ & $\tau$ & EM &
\multicolumn{1}{c}{Abundances}
       & $N_{H\mathrm{\ LMC}}$ & $kT$ & $\tau$ & EM &
\multicolumn{1}{c}{Abundances} & \\
  & ($10^{21}$~cm$^{-2}$) & (keV) & ($10^{11}$ s\,cm$^{-3}$) & ($10^{58}$
cm$^{-3}$) &  & ($10^{21}$~cm$^{-2}$) & (keV) & ($10^{11}$ s\,cm$^{-3}$) &
($10^{58}$ cm$^{-3}$) &  \\
(1) & (2) & (3) & (4) & (5) & \multicolumn{1}{c}{(6)} & (7) & (8) & (9) & (10) &
 \multicolumn{1}{c}{(11)} & (12)\\
\noalign{\medskip}
\hline
\endfirsthead
%
%
\caption[]{(continued)}\\
\hline
\hline
\noalign{\medskip}
  & \multicolumn{5}{c}{Component 1:} &
\multicolumn{5}{c}{Component 2:} & \multirow{2}{*}{$\chi ^2 / \nu$ \ 
$\left(\chi ^2 _{\mathrm{red}}\right)$} \\
\noalign{\medskip}
\cline{2-6} \cline{7-11}
\noalign{\medskip}
 MCSNR & $N_{H\mathrm{\ LMC}}$ & $kT$ & $\tau$ & EM &
\multicolumn{1}{c}{Abundances}
       & $N_{H\mathrm{\ LMC}}$ & $kT$ & $\tau$ & EM &
\multicolumn{1}{c}{Abundances} & \\
  & ($10^{21}$~cm$^{-2}$) & (keV) & ($10^{11}$ s\,cm$^{-3}$) & ($10^{58}$
cm$^{-3}$) &  & ($10^{21}$~cm$^{-2}$) & (keV) & ($10^{11}$ s\,cm$^{-3}$) &
($10^{58}$ cm$^{-3}$) &  \\
(1) & (2) & (3) & (4) & (5) & \multicolumn{1}{c}{(6)} & (7) & (8) & (9) & (10) &
\multicolumn{1}{c}{(11)} & (12)\\
\noalign{\medskip}
\hline
\endhead
%
\hline
\endfoot
%
\hline
\endlastfoot
\noalign{\medskip}
J0449$-$6920 & 6.97$_{-2.08} ^{+2.77}$ & 0.20$_{-0.12} ^{+0.05}$ & CIE &
18.3$_{-16.3} ^{+359}$ & \RD & \multicolumn{5}{c}{---} & 1985.3/1863 (1.07) \\
\noalign{\medskip}
\multirow{3}{*}{J0450$-$7050\tablefootmark{a}} & \multirow{3}{*}{1.31$_{-0.79}
^{+0.95}$} & \multirow{3}{*}{0.24$\pm0.02$} & \multirow{3}{*}{CIE} &
\multirow{3}{*}{33.6$_{-24.1} ^{+36.8}$} & \abund{O}0.20$_{-0.08}
^{+0.14}$ & \multicolumn{5}{c}{\multirow{3}{*}{---}} &
\multirow{3}{*}{4669.4/3294 (1.42)} \\*
 & & & & & \abund{Ne}0.40$_{-0.07} ^{+0.33}$ & & & & & & \\*
 & & & & & \abund{Mg}0.20 $(< 0.48)$ & & & & & & \\
\noalign{\medskip}
\multirow{4}{*}{J0453$-$6655} & \multirow{4}{*}{1.10$_{-0.54} ^{+0.61}$} &
\multirow{4}{*}{0.36$_{-0.08} ^{+0.10}$} &
\multirow{4}{*}{1.38$_{-0.63} ^{+2.07}$} &
\multirow{4}{*}{30.7$_{-15.6} ^{+36.9}$} & \abund{O}0.13$_{-0.04} ^{+0.05}$ &
\multicolumn{5}{c}{\multirow{4}{*}{---}} & \multirow{4}{*}{4460.9/4214 (1.06)} 
\\*
 & & & & & \abund{Ne}0.12$_{-0.05} ^{+0.07}$ & & & & & & \\*
 & & & & & \abund{Mg}0.15$_{-0.01} ^{+0.20}$ & & & & & & \\*
 & & & & & \abund{Fe}0.29$_{-0.08} ^{+0.12}$ & & & & & & \\
\noalign{\medskip}
\multirow{5}{*}{J0453$-$6829\tablefootmark{b}} & \multirow{5}{*}{0.93$_{-0.43}
^{+0.13}$} & \multirow{5}{*}{0.37$_{-0.05} ^{+0.03}$} & 
\multirow{5}{*}{1.58$_{-0.24} ^{+0.51}$} &  \multirow{5}{*}{27.2$_{-3.56}
^{+16.28}$} &
\abund{O}0.17$\pm0.01$ & \multicolumn{5}{c}{\multirow{5}{*}{---}} &
\multirow{5}{*}{2156.2/1860 (1.16)} \\*
 & & & & & \abund{Ne}0.25$_{-0.01} ^{+0.03}$& & & & & & \\*
 & & & & & \abund{Mg}0.37$\pm0.05$ & & & & & & \\*
 & & & & & \abund{Si}0.61$_{-0.23} ^{+0.24}$ & & & & & & \\*
 & & & & & \abund{Fe}0.28$_{-0.02} ^{+0.03}$ & & & & & & \\
\noalign{\medskip}
\multirow{3}{*}{J0454$-$6626} & \multirow{3}{*}{1.15$_{-0.71} ^{+0.76}$} &
\multirow{3}{*}{0.34 $_{-0.07} ^{+0.09}$} & \multirow{3}{*}{1.17$_{-0.70}
^{+1.48}$} & \multirow{3}{*}{1.67$_{-0.99} ^{+1.38}$} & \abund{O}0.14$_{-0.05}
^{+0.07}$  & \multicolumn{5}{c}{\multirow{3}{*}{---}} &
\multirow{3}{*}{2136.8/1811 (1.18)} \\*
 & & & & & \abund{Mg}0.15$(< 0.39)$ & & & & & & \\*
 & & & & & \abund{Fe}0.26$_{-0.08} ^{+0.10}$ & & & & & & \\
\noalign{\medskip}
\multirow{5}{*}{J0505$-$6802\tablefootmark{c}} & \multirow{5}{*}{1.58$_{-0.11}
^{+0.49}$} & \multirow{5}{*}{0.32$_{-0.03} ^{+0.02}$} &
\multirow{5}{*}{3.40$_{-1.62} ^{+2.26}$} & \multirow{5}{*}{107.4$_{-16.3}
^{+76.1}$} & \abund{O}0.11$_{-0.04} ^{+0.03}$ & \multirow{5}{*}{1.58} &
\multirow{5}{*}{1.09$_{-0.09} ^{+0.07}$} & \multirow{5}{*}{2.46$_{-0.48}
^{+0.96}$} & \multirow{5}{*}{6.50$_{-0.60} ^{+12.0}$} &
\multicolumn{1}{c}{\multirow{5}{*}{---}} &
\multirow{5}{*}{454.4/407 (1.12)} \\*
 & & & & & \abund{Ne}0.16$\pm0.02$& & & & & & \\*
 & & & & & \abund{Mg}0.18$\pm0.03$ & & & & & & \\*
 & & & & & \abund{Si}0.37$_{-0.06} ^{+0.05}$ & & & & & & \\*
 & & & & & \abund{Fe}0.20$_{-0.03} ^{+0.06}$ & & & & & & \\
\noalign{\medskip}
J0506$-$6541 & 0.59$(<1.34)$ & 0.18$\pm0.01$ & CIE & 2.27$_{-0.71} ^{+1.35}$ &
\abund{Ne}1.26$_{-0.28} ^{+0.41}$ & \multicolumn{5}{c}{---} & 8549.4/7232 
(1.18) \\
\noalign{\medskip}
\multirow{4}{*}{J0506$-$7026} & \multirow{4}{*}{$< 0.12$} &
\multirow{4}{*}{0.70$_{-0.02} ^{+0.09}$} & \multirow{4}{*}{1.99$_{-0.90}
^{+0.79}$ (1.08)} &
\multirow{4}{*}{0.61$\pm0.09$} & \abund{O}0.33$_{-0.05} ^{+0.07}$ &
\multicolumn{5}{c}{\multirow{4}{*}{---}} & \multirow{4}{*}{9026.4/8365 (1.08)} 
\\*
 & & & & & \abund{Ne}0.42$_{-0.17} ^{+0.19}$ & & & & & & \\*
 & & & & & \abund{Mg}0.22$_{-0.13} ^{+0.14}$ & & & & & & \\*
 & & & & & \abund{Fe}1.49$_{-0.23} ^{+0.17}$ & & & & & & \\*
\noalign{\medskip}
\multirow{2}{*}{J0508$-$6830\tablefootmark{d}} & \multirow{2}{*}{$< 1.80$} &
\multirow{2}{*}{0.71$_{-0.07} ^{+0.06}$} & \multirow{2}{*}{CIE} &
\multirow{2}{*}{1.0$(_{-0.4} ^{+700}) \times 10^{-4}$ } & \abund{O}0  &
\multicolumn{5}{c}{\multirow{2}{*}{---}} & \multirow{2}{*}{1488.3/1554 (0.96)} 
\\*
 & & & & & \abund{Fe}792.9$(> 1.1)$ & & & & & & \\*
\noalign{\medskip}
\multirow{2}{*}{J0508$-$6902\tablefootmark{e}} & \multirow{2}{*}{$< 0.8$} &
\multirow{2}{*}{0.41$_{-0.06} ^{+0.05}$} & \multirow{2}{*}{27.0$_{-9.7}
^{+42.2}$} & \multirow{2}{*}{0.48$_{-0.22} ^{+0.25}$} &
\multicolumn{1}{c}{\multirow{2}{*}{RD92}} & \multirow{2}{*}{0} &
\multirow{2}{*}{0.78$\pm0.03$} & \multirow{2}{*}{CIE} & $< 48.8$ &
\multicolumn{1}{c}{pure O} & \multirow{2}{*}{267.1/219 (1.22)} \\*
 & & & & &  & & & & 0.16$\pm 0.01$ & \multicolumn{1}{c}{pure Fe} & \\
\noalign{\medskip}
\multirow{2}{*}{J0511$-$6759\tablefootmark{d}} & \multirow{2}{*}{$< 2.10$} &
\multirow{2}{*}{0.65$_{-0.04} ^{+0.05}$} & \multirow{2}{*}{CIE} &
\multirow{2}{*}{1.2$(_{-1.0} ^{+8.0}) \times 10^{-2}$ } & \abund{O}0  &
\multicolumn{5}{c}{\multirow{2}{*}{---}} & \multirow{2}{*}{707.5/692 (1.02)} \\*
 & & & & & \abund{Fe}11.4$(>4.7)$ & & & & & & \\
\noalign{\medskip}
J0512$-$6707 & 0.82$_{-0.68} ^{+1.09}$ & 0.24$\pm 0.01$ & CIE &
0.15$_{-0.08} ^{+0.17}$ & \RD & \multicolumn{5}{c}{---} & 3059.4/2083 (1.47)\\
\noalign{\medskip}
\multirow{5}{*}{J0513$-$6912 \textdagger} & \multirow{5}{*}{1.34$_{-1.02}
^{+0.59}$} & \multirow{5}{*}{0.43$\pm0.08$} &
\multirow{5}{*}{10.9$_{-6.66} ^{+37.3}$} & \multirow{5}{*}{1.28$_{-0.89}
^{+1.67}$} & \abund{O}0.27$_{-0.08} ^{+0.23}$ &
\multicolumn{5}{c}{\multirow{5}{*}{---}} &
\multirow{5}{*}{5352.3/4996 (1.07)} \\*
 & & & & & \abund{Ne}0.46$_{-0.13} ^{+0.57}$ & & & & & & \\*
 & & & & & \abund{Mg}0.75$_{-0.22} ^{+0.34}$ & & & & & & \\*
 & & & & & \abund{Si}1.33$_{-0.50} ^{+0.79}$ & & & & & & \\*
 & & & & & \abund{Fe}0.20$_{-0.04} ^{+0.23}$ & & & & & & \\
\noalign{\medskip}
\multirow{2}{*}{J0514$-$6840\tablefootmark{d}} & \multirow{2}{*}{$< 0.09$} &
\multirow{2}{*}{0.30$\pm 0.01$} & \multirow{2}{*}{2.60$_{-0.50} ^{+0.60}$} &
\multirow{2}{*}{0.51$\pm 0.03$} & \abund{O}0.28$\pm 0.03$ &
\multicolumn{5}{c}{\multirow{2}{*}{---}} & \multirow{2}{*}{8005.3/7957 (1.01)} 
\\*
 & & & & & \abund{Fe}0.38$_{-0.11} ^{+0.13}$ & & & & & & \\
\noalign{\medskip}
J0517$-$6759\tablefootmark{d} & 3.5$_{-2.9} ^{+4.8}$ & 0.1$_{-0.02} ^{+0.04}$  &
CIE  & 11.2$_{-9.0} ^{+7.2}$ & \RD & $< 1.7$ & 0.59$_{-0.04} ^{+0.05}$ & 
CIE & 0.19$_{-0.02} ^{+0.002}$ & \RD & 2558.9/2547 (1.00)\\
\noalign{\medskip}
\multirow{5}{*}{J0518$-$6939 \textdagger} & \multirow{5}{*}{$< 1.16$} &
\multirow{5}{*}{0.44$_{-0.09} ^{+0.30}$} & \multirow{5}{*}{4.91$_{-3.62}
^{+15.0}$} & \multirow{5}{*}{1.53$_{-1.08} ^{+1.43}$} & \abund{O}0.14$_{-0.07}
^{+0.06}$ &
\multicolumn{5}{c}{\multirow{5}{*}{---}} & \multirow{5}{*}{2292.1/2190 (1.05)} 
\\*
 & & & & & \abund{Ne}0.23$\pm 0.15$ & & & & & & \\*
 & & & & & \abund{Mg}0.31$_{-0.26} ^{+0.42}$ & & & & & & \\*
 & & & & & \abund{Si}0.96$_{-0.90} ^{+1.63}$ & & & & & & \\*
 & & & & & \abund{Fe}0.27$_{-0.10} ^{+0.34}$ & & & & & & \\
\noalign{\medskip}
\multirow{5}{*}{J0519$-$6926} & \multirow{5}{*}{$< 0.09$} &
\multirow{5}{*}{0.39$_{-0.04} ^{+0.06}$} & \multirow{5}{*}{3.58$_{-1.62}
^{+1.65}$} & \multirow{5}{*}{4.53$_{-0.86} ^{+1.65}$} & \abund{O}0.15$_{-0.02}
^{+0.01}$ &
\multicolumn{5}{c}{\multirow{5}{*}{---}} & \multirow{5}{*}{3602.5/3033 (1.19)} 
\\*
 & & & & & \abund{Ne}0.36$_{-0.05} ^{+0.04}$ & & & & & & \\*
 & & & & & \abund{Mg}0.34$\pm 0.10$ & & & & & & \\*
 & & & & & \abund{Si}1.21$_{-0.36} ^{+0.32}$ & & & & & & \\*
 & & & & & \abund{Fe}0.21$\pm 0.03$ & & & & & & \\
\noalign{\medskip}
\multirow{5}{*}{J0523$-$6753} & \multirow{5}{*}{0.75$_{-0.35} ^{+0.45}$} &
\multirow{5}{*}{0.62$\pm 0.04$} & \multirow{5}{*}{22.7$_{-10.2} ^{+11.0}$} &
\multirow{5}{*}{0.66$_{-0.09} ^{+0.02}$} & \abund{O}1.49$_{-0.80} ^{+0.86}$ &
\multicolumn{5}{c}{\multirow{5}{*}{---}} & \multirow{5}{*}{5075.9/4532 (1.12)} 
\\*
 & & & & & \abund{Ne}3.08$_{-1.63} ^{+2.03}$ & & & & & & \\*
 & & & & & \abund{Mg}1.57$_{-0.76} ^{+0.79}$ & & & & & & \\*
 & & & & & \abund{Si}1.13$\pm 0.43$ & & & & & & \\*
 & & & & & \abund{Fe}0.48$\pm 0.16$ & & & & & & \\
\noalign{\medskip}
\multirow{5}{*}{J0525$-$6559} & \multirow{5}{*}{2.17$_{-0.19} ^{+0.20}$} &
\multirow{5}{*}{0.60$_{-0.04} ^{+0.06}$} & \multirow{5}{*}{2.34$_{-0.56}
^{+0.69}$} &
\multirow{5}{*}{50.3$_{-11.4} ^{+11.3}$} & \abund{O}0.07$\pm 0.05$ &
\multirow{5}{*}{2.17} & \multirow{5}{*}{0.32$\pm 0.07$} &
\multirow{5}{*}{4.34$_{-2.04} ^{+17.7}$} & \multirow{5}{*}{43.6$_{-14.0}
^{+45.58}$} & \multicolumn{1}{c}{\multirow{5}{*}{RD92}} &
\multirow{5}{*}{3981.8/2884 (1.38)} \\*
 & & & & & \abund{Ne}0.27$\pm 0.04$ & & & & & & \\*
 & & & & & \abund{Mg}0.43$\pm 0.09$ & & & & & & \\*
 & & & & & \abund{Si}0.33$\pm 0.07$ & & & & & & \\*
 & & & & & \abund{Fe}0.10$_{-0.06} ^{+0.05}$ & & & & & & \\
\noalign{\medskip}
\multirow{5}{*}{J0526$-$6605\tablefootmark{f}} & \multirow{5}{*}{2.64$_{-0.18}
^{+0.15}$} &
\multirow{5}{*}{0.42$\pm 0.02$} & \multirow{5}{*}{5.26$_{-0.41} ^{+0.75}$} &
\multirow{5}{*}{112.3$_{-3.36} ^{+2.85}$} & \abund{O}0.30$\pm 0.03$ &
\multirow{5}{*}{2.64} & \multirow{5}{*}{1.04$_{0.05} ^{+0.04}$} &
\multirow{5}{*}{$> 35.0$} & \multirow{5}{*}{27.4$_{-1.77}
^{+3.03}$} & \multicolumn{1}{c}{\multirow{5}{*}{---}} &
\multirow{5}{*}{3509.3/2038 (1.72)} \\*
 & & & & & \abund{Ne}0.42$_{-0.05} ^{+0.02}$ & & & & & & \\*
 & & & & & \abund{Mg}0.52$_{-0.05} ^{+0.08}$ & & & & & & \\*
 & & & & & \abund{Si}1.0$\pm 0.11$ & & & & & & \\*
 & & & & & \abund{Fe}0.41$_{-0.05} ^{+0.02}$ & & & & & & \\
\noalign{\medskip}
J0527$-$6714\tablefootmark{g} & 2.0 & 0.18$\pm 0.01$ & CIE &
1.31$\pm 0.23$ & \RD & \multicolumn{5}{c}{---} & 5999.0/5237 (1.15)\\
\noalign{\medskip}
J0527$-$6912\tablefootmark{h} \textdagger& 2.0 & 0.18$\pm 0.01$ & CIE &
1.31$\pm 0.23$ & $0.35_{-0.06} ^{+0.12} \times $ RD92 & \multicolumn{5}{c}{---}
& 12115.5/10771(1.12) \\
\noalign{\medskip}
\multirow{3}{*}{J0527$-$7104} & \multirow{3}{*}{1.33$_{-0.32} ^{+0.97}$} &
\multirow{3}{*}{0.37 $_{-0.04} ^{+0.08}$} & \multirow{3}{*}{1.33$_{-0.55}
^{+4.57}$} & \multirow{3}{*}{0.30$_{-0.13} ^{+0.07}$} & \abund{O}0.05$_{-0.01}
^{+0.04}$  & \multicolumn{5}{c}{\multirow{3}{*}{---}} &
\multirow{3}{*}{6801.8/5788 (1.18)} \\*
 & & & & & \abund{Ne}0.86$_{-0.27} ^{+0.29}$ & & & & & & \\*
 & & & & & \abund{Fe}1.43$_{-0.32} ^{+0.71}$ & & & & & & \\
\noalign{\medskip}
\multirow{3}{*}{J0528$-$6727} & \multirow{3}{*}{$< 0.39$} &
\multirow{3}{*}{0.22$\pm0.02$} & \multirow{3}{*}{CIE} &
\multirow{3}{*}{1.70$_{-0.52} ^{+1.80}$} & \abund{0}0.30$_{-0.19} ^{+0.15}$ &
\multicolumn{5}{c}{\multirow{3}{*}{---}} &
\multirow{3}{*}{5510.0/5265 (1.05)} \\*
 & & & & & \abund{Ne}0.66$_{-0.25} ^{+0.43}$ & & & & & & \\*
 & & & & & \abund{Fe}0.66$_{-0.34} ^{+0.68}$ & & & & & & \\
\noalign{\medskip}
\multirow{5}{*}{J0529$-$6653\tablefootmark{i}} & \multirow{5}{*}{0.48$_{-0.40}
^{+0.37}$} &
\multirow{5}{*}{1.51$\pm 0.28$} & \multirow{5}{*}{0.30$_{-0.10} ^{+0.24}$} &
\multirow{5}{*}{0.83$_{-0.18} ^{+0.22}$} & \abund{O}0.09$\pm 0.02$ &
\multicolumn{5}{c}{\multirow{5}{*}{---}} & \multirow{5}{*}{1961.3/2987 (0.66)} 
\\*
 & & & & & \abund{Ne}0.03$(< 0.05)$ & & & & & & \\*
 & & & & & \abund{Mg}0.07$\pm 0.06$ & & & & & & \\*
 & & & & & \abund{Si}0.21$(< 0.49)$ & & & & & & \\*
 & & & & & \abund{Fe}0.07$_{-0.02} ^{+0.03}$ & & & & & & \\
\noalign{\medskip}
J0530$-$7008 \textdagger & 0.37$_{-0.28} ^{+0.45}$ & 0.18$_{-0.03} ^{+0.02}$  & 
CIE  &
1.34$_{-0.34} ^{+0.45}$ & \RD & 0.37 & 0.74$\pm 0.06$ & CIE &
0.36$\pm 0.06$ & \RD & 9284.5/7105 (1.31) \\*
\noalign{\medskip}
J0531$-$7100 & $< 0.12$ & 0.52$_{-0.04} ^{+0.05}$ & 2.50$_{-0.52} ^{+1.70}$ &
1.63$_{-0.16} ^{+0.21}$ & \RD & \multicolumn{5}{c}{---} & 2864.9/2361 (1.20) \\
\noalign{\medskip}
\multirow{5}{*}{J0532$-$6732} & \multirow{5}{*}{0.94$_{-0.51} ^{+0.68}$} &
\multirow{5}{*}{0.53$_{-0.11} ^{+0.29}$} & \multirow{5}{*}{0.83$_{-0.24}
^{+0.46}$} &
\multirow{5}{*}{2.30$_{-1.34} ^{+3.53}$} & \abund{O}0.25$_{-0.05} ^{+0.06}$ &
\multicolumn{5}{c}{\multirow{5}{*}{---}} & \multirow{5}{*}{3824.0/3562 (1.10)} 
\\*
 & & & & & \abund{Ne}0.24$_{-0.07} ^{+0.12}$ & & & & & & \\*
 & & & & & \abund{Mg}0.17$_{-0.14} ^{+0.20}$ & & & & & & \\*
 & & & & & \abund{Si}1.44$_{-0.83} ^{+1.23}$ & & & & & & \\*
 & & & & & \abund{Fe}0.34$_{-0.09} ^{+0.14}$ & & & & & & \\
\noalign{\medskip}
\multirow{4}{*}{J0533$-$7202} & \multirow{4}{*}{0.47$_{-0.41} ^{+0.44}$} &
\multirow{4}{*}{0.31$_{-0.09} ^{+0.06}$} & \multirow{4}{*}{1.39$_{-0.54}
^{+1.16}$} & \multirow{4}{*}{1.13$_{-0.63} ^{+0.87}$} & \abund{O}0.27$_{-0.05}
^{+0.03}$ & \multicolumn{5}{c}{\multirow{4}{*}{---}} &
\multirow{4}{*}{4128.0/3265 (1.26)} \\*
 & & & & & \abund{Ne}0.37$_{-0.09} ^{+0.07}$ & & & & & & \\*
 & & & & & \abund{Mg}$< 0.18$ & & & & & & \\*
 & & & & & \abund{Fe}0.37$_{-0.10} ^{+0.17}$ & & & & & & \\
\noalign{\medskip}
\multirow{4}{*}{J0534$-$6955\tablefootmark{j} \textdagger} & 
\multirow{4}{*}{2.27$_{-0.26}
^{+0.18}$} & \multirow{4}{*}{0.31$_{-0.02} ^{+0.03}$} &
\multirow{4}{*}{3.04$_{-1.06} ^{+0.66}$} &
\multirow{4}{*}{26.1$_{-4.81} ^{+10.83}$} & \abund{O}0.14$\pm 0.01$ &
\multirow{4}{*}{2.27} &
\multirow{4}{*}{1.32$_{-0.20} ^{+0.09}$} &
\multicolumn{3}{l}{pure Si: $\tau=0.05\pm 0.01$; \ EM $= 130.6_{-94.6} 
^{+133.3}$ } & \multirow{4}{*}{3400.8/2727 (1.25)} \\*
&&&&& \abund{Ne}0.23$_{-0.03} ^{+0.02}$ &&&
\multicolumn{3}{l}{pure S\,: \,$\tau = 0.75_{-0.51}^{+4.44}$; \ EM 
$=1.04_{-0.51} ^{+4.69}$} & \\*
&&&&& \abund{Mg}0.27$_{-0.03} ^{+0.04}$ &&&
\multicolumn{3}{l}{pure Fe: $\tau = 0.57_{-0.04}^{+0.06}$; \ EM $=1.26_{-0.17} 
^{+0.25}$} & \\*
 & & & & & \abund{Fe}0.34$_{-0.03} ^{+0.04}$ & & & & & & \\
\noalign{\medskip}
\multirow{3}{*}{J0534$-$7033\tablefootmark{k}} & \multirow{3}{*}{$< 0.03)$} &
\multirow{3}{*}{0.78$\pm0.01$} & \multirow{3}{*}{208$(> 63.1)$} &
\multirow{3}{*}{0.87$_{-0.16} ^{+0.14}$} & \abund{Mg}0.30$\pm 0.18$ &
\multirow{3}{*}{0} & \multirow{3}{*}{0.27$_{-0.04} ^{+0.07}$} &
\multirow{3}{*}{2.65$_{-1.35} ^{+35.3}$} & \multirow{3}{*}{0.46$_{-0.15}
^{+0.58}$} & \multicolumn{1}{c}{ \multirow{3}{*}{RD92}} &
\multirow{3}{*}{3496.61/2922 (1.20)} \\*
 & & & & & \abund{Si}0.08$(< 0.27)$ & & & & & & \\*
 & & & & & \abund{Fe}1.26$_{-0.24} ^{+0.29}$ & & & & & & \\*
\noalign{\medskip}
\multirow{7}{*}{J0535$-$6602\tablefootmark{l}} & \multirow{7}{*}{0.68$_{-0.04}
^{+0.08}$} & \multirow{7}{*}{0.52$\pm 0.01$} & \multirow{7}{*}{7.49$_{-0.90}
^{+2.34}$} & \multirow{7}{*}{251.1$_{-10.6} ^{+31.0}$} &   &
\multirow{7}{*}{2.48$_{-0.81} ^{+1.10}$} & \multirow{7}{*}{1.10$_{-0.01}
^{+0.04}$} & \multirow{7}{*}{10.4$_{-0.90} ^{+3.80}$} &
\multirow{7}{*}{99.1$_{-8.63} ^{+6.05}$} & \abund{O}0.71$_{-0.31} ^{+0.23}$ &
\multirow{7}{*}{1962.4/1634 (1.20)} \\*
&&&&& \abund{O}0.23$_{-0.05} ^{+0.04}$  &&&&& \abund{Ne}$< 0.12$& \\*
&&&&& \abund{Ne}0.57$_{-0.05} ^{+0.02}$ &&&&& \abund{Mg}0.85$_{-0.07} ^{+0.05}$&
\\*
&&&&& \abund{Mg}0.12$_{-0.03} ^{+0.04}$ &&&&& \abund{Si}0.76$_{-0.07}^{+0.05}$ &
\\*
&&&&& \abund{Si}$<0.13$ &&&&& \abund{S}0.48$\pm 0.05$ & \\*
&&&&& \abund{Fe}0.28$\pm 0.01$ &&&&& \abund{Fe}$< 0.02$ & \\*
 & & & & & & & & & & \abund{Ar}0.14$(< 0.34)$ & \\
\noalign{\medskip}
\multirow{5}{*}{J0535$-$6918} & \multirow{5}{*}{2.31$\pm 0.03$} &
\multirow{5}{*}{0.31$_{-0.03} ^{+0.05}$} & \multirow{5}{*}{1.71$_{-0.71}
^{+1.09}$} & \multirow{5}{*}{1.85$_{-0.71} ^{+0.92}$} & \abund{O}0.15$_{-0.03}
^{+0.04}$ & \multicolumn{5}{c}{\multirow{5}{*}{---}} &
\multirow{5}{*}{1999.9/1589 (1.26)} \\*
 & & & & & \abund{Ne}0.34$_{-0.07} ^{+0.09}$ & & & & & & \\*
 & & & & & \abund{Mg}0.34$_{-0.14} ^{+0.18}$ & & & & & & \\*
 & & & & & \abund{Si}1.60$_{-0.86} ^{+1.13}$ & & & & & & \\*
 & & & & & \abund{Fe}0.30$_{-0.07} ^{+0.10}$ & & & & & & \\
\noalign{\medskip}
\multirow{4}{*}{J0536$-$6735\tablefootmark{m}} & \multirow{4}{*}{2.26$\pm 0.20$}
&
\multirow{4}{*}{0.56$_{-0.03} ^{+0.04}$} & \multirow{4}{*}{7.59$_{-1.00}
^{+1.12}$} & \multirow{4}{*}{2.00$_{-0.24} ^{+0.67}$} & \abund{O}1.57$_{-0.19}
^{+0.61}$ & \multicolumn{5}{c}{\multirow{4}{*}{---}} &
\multirow{4}{*}{6015.7/4816 (1.25)} \\*
 & & & & & \abund{Ne}2.52$_{-0.18} ^{+0.87}$ & & & & & & \\*
 & & & & & \abund{Mg}1.35$_{-0.11} ^{+0.27}$ & & & & & & \\*
 & & & & & \abund{Fe}0.04$_{-0.05} ^{+0.09}$ & & & & & & \\
\noalign{\medskip}
\multirow{4}{*}{J0536$-$6913\tablefootmark{n}} & \multirow{4}{*}{4.89$_{-3.30}
^{+0.50}$} & \multirow{4}{*}{0.75$_{-0.07} ^{+0.27}$} &
\multirow{4}{*}{10.5$_{-8.19} ^{+2.34}$} & \multirow{4}{*}{0.11$_{-0.34}
^{+0.02}$} & \multicolumn{1}{c}{\multirow{4}{*}{RD92}} &
\multirow{4}{*}{4.89} & \multirow{4}{*}{4.22$_{-2.06} ^{+0.40}$} &
\multirow{4}{*}{0.08$_{-0.02} ^{+0.03}$} & \multirow{4}{*}{0.06$\pm 0.02$ } &
 \abund{O}1.68$_{-0.39} ^{+0.92}$ & \multirow{4}{*}{1107.4/1139 (0.97)} \\*
&&&&&&&&&& \abund{Ne}0.59$_{-0.26} ^{+0.31}$& \\*
&&&&&&&&&& \abund{Mg}1.51$_{-1.1} ^{+2.35}$& \\*
&&&&&&&&&& \abund{Si}7.9$_{-5.20}^{+17.83}$ & \\
\noalign{\medskip}
\multirow{4}{*}{J0536$-$7039\tablefootmark{o}} & \multirow{4}{*}{$< 0.07$} & 
\multirow{4}{*}{0.27$_{-0.08} ^{+0.14}$} &
\multirow{4}{*}{3.85$_{-2.60} ^{+52.9}$} & \multirow{4}{*}{0.56$_{-0.26}
^{+0.26}$} & \multicolumn{1}{c}{\multirow{4}{*}{RD92}} &
\multirow{4}{*}{0} & \multirow{4}{*}{0.73$_{-0.05} ^{+0.07}$} &
\multirow{4}{*}{7.26$_{-4.00} ^{+500}$} & \multirow{4}{*}{0.50$_{-0.08}
^{+0.18}$ } &
 \abund{Ne}0.12$(< 1.12)$ & \multirow{4}{*}{3555.7/3441 (1.03)} \\*
&&&&&&&&&& \abund{Mg}0.37$_{-0.21} ^{+0.57}$& \\*
&&&&&&&&&& \abund{Si}0.37$(< 0.77)$ & \\*
&&&&&&&&&& \abund{Fe}1.89$_{-0.79}^{+0.69}$ & \\*
\noalign{\medskip}
\multirow{4}{*}{J0537$-$6628} & \multirow{4}{*}{0.56$(< 1.13)$} &
\multirow{4}{*}{0.42$_{-0.09} ^{+0.11}$} & \multirow{4}{*}{10.2$_{-6.7}
^{+23.4}$} & \multirow{4}{*}{0.60$_{-0.55} ^{+1.30}$} & \abund{O}0.32$_{-0.13}
^{+0.75}$ & \multicolumn{5}{c}{\multirow{4}{*}{---}} &
\multirow{4}{*}{3489.1/2983 (1.17)} \\*
 & & & & & \abund{Ne}0.34$_{-0.19} ^{+0.46}$ & & & & & & \\*
 & & & & & \abund{Mg}0.07$(< 0.31)$ & & & & & & \\*
 & & & & & \abund{Fe}0.26$_{-0.10} ^{+0.23}$ & & & & & & \\
\noalign{\medskip}
J0537$-$6910\tablefootmark{p} & 9.11$_{-0.29} ^{+0.33}$ & 4.92$\pm 0.35$ &
0.19$_{-0.06} ^{+0.07}$  & 0.53$_{-0.14} ^{+0.18}$ & \RD &
\multicolumn{5}{c}{---} & 3866.8/3452 (1.12)\\
\noalign{\medskip}
J0540$-$6920\tablefootmark{q} & 6.85$_{-0.16} ^{+0.18}$ & 0.49$\pm 0.06$ &
1.87$_{-0.71} ^{+1.06}$  & 6.03$_{-1.08} ^{+1.32}$ & \RD &
\multicolumn{5}{c}{---} & 5528.0/4763 (1.16)\\
\noalign{\medskip}
J0540$-$6944\tablefootmark{r} & 11.4$_{-0.90} ^{+1.21}$ & 0.20$\pm 0.02$ & CIE
& 
30.64$_{-16.04} ^{+18.70}$ & \RD & \multicolumn{5}{c}{---} & 1323.4/770 (1.72)\\
\noalign{\medskip}
J0541$-$6659 & $< 0.07$ & 0.40$_{-0.07} ^{+0.15}$ & 0.36$_{-0.16} ^{+0.19}$
& 
0.38$_{-0.16} ^{+0.10}$ & \RD & \multicolumn{5}{c}{---} & 6933.5/5886 (1.18)\\
\noalign{\medskip}
\multirow{5}{*}{J0543$-$6858} & \multirow{5}{*}{2.09$_{-0.24} ^{+0.65}$} &
\multirow{5}{*}{1.12$_{-0.51} ^{+0.25}$} & \multirow{5}{*}{0.35$_{-0.08}
^{+0.11}$} & \multirow{5}{*}{1.14$_{-0.59} ^{+0.76}$} & \abund{O}0.37$\pm 0.10$
&
\multicolumn{5}{c}{\multirow{5}{*}{---}} &
\multirow{5}{*}{7211.4/6717 (1.07)} \\*
 & & & & & \abund{Ne}0.36$_{-0.12} ^{+0.15}$ & & & & & & \\*
 & & & & & \abund{Mg}0.29$_{-0.16} ^{+0.23}$ & & & & & & \\*
 & & & & & \abund{Si}0.36$(< 0.98)$ & & & & & & \\*
 & & & & & \abund{Fe}0.32$_{-0.13} ^{+0.16}$ & & & & & & \\
\noalign{\medskip}
\multirow{5}{*}{J0547$-$6941} & \multirow{5}{*}{4.91$_{-1.08} ^{+1.31}$} &
\multirow{5}{*}{1.25$\pm 0.18$} & \multirow{5}{*}{2.38$_{-0.30} ^{+0.67}$} &
\multirow{5}{*}{0.83$_{-0.25} ^{+0.24}$} & \abund{O}0.26$_{-0.15} ^{+0.28}$ &
\multicolumn{5}{c}{\multirow{5}{*}{---}} &
\multirow{5}{*}{3451.2/3196 (1.08)} \\*
 & & & & & \abund{Ne}$< 0.37$ & & & & & & \\*
 & & & & & \abund{Mg}0.39$_{-0.14} ^{+0.21}$ & & & & & & \\*
 & & & & & \abund{Si}0.71$_{-0.27} ^{+0.33}$ & & & & & & \\*
 & & & & & \abund{Fe}2.30$\pm 0.70$ & & & & & & \\
\noalign{\medskip}
\multirow{6}{*}{J0547$-$6943\tablefootmark{s}} & \multirow{6}{*}{3.73$_{-0.72}
^{+1.15}$} &
\multirow{6}{*}{0.27$_{-0.05} ^{+0.26}$} & \multirow{6}{*}{493$(> 5.32)$} &
\multirow{6}{*}{3.67$_{-2.72} ^{+4.78}$} & \abund{O}0.22$_{-0.09} ^{+0.40}$ &
\multirow{6}{*}{3.73} & \multirow{6}{*}{2.16$_{-0.27} ^{+0.45}$} &
\multirow{6}{*}{2.33$_{-0.66} ^{+1.23}$} &
\multirow{6}{*}{0.89$_{-0.25} ^{+0.22}$} &
\multicolumn{1}{c}{\multirow{6}{*}{ --- }} & 
\multirow{6}{*}{4468.2/4356 (1.03)} \\*
 & & & & & \abund{Ne}0.27$_{-0.22} ^{+0.33}$ & & & & & & \\*
 & & & & & \abund{Mg}0.76$_{-0.28} ^{+0.45}$ & & & & & & \\*
 & & & & & \abund{Si}0.58$_{-0.24} ^{+0.35}$ & & & & & & \\*
 & & & & & \abund{S}0.33$_{-0.31} ^{+0.34}$ & & & & & & \\*
 & & & & & \abund{Fe}0.58$_{-0.19} ^{+0.26}$ & & & & & & \\
\noalign{\medskip}
\multirow{6}{*}{J0547$-$7025 \textdagger} & \multirow{6}{*}{3.04$_{-0.35} 
^{+0.37}$} & 
\multirow{6}{*}{0.31$_{-0.05} ^{+0.01}$} & \multirow{6}{*}{3.12$_{-3.66}
^{+11.7}$} & \multirow{6}{*}{4.72$_{-1.81} ^{+3.14}$} &
\multicolumn{1}{c}{\multirow{6}{*}{RD92}} &
\multirow{6}{*}{3.04} & \multirow{6}{*}{0.80$_{-0.10} ^{+0.13}$} &
\multirow{6}{*}{1.18$_{-0.18} ^{+0.90}$} & \multirow{6}{*}{1.65$_{-0.52}
^{+0.35}$ } & \abund{O}$< 0.08$ & \multirow{6}{*}{2949.1/2559 (1.15)} \\*
&&&&&&&&&& \abund{Ne}0.32$\pm 0.01$& \\*
&&&&&&&&&& \abund{Mg}0.07$(< 0.12)$& \\*
&&&&&&&&&& \abund{Si}0.28$\pm 0.25$ & \\*
&&&&&&&&&& \abund{S}1.10$_{-0.57}^{+0.50}$ & \\*
&&&&&&&&&& \abund{Fe}0.92$_{-0.19}^{+0.23}$ & \\
\noalign{\medskip}
%
%
\end{longtable}
\tablefoot{Columns are described in
Sect.\,\ref{results_spectra_general}.\\
\tablefoottext{a}{Only MOS data available;\ }
\tablefoottext{b}{Fit includes a power-law component for the central PWN;\ }
\tablefoottext{c}{Same absorption column and abundances in the two components;
Nitrogen abundance is also fitted to 0.07 solar;\ }
\tablefoottext{d}{Results from \citetalias{2014A&A...561A..76M};\ }
\tablefoottext{e}{Results from \citetalias{2014MNRAS.439.1110B}. The first
component is a Sedov model. The second is split into two pure-metal components
(O and Fe) with the corresponding emission measures given as EM$_{X} \times (n_X
/ n_H)$; }
\tablefoottext{f}{Same absorption column and abundances in the two components;
Fit includes a power-law component for SGR 0526$-$66
\citep{2012ApJ...748..117P};\ }
\tablefoottext{g}{$N_{H\mathrm{\ LMC}}$ fixed to the value from the \ion{H}{i}
map of the LMC;\ }
\tablefoottext{h}{Abundances from RD92 are scaled by a common factor;\ }
\tablefoottext{i}{Affected by strong background flare;\ }
\tablefoottext{j}{The second component comprises three pure-metals NEI models
with a common temperature and distinct $\tau$ and EM, for each Si, S, and Fe;\ }
\tablefoottext{k}{C, N, O, and Ne abundances of the iron-rich component
(Component 1) are fixed to 0;\ }
\tablefoottext{l}{Abundance of Ca fixed at 0 in the second component; fit
includes an Fe~K (Gaussian) line, see Sect.\,\ref{results_spectra_FeK};\ }
\tablefoottext{m}{Fit includes a power-law component for the interior HMXB
\citep{2012ApJ...759..123S};\ }
\tablefoottext{n}{Elements other than O, Ne, Mg, and Si in the second component
are set to 0;\ }
\tablefoottext{o}{C, N, and O abundances of the iron-rich component (Component
2) are fixed to 0;\ }
\tablefoottext{p}{Fit includes a power-law component for the interior PWN;\ }
\tablefootmark{q}{Strongly affected by the emission from PSR~0540$-$69.3
(modeled by a power law);\ }
\tablefoottext{r}{Dominated by the contamination from LMC~X-1 (modeled by a
power law);\ }
\tablefoottext{s}{Same absorption column and abundances in the two 
components.}\\
\revision{\textdagger\ These SNRs were observed with \xmm\ EPIC instruments for 
the first time over the course of this work. Their spectra are shown in 
Fig.\,\ref{fig_data_spectra_newSNRs}.}
}

\begin{table}
\caption{X-ray spectral results for bright LMC SNRs.}
\label{appendix_table_spectra_brightest}
\begin{center}
\begin{tabular}{@{}l @{\hspace{0.05cm}} @{\hspace{0.05cm}} c @{\hspace{0.05cm}}
@{\hspace{0.05cm}} c @{\hspace{0.05cm}} @{\hspace{0.05cm}} c @{\hspace{0.15cm}}
@{\hspace{0.15cm}} c @{\hspace{0.05cm}} @{\hspace{0.05cm}} c @{\hspace{0.05cm}}
@{\hspace{0.05cm}} c @{\hspace{0.05cm}} @{\hspace{0.05cm}} c @{\hspace{0.05cm}}
@{\hspace{0.05cm}} c @{\hspace{0.05cm}} @{\hspace{0.05cm}} c @{\hspace{0.05cm}}
@{\hspace{0.05cm}} c @{\hspace{0.05cm}} @{\hspace{0.05cm}} c @{\hspace{0.05cm}}
@{\hspace{0.05cm}} c @{}}
\hline\hline
\noalign{\smallskip}
\noalign{\smallskip}
Component & $N_{H\mathrm{\ LMC}}$ & $kT$ & $\tau$ & EM &
O & Ne & Mg & Si & S & Ar & Ca & Fe \\
 & ($10^{21}$~cm$^{-2}$) & (keV) & ($10^{11}$ s\,cm$^{-3}$) & ($10^{58}$
cm$^{-3}$) & & & & & & & &  \\
\noalign{\smallskip}
\noalign{\smallskip}
\hline
\noalign{\smallskip}
\multicolumn{13}{c}{DEM~L71 --- MCSNR J0505$-$6753 : 
$\chi ^2 / \nu$ = 3300.5/1784 (1.85)}\\
\noalign{\smallskip}
\hline
\noalign{\smallskip}
CSM/ISM & \multirow{3}{*}{0.28$\pm 0.001$} & 0.46$\pm0.01$ &
2.59$_{-0.02}^{+0.03}$ &
38.2$_{-0.17} ^{+2.88}$ & 0.26$\pm0.001$ & 0.40$\pm 0.001$ &
0.49$_{-0.02}^{+0.01}$ & 0 & 0 & --- & --- & 0 \\
\noalign{\smallskip}
Ejecta$_{\mathrm{\ cool}}$ &  & 0.42$_{-0.009}^{+0.004}$ &
3.56$_{-0.40}^{+0.98}$ & 11.3$_{-0.14}^{+0.12}$ & 0 & 0 & 0 &
\multirow{2}{*}{1.03$_{-0.03}^{+0.05}$} &
\multirow{2}{*}{1.27$_{-0.07}^{+0.09}$} & 0 & 0 & \multirow{2}{*}{1.21$\pm
0.001$} \\
Ejecta$_{\mathrm{\ hot}}$ &  & 0.88$_{-0.001}^{+0.007}$ &
1.83$_{-0.04}^{+0.06}$ & 7.30$_{-0.63}^{+0.36}$ & 0 & 0 & 0 &
 &  & 0 & 0 &  \\
\noalign{\smallskip}
\hline
\noalign{\smallskip}
\multicolumn{13}{c}{N103B --- MCSNR J0509$-$6844 : $\chi ^2 / \nu$ = 
694.2/583 (1.19)}\\
\noalign{\smallskip}
\hline
\noalign{\smallskip}
CSM/ISM & \multirow{3}{*}{3.09$_{-0.11}^{+0.20}$} & 0.33$_{-0.10}^{+0.03}$ &
34.9$_{-19.7}^{+137.5}$ & 40.73$_{-4.68}^{+4.27}$
& --- & --- & --- & --- & --- & --- & --- & --- \\
\noalign{\smallskip}
Ejecta$_{\mathrm{\ cool}}$ & & 0.71$\pm 0.02$ &
$> 41.4$ & 25.9$_{-2.84}^{+1.47}$ & \multirow{2}{*}{$<0.25$} &
\multirow{2}{*}{1.71$_{-0.32}^{+0.49}$} &
\multirow{2}{*}{0.33$_{-0.14}^{+0.10}$} &
\multirow{2}{*}{2.84$_{-0.30}^{+0.56}$} &
\multirow{2}{*}{3.48$_{-0.28}^{+0.33}$} &
\multirow{2}{*}{5.26$_{-0.59}^{+1.0}$}  &
\multirow{2}{*}{9.51$_{-2.05}^{+2.43}$} &
\multirow{2}{*}{1.10$_
{-0.14}^{+0.36}$} \\
Ejecta$_{\mathrm{\ hot}}$ & & 1.62$_{-0.16}^{+0.10}$ &
8.59$_{-2.64}^{+4.27}$ & 9.95$_{-0.57}^{+1.50}$ & & & & & & & &  \\
\noalign{\smallskip}
\hline
\noalign{\smallskip}
\multicolumn{13}{c}{N132D --- MCSNR J0525$-$6938 : $\chi ^2 / \nu$ = 
1288.4/1116 (1.15)}\\
\noalign{\smallskip}
\hline
\noalign{\smallskip}
CSM/ISM & \multirow{3}{*}{0.74$\pm 0.03$}& 0.64$_{-0.01}^{+0.02}$ &
5.97$_{-0.35}^{+0.44}$ &
358.6$_{-3.33}^{+4.78}$ & 0.05$_{-0.01}^{+0.02}$ & 0.46$\pm 0.02$ & 0.19$\pm
0.01$ & 0.39$\pm 0.03$ & --- & --- & --- & 0.28$\pm0.01$ \\
O-rich & & 1.46$_{-0.02}^{+0.04}$ & 10.0$_{-4.030}^{+1.13}$ &
33.2$_{-1.81}^{+10.1}$ & 9.03$_{-3.48}^{+1.44}$ & $< 0.25$ &
2.59$_{-0.17}^{+0.19}$ & 1.37$_{-0.15}^{+0.19}$ & 1.03$_{-0.10}^{+0.11}$ &
--- & --- & 0.15$_{-0.02}^{+0.05}$ \\
Si-S-Fe &  & 5.12$_{-0.72}^{+0.93}$ &
500 & 7.84$_{-0.86}^{+1.33}$ & 0 & 0 & 0 & 0 & 0 & 5.29$_{-3.26}^{+4.05}$ &
2.20$(<4.86)$ & 1.0 \\
\noalign{\smallskip}
\hline
\noalign{\smallskip}
\multicolumn{13}{c}{0509$-$67.5 --- MCSNR J0509$-$6731 : $\chi ^2 / \nu$ = 
871.1/419 (2.08)}\\
\noalign{\smallskip}
\hline
\noalign{\smallskip}
CSM/ISM & \multirow{4}{*}{1.64$\pm 0.07$}& 0.27$_{-0.01}^{+0.09}$ &
9.44$_{-3.02}^{+1.69}$ & 6.26$_{-2.70}^{+4.42}$ & 1.14$_{-0.03}^{+1.58}$ &
2.37$_{-0.05}^{+0.06}$ & 2.45$_{-0.56}^{+0.29}$ & 0 & 0 & 0 & 0 &
2.84$_{-0.15}^{+5.78}$ \\
Fe$_{\mathrm{\ cool}}$ & & 1.42$_{-0.05}^{+0.09}$ & 0.08$\pm 0.01$ &
10.5$_{-4.46}^{+2.20}$ & 0 & 0 & 0 & 0 & 0 & 0 & 0 & pure \\
Fe$_{\mathrm{\ hot}}$-Ca & & 11.7$_{-1.5}^{+9.9}$ & 11.4$_{-4.4}^{+7.6}$ &
1.75$_{-0.24}^{+0.29}$ & 0 & 0 & 0 & 0 & 0 & 0 & 4.51$_{-1.64}^{+2.05}$ & 1.0 \\
Si-S-Ar & & 1.18$_{-0.12}^{+0.11}$ & 0.23$_{-0.003}^{+0.014}$ &
57.3$_{-3.0}^{+3.5}$ & 0 & 0 & 0 & 1.0 & 3.28$_{-0.11}^{+0.19}$ &
3.38$_{-0.10}^{+0.25}$ & 0 & 0 \\
\noalign{\smallskip}
\hline
\noalign{\smallskip}
\multicolumn{13}{c}{0519$-$69.0 --- MCSNR J0519$-$6902 : $\chi ^2 / \nu$ = 
5971.7/2438 (2.45)}\\
\noalign{\smallskip}
\hline
\noalign{\smallskip}
CSM/ISM & \multirow{6}{*}{0.96$\pm 0.04$}& 0.60$_{-0.01}^{+0.02}$ &
29.4$_{-8.9}^{+29.6}$ &
22.5$_{-0.90}^{+0.53}$ & --- & --- & 0.35$_{-0.01}^{+0.03}$ & --- & --- & --- &
--- & --- \\
Fe$_{\mathrm{\ cool}}$ & & 1.37$_{-0.02}^{+0.01}$ &
0.88$\pm0.02$ &
10.0$_{-0.23}^{+0.30}$ & 0 & 0 & 0 & 0 & 0 & 0 & 0 & pure \\
Fe$_{\mathrm{\ hot}}$ & & 8.12 & 3.60 & 8.61 & 0 & 0 & 0
& 0 & 0 & 0 & 0 & pure \\
Si-S & & 3.43$_{-0.45}^{+0.50}$ & 1.12$_{-0.06}^{+0.12}$ &
10.6$_{-30}^{+0.67}$ & 0 & 0 & 0 & 1.0 & 1.26$\pm 0.06$ & 0 & 0 & 0 \\
Ar-Ca & & 4.42$\pm 1.20$ & 2.03$_{-0.46}^{+0.51}$ &
15.8$_{-5.5}^{+4.6}$ & 0 & 0 & 0 & 0 & 0 & 1.0 & 1.33 $_{-0.18}^{+0.25}$ & 0 \\
O & & 4.32$_{-4.03}^{+1.47}$ & 0.04$_{-0.02}^{+0.17}$ & 
0.10$_{-0.03}^{+0.08}$ & pure & 0 & 0 & 0 & 0 & 0 & 0 & 0 \\
\noalign{\smallskip}
\noalign{\smallskip}
\hline
\end{tabular}
\end{center}
\tablefoot{Details of the spectral models used are given in
Sect.\,\ref{results_spectra_brightest}.}
\end{table}

\begin{table}
\caption{Spectral results, abundances, fluxes, and Fe K line properties of
SNR~1987A}
\label{appendix_table_spectra_1987A}
\begin{center}
\begin{tabular}{c c c c c c c c c}
\hline\hline
\noalign{\smallskip}
Epoch & Age  & \multicolumn{2}{c}{low-$kT$ component} & 1.15~keV component &
\multicolumn{2}{c}{high-$kT$ component} & \multicolumn{2}{c}{ionisation age} \\
\noalign{\smallskip}
\cline{3-7}
\noalign{\smallskip}
 &  & $kT$ & EM & EM & $kT$ & EM & \multicolumn{2}{c}{}\\
 & (days) & keV & ($10^{58}$~cm$^{-3}$) & ($10^{58}$~cm$^{-3}$) & (keV) &
($10^{58}$~cm$^{-3}$) & \multicolumn{2}{c}{($10^{11}$ s\,cm$^{-3}$)} \\
\noalign{\smallskip}
\hline
\noalign{\smallskip}
2007 Jan & 7267 & 0.35$\pm 0.01$ & 35.69$_{-2.07}^{+2.34}$ &
7.31$_{-0.27}^{+0.17}$ & 13.89$_{-4.80}^{+7.11}$ & 0.55$_{-0.03}^{+0.05}$ &
\multicolumn{2}{c}{4.42$_{-0.56}^{+0.41}$} \\
\noalign{\smallskip}
2008 Jan & 7626 & 0.36$_{-0.01}^{+0.02}$ & 42.63$_{-3.51}^{+4.09}$ &
7.80$_{-0.37}^{+0.75}$ & 5.82$_{-0.52}^{+4.69}$ & 1.06$_{-0.20}^{+0.15}$ &
\multicolumn{2}{c}{4.60$_{-0.53}^{+0.57}$} \\
\noalign{\smallskip}
2009 Jan & 8013 & 0.37$\pm 0.01$ & 49.66$_{-4.57}^{+3.82}$ &
11.89$_{-1.24}^{+0.42}$ & 10.86$_{-4.71}^{+3.64}$ & 0.90$_{-0.05}^{+0.02}$ &
\multicolumn{2}{c}{5.27$_{-0.67}^{+0.39}$} \\
\noalign{\smallskip}
2009 Dec & 8328 & 0.40$\pm 0.01$ & 47.29$_{-4.37}^{+2.64}$ &
12.97$_{-1.14}^{+0.40}$ & 5.98$_{-1.14}^{+1.47}$ & 1.40$_{-0.19}^{+0.34}$ &
\multicolumn{2}{c}{4.77$_{-0.25}^{+0.26}$} \\
\noalign{\smallskip}
2010 Dec & 8693 & 0.42$_{-0.01}^{+0.02}$ & 48.13$_{-1.73}^{+2.03}$ &
13.40$_{-0.90}^{+1.28}$ & 3.80$_{-0.49}^{+0.99}$ & 2.43$_{-0.53}^{+0.34}$ &
\multicolumn{2}{c}{4.98$_{-0.43}^{+0.71}$} \\
\noalign{\smallskip}
2011 Dec & 9048 & 0.44$_{-0.01}^{+0.02}$ & 45.72$_{-1.38}^{+1.66}$ &
15.92$_{-0.63}^{1.10+}$ & 5.66$_{-0.70}^{+0.41}$ & 2.20$_{-0.34}^{+0.27}$ &
\multicolumn{2}{c}{4.84$_{-0.37}^{+0.33}$} \\
\noalign{\smallskip}
2012 Dec & 9423 & 0.44$_{-0.01}^{+0.02}$ & 44.56$_{-2.41}^{+0.99}$ &
19.90$_{-1.26}^{+0.44}$ & 7.66$_{-0.86}^{+1.81}$ & 1.95$_{-0.07}^{+0.15}$ &
\multicolumn{2}{c}{5.76$_{-0.26}^{+0.54}$} \\
\noalign{\smallskip}
\hline\hline
\noalign{\smallskip}
 & N & O & Ne & Mg & Si & S & Fe & \\ 
\noalign{\smallskip}
\hline
\noalign{\smallskip}
 & 1.35$\pm 0.06$ & 0.08$\pm 0.01$ & 0.29$\pm 0.01$ & 0.26$\pm 0.01$ & 0.51$\pm
0.01$ & 0.48$\pm 0.02$ & 0.24$\pm 0.01$ & \\ 
\noalign{\smallskip}
\noalign{\smallskip}
\hline\hline
\noalign{\smallskip}
\noalign{\smallskip}
Epoch & Age  & Flux (0.5--2 keV) & Flux (3--10 keV) & 
$\Delta{F_X}$ & $E_{\mathrm{line}}$ & $\sigma$-width & Flux & EW \\
 & (days) & \multicolumn{2}{c}{$\left(10^{-13}\right.$
erg\,s$^{-1}$\,cm$\left.^{-2}\right)$} & (\%) & (keV) & (eV) & ($10^{-6}$
ph\,cm$^{-2}$\,s$^{-1}$) & (eV) \\
\noalign{\smallskip}
\hline
\noalign{\smallskip}
2007 Jan & 7267 & 33.51$^{+0.46}_{-0.49}$ & 4.09$^{+0.86}_{-1.71}$ & --- &
--- & --- &  --- & --- \\
\noalign{\smallskip}
2008 Jan & 7626& 43.34$^{+0.56}_{-0.57}$ & 5.22$^{+0.67}_{-1.54}$ & 29.8 &
6.58$^{+0.05}_{-0.07}$&46 $(< 146)$&1.07$^{+0.57}_{-0.46}$&174 \\
\noalign{\smallskip}
2009 Jan &  8013 & 52.77$^{+0.71}_{-0.64}$ & 6.31$^{+0.76}_{-1.86}$ & 20.5 &
6.55$^{+0.15}_{-0.14}$&125 $(< 433)$&0.94$^{+1.50}_{-0.65}$&169 \\
\noalign{\smallskip}
2009 Dec & 8328 & 59.04$^{+0.66}_{-0.69}$ & 7.42$^{+0.89}_{-2.37}$ & 13.8 & 
6.63$^{+0.11}_{-0.09}$&229 $(< 424)$&2.64$^{+1.69}_{-1.78}$&432 \\
\noalign{\smallskip}
2010 Dec & 8693 & 65.90$^{+1.08}_{-1.17}$ & 8.16$^{+0.96}_{-2.24}$  & 11.6  & 
6.61$^{+0.06}_{-0.06}$&105 $(< 227)$&2.51$^{+1.39}_{-0.99}$&344 \\
\noalign{\smallskip}
2011 Dec & 9048 & 70.69$^{+0.71}_{-0.76}$ & 11.31$^{+1.08}_{-2.16}$ & 7.5 &
6.61$^{+0.06}_{-0.06}$& 83 $(< 178)$&2.08$^{+1.01}_{-0.86}$&238 \\
\noalign{\smallskip}
2012 Dec & 9423 & 74.60$^{+0.79}_{-0.87}$ & 11.65$^{+1.22}_{-2.33}$ & 5.4 &
6.78$^{+0.06}_{-0.05}$ & 84 $(< 155)$ & 2.11$^{+0.97}_{-0.85}$& 230 \\
\noalign{\smallskip}
\hline
\end{tabular}
\end{center}
\tablefoot{Top panel: best-fit values and 90\,\% C.\,L. uncertainties for the
parameters of the three-components spectral model described in 
Sect.\,\ref{results_spectra_1987A}, in spectra obtained from 2007 to 2012.
Middle panel: Best-fit abundances relative to the solar values as listed in
\citet{2000ApJ...542..914W}.
Bottom panel: Fluxes and Fe~K line properties. Columns (3) and (4) list the soft
and hard X-ray fluxes with 3$\sigma$ errors (99.73\,\% C.L.). Column (5) gives
the increase rate of the flux (in \%) since previous measurement, normalised to
one year. Columns (6) to (9) give the central energy, $\sigma$-width, total
photon flux and equivalent width (EW) of the Gaussian used to characterise the
Fe~K feature in the spectra of SNR~1987A (with 90\,\% C.\,L. uncertainties).
}
\end{table}

\end{landscape}

}

\onlfig{
\section{X-ray images, spectral extraction regions, and SFH for all LMC SNRs}
    \label{appendix_images}

This Appendix presents, for each SNR, an X-ray image (top), the regions used for 
spectral analysis (middle), and the SFH of the cell including the remnant 
(bottom panel). The images are using the (0.3--0.7 keV), medium (0.7--1.1 keV), 
and hard (1.1--4.2 keV) bands as red, green, and blue components, respectively, 
\revision{and have been adaptively smoothed (see 
Sect.\,\ref{observations_reduction}}. The white bars indicate the scale of 
1\arcmin. North is up and east is left. A linear scale is used to display the 
pixel values, but the cut levels are adapted for each SNR. The extraction 
regions used for spectral analysis are shown for pn, MOS1, and MOS2 detectors 
(left to right), as in Fig.\,\ref{fig_data_spectra_extraction}. The star 
formation history plots are shown as in Fig.\,\ref{fig_sfh_examples}.
\revision{For MCSNR J0509$-$6844, J0509$-$6731, and J0509$-$6902, which are 
highly unresolved with \xmm, we show \chandra\ images, using processed data 
obtained from the \chandra\ SNR catalogue.}

\begin{figure}[ht]
    \begin{center}
    \includegraphics[width=0.253\hsize]
{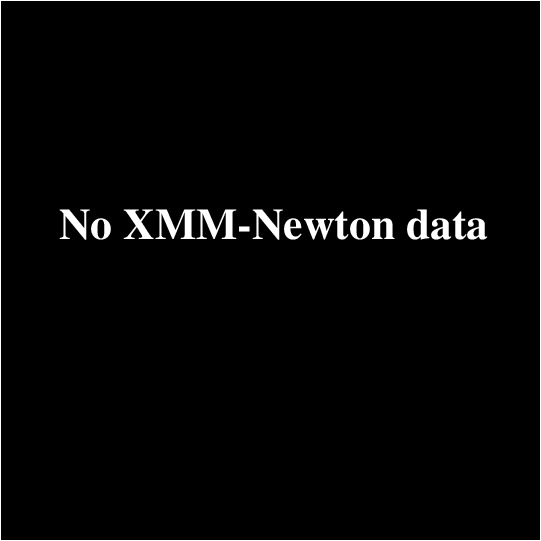}
    \includegraphics[width=0.253\hsize]
{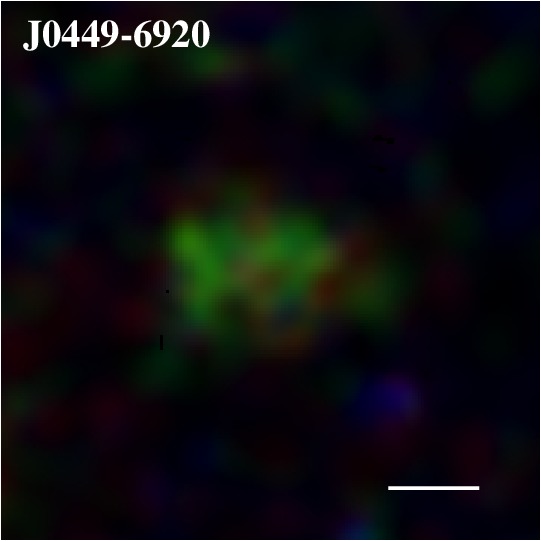}
    \includegraphics[width=0.253\hsize]
{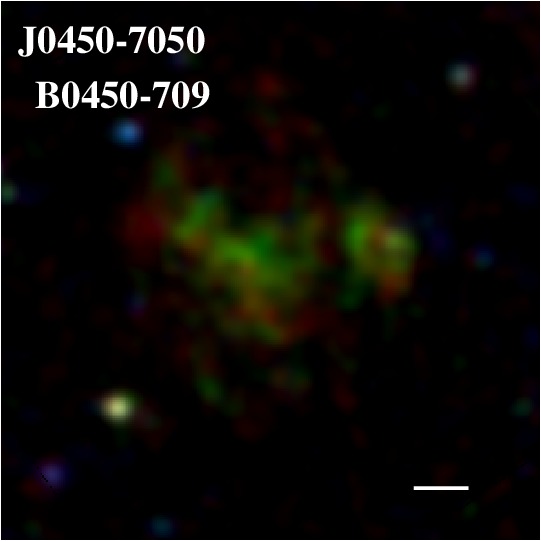}

    \includegraphics[width=0.253\hsize]
{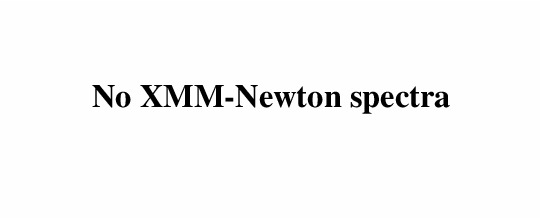}
    \includegraphics[width=0.253\hsize]
{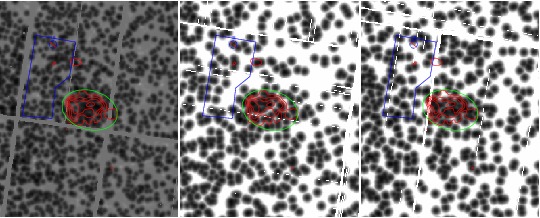}
    \includegraphics[width=0.253\hsize]
{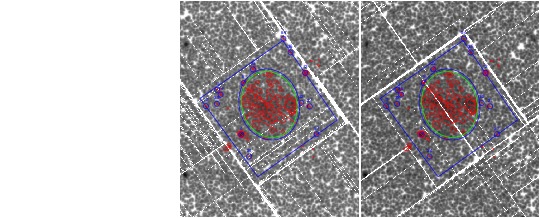}

    \includegraphics[width=0.253\hsize]
{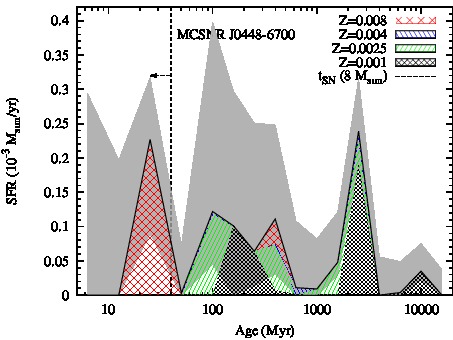}
    \includegraphics[width=0.253\hsize]
{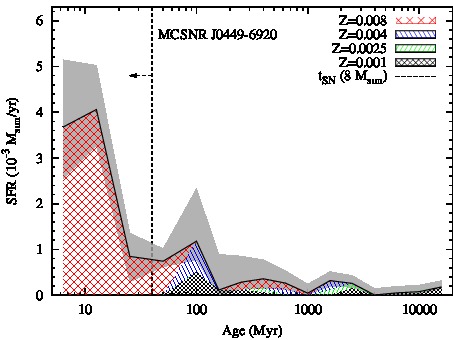}
    \includegraphics[width=0.253\hsize]
{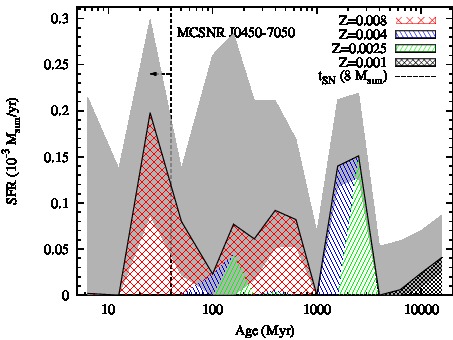}

    \includegraphics[width=0.253\hsize]
{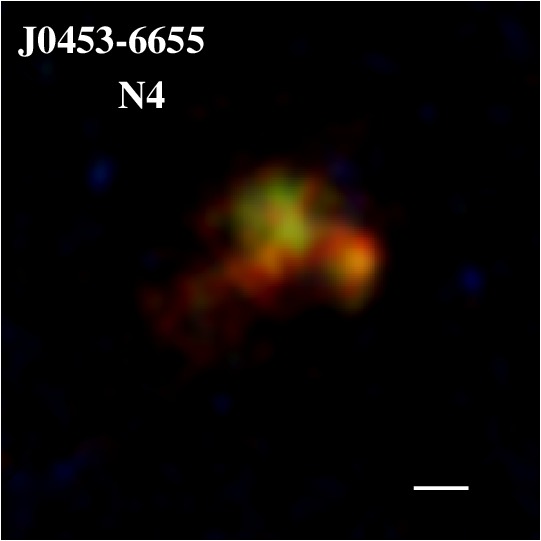}
    \includegraphics[width=0.253\hsize]
{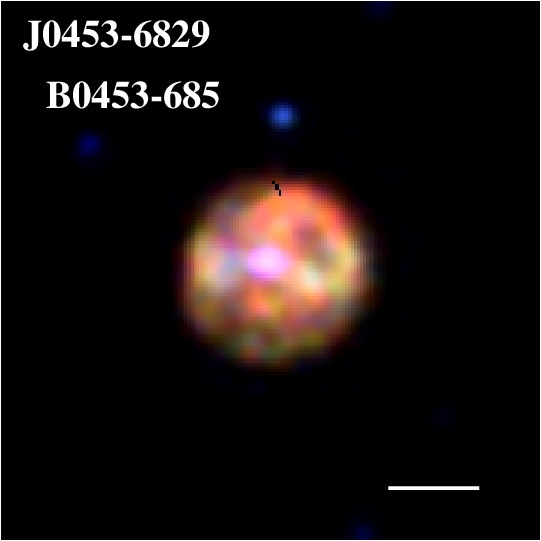}
    \includegraphics[width=0.253\hsize]
{noxmmdata.jpg}

    \includegraphics[width=0.253\hsize]
{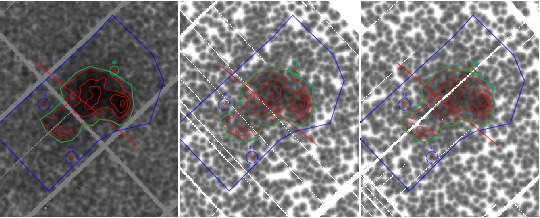}
    \includegraphics[width=0.253\hsize]
{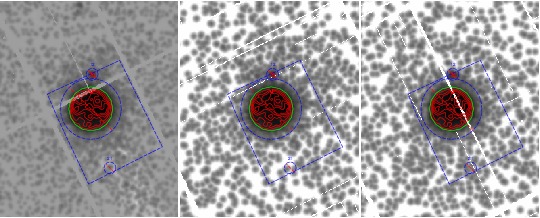}
    \includegraphics[width=0.253\hsize]
{noxmmspectra.jpg}

    \includegraphics[width=0.253\hsize]
{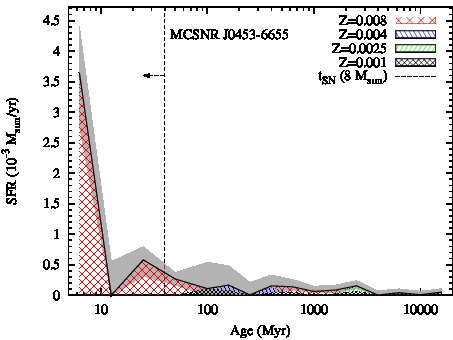}
    \includegraphics[width=0.253\hsize]
{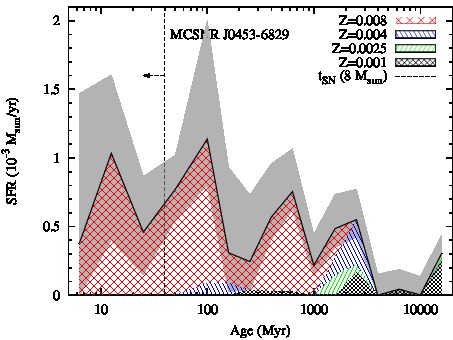}
    \includegraphics[width=0.253\hsize]
{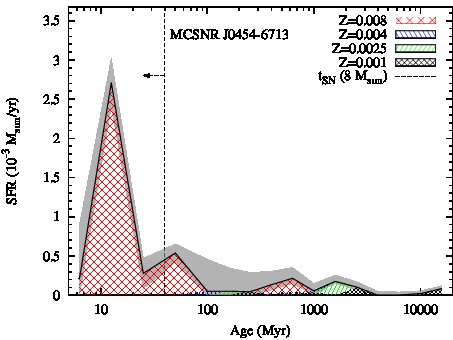}
    \end{center}
\end{figure}

\begin{figure}[ht]
    \begin{center}
    \includegraphics[width=0.312\hsize]
{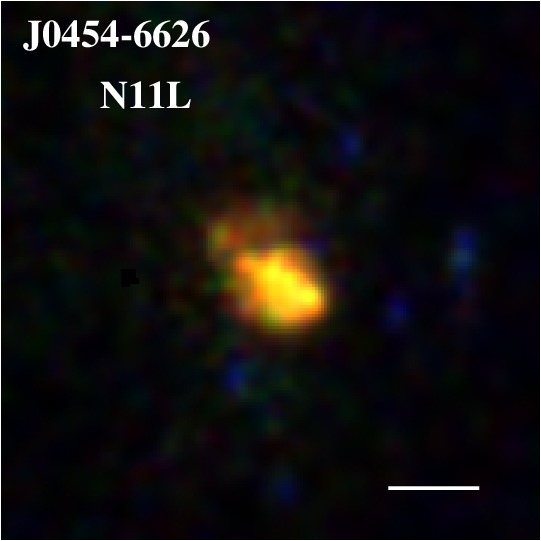}
    \includegraphics[width=0.312\hsize]
{noxmmdata.jpg}
    \includegraphics[width=0.312\hsize]
{noxmmdata.jpg}

    \includegraphics[width=0.312\hsize]
{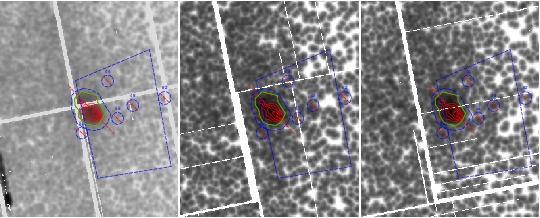}
    \includegraphics[width=0.312\hsize]
{noxmmspectra.jpg}
    \includegraphics[width=0.312\hsize]
{noxmmspectra.jpg}

    \includegraphics[width=0.312\hsize]
{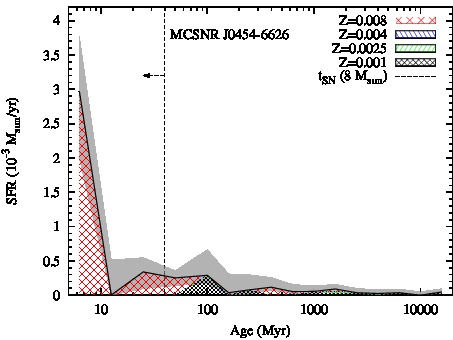}
    \includegraphics[width=0.312\hsize]
{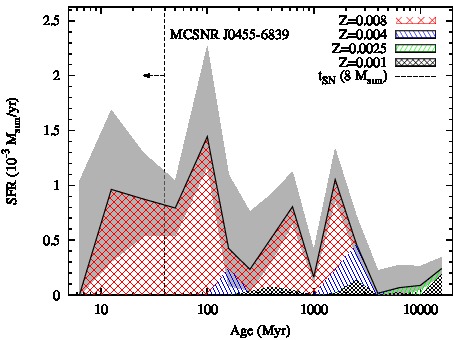}
    \includegraphics[width=0.312\hsize]
{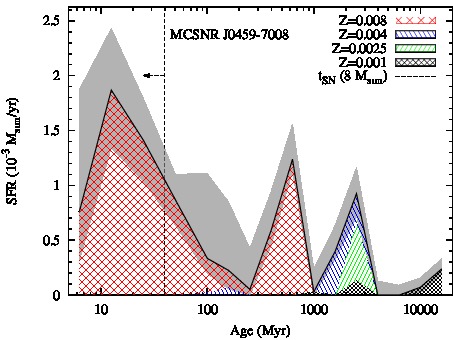}

\vspace{1cm}

\includegraphics[width=0.312\hsize]
{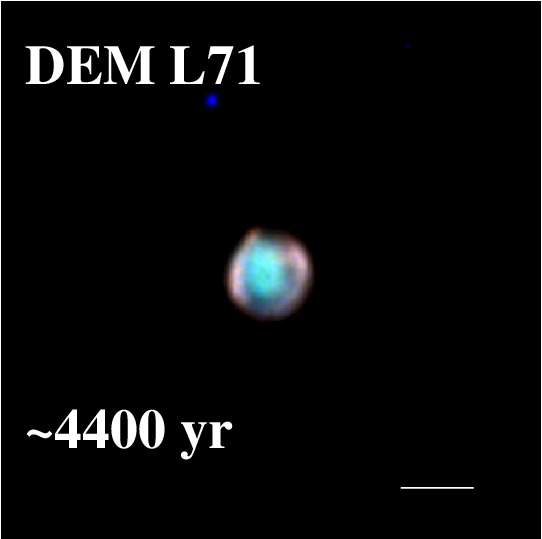}
    \includegraphics[width=0.312\hsize]
{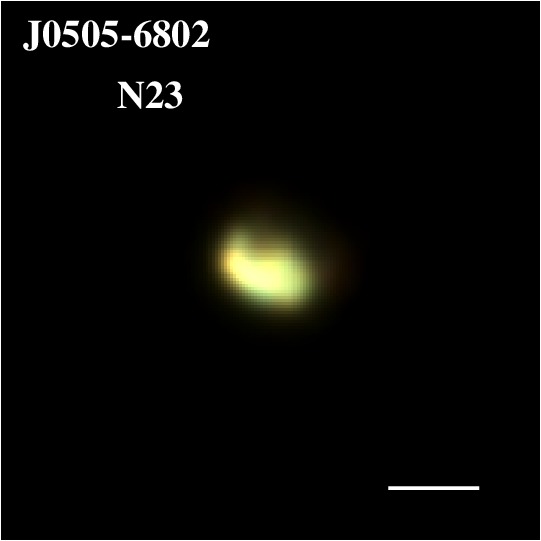}
    \includegraphics[width=0.312\hsize]
{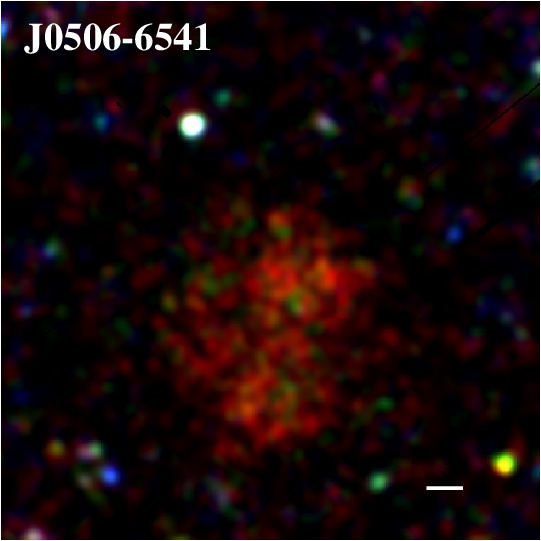}

    \includegraphics[width=0.312\hsize]
{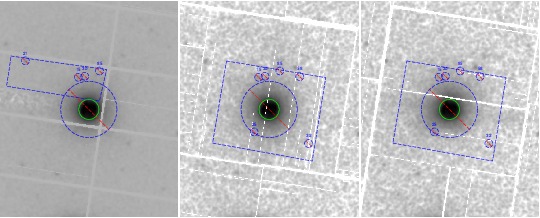}
    \includegraphics[width=0.312\hsize]
{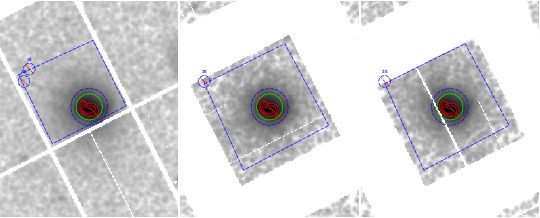}
    \includegraphics[width=0.312\hsize]
{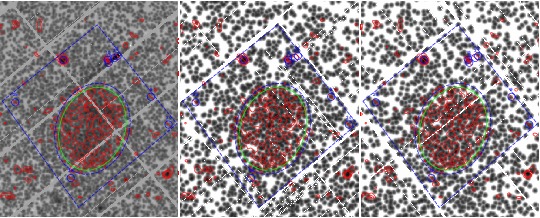}

    \includegraphics[width=0.312\hsize]
{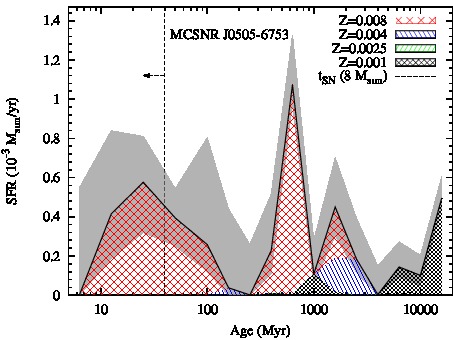}
    \includegraphics[width=0.312\hsize]
{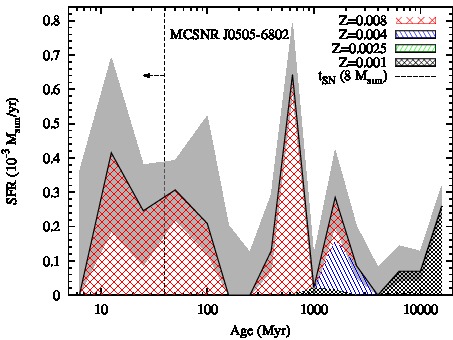}
    \includegraphics[width=0.312\hsize]
{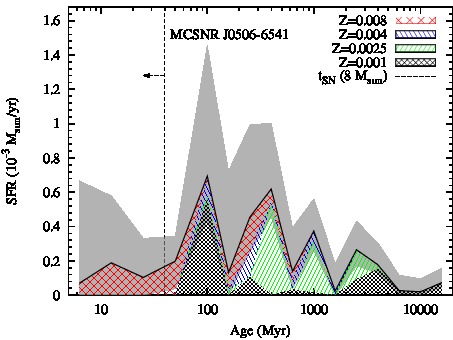}
    \end{center}
\end{figure}

\begin{figure}[ht]
    \begin{center}
    \includegraphics[width=0.312\hsize]
{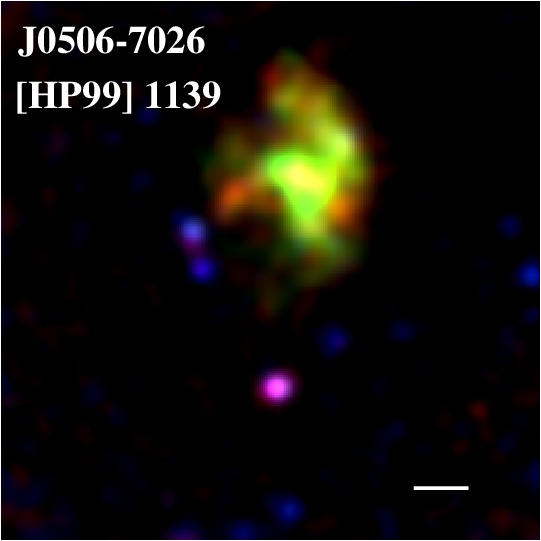}
    \includegraphics[width=0.312\hsize]
{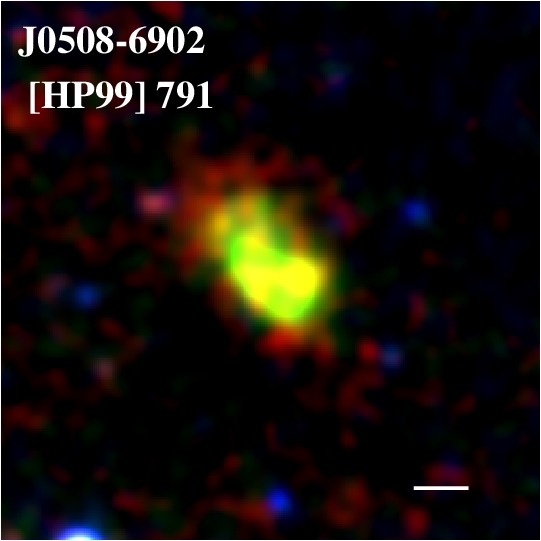}
    \includegraphics[width=0.312\hsize]
{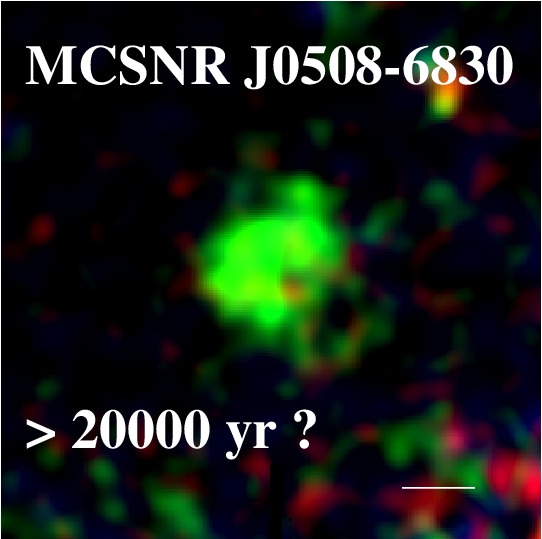}

    \includegraphics[width=0.312\hsize]
{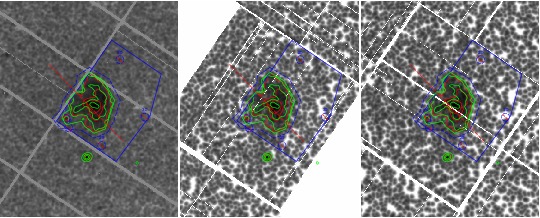}
    \includegraphics[width=0.312\hsize]
{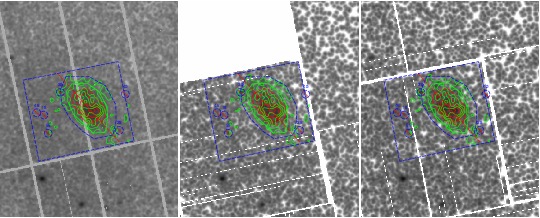}
    \includegraphics[width=0.312\hsize]
{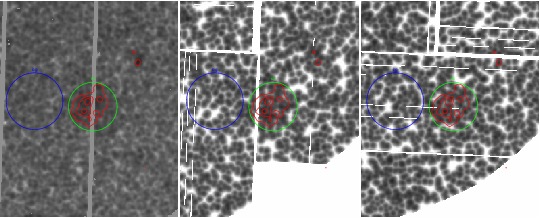}

    \includegraphics[width=0.312\hsize]
{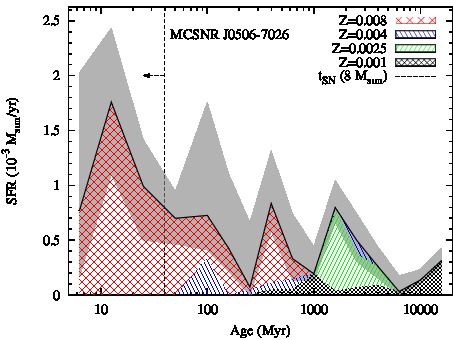}
    \includegraphics[width=0.312\hsize]
{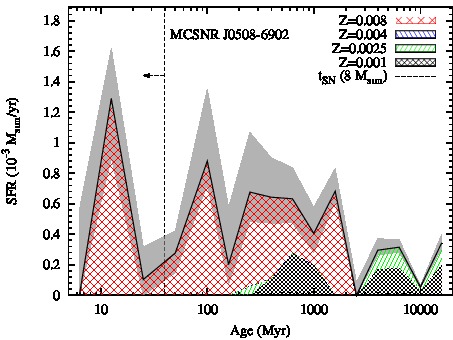}
    \includegraphics[width=0.312\hsize]
{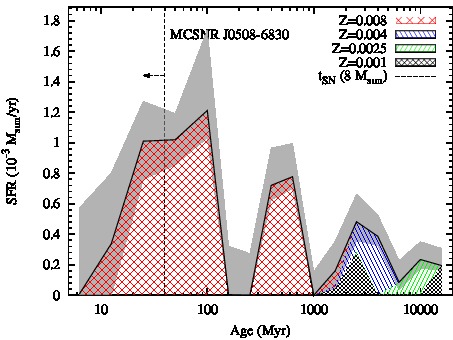}

\vspace{1cm}

\includegraphics[width=0.312\hsize]
{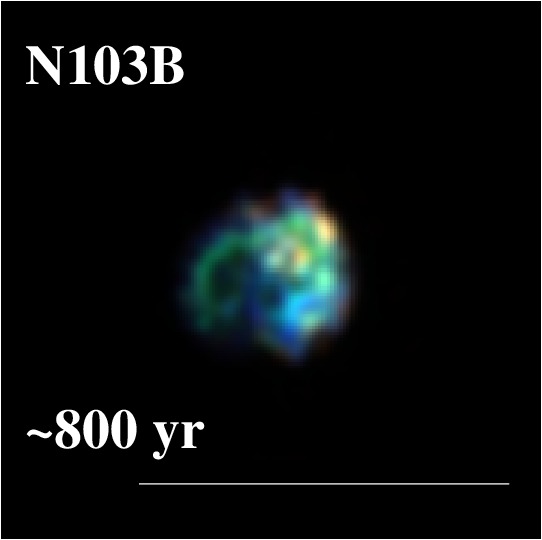}
\includegraphics[width=0.312\hsize]
{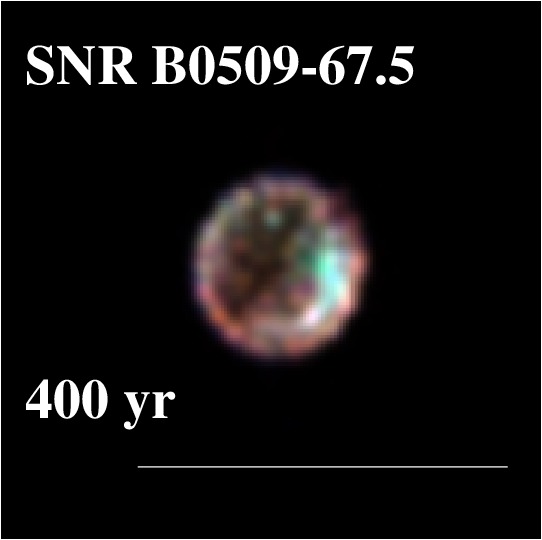}
    \includegraphics[width=0.312\hsize]
{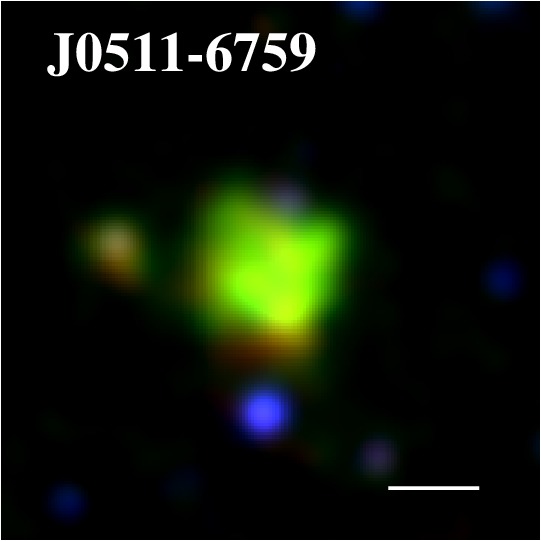}

    \includegraphics[width=0.312\hsize]
{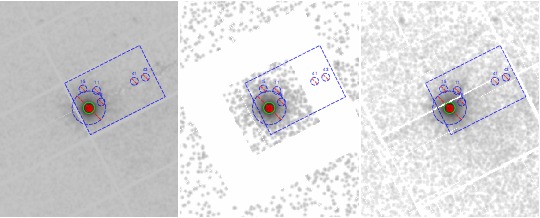}
    \includegraphics[width=0.312\hsize]
{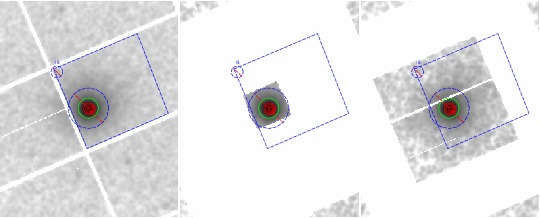}
    \includegraphics[width=0.312\hsize]
{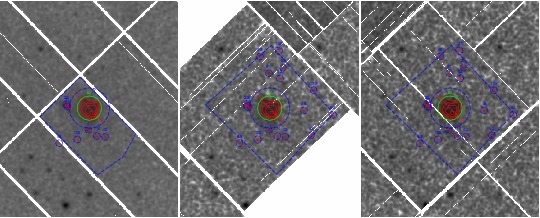}

    \includegraphics[width=0.312\hsize]
{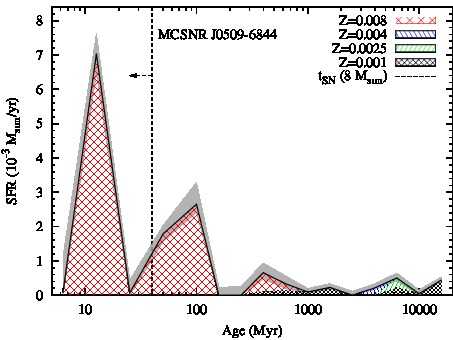}
    \includegraphics[width=0.312\hsize]
{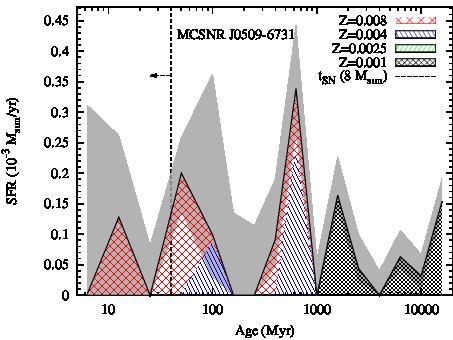}
    \includegraphics[width=0.312\hsize]
{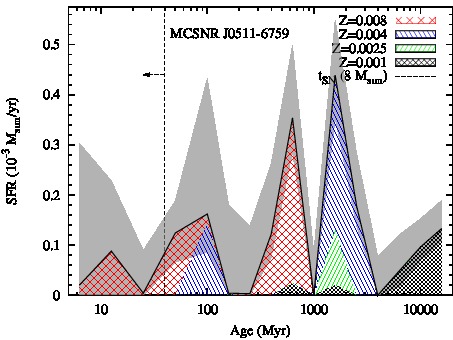}
    \end{center}
\end{figure}

\begin{figure}[ht]
    \begin{center}
    \includegraphics[width=0.312\hsize]
{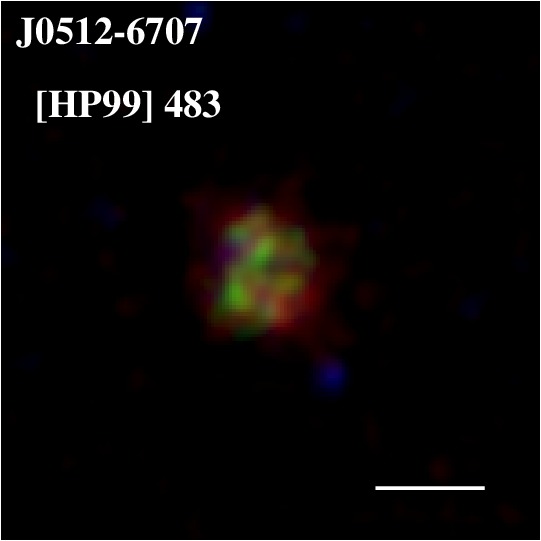}
    \includegraphics[width=0.312\hsize]
{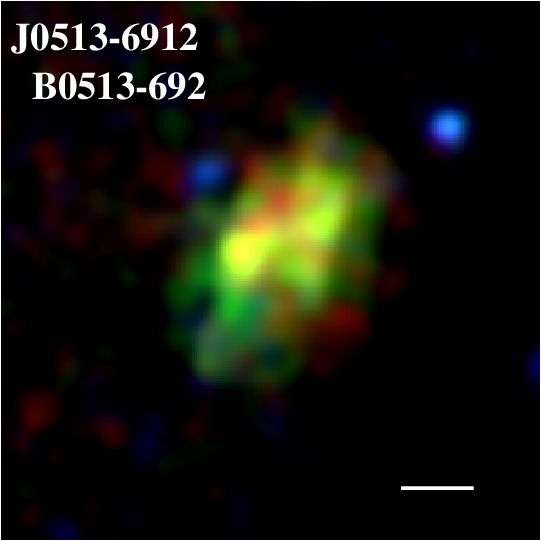}
    \includegraphics[width=0.312\hsize]
{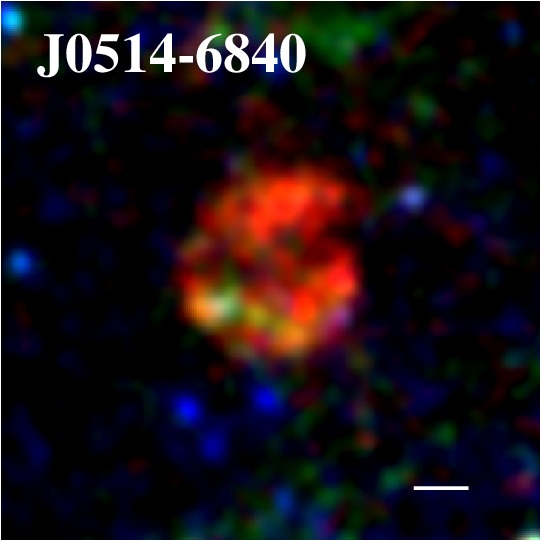}

    \includegraphics[width=0.312\hsize]
{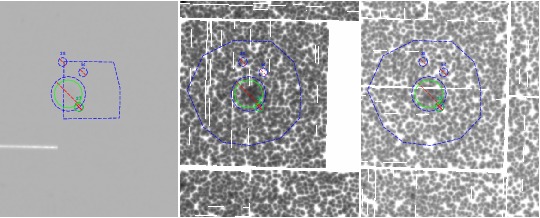}
    \includegraphics[width=0.312\hsize]
{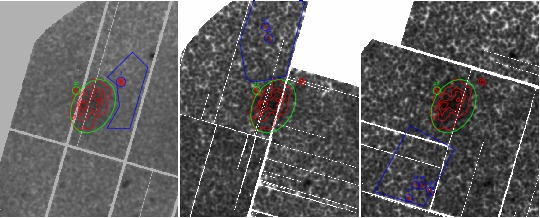}
    \includegraphics[width=0.312\hsize]
{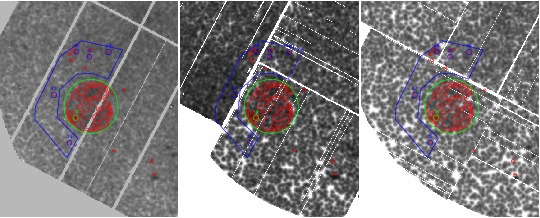}

\includegraphics[width=0.312\hsize]
{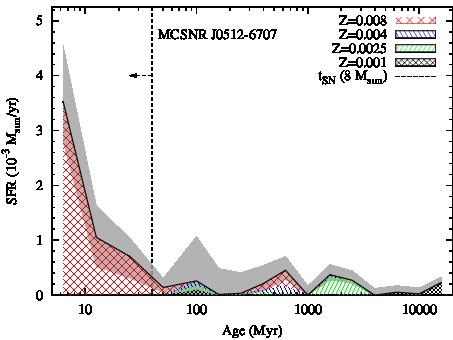}
    \includegraphics[width=0.312\hsize]
{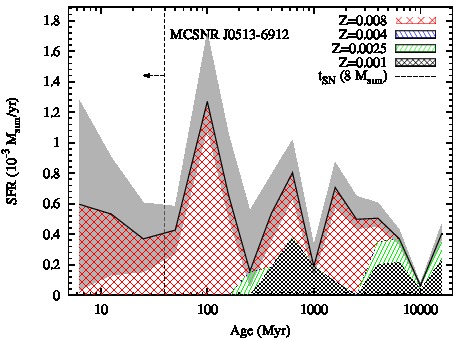}
    \includegraphics[width=0.312\hsize]
{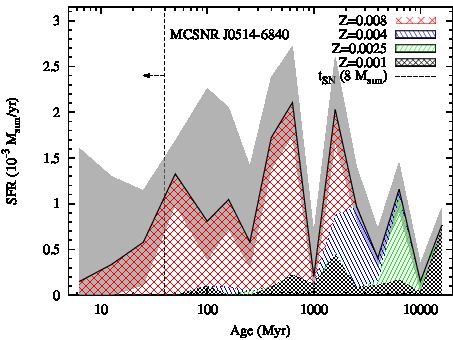}

\vspace{1cm}

\includegraphics[width=0.312\hsize]
{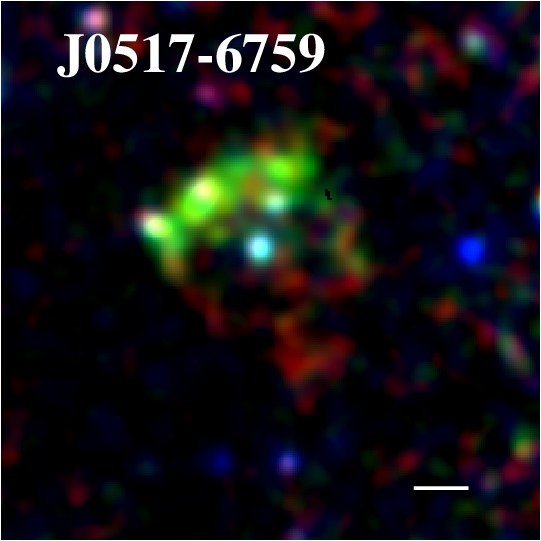}
    \includegraphics[width=0.312\hsize]
{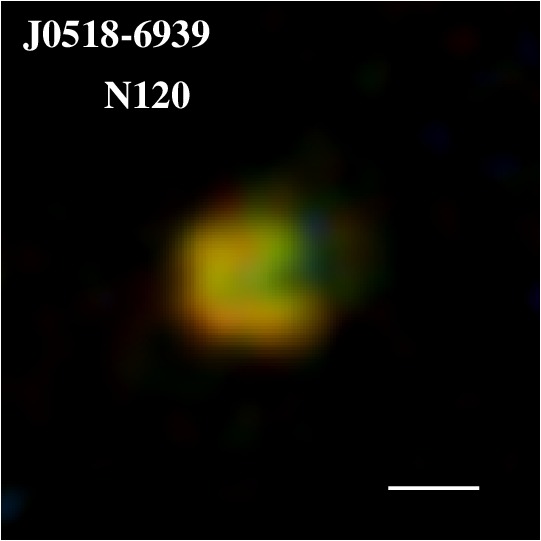}
\includegraphics[width=0.312\hsize]
{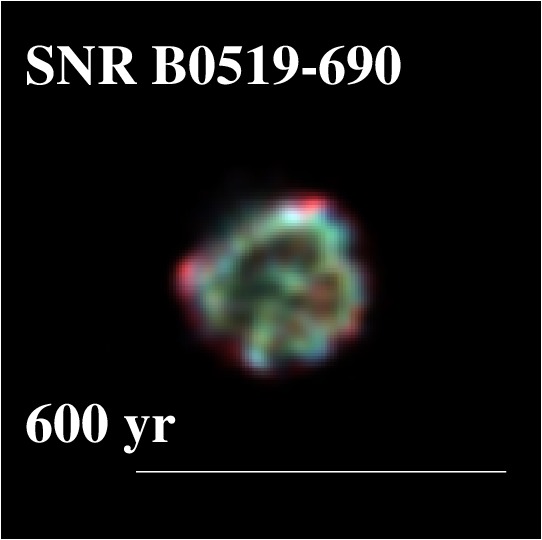}

    \includegraphics[width=0.312\hsize]
{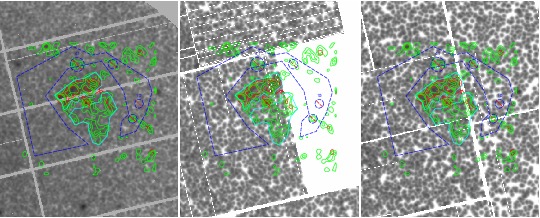}
    \includegraphics[width=0.312\hsize]
{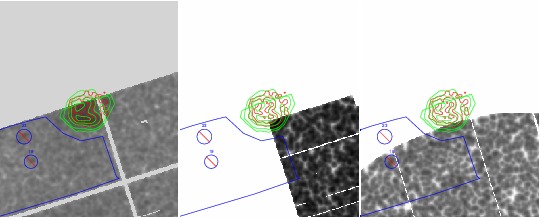}
    \includegraphics[width=0.312\hsize]
{J0519-6902_extraction_region.jpg}

    \includegraphics[width=0.312\hsize]
{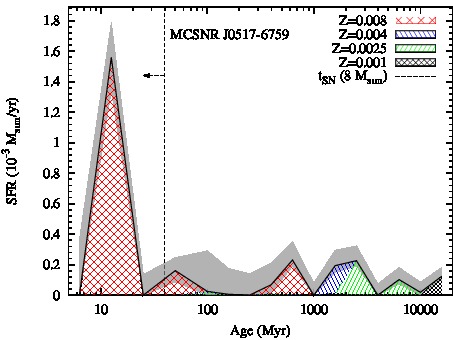}
    \includegraphics[width=0.312\hsize]
{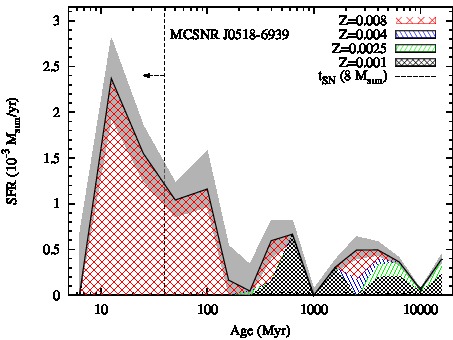}
    \includegraphics[width=0.312\hsize]
{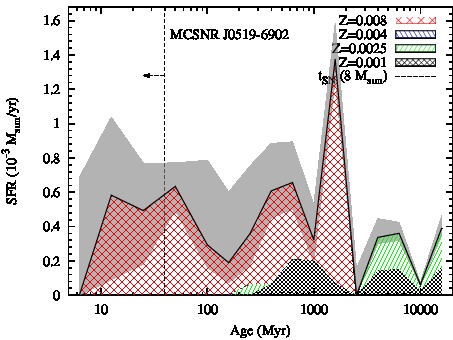}
    \end{center}
\end{figure}

\begin{figure}[ht]
    \begin{center}
\includegraphics[width=0.312\hsize]
{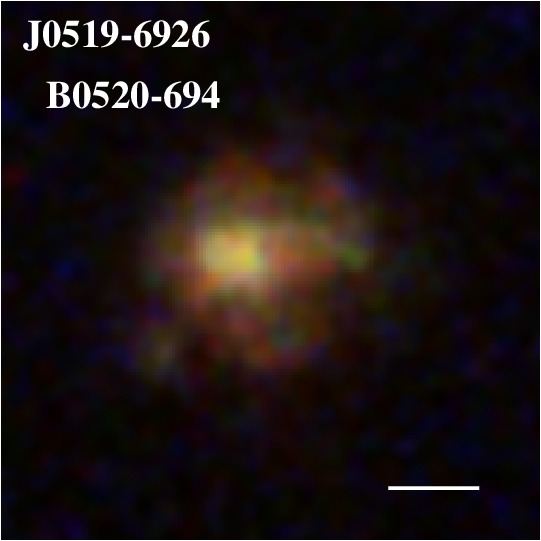}
    \includegraphics[width=0.312\hsize]
{noxmmdata.jpg}
\includegraphics[width=0.312\hsize]
{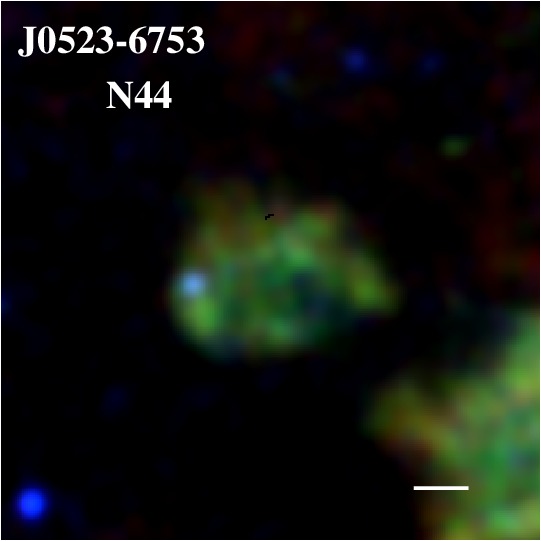}

    \includegraphics[width=0.312\hsize]
{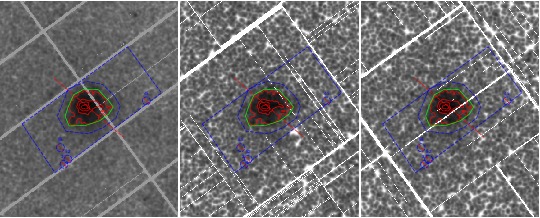}
    \includegraphics[width=0.312\hsize]
{noxmmspectra.jpg}
    \includegraphics[width=0.312\hsize]
{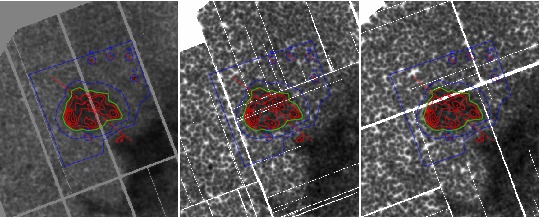}

    \includegraphics[width=0.312\hsize]
{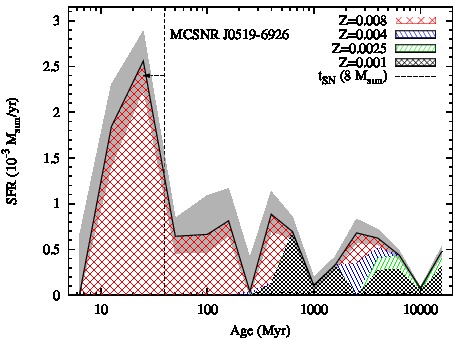}
    \includegraphics[width=0.312\hsize]
{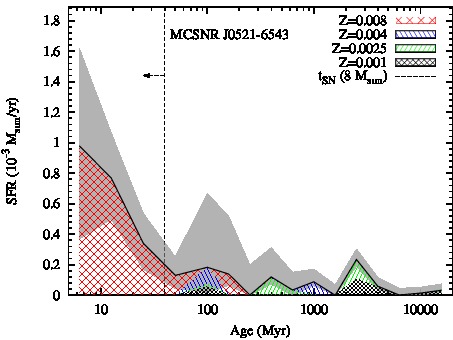}
    \includegraphics[width=0.312\hsize]
{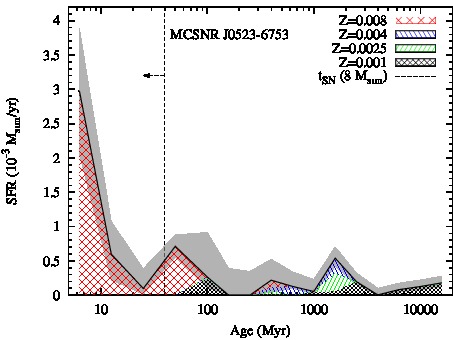}

\vspace{1cm}

\includegraphics[width=0.312\hsize]
{noxmmdata.jpg}
    \includegraphics[width=0.312\hsize]
{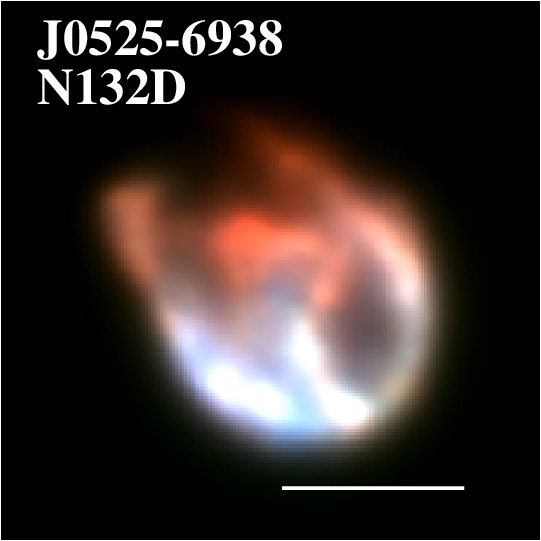}
\includegraphics[width=0.312\hsize]
{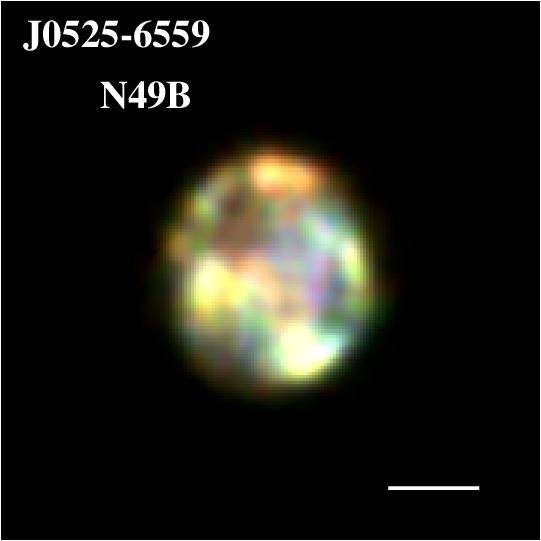}

    \includegraphics[width=0.312\hsize]
{noxmmspectra.jpg}
    \includegraphics[width=0.312\hsize]
{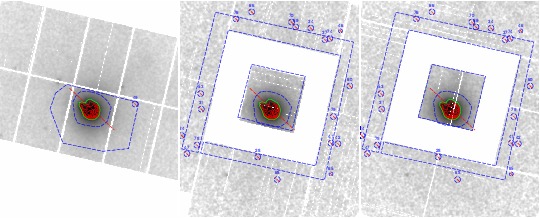}
    \includegraphics[width=0.312\hsize]
{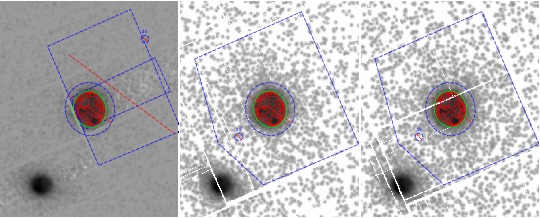}

    \includegraphics[width=0.312\hsize]
{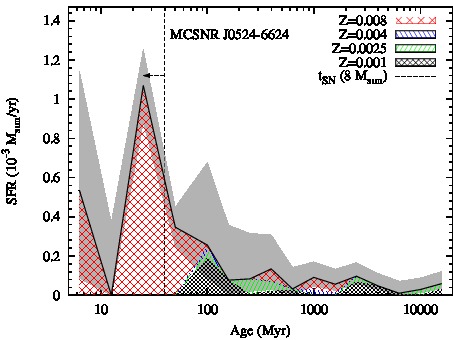}
    \includegraphics[width=0.312\hsize]
{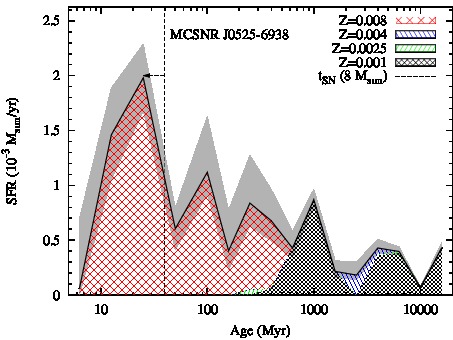}
    \includegraphics[width=0.312\hsize]
{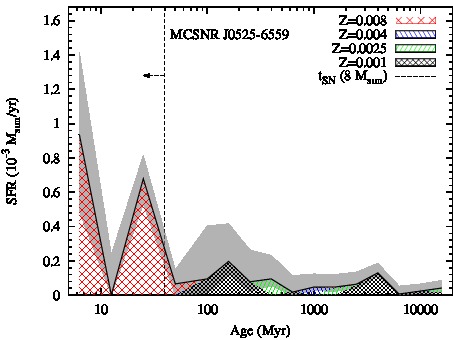}
    \end{center}
\end{figure}

\begin{figure}[ht]
    \begin{center}
\includegraphics[width=0.312\hsize]
{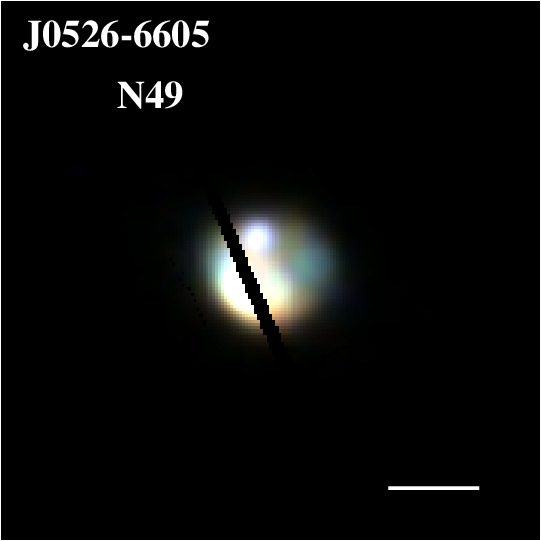}
    \includegraphics[width=0.312\hsize]
{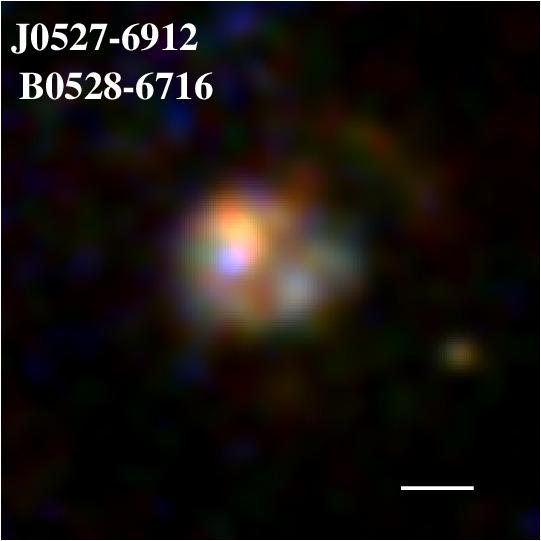}
\includegraphics[width=0.312\hsize]
{noxmmdata.jpg}

    \includegraphics[width=0.312\hsize]
{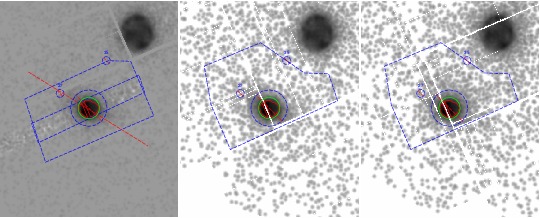}
    \includegraphics[width=0.312\hsize]
{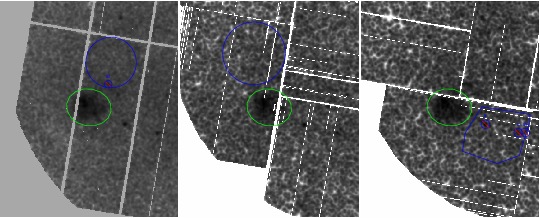}
    \includegraphics[width=0.312\hsize]
{noxmmspectra.jpg}

    \includegraphics[width=0.312\hsize]
{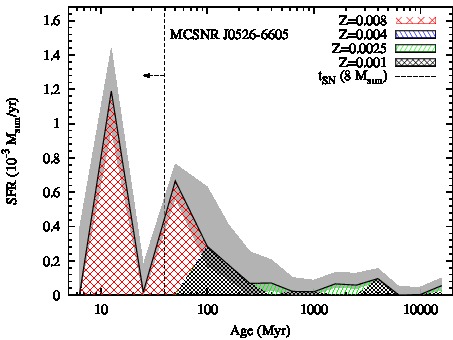}
    \includegraphics[width=0.312\hsize]
{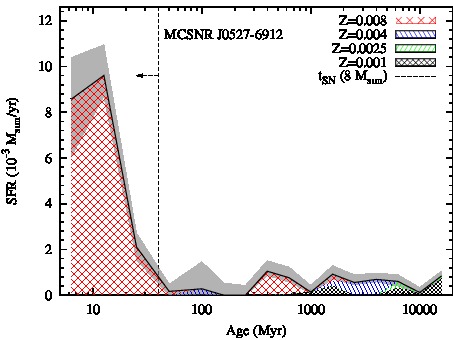}
    \includegraphics[width=0.312\hsize]
{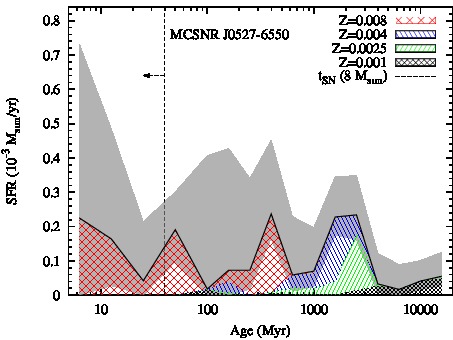}

\vspace{1cm}

\includegraphics[width=0.312\hsize]
{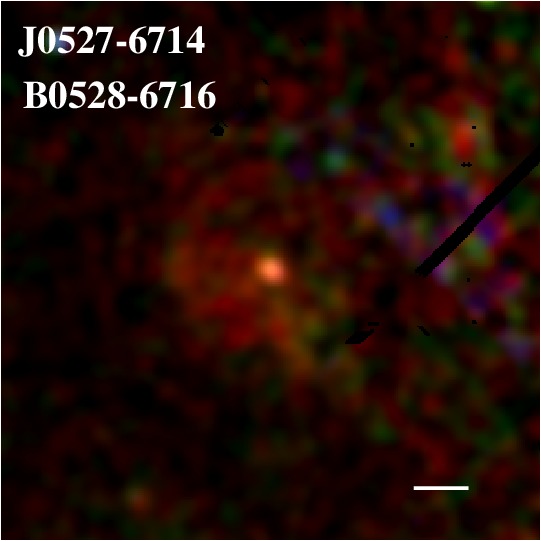}
    \includegraphics[width=0.312\hsize]
{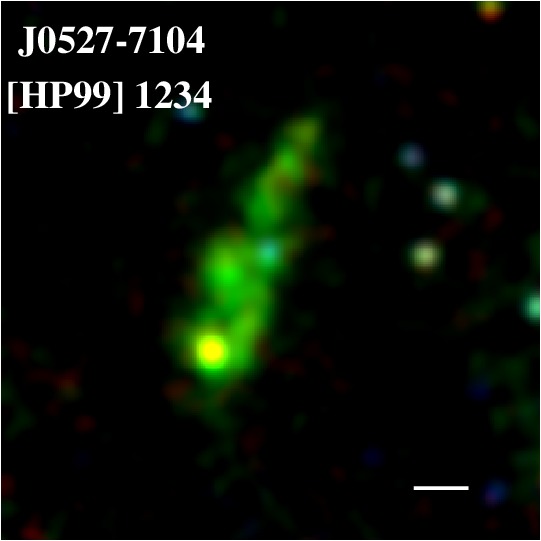}
\includegraphics[width=0.312\hsize]
{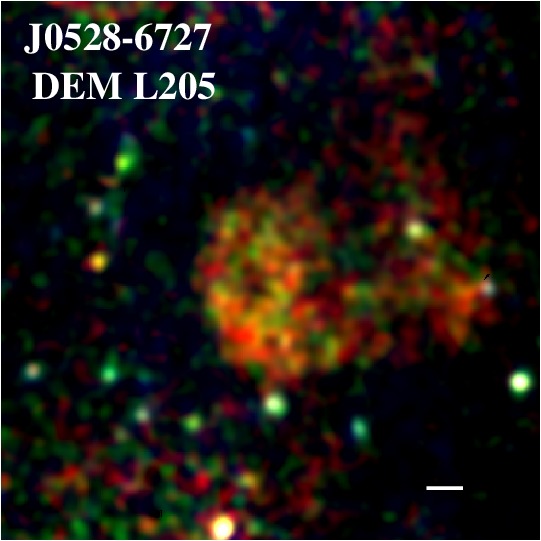}

    \includegraphics[width=0.312\hsize]
{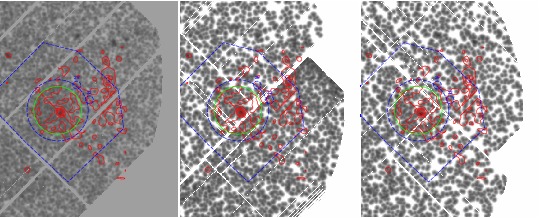}
    \includegraphics[width=0.312\hsize]
{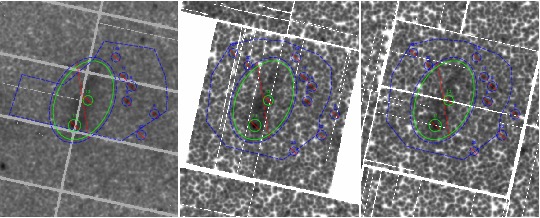}
    \includegraphics[width=0.312\hsize]
{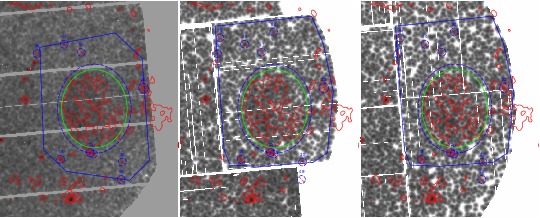}

    \includegraphics[width=0.312\hsize]
{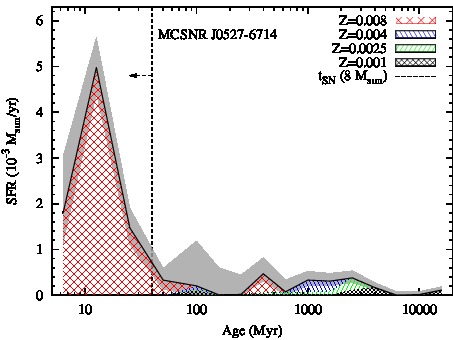}
    \includegraphics[width=0.312\hsize]
{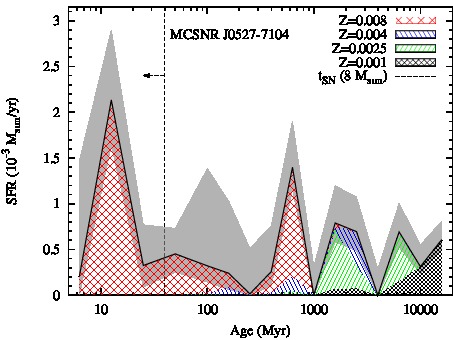}
    \includegraphics[width=0.312\hsize]
{closestsfh_J0528-6727.jpg}
    \end{center}
\end{figure}

\begin{figure}[ht]
    \begin{center}
\includegraphics[width=0.312\hsize]
{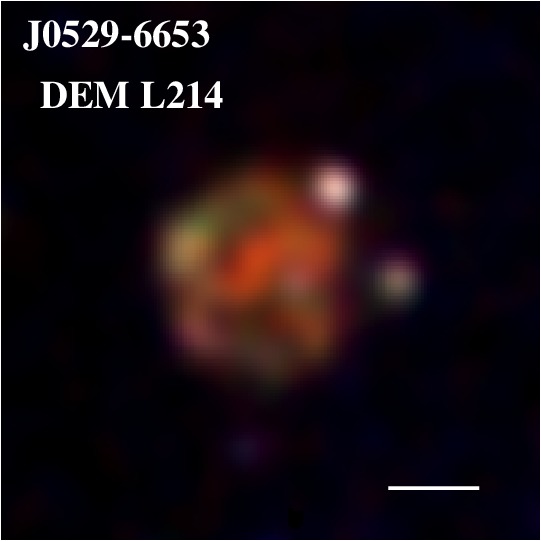}
    \includegraphics[width=0.312\hsize]
{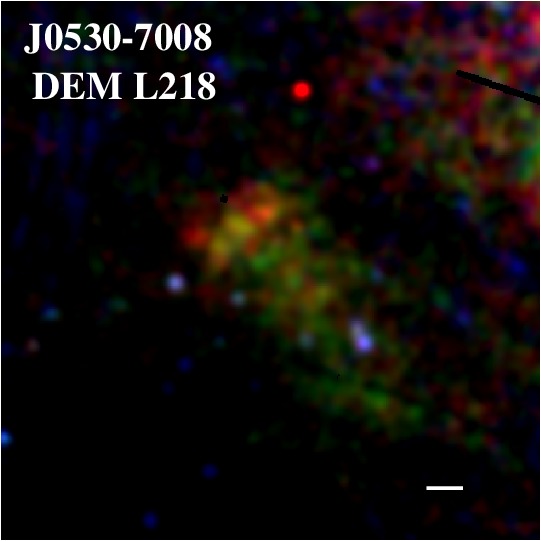}
\includegraphics[width=0.312\hsize]
{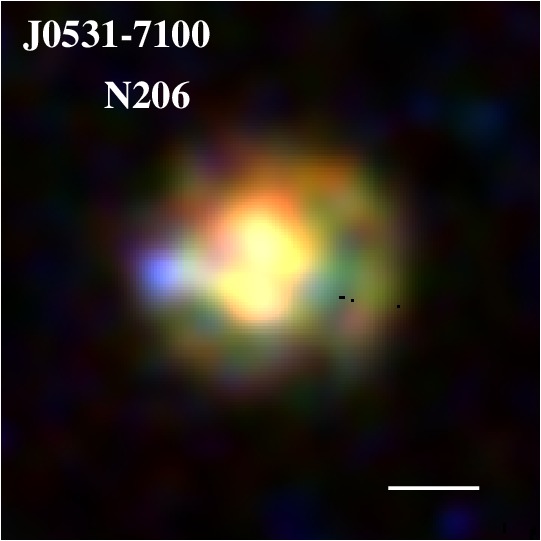}

    \includegraphics[width=0.312\hsize]
{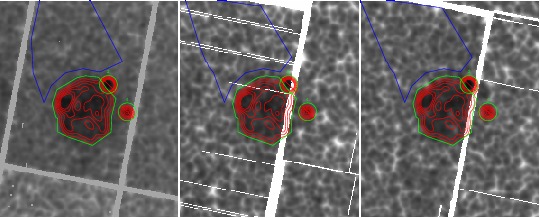}
    \includegraphics[width=0.312\hsize]
{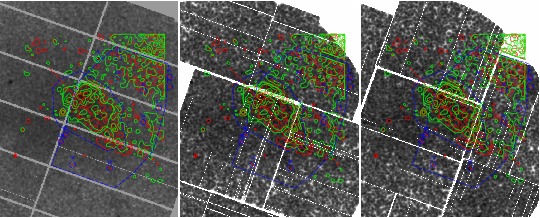}
    \includegraphics[width=0.312\hsize]
{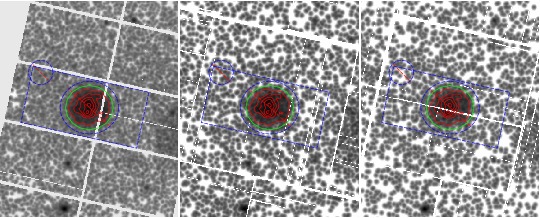}

    \includegraphics[width=0.312\hsize]
{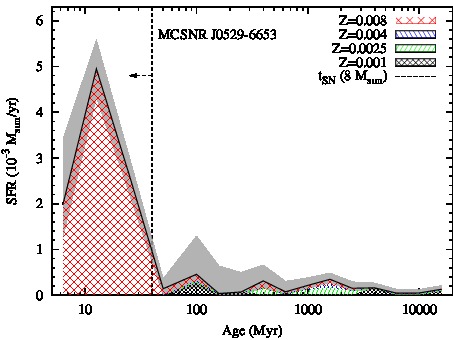}
    \includegraphics[width=0.312\hsize]
{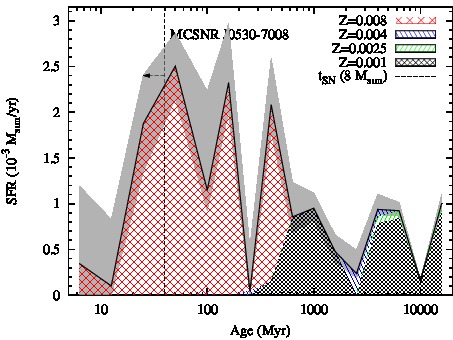}
    \includegraphics[width=0.312\hsize]
{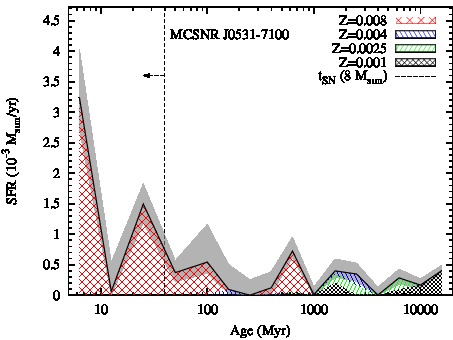}

\vspace{1cm}

\includegraphics[width=0.312\hsize]
{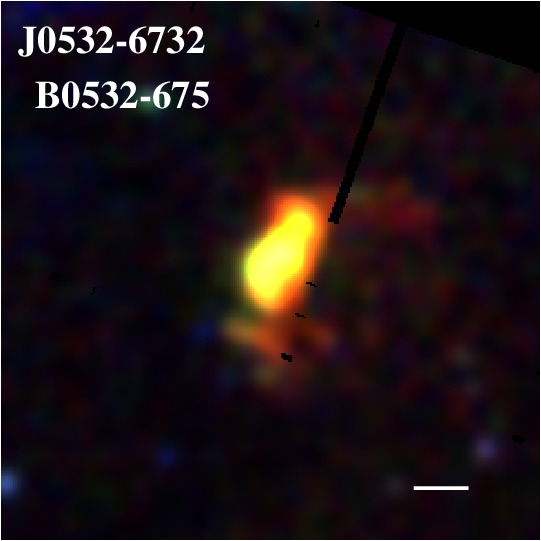}
    \includegraphics[width=0.312\hsize]
{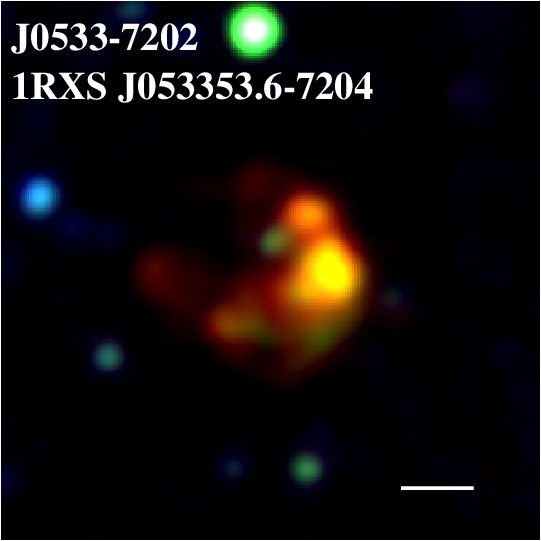}
\includegraphics[width=0.312\hsize]
{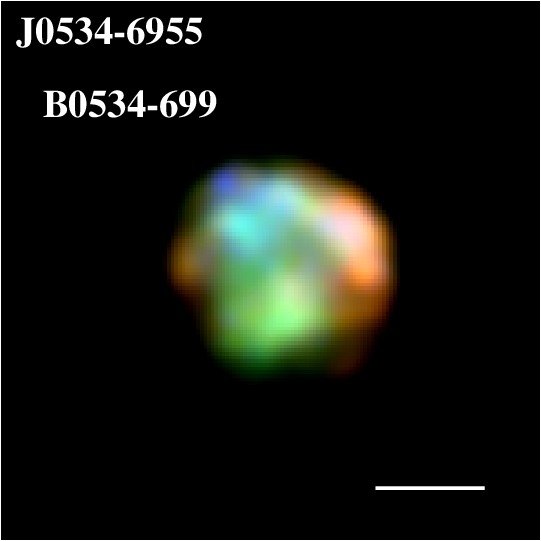}

    \includegraphics[width=0.312\hsize]
{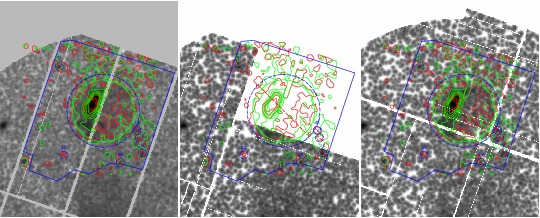}
    \includegraphics[width=0.312\hsize]
{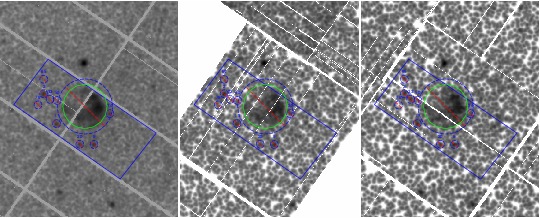}
    \includegraphics[width=0.312\hsize]
{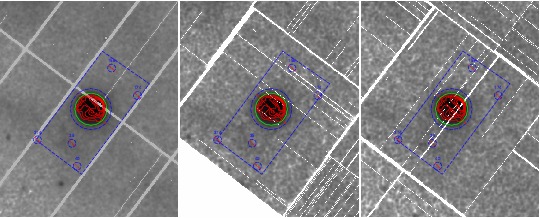}

    \includegraphics[width=0.312\hsize]
{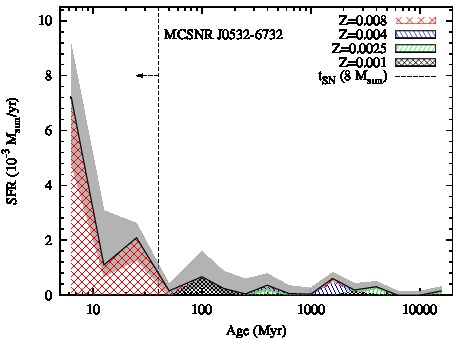}
    \includegraphics[width=0.312\hsize]
{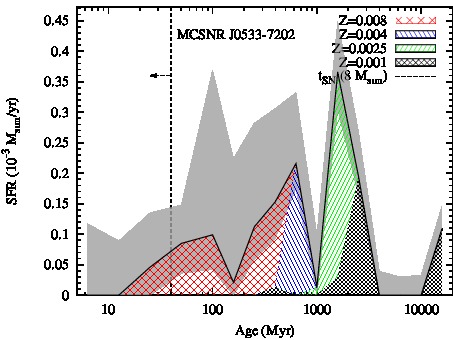}
    \includegraphics[width=0.312\hsize]
{closestsfh_J0534-6955.jpg}
    \end{center}
\end{figure}

\begin{figure}[ht]
    \begin{center}
\includegraphics[width=0.312\hsize]
{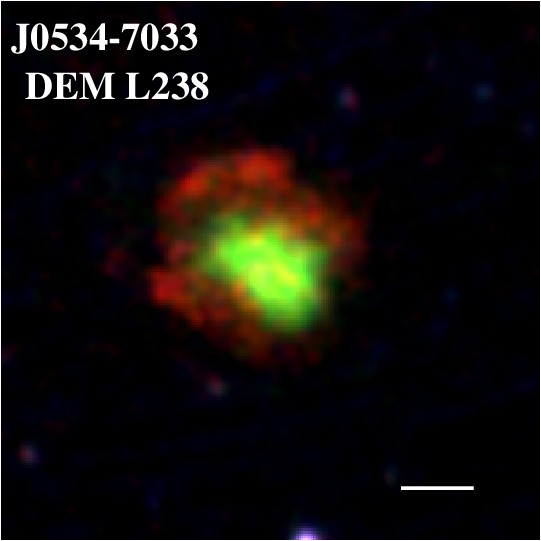}
    \includegraphics[width=0.312\hsize]
{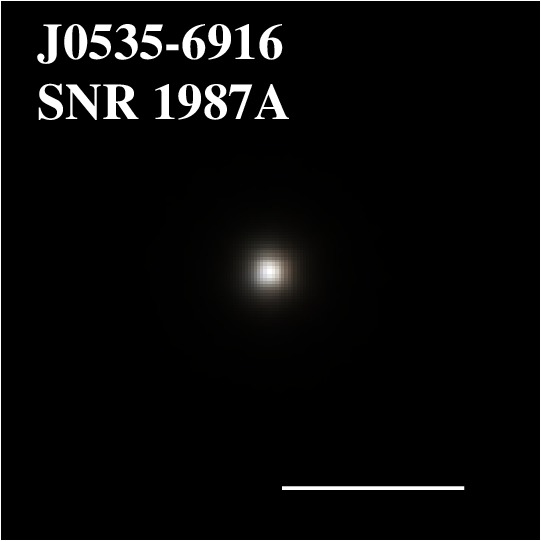}
\includegraphics[width=0.312\hsize]
{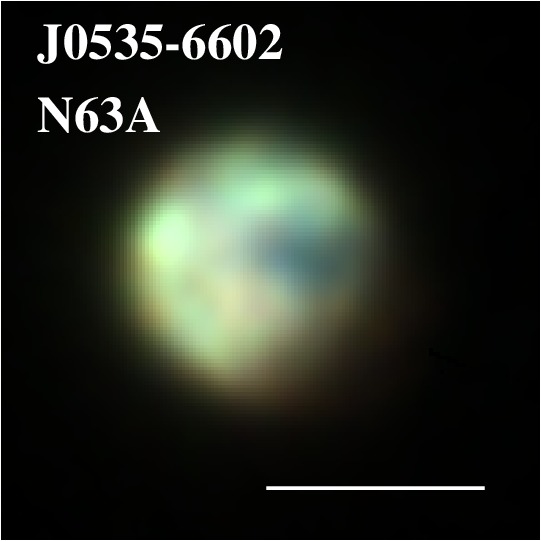}

    \includegraphics[width=0.312\hsize]
{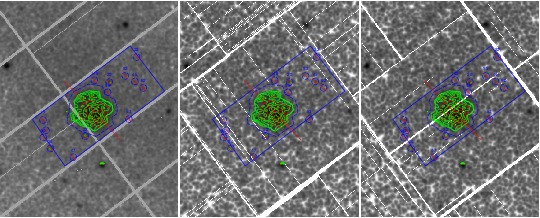}
    \includegraphics[width=0.312\hsize]
{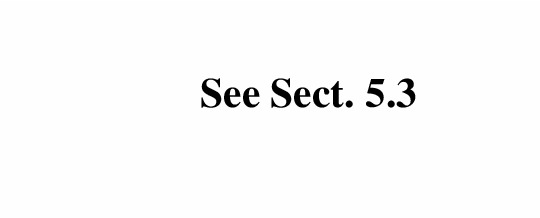}
    \includegraphics[width=0.312\hsize]
{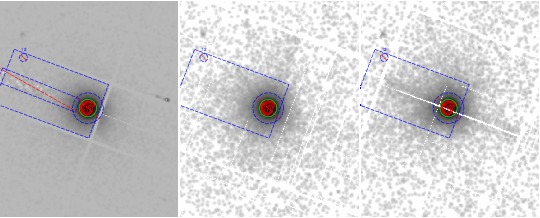}

    \includegraphics[width=0.312\hsize]
{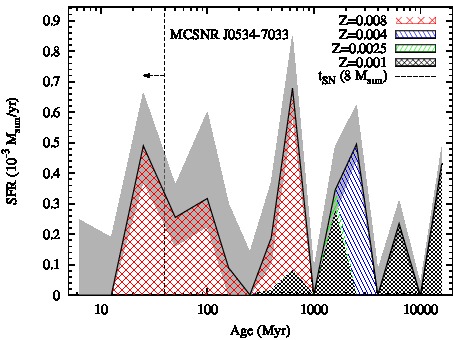}
    \includegraphics[width=0.312\hsize]
{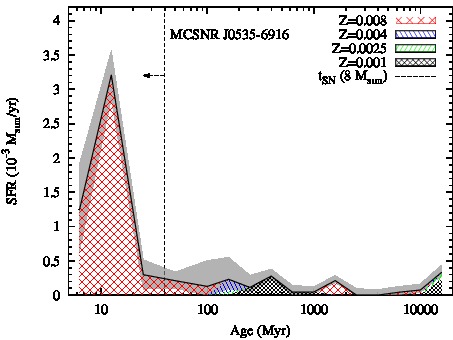}
    \includegraphics[width=0.312\hsize]
{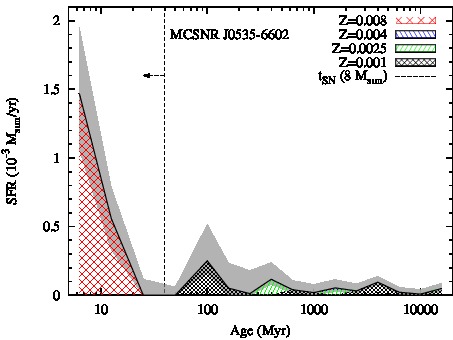}

\vspace{1cm}

\includegraphics[width=0.312\hsize]
{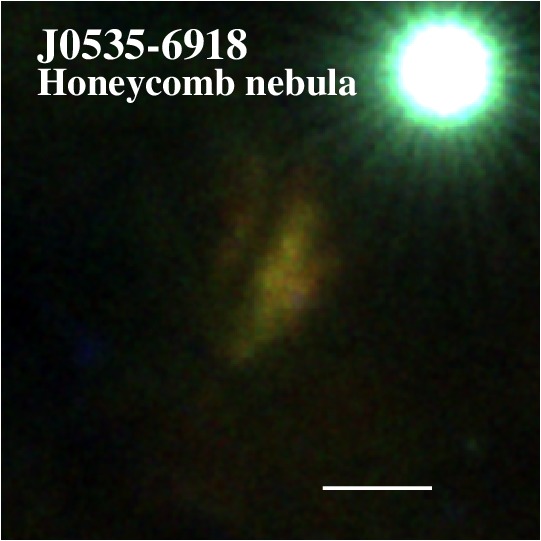}
    \includegraphics[width=0.312\hsize]
{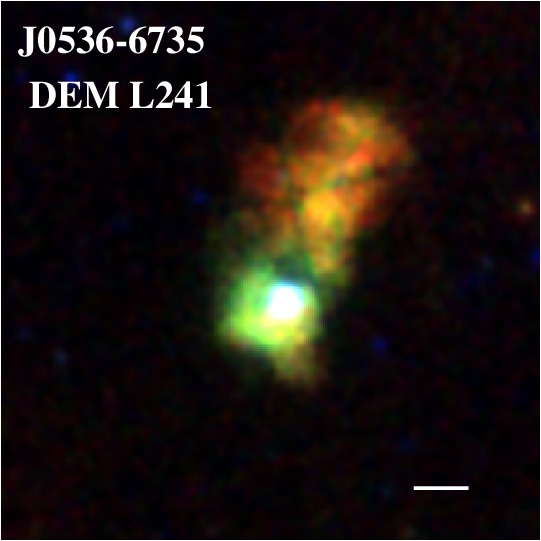}
\includegraphics[width=0.312\hsize]
{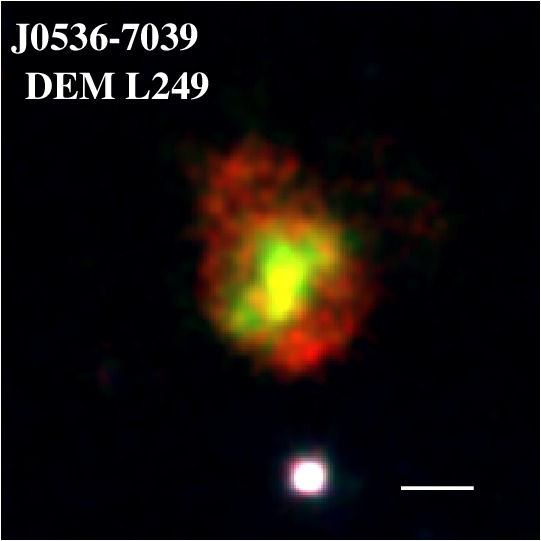}

    \includegraphics[width=0.312\hsize]
{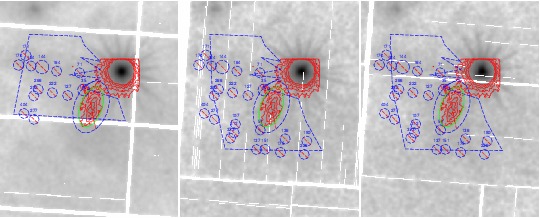}
    \includegraphics[width=0.312\hsize]
{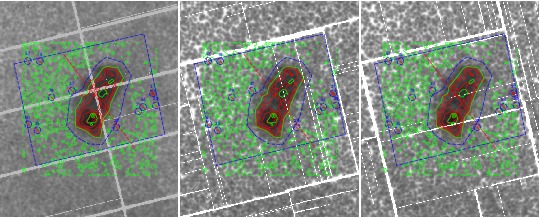}
    \includegraphics[width=0.312\hsize]
{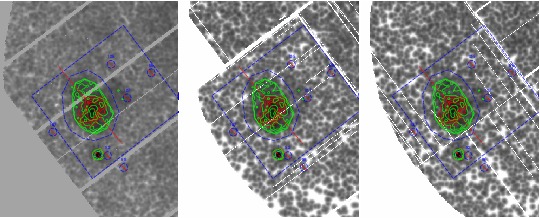}

    \includegraphics[width=0.312\hsize]
{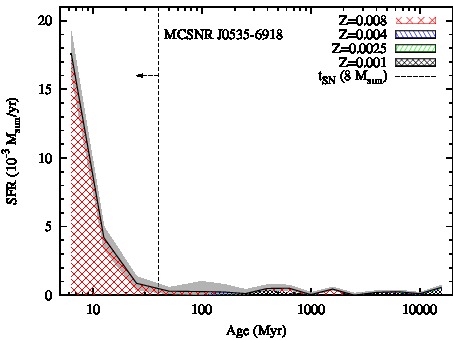}
    \includegraphics[width=0.312\hsize]
{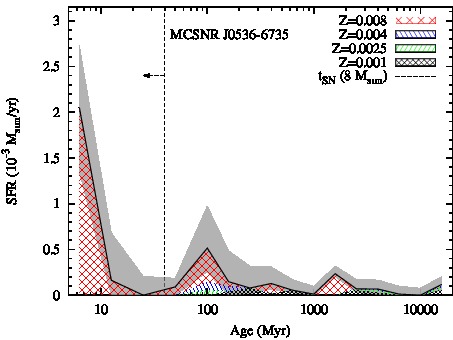}
    \includegraphics[width=0.312\hsize]
{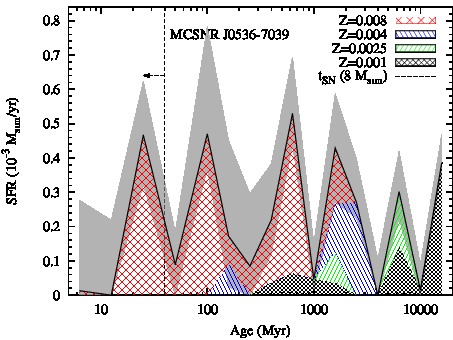}
    \end{center}
\end{figure}

\begin{figure}[ht]
    \begin{center}
    \includegraphics[width=0.312\hsize]
{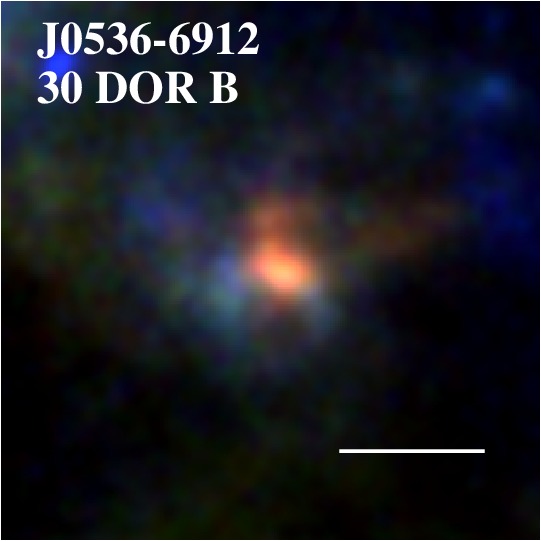}
    \includegraphics[width=0.312\hsize]
{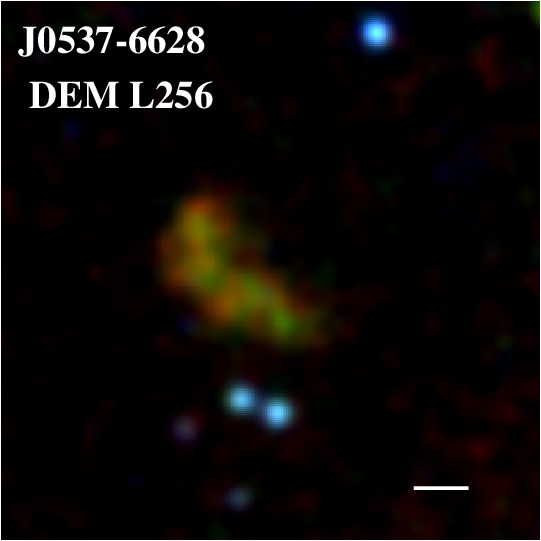}
\includegraphics[width=0.312\hsize]
{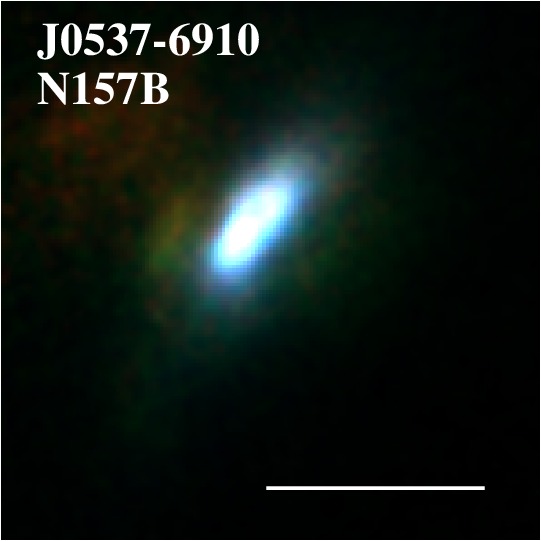}

    \includegraphics[width=0.312\hsize]
{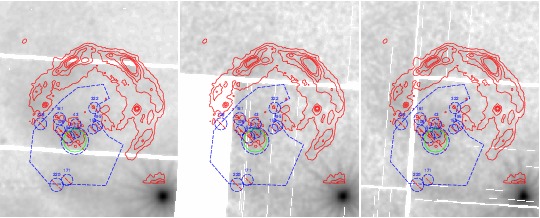}
    \includegraphics[width=0.312\hsize]
{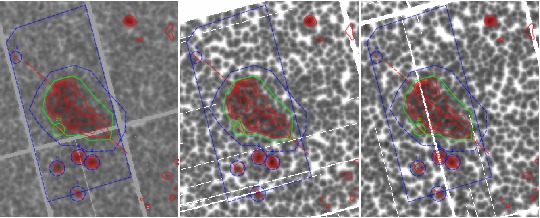}
    \includegraphics[width=0.312\hsize]
{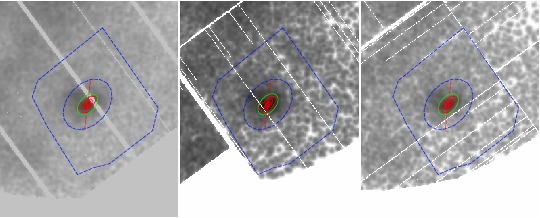}

    \includegraphics[width=0.312\hsize]
{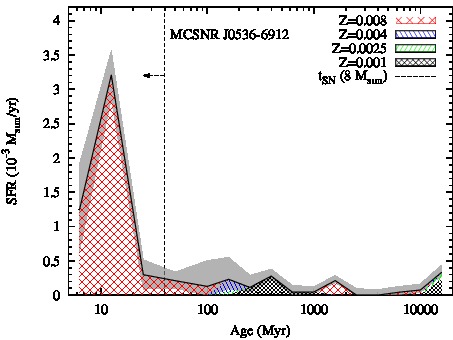}
    \includegraphics[width=0.312\hsize]
{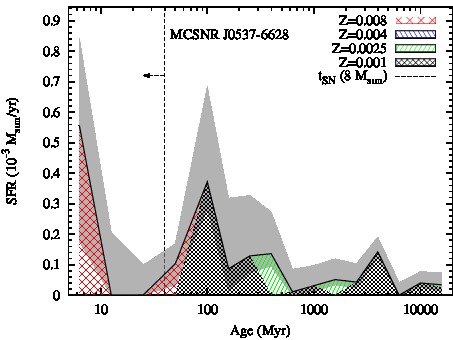}
    \includegraphics[width=0.312\hsize]
{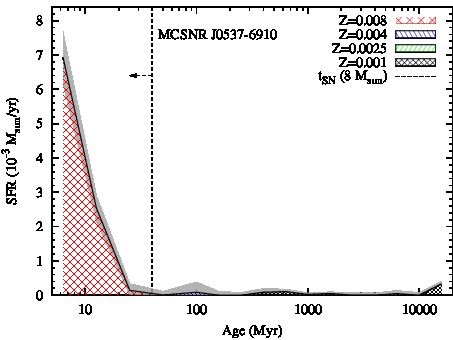}

\vspace{1cm}

\includegraphics[width=0.312\hsize]
{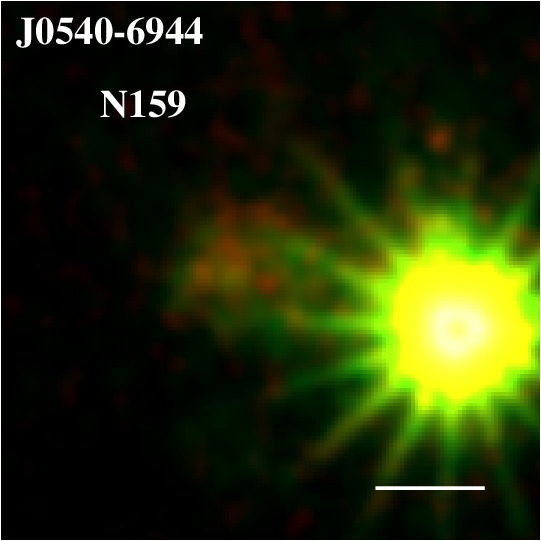}
    \includegraphics[width=0.312\hsize]
{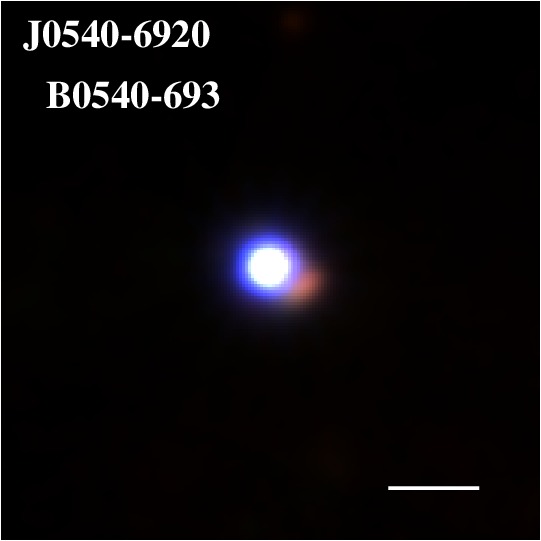}
\includegraphics[width=0.312\hsize]
{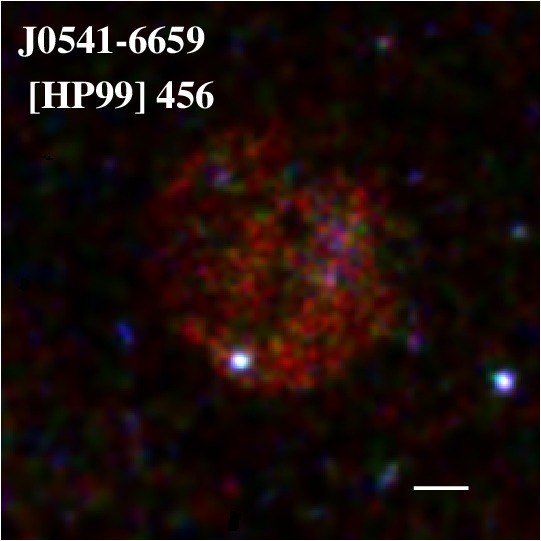}

    \includegraphics[width=0.312\hsize]
{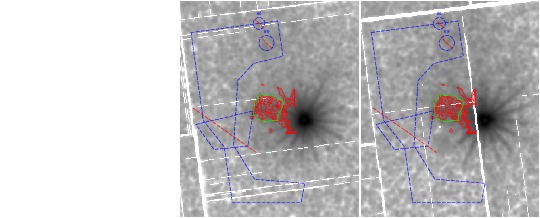}
    \includegraphics[width=0.312\hsize]
{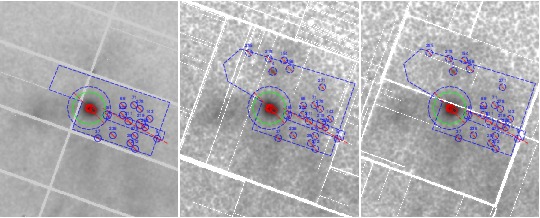}
    \includegraphics[width=0.312\hsize]
{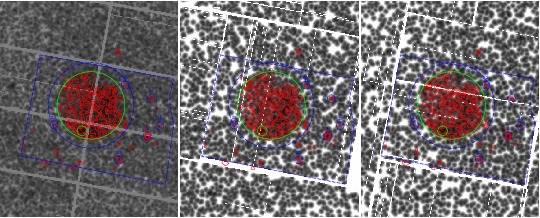}

    \includegraphics[width=0.312\hsize]
{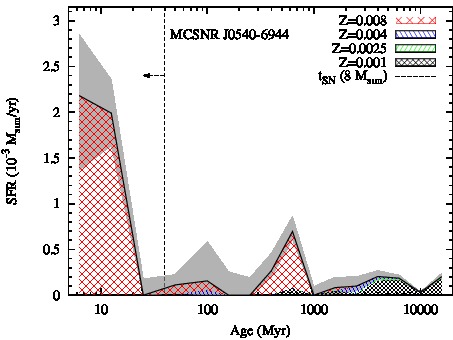}
    \includegraphics[width=0.312\hsize]
{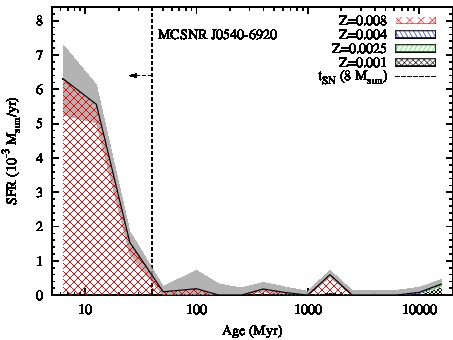}
    \includegraphics[width=0.312\hsize]
{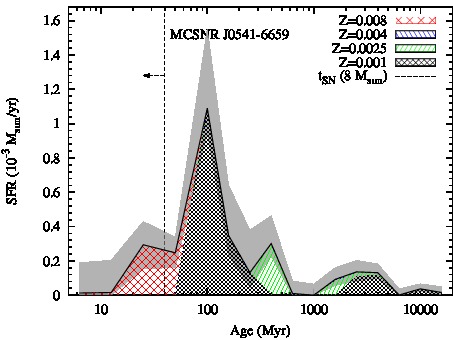}
    \end{center}
\end{figure}

\begin{figure}[ht]
    \begin{center}
\includegraphics[width=0.312\hsize]
{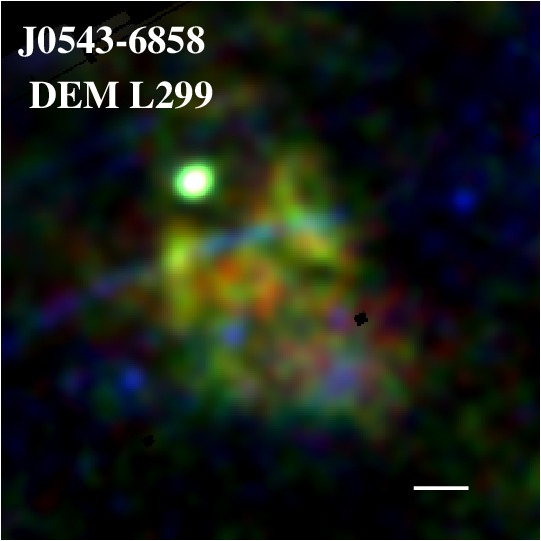}
    \includegraphics[width=0.312\hsize]
{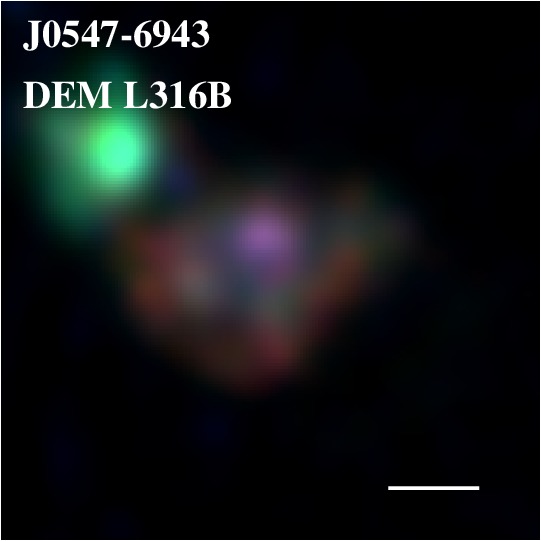}
\includegraphics[width=0.312\hsize]
{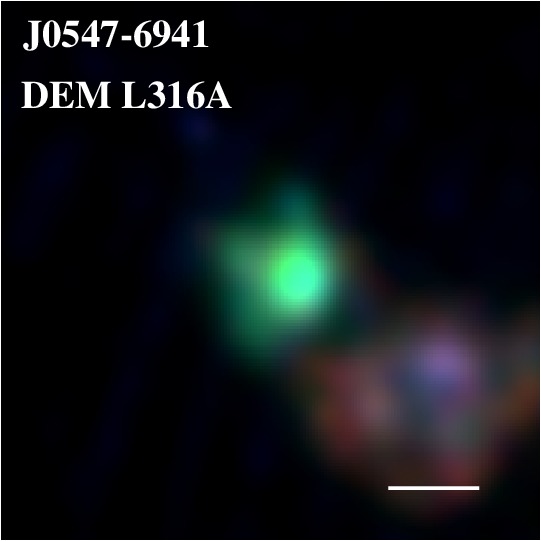}

    \includegraphics[width=0.312\hsize]
{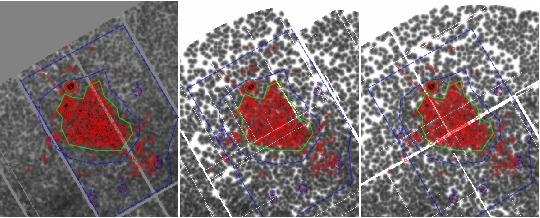}
    \includegraphics[width=0.312\hsize]
{J0547-6943_extraction_region.jpg}
    \includegraphics[width=0.312\hsize]
{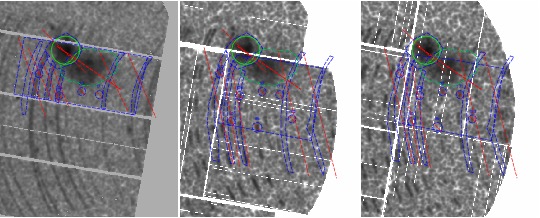}

    \includegraphics[width=0.312\hsize]
{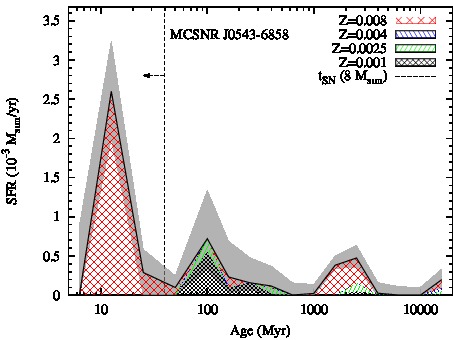}
    \includegraphics[width=0.312\hsize]
{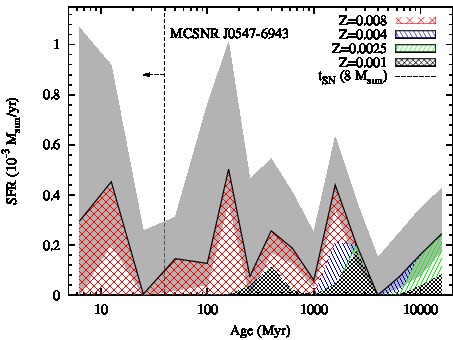}
    \includegraphics[width=0.312\hsize]
{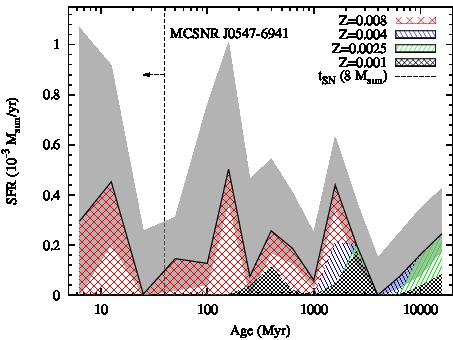}

\vspace{1cm}

\includegraphics[width=0.312\hsize]
{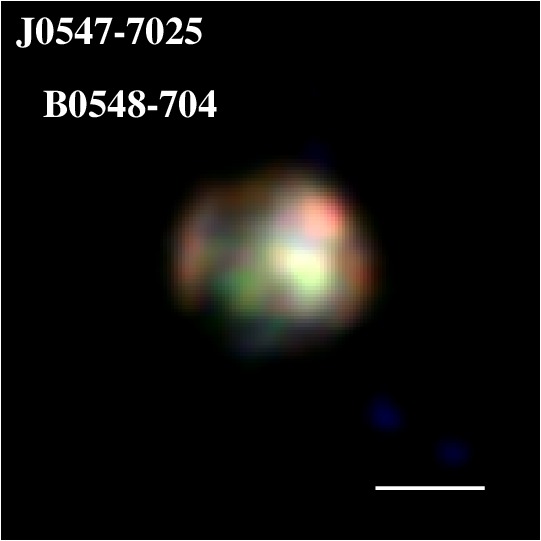}
    \includegraphics[width=0.312\hsize]
{noxmmdata.jpg}

    \includegraphics[width=0.312\hsize]
{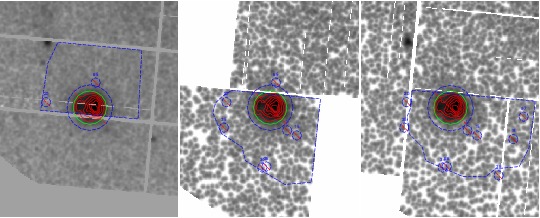}
    \includegraphics[width=0.312\hsize]
{noxmmspectra.jpg}

    \includegraphics[width=0.312\hsize]
{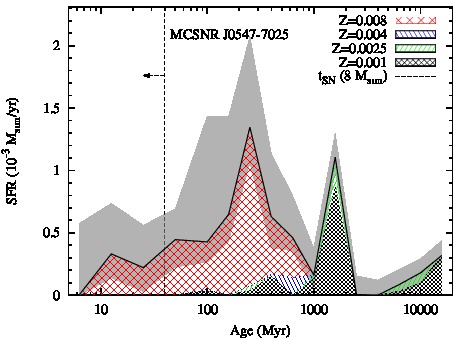}
    \includegraphics[width=0.312\hsize]
{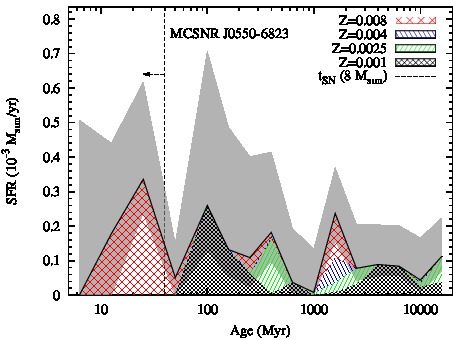}
    \end{center}
\end{figure}

}

\end{document}